\documentclass[11pt]{report}
\usepackage{epsfig,cite,rotate}
\evensidemargin \oddsidemargin
\input paperdef

\makeatletter
\renewcommand\section{\@startsection {section}{1}{\z@}%
                           {-3.5ex \@plus -1ex \@minus -.2ex}%
                           {2.3ex \@plus.2ex}%
                           {\mathversion{bold}\normalfont\Large\bfseries}}
\renewcommand\subsection{\@startsection{subsection}{2}{\z@}%
                           {-3.25ex\@plus -1ex \@minus -.2ex}%
                           {1.5ex \@plus .2ex}%
                           {\mathversion{bold}\normalfont\large\bfseries}}
\renewcommand\subsubsection{\@startsection{subsubsection}{3}{\z@}%
                           {-3.25ex\@plus -1ex \@minus -.2ex}%
                           {1.5ex \@plus .2ex}%
                           {\mathversion{bold}\normalfont\normalsize\bfseries}}
\makeatother

\begin{document}

\thispagestyle{empty}
\setcounter{page}{0}
\def\thefootnote{\fnsymbol{footnote}}

\begin{flushright}
CERN--PH--TH/2005--228\\
hep-ph/0511332\\
\end{flushright}

\vspace{1cm}

\begin{center}
{\Large {\bf Toward High Precision Higgs-Boson Measurements at the\\[.5em]
International Linear $\ee$ Collider}} \\

\vspace{4mm}
The Higgs Working Group at Snowmass '05\\[1em]

S.~Heinemeyer$^{ay}$%
\footnote{email: Sven.Heinemeyer@cern.ch}%
, S.~Kanemura$^b$, H.~Logan$^c$, A.~Raspereza$^d$, T.~Tait$^e$\\[.2em]
H.~Baer$^f$,
E.L.~Berger$^e$,
A.~Birkedal$^g$,
J.-C.~Brient$^h$,
M.~Carena$^i$
J.A.R.~Cembranos$^j$,
S.~Choi$^k$,
S.~Godfrey$^c$, 
J.~Gunion$^l$,
H.E.~Haber$^g$, 
T.~Han$^m$,
H.~Heath$^n$,
S.~Hesselbach$^o$,
J.~Kalinowski$^p$, 
W.~Kilian$^q$,
G.~Moortgat-Pick$^a$, 
S.~Moretti$^r$,
S.~Mrenna$^i$, 
M.~Muhlleitner$^s$,
F.~Petriello$^m$,
J.~Reuter$^q$,
M.~Ronan$^t$,
P.~Skands$^i$,
A.~Sopczak$^u$,
M.~Spira$^v$,
Z.~Sullivan$^e$,
C.~Wagner$^{ew}$,
G.~Weiglein$^x$,
P.~Zerwas$^o$

\vspace{1em}

\end{center}
\BC
{\bf Abstract}
\EC
This report reviews the properties of Higgs bosons in the Standard 
Model (SM) and its various extensions. We give an extensive overview of
the potential of the ILC operated at center-of-mass
energies up to 1~TeV (including the $\ga\ga$ collider option) for the
determination of the Higgs boson properties.
This includes the measurement of the Higgs boson mass,
its couplings to SM fermions and gauge bosons, and the determination of
the spin and the CP quantum numbers of the Higgs.  We also discuss
extensions of the SM, including heavy SM-like Higgs
bosons, heavy Higgs bosons in the framework of Supersymmetry (SUSY)
and more exotic scenarios.
We review recent theoretical developments in the field of Higgs boson 
physics, and the 
impact of Higgs boson physics on cosmology in several SUSY
frameworks is considered.
The important questions as to what the ILC can contribute to Higgs boson
physics after the LHC, the LHC/ILC interplay and synergy, are addressed.
The impact of the accelerator and detector performance on the precision
of measurements are discussed in detail and we
propose a strategy to optimize future analyses.
Open questions arising for the
various topics are listed, and further topics of study
and corresponding roadmaps are suggested. 

\newpage
\thispagestyle{empty}
\setcounter{page}{0}

\mbox{}\vspace{1cm}
\BC

$^a$CERN, TH division, Dept.\ of Physics, 1211 Geneva 23, Switzerland

$^b$Department of Physics, Osaka University, 
    Toyonaka, Osaka 560-0043, Japan

$^c$Ottawa Carleton Inst. for Physics, Carleton University, Ottawa, Ontario K1S 5B6, Canada

$^d$Max-Planck-Institut für Physik, F\"ohringer Ring 6, 80805
    M\"unchen, Germany

$^e$High Energy Physics Division,  Argonne National Laboratory, 
    Argonne, IL 60439, USA

$^f$Dept. of Physics, Florida State University, Tallahassee, Florida 32306,
    USA

$^g$Santa Cruz Institute for Particle Physics, 
    Santa Cruz, CA 95064 USA

$^h$Laboratoire Leprince-Ringuet, Ecole polytechnique, Palaiseau, France

$^i$Theoretical Physics Department, Fermilab, Batavia, IL 60510, USA

$^j$Department of Physics and Astronomy,
    University of California, Irvine, CA 92697, USA

$^k$Department of Physics, Chonbuk National University,
    Chonju 561-756, Korea.

$^l$Department of Physics. University of California at Davis, Davis
    CA 95616, USA

$^m$Department of Physics, University of Wisconsin, Madison, WI,
    53706, USA

$^n$Physics Department, University of Bristol, Tyndall Ave, Bristol
    BS8 1TL, UK

$^o$High Energy Physics, Uppsala University, Box 535, S-751 21 Uppsala, Sweden

$^p$Institute of Theoretical Physics, Warsaw University,   
    Hoza 69, 00681 Warsaw, Poland

$^q$Deutsches Elektronen-Synchrotron DESY, D--22603 Hamburg, Germany

$^r$School of Physics \& Astronomy, Southampton University,
    Highfield SO17 1BJ, UK

$^s$Laboratoire d'Annecy-Le-Vieux de Physique Th\'eorique, LAPTH,
    Annecy-Le-Vieux, France

$^t$Lawrence Berkeley National Laboratory, USA

$^u$Department of Physics, Lancaster University, Lancaster, LA1 4YW, UK

$^v$Paul Scherrer Institut, CH-5232 Villigen PSI, Switzerland

$^w$Enrico Fermi Institute, University of Chicago, 5640 Ellis Ave., 
    Chicago, IL 60637, USA

$^x$Institute for Particle Physics Phenomenology, University of Durham,
    Durham DH1~3LE, UK

$^y$Depto.\ de F\'isica Te\'orica, Universidad de Zaragoza, 50009 Zaragoza,
    Spain

\EC

\def\thefootnote{\arabic{footnote}}
\setcounter{footnote}{0}


\newpage
\tableofcontents
\newpage


\chapter{Introduction}

Elucidating the mechanism of electroweak symmetry breaking (EWSB), by
which the electroweak gauge bosons and the fermions acquire masses,
will be one of the main tasks of future collider experiments.
Ultimately, all theoretical models rely on some form of the Higgs
mechanism to explain EWSB through spontaneous symmetry breaking. 
In the Standard Model (SM)~\cite{sm}, this is postulated
by introducing one fundamental scalar doublet field whose potential leads it
to acquire a non-zero vacuum expectation value.  Three of the real 
scalar components of the doublet are eaten by the gauge bosons, leaving
one physical scalar particle referred to as Higgs boson~\cite{higgs}.
Although this particle has eluded experimental observation so far, 
high precision measurements of electroweak observables provide indirect 
indications for a light Higgs boson. Updated electroweak 
data set an upper limit on the Higgs mass of about 200~GeV 
at 95\% confidence level~\cite{eweak_higgs}. 
The preferred value of the Higgs boson mass
as extracted from the global fits of electroweak observables is
91~GeV, with an experimental uncertainty of +45 and 
$-$32 GeV~\cite{eweak_higgs}.  
This result strongly depends on the assumption that the SM describes all 
physics up to high energy scales, and thus is not a proof that the SM 
Higgs boson actually exists.  Rather,
the results of electroweak fits serve as guidelines as to what
mass range is to be expected.
The results of experimental searches for the Higgs boson at 
LEP set a lower limit on the SM Higgs boson mass of 
about 114 GeV~\cite{lep_higgs}. 

If a SM-like Higgs boson exists, it is expected to be seen at the
Large Hadron Collider (LHC) at CERN~\cite{lhc_higgs}. To be certain
the state observed is the Higgs boson, it is 
necessary to measure the couplings of this state to
the $W$~and $Z$~gauge bosons, and to fermions such as the top and
bottom quarks and the tau leptons. 
Progress along these lines will be achieved at the
LHC~\cite{lhc_higgs,zeppi,HcoupLHCSM}, but the international linear
$e^+e^-$-collider (ILC) 
is absolutely essential for making precise measurements 
of a full set of these couplings and for establishing the
nature of EWSB~\cite{tdr,nlc,jlc,gagatdr}.

Many models beyond the SM, e.g., supersymmetric
extensions of the SM, predict extended Higgs sectors containing
more than one physical scalar state and therefore enriching Higgs boson
phenomenology. 
A common feature of the models with such extended Higgs 
sectors is the existence 
of additional scalar bosons, such as charged Higgs bosons 
and CP-odd Higgs boson(s). 
Discovery of these extra scalar particles would be direct evidence
for physics beyond the SM and be vital for our understanding of the 
structure of EWSB.
Once these additional Higgs bosons are found, 
we would be able to test each new physics model 
by measuring details of their properties accurately at ILC.
A $\gamma\gamma$ collider option provides further opportunity 
for the heavy Higgs search, where 
various properties of the heavy Higgs bosons, 
such as the CP-parity, can be explored.

\bigskip
In this report we address the following questions:
\begin{itemize}

\item 
What are the most important measurements that the ILC should perform
in the subject of Higgs boson physics?

\item
What are the key measurements by which the ILC will add to what will
be known already from the LHC?

\item
For each of these measurements, what criteria for the detectors are
necessary to allow the measurements with the required precision?

\item
What are the open questions in the fields of theory, experimental
analyses and detector developments that have to be addressed for a
successful ILC program?

\item
For these open questions, what is the best way to answer them in a
comprehensive and systematic way?

\end{itemize}
Consequently, we review the properties of Higgs bosons in various
models in \refse{sec:productiondecay}. This includes the possible
mass ranges, as well as a general overview about production and decay
modes. In \refse{sec:ILCmeasurements} we give an extensive overview about 
the potential of the ILC operated at center-of-mass
energies up to 1~TeV (including the $\ga\ga$ option) for the
determination of the Higgs boson properties 
within the framework of the SM and its various extensions. 
This comprises the measurement of the Higgs boson mass,
its couplings to SM fermions and gauge bosons, and the determination of
the spin and the CP quantum numbers of the Higgs. The extensions of
the SM that are analyzed in more detail are heavy SM-like Higgs
bosons, heavy Higgs bosons in the framework of supersymmetry (SUSY)
and further exotic scenarios. We review in \refse{sec:theory} recent
theoretical developments in the field of Higgs boson physics. This
includes higher-order corrections to Higgs boson masses and production
cross sections in the SM and the Minimal Supersymmetric Standard Model
(MSSM), the possibility of CP-violation, as well as extensions of
the MSSM such as the NMSSM, gauge-extended models, and Fat Higgs. 
We also summarize work on non-supersymmetric models,
such as Little Higgs models, models with extra dimensions,
and dimension-six Higgs operators. A list of Higgs related computer
tools (spectrum generators, decay packages, event generators) is
provided in \refse{sec:tools}. 
The important question of what the ILC can contribute to Higgs boson
physics after the LHC, the LHC/ILC interplay and synergy, is discussed
in \refse{sec:LHCILC}. It is emphasized that if a Higgs-like state is
discovered at the LHC and the ILC, 
independent of the realization of the Higgs mechanism, important
synergistic effects arise from the interplay of LHC and
ILC~\cite{lhcilc}.
Higgs boson physics can also have an important impact on our understanding of
cosmology. This is analyzed for the MSSM and the NMSSM case in
\refse{sec:cosmo}. 
The above listed issues and analyses impose certain detector
requirements to achieve the necessary precision.
The impact of the accelerator and detector performance on the precision
of measurements are analyzed in detail in \refse{sec:detector}. We
propose a strategy to optimize future analyses.
Open questions arising for the
topics in the various chapters are listed, and further topics of study
and corresponding roadmaps are suggested. 

\chapter{Production and Decays of Higgs Bosons}
\label{sec:productiondecay}

In this section we review the experimental status of Higgs boson
production and decay at the ILC and the photon collider ($\ga$C). 
For the corresponding
theoretical results and developments, see \refse{sec:theory}.


\section{The Higgs boson mass}

\subsection{Limits from theory}

\begin{figure}[htb]
\begin{minipage}[c]{0.50\textwidth}
\epsfysize=3.0in \rotatebox{90}{\epsffile{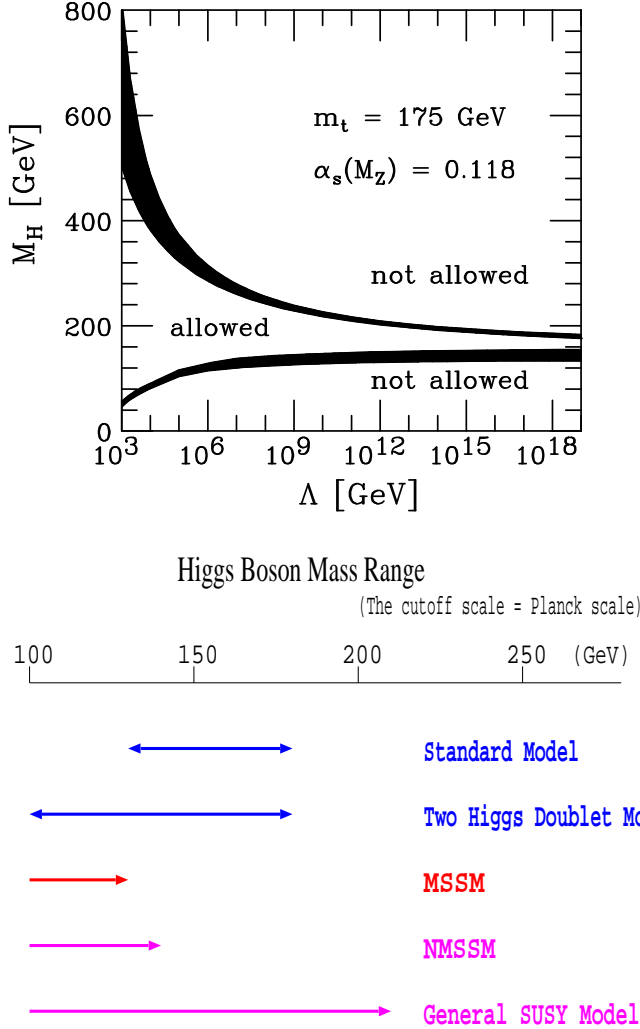}}
\end{minipage}
\begin{minipage}[c]{0.03\textwidth}
$\phantom{0}$
\end{minipage}
\begin{minipage}[c]{0.45\textwidth}
\caption{Summary of the uncertainties connected to the bounds on $\mh$.
  The upper solid area indicates the sum of theoretical uncertainties
  in the $\mh$ upper bound for $m_t=175 \gev$~\cite{LambdaMH}. The
  upper edge corresponds to Higgs masses for which the SM Higgs sector
  ceases to be meaningful at scale $\Lambda$, and the lower
  edge indicates a value of $\mh$ for which perturbation theory is
  certainly expected to be reliable at scale $\Lambda$.  The lower
  solid area represents the theoretical uncertainties in the $\mh$
  lower bounds derived from stability requirements
  using $\mt = 175 \gev$ and $\al_s=0.118$.} 
\label{fig:LambdaMH}
\end{minipage}
\end{figure}

The Higgs boson $h$ is yet to be discovered, 
and its mass $\mh$ remains unknown. 
In the SM, $\mh$ is a parameter which 
characterizes the property of the Higgs dynamics. 
Since $m_h \propto \sqrt{\lambda} v$ ($\la$ is the quartic term in the
Higgs potential, $v$ denotes the vacuum expectation value), 
a light $h$ means that the Higgs sector is weakly-interacting, 
while a heavy $h$ corresponds to strong coupling.
Although $\mh$ is an unknown parameter, 
by imposing the requirement that the theory must be consistent 
up to a given value of the cutoff scale $\Lambda$,  
the allowed region of $\mh$ can be predicted 
as a function of $\Lambda$~\cite{lindner}.  
Based on this requirement, a renormalization group 
analysis for the coupling constant $\lambda$ gives  
upper and lower bounds on $\mh$ for a given $\Lambda$. 
In the SM, for $\Lambda=10^{19} \gev$, the allowed region of $\mh$ 
is evaluated as about $135 < m_h < 180 \gev$, 
while for $\Lambda=10^{3} \gev$ it is about $m_h < 500 \gev$.
The results are summarized in \reffi{fig:LambdaMH}~\cite{LambdaMH},
evaluated for $\mt = 175 \gev$.
%

\begin{figure}[htb!]
\begin{minipage}[c]{0.55\textwidth}
\psfig{figure=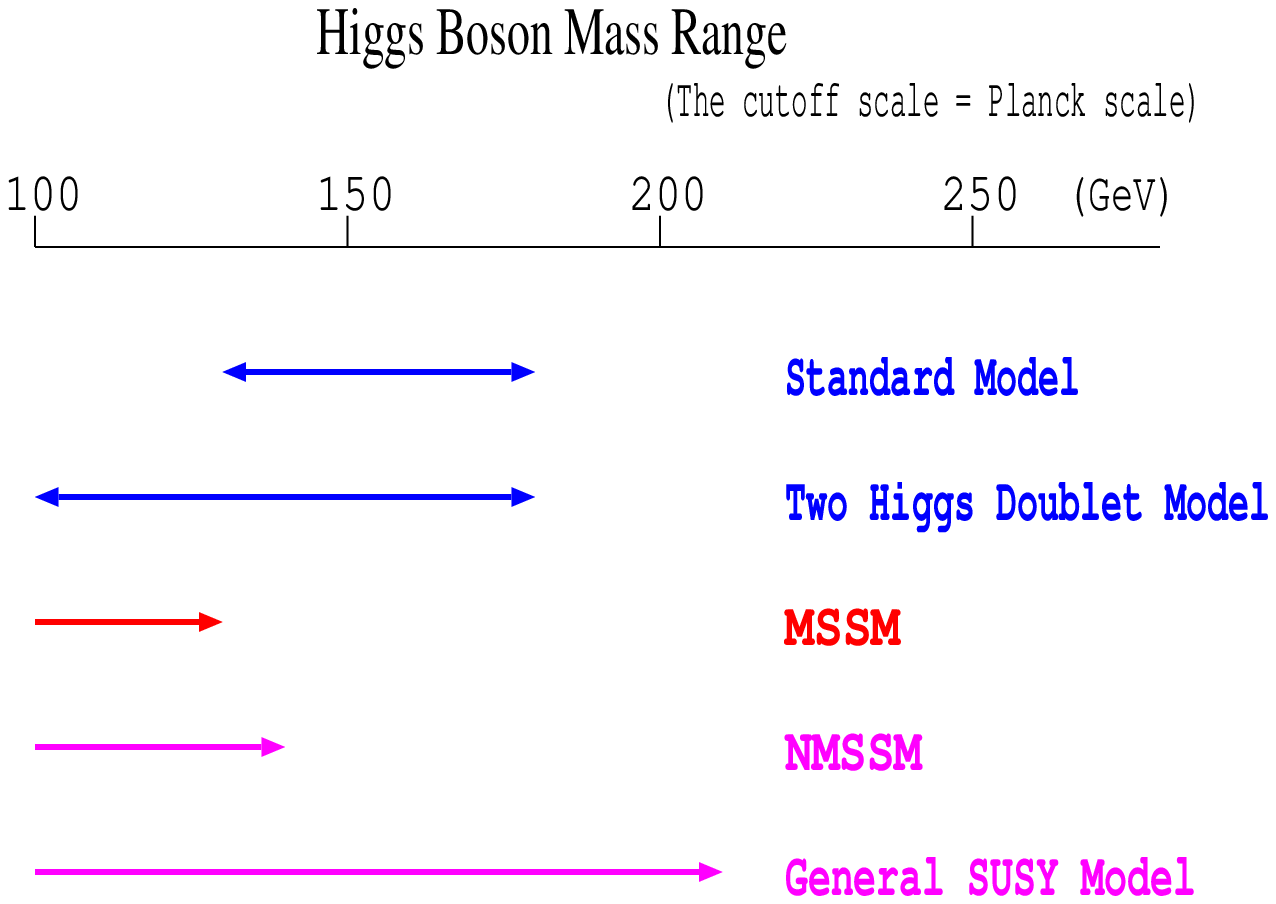,width=0.99\textwidth}
\end{minipage}
\begin{minipage}[c]{0.03\textwidth}
$\phantom{0}$
\end{minipage}
\begin{minipage}[c]{0.40\textwidth}
\caption{Predictions of the mass range 
of a Higgs boson for various models\cite{okada}. 
The upper and lower
bound of the Higgs boson mass is derived from an assumption that each model
is valid up to the cut-off scale of the theory, which is taken to be 
$10^{19} \gev$. For the MSSM case, the mass bound is obtained without
reference to the cutoff scale.
 }
\label{fig:h-mass-bounds}
\end{minipage}
\end{figure}

Hence, as long as the SM Higgs sector is assumed, 
a light Higgs boson would indicate 
a weakly coupled theory with a high cut-off scale. 
Scenarios based on grand unified theories
might correspond to this case. 
In such cases, supersymmetry would necessarily be required 
to reduce the problem of the 
large hierarchy between the weak scale and the scale $\Lambda$.
On the contrary, a heavy Higgs boson with $m_h \sim$ several hundred GeV 
would imply a strongly-coupled Higgs sector with a low cutoff 
$\Lambda \sim {\cal O}(1)$ TeV.
In such a case, the Higgs sector should be considered 
as an effective theory of a new dynamics at TeV scales. 
Therefore, from knowing the mass of the Higgs boson, a useful hint 
for new physics beyond the SM can be obtained.
Figure~\ref{fig:h-mass-bounds} shows the mass range of the 
(lightest) Higgs boson under the assumption of $\Lambda=10^{19} \gev$ 
in the SM\cite{lindner}, the minimal supersymmetric SM
(MSSM)\cite{ERZ,mhiggslong,mhiggsAEC} (see also \refse{sec:mhMSSM}),  
some extended SUSY models \cite{quiros} and 
the general two Higgs doublet model \cite{kko}.
Due to its strong dependence on the other model parameters, the
lightest Higgs boson mass in the MSSM and its extensions is one of the
most important observables in the Higgs sector.

\subsection{Experimental limits on the SM Higgs boson mass}
\label{subsec:MHexp}

The mass of the SM Higgs boson is bounded from below by direct
searches at LEP to be $\MH > 114.4 \gev$ at the
95\%~C.L.~\cite{lep_higgs}. 
%
\begin{figure}[htb!]
\begin{minipage}[c]{0.50\textwidth}
\epsfysize=3.0in{\epsffile{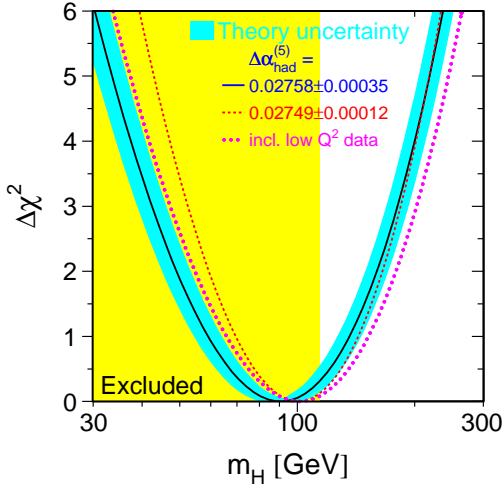}}
\end{minipage}
\begin{minipage}[c]{0.03\textwidth}
$\phantom{0}$
\end{minipage}
\begin{minipage}[c]{0.45\textwidth}
\caption{Global fit to all electroweak precision data within the
  SM~\cite{eweak_higgs} (including the new $\mt$ value of 
  $\mt = 172.7 \gev$). The fit yields $\MH = 91^{+45}_{-32} \gev$
  with an upper limit of $\MH < 186 \gev$ at the 95~\% C.L. The fit
  was performed under the assumption that the SM including the Higgs
  sector gives a correct description of the experimental data.
} 
\label{fig:MH_EWfit}
\end{minipage}
\end{figure}
%
An upper limit can be obtained from a global fit to all electroweak
precision data~\cite{eweak_higgs}. Including the new experimental
$\mt$ value of $\mt = 172.7 \gev$~\cite{newmt}, 
the fit yields $\MH = 91^{+45}_{-32} \gev$
with an upper limit of $\MH < 186 \gev$ at the 95\% C.L. The fit
was performed under the assumption that the SM including the Higgs
sector gives a correct description of the experimental data. 
Therefore it does {\em not} confirm the SM Higgs sector.
Higgs masses larger than about $200 \gev$ are only possible if other
new physics effects would compensate the effect of the heavy Higgs on
the electroweak precision data, see e.g.~\cite{peskinwells,Choudhury:2002qb}.
In most cases, some hints of these new physics effects are expected to be 
visible at either the LHC or ILC (or both).


\section{Standard Model Higgs Production and Decays}
\label{sec:SMproddecay}

\begin{figure}[htb!]
\vspace{1em}
\begin{minipage}[c]{0.47\textwidth}
\psfig{figure=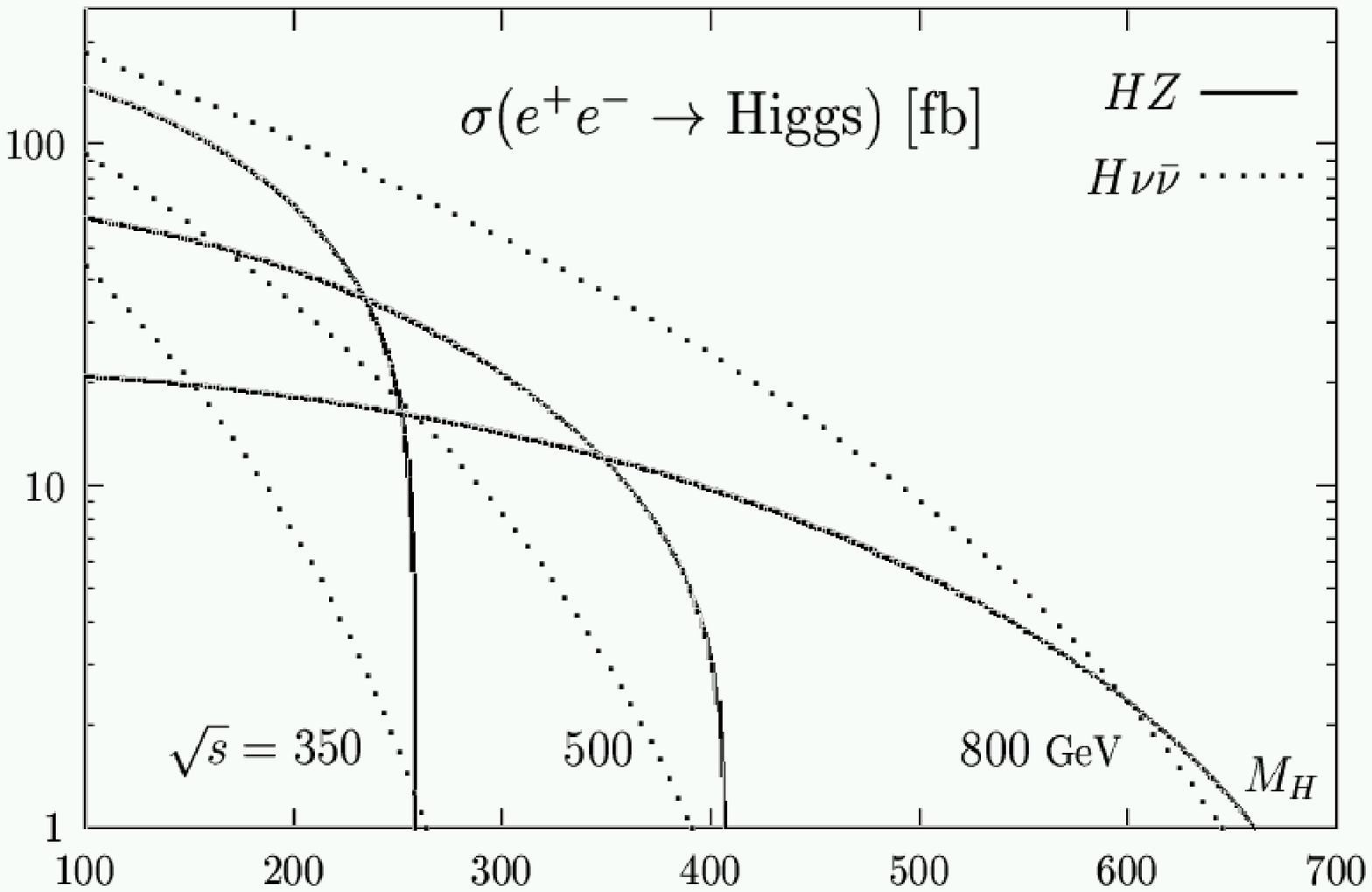,width=0.99\textwidth}
\end{minipage}
\begin{minipage}[c]{0.47\textwidth}
\psfig{figure=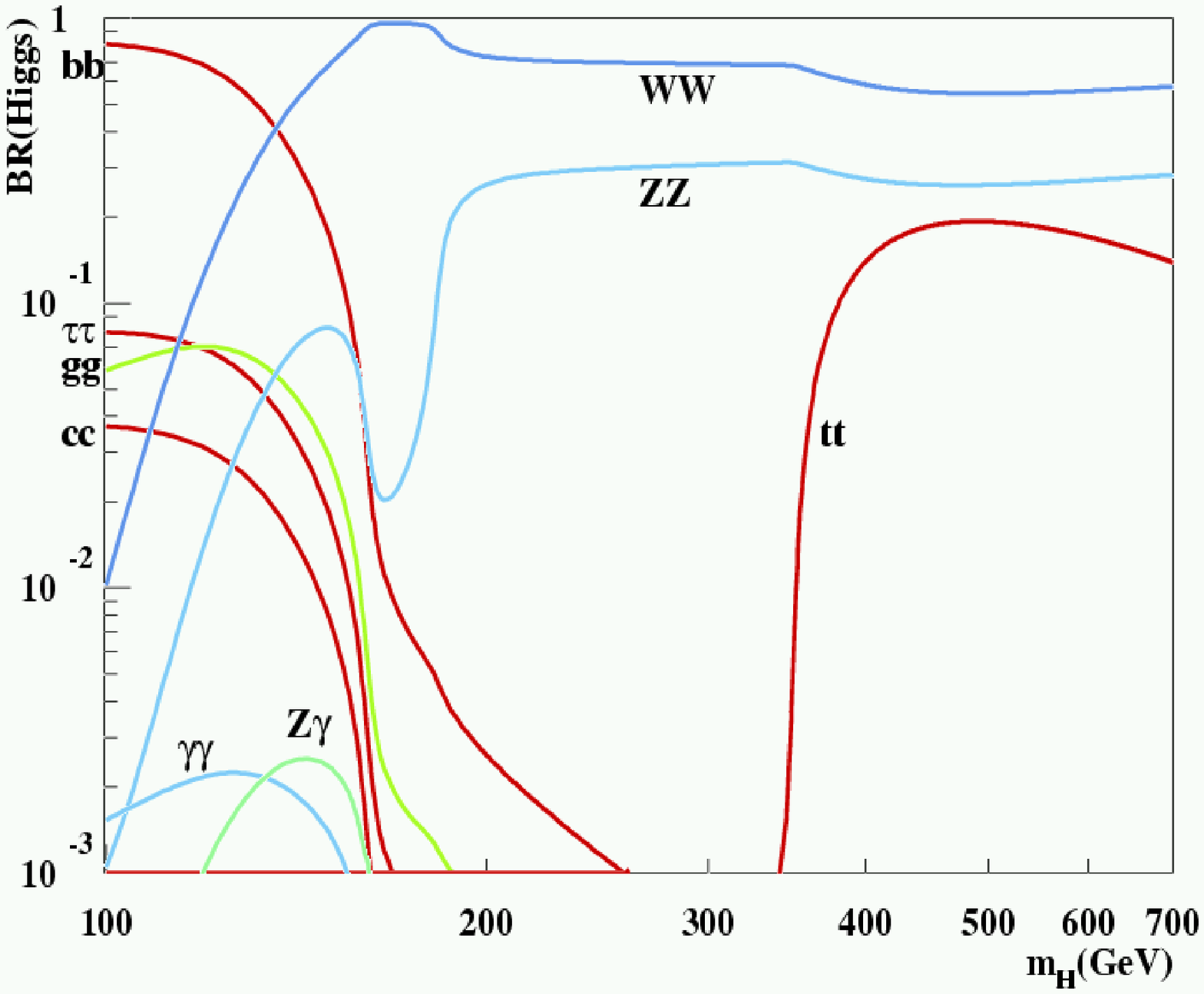,width=0.85\textwidth}
\end{minipage}
\begin{minipage}[c]{0.45\textwidth}
\caption{SM Higgs boson production cross sections 
as a functions of the Higgs boson mass.
\label{fig:xsec_sm}
}
\end{minipage}
\begin{minipage}[c]{0.08\textwidth}
$\phantom{0}$
\end{minipage}
\begin{minipage}[c]{0.45\textwidth}
\caption{SM Higgs boson branching ratios as a
functions of the Higgs boson mass.
\label{fig:higgs_branching}
}
\end{minipage}
\end{figure}

In $\EE$ collisions, the SM Higgs boson is predominantly 
produced through the Higgs-strahlung process, 
$\EEHZ$, and through vector boson fusion 
processes $\EEHnnee$.
The cross section of the main production mechanisms as a function of 
the Higgs boson mass is presented
in \reffi{fig:xsec_sm} for the three representative 
centre-of-mass energies,
350, 500 and 800 GeV. 

The Higgs boson can be detected and
its profile can be studied in detail over a wide mass range by exploiting 
various decay modes. Being responsible for generation of fermion and weak boson
masses, the Higgs boson decays preferentially into the heaviest kinematically
accessible final states. The dependence of the branching ratios of 
the SM Higgs boson on its mass is illustrated in 
\reffi{fig:higgs_branching}.

\section{Higgs Bosons in General Two Higgs Doublet Models}

Among models with an extended Higgs sector, two Higgs doublet models 
are of particular interest, since this structure of the Higgs sector
is required in MSSM~\cite{mssm,higgs_hunters}.
In CP-conserving two Higgs doublet models, spontaneous symmetry 
breaking gives rise to five physical scalar states: two CP-even Higgs bosons, 
the lighter one denoted $h$ and heavier $H$, one CP-odd boson $A$ and 
two charged bosons, $\HPM$. 
At tree level in the MSSM, the mass spectrum is determined by $\tanb$
and the mass of CP-odd boson $\mA$. The mass of the 
light Higgs boson is constrained to be smaller than the mass of the 
$Z$ boson, $\mh < \mZ$. Higher order corrections, in particular from 
loops involving third generation fermions and their superpartners,
move the upper limit on $\mh$ up to about 135 GeV~\cite{mhiggslong,mhiggsAEC}.
 
In addition to supersymmetry, 
there are lots of motivations to consider models 
with two Higgs doublets at the electroweak scale, such as 
top-color\cite{topcolor}, 
Little Higgs models\cite{little}, some extra dimension models,   
and the model of 
gauge-Higgs unification\cite{hosotani}. 
Tiny neutrino masses ({\it e.g.}, the Zee model\cite{zee}),
and extra CP violating phases which may be required for 
the realization of electroweak baryogenesis\cite{EWBG} 
also can be studied by introducing multi scalar doublets 
(plus singlets) at the electroweak scale. 
The analysis of phenomenology in the framework 
of the general two Higgs doublet model can make it possible 
to distinguish these nonstandard models.

In the CP-conserving MSSM and more general two Higgs doublet model, 
the cross sections of the Higgs-strahlung and weak boson fusion processes
involving CP-even bosons scale with the coupling of the appropriate Higgs 
to gauge bosons relative to that in the SM. These couplings are given by
\BEQ
g_{hZZ,hWW} \sim \sin(\beta-\alpha), \hspace{2cm}
g_{HZZ,HWW} \sim \cos(\beta-\alpha),
\EEQ
where $\alpha$ is the mixing angle in the CP-even sector.
The set of tree level couplings between
neutral Higgs particles and weak bosons is extended by 
two additional couplings,
\BEQ
g_{hAZ} \sim \cos(\beta-\alpha),\hspace{2cm}
g_{HAZ} \sim \sin(\beta-\alpha).
\EEQ
As a consequence, in $\EE$ collisions, the Higgs-strahlung 
and fusion processes will be complemented by the associated 
Higgs boson pair production mechanisms, $\EEhHA$. Neutral
Higgs bosons can be also accessed via Yukawa processes, 
$\EEHAbb$, $\EEHAtt$.
The former reaction is of particular importance, since it 
allows to extend the mass reach for neutral Higgs bosons 
up to values close to the collision centre-of-mass energy.

The MSSM exhibits a so-called decoupling limit as $\mA$ becomes
large, $\mA \gsim 200 \gev$. In this scenario, $h$ approaches 
the properties of the SM Higgs boson, while $H$ decouples 
from the weak bosons. As a consequence the $H$ boson production 
via Higgs-strahlung or weak boson fusion processes gets suppressed, whereas  
the cross section of the process $\EEHA$ reaches its maximal value.
Another distinct feature of the decoupling limit is a tiny 
mass splitting between $H$ and $A$, making it practically 
impossible to distinguish them by mass.
For a large portion of MSSM parameter space, decoupling is realized
and the decay properties
of $h$ are close to those of the SM Higgs boson, 
while $H$ and $A$ decay predominantly into the heaviest kinematically 
accessible fermions, $\tautau$, $\bb$ or $\toptop$.
However, scenarios are possible in which decay rates of heavy 
neutral Higgs bosons into supersymmetric particles become significant.

Charged Higgs bosons can be produced in $\EE$ collisions 
in pairs, $\EEHPM$. In this process, the mass 
reach for the charged Higgs boson is limited to half of centre-of-mass
energy. Hence, it would be desirable to investigate also  
the rare processes of single charged Higgs boson production. The dominant
processes of the single charged boson production are: 
$\EEHPMbt$, $\EEHPMtaunu$, $\EEHPMW$, $e^+e^- \to H^+e^-\bar\nu_e$.
Recent theoretical
calculations~\cite{single_hpm_xsec,single_hpm_xsec2,eeHW,eeenH} 
showed that, in general, the parameter regions for which single charged Higgs 
boson production cross sections exceeds 0.1~fb are rather small beyond
the pair production threshold.
The main decay channels of the charged Higgs bosons are
$\HPMcs$, $\HPMtaunu$ and, if kinematically allowed, $\HPMtb$.

\begin{figure}[htb!]
\begin{minipage}[c]{\textwidth}
\psfig{figure=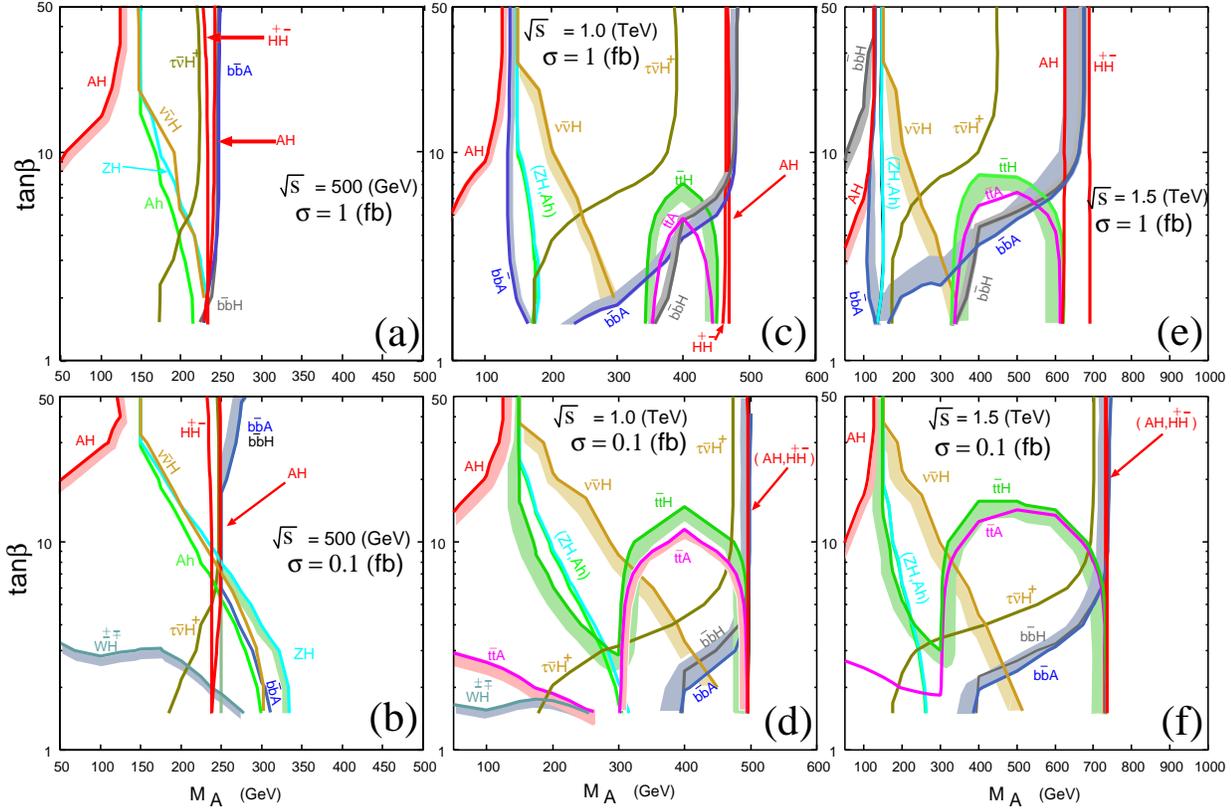,width=0.99\textwidth}
\end{minipage}
\caption{
The cross-section contours: (a) for
$\sqrt{s}$ = 500 GeV and $\sigma$ = 1 fb, (b) for $\sqrt{s}$ = 500 GeV
and $\sigma$ = 0.1 fb, (c) for $\sqrt{s}$ = 1.0 TeV and $\sigma$ =
1 fb, (d) for $\sqrt{s}$ = 1.0 TeV and $\sigma$ = 0.1 fb, (e) for
$\sqrt{s}$ = 1.5 TeV and $\sigma$ = 1 fb, and (f) for $\sqrt{s}$ = 1.5
TeV and $\sigma$ = 0.1 fb. GRACE/SUSY\cite{grace} is used to calculate 
the production cross-sections of the tree-level processes.
One-loop induced production of $W^+H^-$ is calculated as in 
\citeres{single_hpm_xsec,single_hpm_xsec2}.
Masses of the Higgs bosons and the mixing angle of the
neutral Higgs bosons are obtained by using
FeynHiggs\cite{feynhiggs},
where we assume the diagonal masses to be (1~TeV)${}^2$ in the stop
mass matrix and maximal stop mixing.
HDECAY\cite{Djouadi:1997yw} is used to calculate decay widths of the Higgs
bosons.
}
\label{fig:heavy-higgs-contour}
\end{figure}

Figure~\ref{fig:heavy-higgs-contour} shows  
contour plots of the production cross-sections   
in the $m_A$-$\tb$ plane for the following processes:
$e^+e^- \to$ $ZH$, $Ah$, $AH$, $H^+H^-$, $W^{\pm}H^{\mp}$,
$b\overline{b}A$, $b\overline{b}H$, $t\overline{t}A$,
$t\overline{t}H$,
and $\nu\overline{\nu}H$\cite{kiyoura}.
In Figures~\ref{fig:heavy-higgs-contour} (a), (c), (d), (e) and (f), 
the mass reach for $A$, $H$, and $H^{\pm}$ 
at the ILC is determined by half of $\sqrt{s}$.
In Figure~\ref{fig:heavy-higgs-contour}(b), 
if the sensitivity reaches to 0.1fb, 
the discovery contours for the $b\overline{b}A$ and $b\overline{b}H$
modes go beyond $\sqrt{s}/2$ for large $\tb$.
For $\tb\lsim$10, the $ZH$ ($Ah$) mode is
available above $m_A > \sqrt{s}/2$.
In Figure~\ref{fig:heavy-higgs-contour}(c)-(f), 
cross-section contours of $e^+e^-
\to t\overline{t}A$, $t\overline{t}H$, $b\overline{b}A$, and
$b\overline{b}H$ exhibit a dependence on $\tb$ for
350~GeV$\lsim m_A\lsim \sqrt{s}/2$.
These processes include $e^+e^- \to AH$ followed by $A$ or $H$
decaying into a $b\overline{b}$ or $t\overline{t}$ quark pair.
The $\tb$ dependence shown in 
Figure~\ref{fig:heavy-higgs-contour}(c)-(f) comes from 
the branching ratio of the heavy Higgs bosons.
Figure~\ref{fig:heavy-higgs-contour}(c)-(e) 
also shows that the ILC will cover the region of moderate
$\tb\lsim10$ and $M_A\lsim\sqrt{s}/2$ where the detection
of the heavy Higgs bosons at LHC is expected to be difficult.
If kinematically allowed, the heavy Higgs bosons are expected to be
found in several modes at the ILC. More details can be found in
\refse{sec:2to3}. 

\section{Higgs Boson Production at a $\GG$ collider}

\begin{figure}[htb!]
\begin{minipage}[c]{0.28\textwidth}
\psfig{figure=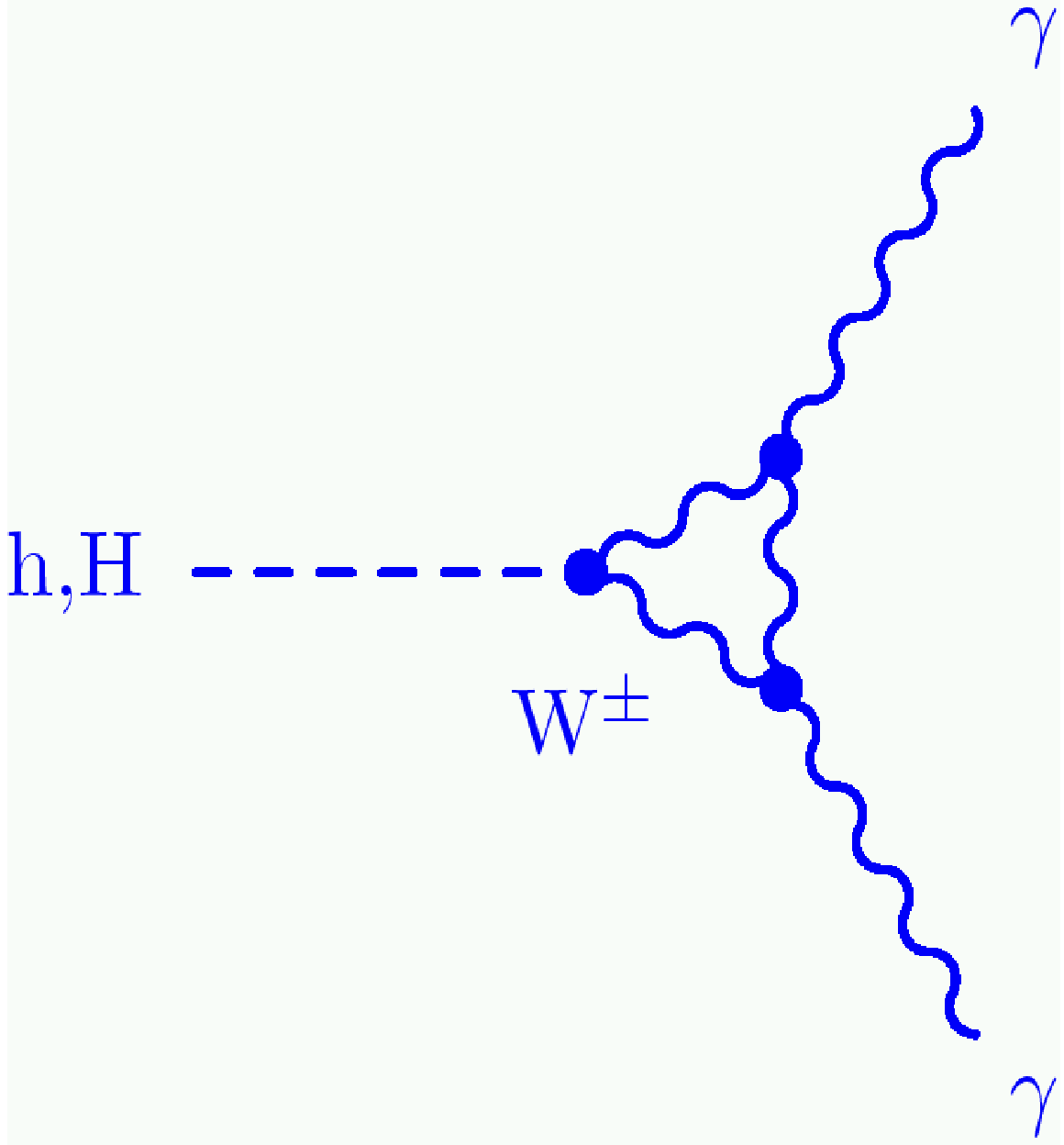,width=0.99\textwidth}
\end{minipage}
\begin{minipage}[c]{0.28\textwidth}
\psfig{figure=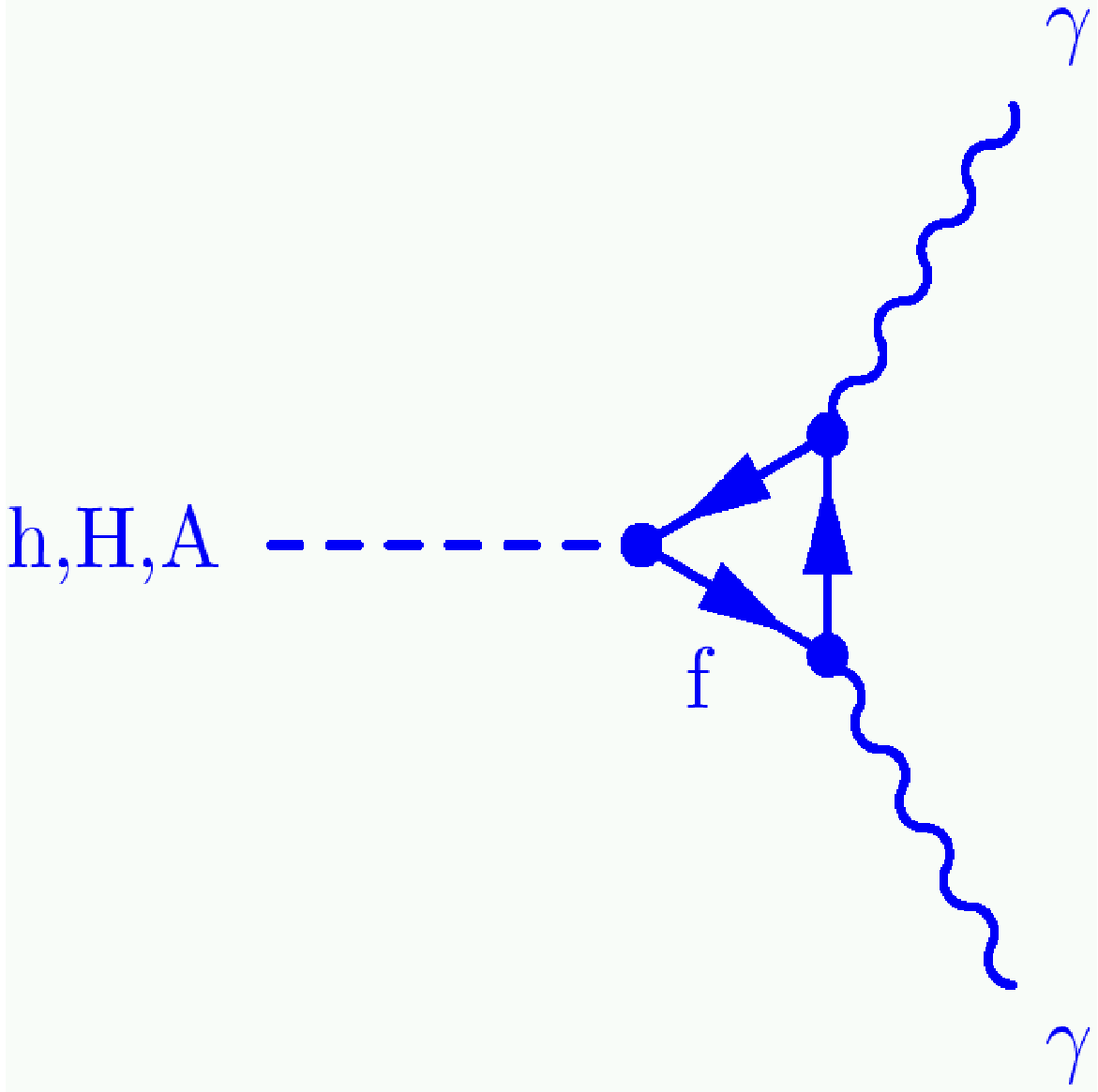,width=0.99\textwidth}
\end{minipage}
\begin{minipage}[c]{0.35\textwidth}
\caption{
Loop-induced couplings of Higgs bosons to photons.
Feynman diagram involving W boson loop (left plot).
Feynman diagram involving charged fermion loop (right plot).
\label{fig:higgs_to_photons}
}
\end{minipage}
\end{figure}

At a $\GG$ collider, neutral Higgs bosons can be produced resonantly in the
s-channel. Although Higgs bosons do not couple to photons at tree level,
Higgs boson production in $\GG$ collisions
becomes possible due to loop-induced couplings as 
illustrated in \reffi{fig:higgs_to_photons}.  
As compared to the $\EE$ collision option, 
the $\GG$ collider, operated at the 
same centre-of-mass energy, has higher mass reach for the Higgs bosons,
as the whole collision energy is available for the resonant production of 
the Higgs particles in the
s-channel~\cite{photon-higgs1,photon-higgs2,photon-higgs3,photon-higgs4}.  
In addition, we can determine the CP parity of the heavy Higgs boson
through the process $\ga\ga \to t\bar{t}$ by
measuring the helicity of the top quark\cite{photon-CP}.
Along with other nice features, described later on in this report, 
the high-mass discovery potential of the $\GG$ collider
makes it a particularly attractive running option of the 
International Linear Collider facility. 

Technical aspects related to the $\GG$ collision option are highlighted in 
References~\cite{gg_telnov1,gg_telnov2,gg_rosca}.

\section{The role of polarization}

For the ILC the possibility to have a polarized $e^-$ beam is in the
baseline design~\cite{talkWalker}. A polarization of up to 80\% is
foreseen. The possibility to have polarized $e^+$ as well is still
under discussion. Here in principle a polarization of $\sim 60\%$
seems to be achievable. Polarization can be helpful in various
aspects of Higgs boson physics.

\begin{itemize}

\item
One of the major physics goals at the ILC is the precise
analysis of all the properties of the Higgs particle.  For a light Higgs the
two major production processes, Higgs-strahlung, $e^+e^- \to H Z$,
and $WW$ fusion, $e^+e^-\to H \nu \bar{\nu}$, will have similar
rates at $\sqrt{s}=500 \gev$; see \refse{sec:SMproddecay}.
Beam polarization will be
important for background suppression and a better separation of the two
processes. Furthermore, the determination of the general Higgs couplings is
greatly improved when both beams are polarized~\cite{power}. 

\item
Searches for heavy SUSY  Higgs particles can be
extremely challenging for both the LHC and the ILC.  Exploiting
single Higgs-boson production in $e^+e^- \to \nu \bar{\nu} H$ extends
the kinematical reach considerably.  However, in the decoupling
region, $M_A\gg M_Z$, the  
suppressed couplings of the heavy Higgs boson to SM gauge bosons
lead to very small rates. This difficulty could be attenuated by
accumulating a very high integrated 
luminosity, together with a further enhancement of the signal cross
section by polarizing both beams, see also \refse{sec:2to3}. 
Polarization can be decisive to discover MSSM Higgs bosons with low
production cross sections~\cite{eennH}.
Polarization of both the $e^-$ and $e^+$ beams is important in this case.

\item
Compared with the case where only the electron beam is
polarized, the polarization of both beams leads to a gain of about one
order of magnitude in the accuracy of the effective weak mixing angle,
$\sweff$~\cite{sweff}.  Within the SM, this has a dramatic effect on
the indirect 
determination of the Higgs-boson mass, providing a highly sensitive
consistency test of the model that may possibly point towards large
new-physics scales~\cite{blueband}. Within the MSSM, the large
increase in the 
precision of $\sweff$ will allow to obtain stringent indirect bounds
on SUSY parameters~\cite{power}. This will constitute, in analogy to
the SM case, a 
powerful consistency test of supersymmetry at the quantum level and
may be crucial to constrain SUSY parameters that are not directly
experimentally accessible~\cite{PomssmRep,gigaz}.

\end{itemize}

In order to fully exploit the possibility of beam polarization,
several questions should be addressed in the above scenarios:
\begin{itemize}
\item
What are the detector requirements for these measurements?
\item
What are the accelerator requirements?
\item
What are the theory requirements?
\end{itemize}
Partial answers are already available in \citere{power}.

\chapter{Measurements in the Higgs Sector at $\EE$ and $\GG$ Colliders}
\label{sec:ILCmeasurements}

In the following we review the Higgs boson sector measurements that
can be performed at the ILC and the $\ga$C. The description of a
SM-like Higgs boson (with mass below $\lsim 140 \gev$) also applies to
the lightest MSSM Higgs boson for large parts of the MSSM parameter
space. The issue of heavy MSSM Higgs bosons is addressed separately. 


\section{Model Independent Determination of the 
Higgs-strahlung cross section}
\label{sec:hstrahl}

\begin{figure}[htb!]
\begin{minipage}[c]{0.47\textwidth}
\psfig{figure=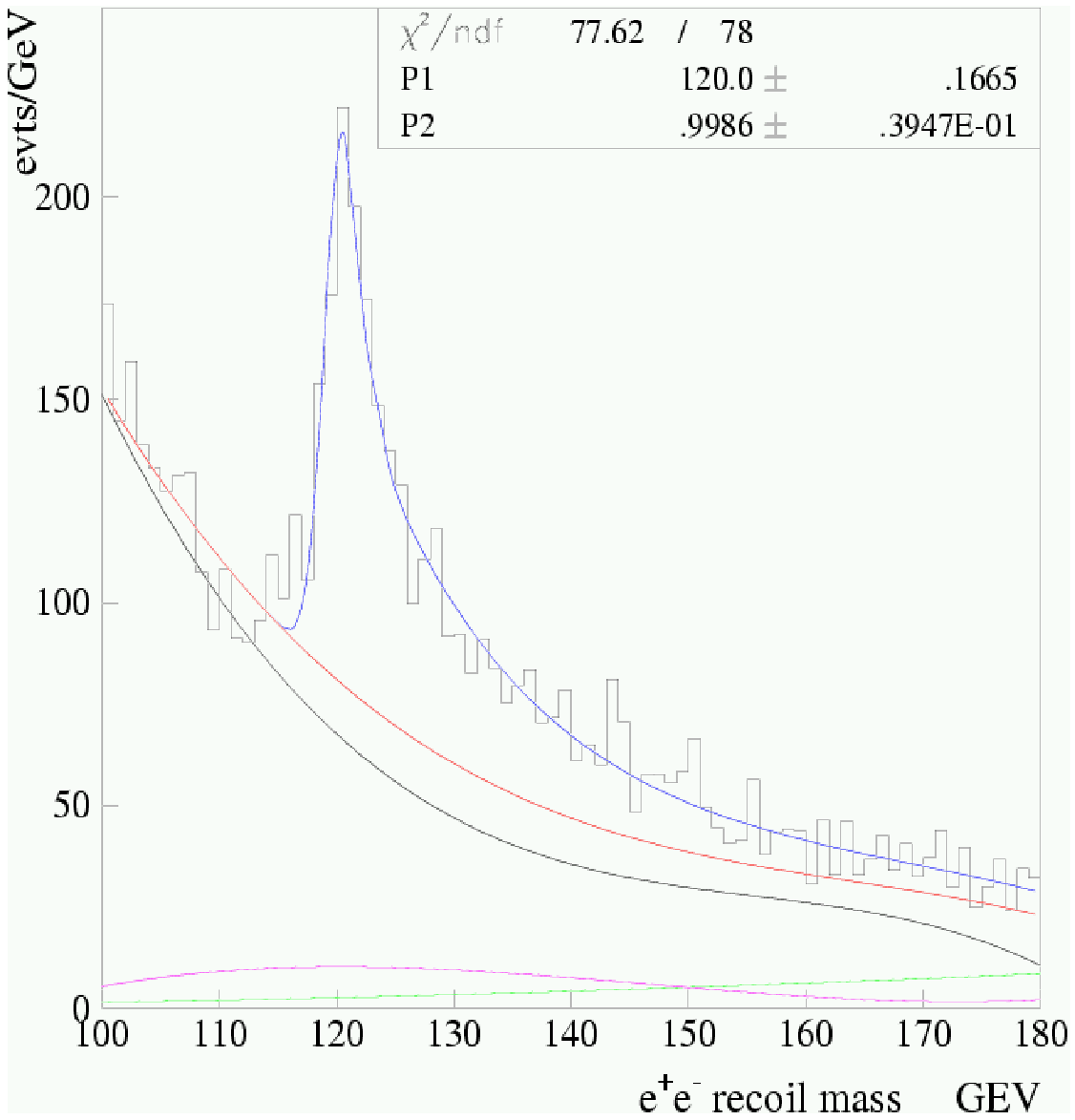,width=0.88\textwidth}
\end{minipage}
\begin{minipage}[c]{0.47\textwidth}
\psfig{figure=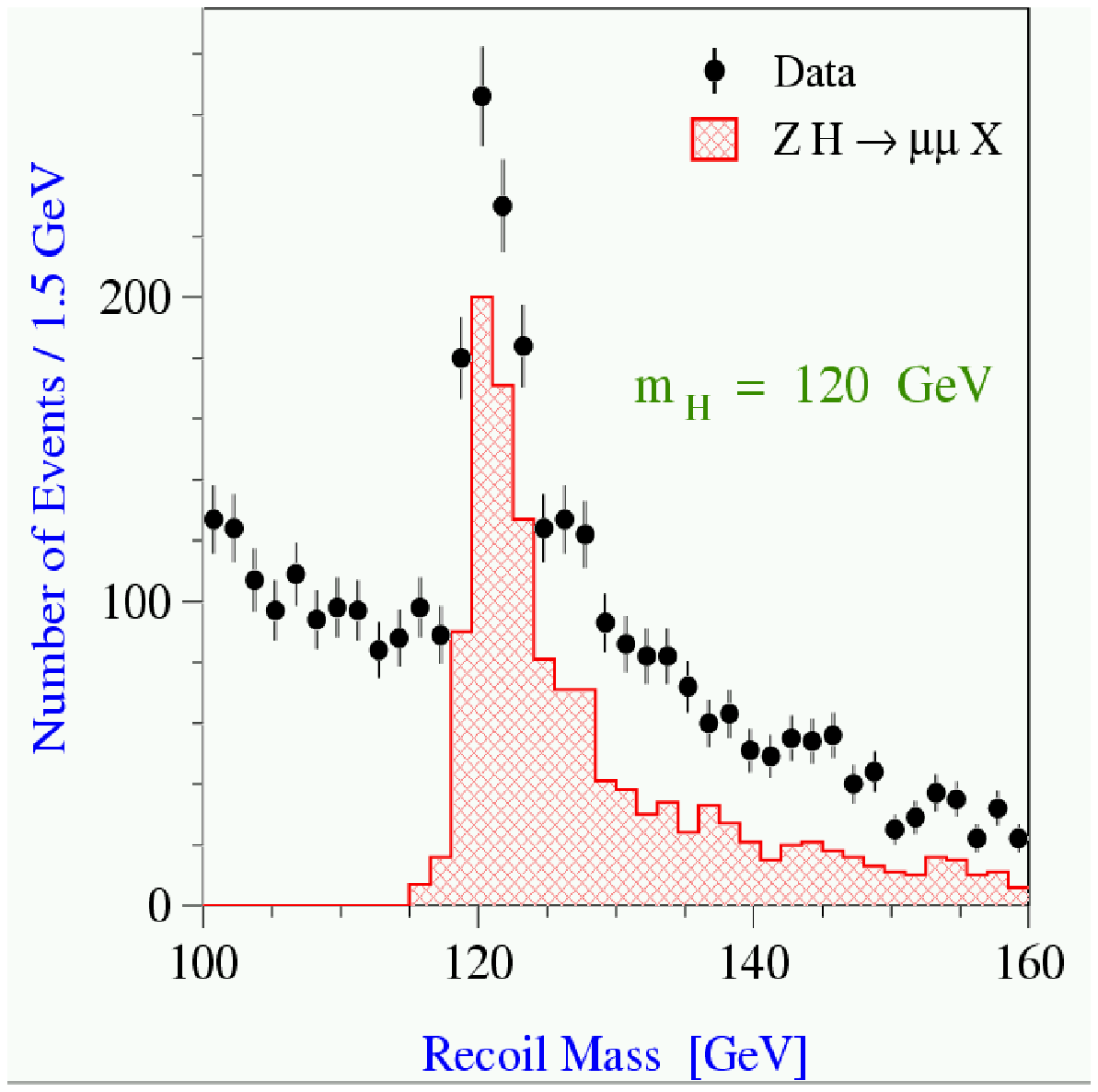,width=0.99\textwidth}
\end{minipage}
\begin{minipage}[c]{1.0\textwidth}
\caption{Distribution of the invariant mass
of the system recoiling against a pair leptons in 
the channel $HZ\ra Xe^+e^-$~\cite{hzxll_brient} (left plot) 
and in the channel $HZ\ra X\mu^+\mu^-$~\cite{hzxll_wolfgang} (right plot).
The Higgs boson mass is 120 GeV. The left plot corresponds 
to the centre-of-mass energy of 340 and the right plot to 350 GeV.
Assumed integrated luminosity is 500 fb$^{-1}$ for both cases.
\label{fig:HZXll}
}
\end{minipage}
\end{figure}

One of the nice features of 
the linear collider is its capability to detect the Higgs boson
independent of its decay mode. Even if the Higgs boson decays
into invisible particles\footnote{Scenarios of the Higgs boson 
decays into invisible particles are discussed in \refse{subsec:invisHiggs}.}, 
it can still be detected by exploiting the Higgs-strahlung production
mechanism with subsequent decay of the $Z$ into electron or muon pairs. 
The signal manifests itself as a peak in the distribution of 
invariant mass of the system recoiling against the electron or muon pair 
stemming from $Z$ boson decay. This is illustrated in \reffi{fig:HZXll}.

The results of the dedicated study~\cite{hzxll_wolfgang}
on model-independent measurement of 
the Higgs-strahlung cross-section, exploiting the $\HZXll$ channel,
are presented in Table~\ref{tab:HZXll_xsec}.
Combining the two final states $\HZXee$ and $\HZXmm$,
the cross sections can be measured with statistical errors of 
2.6 to 3.1\% for Higgs masses from 120 to 160 GeV.

\begin{table}[htb!]
\begin{minipage}[c]{0.50\textwidth}
\begin{tabular}{ccc}
$\mH$ (GeV) & $\sigma$ (fb) $\Zee$ & $\sigma$ (fb) $\Zmm$ \\
\hline
120 & 5.26$\pm$0.18$\pm$0.13 & 5.35$\pm$0.21$\pm$0.13 \\
140 & 4.38$\pm$0.18$\pm$0.11 & 4.39$\pm$0.17$\pm$0.10 \\
160 & 3.68$\pm$0.17$\pm$0.09 & 3.52$\pm$0.15$\pm$0.08 \\
\end{tabular}
\end{minipage}
\begin{minipage}[c]{0.05\textwidth}
$\phantom{0}$
\end{minipage}
\begin{minipage}[c]{0.4\textwidth}
\caption{
The results for the Higgs-strahlung cross-section 
measurement in the $\HZXee$ and 
$\HZXmm$ channels at $\sqrt{s}$ = 350 GeV with 
an integrated luminosity of 500 fb$^{-1}$.
The first error is statistical and the second is systematic.
\label{tab:HZXll_xsec}
}
\end{minipage}
\end{table}


\section{Determination of Higgs Boson Mass}

\begin{figure}[htb!]
\begin{minipage}[c]{0.47\textwidth}
\psfig{figure=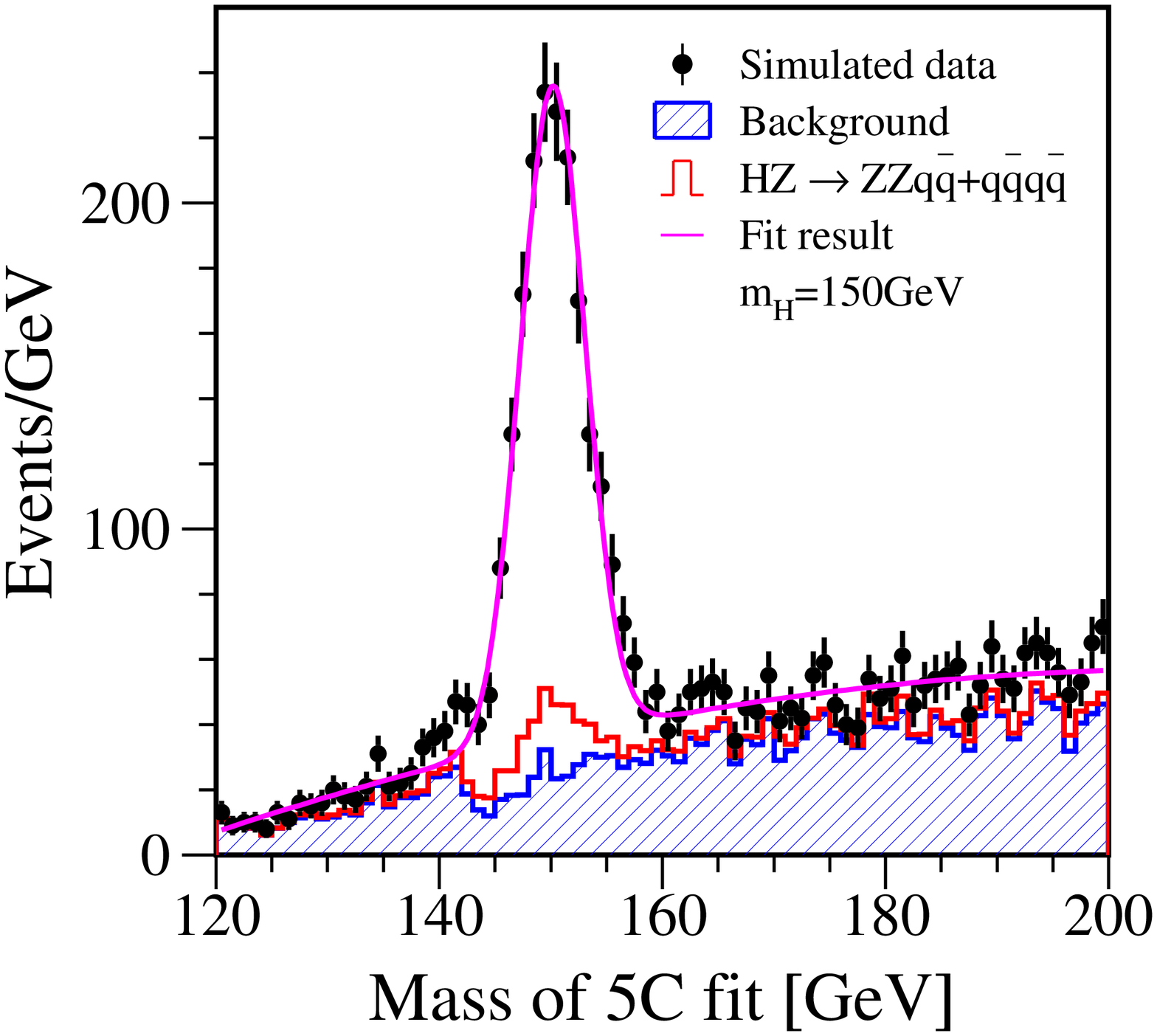,width=0.99\textwidth}
\end{minipage}
\begin{minipage}[c]{0.47\textwidth}
\psfig{figure=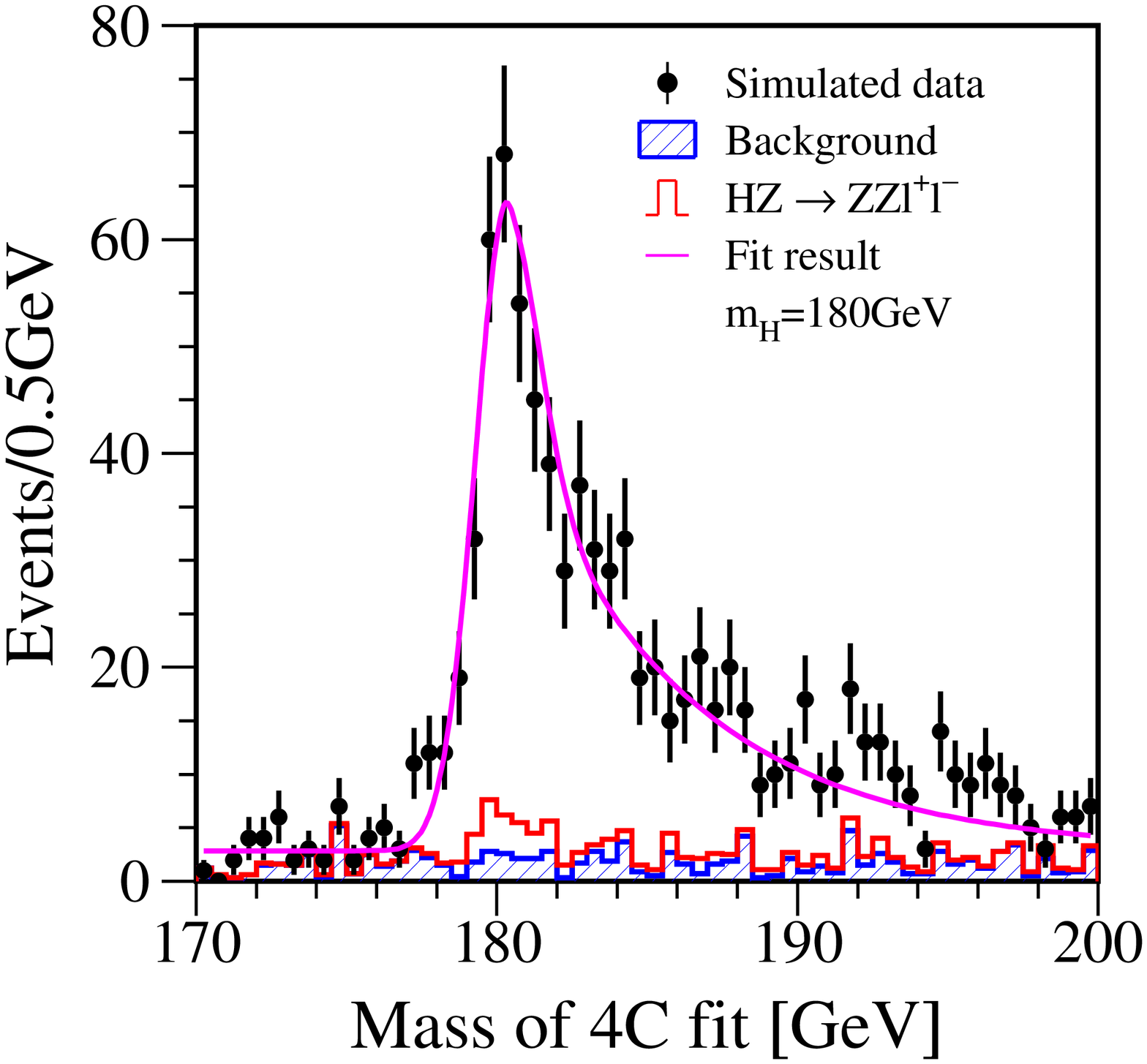,width=0.99\textwidth}
\end{minipage}
\begin{minipage}[c]{1.0\textwidth}
\caption{
Distribution of the reconstructed Higgs boson mass
in the $\HZWWqq$ channel for $\mH$ = 150 GeV (left plot)
and in the $\HZWWll$ channel for $\mH$ = 180 GeV (right plot).
\label{fig:higg_mass}
}
\end{minipage}
\end{figure}

The SM Higgs boson mass is a free parameter of the model and of 
great importance for the exploration of the EWSB mechanism. 
The prospects of Higgs mass measurement 
at a linear $\EE$ collider are investigated in 
References~\cite{higgs_mass,higgs_lineshape}. 
The study presented in~\cite{higgs_mass} focuses on the low and 
intermediate Higgs boson masses in the range from 120 to 180 GeV.
The analysis utilizes Higgs boson decays into $b$-quark and $W$-boson 
pairs and $Z$ boson decays into quark and charged lepton pairs. 
Considering hadronic decays of $W$ bosons, 
four distinct topologies are covered: 
2jet$+2\ell$ final states resulting from $\HZqqll$, 4jet from $\HZbbqq$,
4jet$+2\ell$ from $\HZWWll$ and 6jet from $\HZWWqq$. 
The study is performed for centre-of-mass energy of 350 GeV, 
assuming an integrated luminosity of 500 fb$^{-1}$.
As an example, \reffi{fig:higg_mass} shows the reconstructed 
Higgs boson mass spectra in the $\HZWWqq$ channel for $\mH$ = 150 GeV and
the $\HZWWll$ channel for $\mH$ = 180 GeV. 

From the fit of the resulting mass spectra, the Higgs boson mass can be 
extracted.  Results of the analysis are summarized in 
Table~\ref{tab:higgs_mass}.

\begin{table}[htb!]
\begin{center}
\begin{tabular}{lccc}
                          & \multicolumn{3}{c}{$\Delta(\mH)$ in MeV}    \\ 
\cline{2-4} 
   Decay mode \hfill      &         120  &             150  &             180  \\ 
\hline
   $\HZqqll$ &          $\phantom{0}$85  &             100  & $\phantom{0}$--  \\
   $\HZbbqq$ &          $\phantom{0}$45  &             170  & $\phantom{0}$--  \\
   $\HZWWll$ &          $\phantom{0}$--  & $\phantom{0}$90  & $\phantom{0}$80  \\
   $\HZWWqq$ &          $\phantom{0}$--  &             100  &             150  \\
\hline
   Combined  &          $\phantom{0}$40  & $\phantom{0}$65  & $\phantom{0}$70  \\
\end{tabular}
\caption{Uncertainties on the  determination  of $\mH$ for $\mH$ =
         120, 150 and 180 GeV. For $\mH$ = 150 GeV, 
         the combination is done using the $\HZWWll$ and $\HZWWqq$ channels.}
\label{tab:higgs_mass}
\end{center}
\end{table}

The study presented in Reference~\cite{higgs_lineshape} focuses
on a heavy Higgs boson with mass greater than 200 GeV. Exploiting 
4jet$+2\ell$ final states stemming from $\HZWWZZll$ 
decay modes, the Higgs boson mass is extracted from a lineshape
analysis of the invariant mass of the hadronic system stemming from Higgs 
decays to weak boson pairs. The relative accuracy on the Higgs boson mass 
varies from 0.11 to 0.36\% for $\mH$ between 200 and 320 GeV.
As discussed in Sect.~\ref{sec:heavySMHiggs},
in the case of a heavy Higgs boson  
not only its mass but also its width can be determined 
with good accuracy from the mass lineshape analysis.


\section{Measurements of the Higgs Boson Couplings 
to Gauge Bosons, Fermions and Itself}
\label{sec:HiggsCouplings}

\begin{figure}[htb!]
\begin{minipage}[c]{0.55\textwidth}
\epsfig{figure=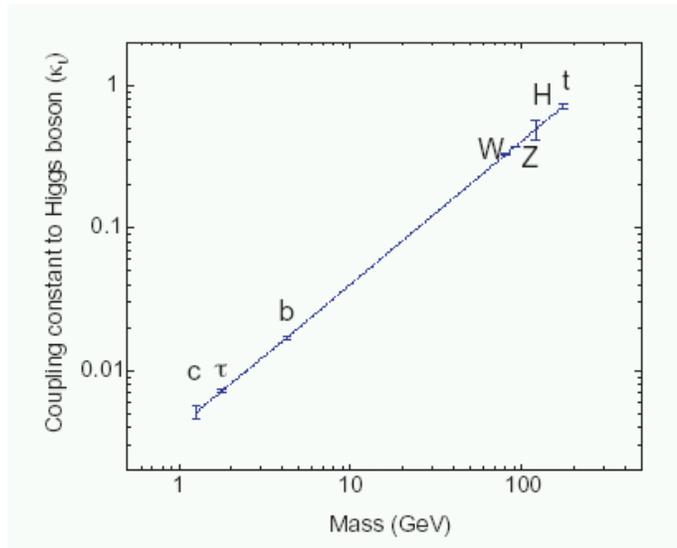,width=0.99\textwidth}
\end{minipage}
\begin{minipage}[c]{0.03\textwidth}
$\phantom{0}$
\end{minipage}
\begin{minipage}[c]{0.40\textwidth}
\caption{The couplings of the SM particles to the Higgs boson are
 shown as a function of their respective mass. Also shown are the 
prospective ILC precisions as discussed in the next sections.
 }
\label{fig:HiggsCouplings}
\end{minipage}
\end{figure}

In the SM, the electroweak symmetry is spontaneously broken by 
introducing the Higgs doublet field. 
Its neutral component receives the vacuum expectation value ($v$).  
The weak gauge bosons then obtain their masses 
through the Higgs mechanism.
At the same time, all quarks and charged leptons receive their 
masses from the Yukawa interactions with the Higgs field. 
Moreover, the Higgs boson ($h$) itself is 
also given its mass ($m_h$) by $v$
through the self-interaction of the Higgs boson. 
All these masses of the SM particles 
are expressed as multiplication of coupling constants with $v$. 
Therefore, there is a universality 
between masses and coupling constants:  
\begin{eqnarray}
\frac{2 m_W}{g} = \frac{\sqrt{2} m_t}{y_t} 
= \frac{\sqrt{2} m_b}{y_b} 
= \frac{\sqrt{2} m_{\tau}}{y_\tau} = ... = \frac{ m_h}{\sqrt{2 \lambda}} 
= v \simeq 246 \gev, 
\end{eqnarray}
where $g$ is the weak gauge coupling, $y_f$ is the Yukawa coupling 
constant to the fermion $f$, $\lambda$ is the self-coupling 
constant of $h$, and $m_i$ is the mass of the field $i$. 
Experimental validation of these relations would be a crucial 
step in establishment of 
the Higgs mechanism, the fermion mass generation mechanism,
and the structure of the Higgs potential. This is visualized in
\reffi{fig:HiggsCouplings}, where the coupling of SM particles to the
Higgs boson is shown as a function of their respective mass. Also
shown are the prospective ILC precisions as discussed in the next
sections.
Information on the triple Higgs coupling is crucial for
the reconstruction of the Higgs potential by which
the breaking of the electroweak symmetries is induced.
The accurate measurement of the self-couplings can also be
important to discriminate models.

\subsection{Couplings to heavy SM gauge bosons}

\begin{figure}[htb!]
\begin{minipage}[c]{0.47\textwidth}
\psfig{figure=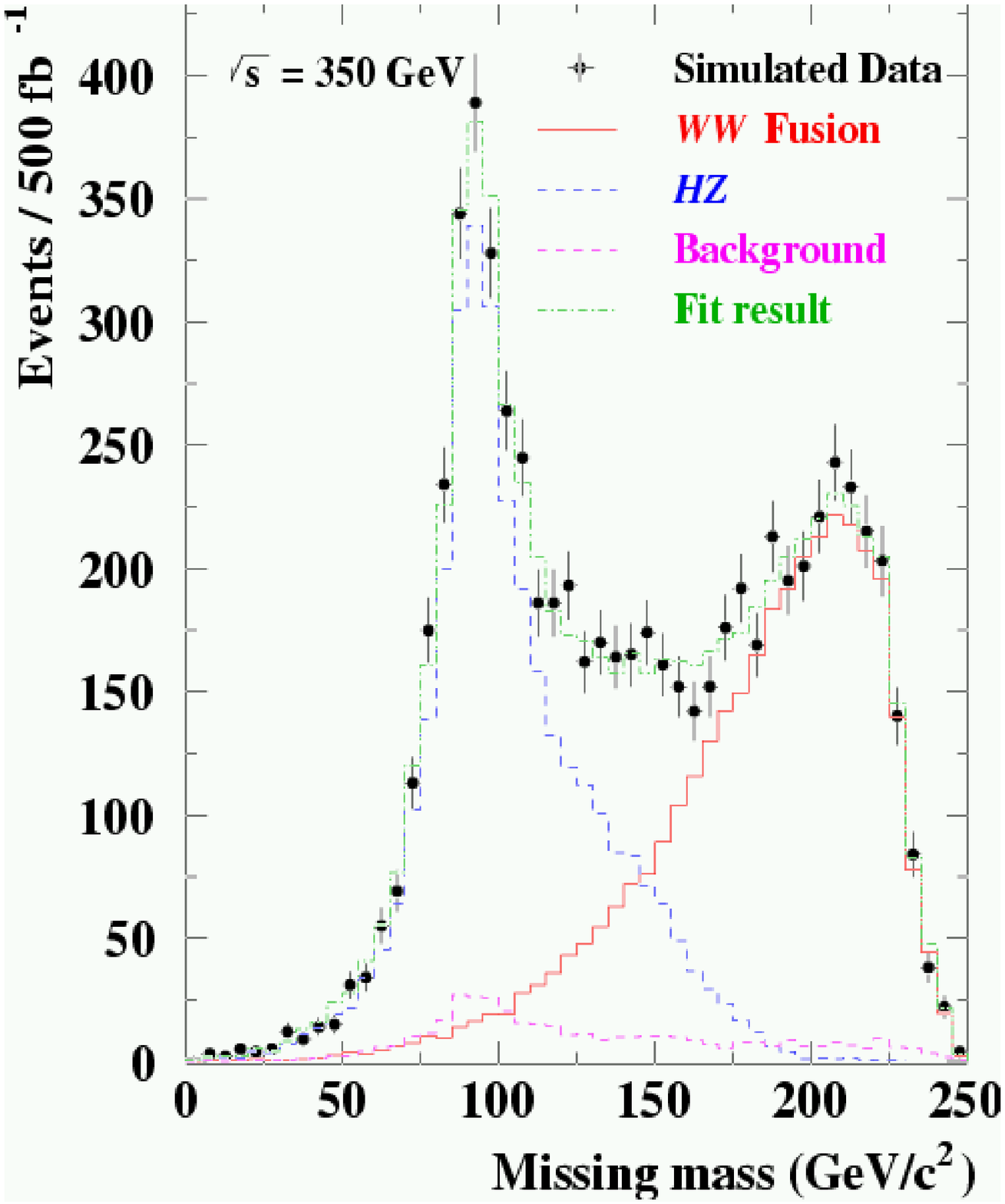,width=0.99\textwidth}
\end{minipage}
\begin{minipage}[c]{0.47\textwidth}
\psfig{figure=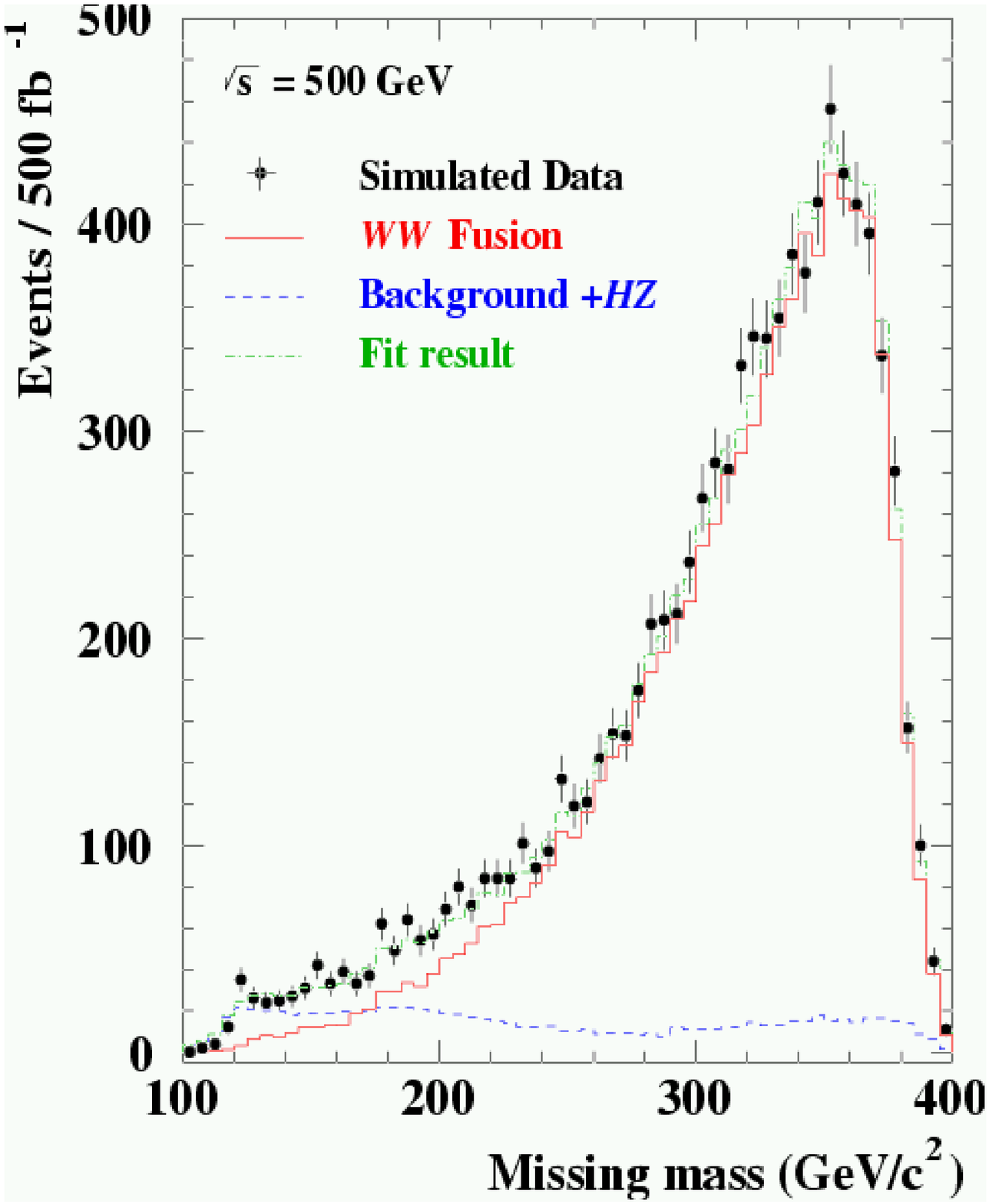,width=0.99\textwidth}
\end{minipage}
\begin{minipage}[c]{0.99\textwidth}
\caption{The missing mass spectrum in the $e^+e^-\ra
  H(\nu_e\bar{\nu}_e)\ra b\bar{b}\nu_e\bar{\nu}_e$ channel at
$\sqrt{s}$ = 350~GeV (left plot) and  
500~GeV (right plot). Assumed integrated luminosity is 500 fb$^{-1}$.
\label{fig:wwfusion}
}
\end{minipage}
\end{figure}

Couplings to $Z$ and $W$ bosons
can be directly accessed by measuring Higgs production rates in 
the Higgs-strahlung and $W$ boson fusion processes. The prospects 
for measuring the Higgs-strahlung cross section at ILC was discussed in 
Sect.~\ref{sec:hstrahl}. The feasibility of measuring $W$ fusion cross section 
at centre-of-mass energies 350 and 500 GeV is investigated in
\citere{wwfusion}. 
The analysis utilizes the $\Hbb$ decay mode and is based on the selection 
of final states characterized by two jets with the invariant mass compatible
with $\mH$ and missing energy carried by a pair of 
neutrinos\footnote{This analysis assumes that values for the 
Higgs boson mass and $\Hbb$ branching ratio will be already known
from other channels.}. \reffi{fig:wwfusion} presents
the spectrum of missing mass after applying a dedicated selection in the 
$e^+e^-\ra H\nu_e\bar{\nu}_e\ra b\bar{b}\nu_e\bar{\nu}_e$ channel. 

The selected samples contain a large contribution from the 
Higgs-strahlung process followed by Higgs decays into $b$ quarks and
$Z$ decays into neutrinos. However, the reconstructed missing mass 
clearly distinguishes between Higgs-strahlung and $WW$-fusion processes.
The missing mass spectrum is then fitted with two normalization factors, 
quantifying contributions from the Higgs-strahlung and fusion 
processes, as free parameters.
Results of the fits are given in Table~\ref{tab:wwfusion}.  

\begin{table}[htb!]
\begin{center}
\begin{tabular}{|c|c||c|}
\hline
$\sqrt{s}$ (GeV) & $\mH$ (GeV) & 
$\Delta\sigma_{WW-{\rm fusion}}$ \\
     & 120 & 3.3\% \\
350  & 140 & 4.7\% \\
\hline
     & 120 & 2.8\% \\
500  & 140 & 3.7\% \\
     & 160 & 13.0\% \\
\hline
\end{tabular}
\end{center}
\caption{
Results on the accuracy on the $WW$-fusion cross-section measurements in the 
$e^+e^-\ra H\nu_e\bar{\nu}_e\ra b\bar{b}\nu_e\bar{\nu}_e$ channel.
An integrated luminosity of 500 fb$^{-1}$ is assumed for each 
center-of-mass energy.
\label{tab:wwfusion}
}
\end{table}
%
Using a polarized electron beam will considerably enhance the
Higgs production rate in the $WW$-fusion process and allow for improvement 
of the precision on the cross-section by a factor of 2 or better for 
electron beam 
polarization of 80\%~\cite{wwfusion}. Polarization of the positron
beam could lead to a further improvement; see also \citere{power}.


\subsection{Higgs-photon coupling}

\begin{figure}[htb!]
\begin{minipage}[l]{0.47\textwidth}
\psfig{figure=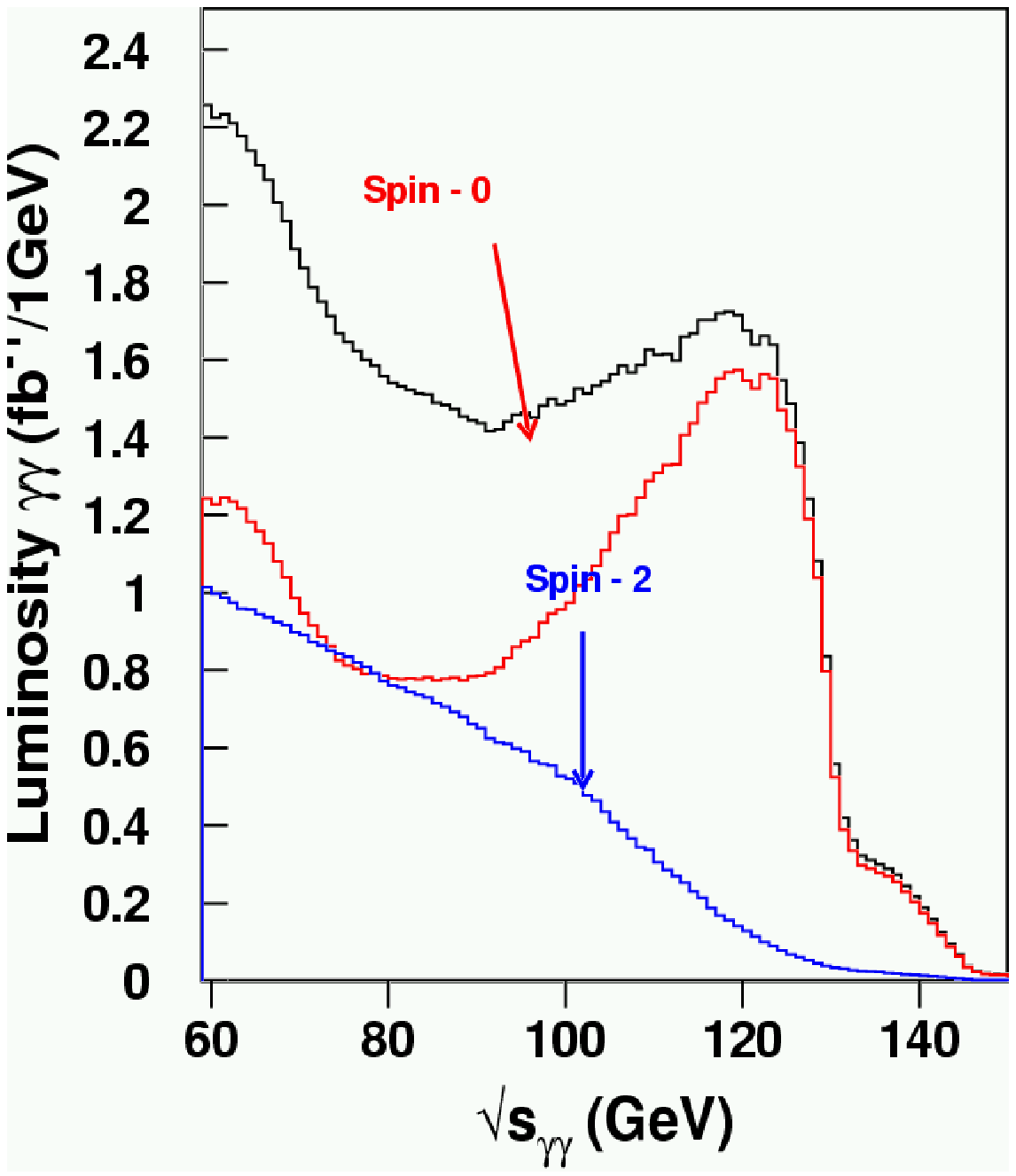,width=0.9042\textwidth}
\end{minipage}
\begin{minipage}[c]{0.03\textwidth}
$\phantom{0}$
\end{minipage}
\begin{minipage}[c]{0.47\textwidth}
\psfig{figure=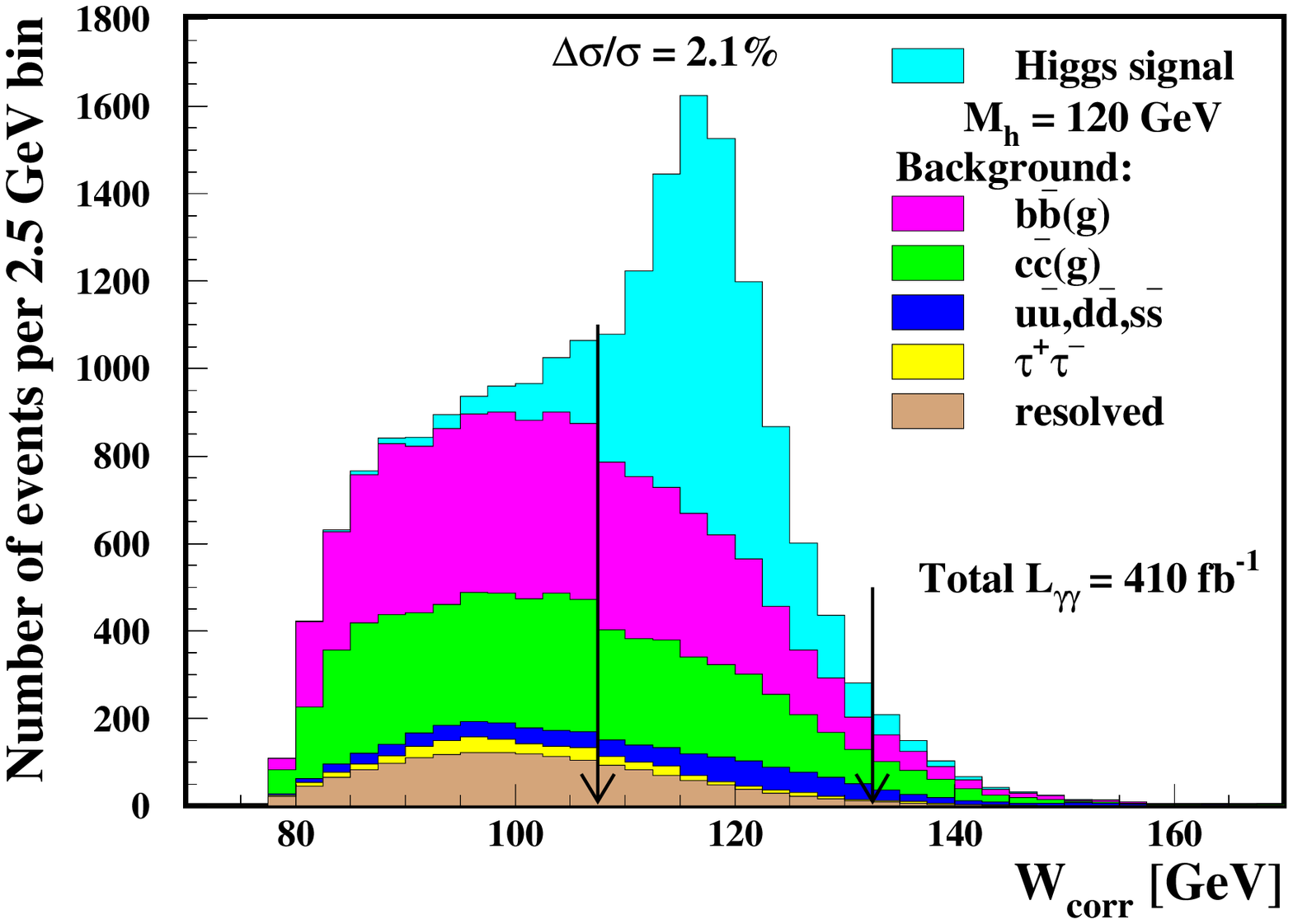,width=0.9042\textwidth}
\end{minipage}
\begin{minipage}[c]{0.47\textwidth}
\caption{
Luminosity spectra for the spin of the two colliding
photons $J_z$ = 0 and $J_z$ = 2. Distribution 
is obtained with the CIRCE2 program for a linear $e^-e^-$ collider 
operated at center-of-mass energy of 210 GeV.
\label{fig:gg_lumi} 
}
\end{minipage}
\begin{minipage}[c]{0.03\textwidth}
$\phantom{0}$
\end{minipage}
\begin{minipage}[c]{0.47\textwidth}
\caption{
Reconstructed Higgs boson mass in the process 
$\ga\ga\ra H\ra b\bar{b}$ assuming 
integrated photon-photon luminosity of 410 fb$^{-1}$ collected
at a linear $e^-e^-$ collider operated at center-of-mass 
energy 210 GeV.
\label{fig:hbb_gg}
}
\end{minipage}
\end{figure}

The coupling of the Higgs boson to photons can be determined through
the measurement of the resonant Higgs production rate in $\GG$ collisions. 
Colliding photons of high energy are produced through the Compton 
back-scattering
of laser light on high energy electron beams. The high energy photon beams
resulting from this process are not monochromatic. A number of 
simulation tools have been developed, allowing for estimation 
of the two-photon luminosity spectrum~\cite{CompAZ,CIRCE2}. 
As an example, \reffi{fig:gg_lumi} shows the 
resulting two-photon luminosity spectrum for a linear 
$e^-e^-$ collider operated at 
$\sqrt{s}$ = 210 GeV as simulated using the CIRCE2 program~\cite{CIRCE2}. 
A Higgs boson produced in $\GG$ collisions can be identified via its 
decays into $b$-quarks~\cite{higgs_gg_rm,higgs_gg_p1,higgs_gg_p2}. 
\reffi{fig:hbb_gg} presents the spectrum of the reconstructed
Higgs boson mass in the channel $\ga\ga\ra H\ra b\bar{b}$
for $\mH$ = 120 GeV and assuming integrated photon-photon 
luminosity of 410 fb$^{-1}$ collected at a linear $e^-e^-$ 
collider operated at center-of-mass 
energy 210 GeV~\cite{higgs_gg_p2}. Dedicated 
analyses~\cite{higgs_gg_rm,higgs_gg_p1,higgs_gg_p2} showed that 
the quantity $\Gamma(H\ra\ga\ga){\rm BR}(H\ra b\bar{b})$
can be measured with an accuracy of 2.1$-$7.8\% for 
$\mH$ = 120$-$160 GeV, respectively. 
Combining this with the measurements of $\Hbb$ in $\EE$ collisions,
the partial decay width $\Gamma(H\ra\ga\ga)$,
and therefore the photonic coupling, of the Higgs boson
can be extracted.

One potential caveat is that the reach analyses performed to date have assumed
that the QCD background is well-described by current theoretical estimates.  A
study during this Workshop (see \cite{Sullivan:2005pb} for details) of the QCD
structure in the resolved portion of the background in \reffi{fig:hbb_gg} has
demonstrated a factor of 5 uncertainty in that term, which raises the overall
background uncertainty by a factor of 2.  The origin of this large uncertainty
is that the gluonic component of the photon is nearly unconstrained at these
large energies.  The gluon contribution will be well-measured by an \textit{in
situ} determination once the collider is running.  Hence, at worst, a slightly
larger luminosity may be required to obtain the high precision measurements of
the photon coupling to the Higgs.


\subsection{Higgs couplings to SM fermions}

Information on Higgs couplings to SM fermions can be extracted through the 
measurements of the Higgs boson decay branching fractions. Two approaches
for the measurements of Higgs branching fractions can be used:
\begin{itemize}
\item
Extract topological cross sections, 
$\sigma(e^+e^-\ra H\nu_e\bar{\nu}_e)\br(H\ra X)$ or
$\sigma(e^+e^-\ra HZ) \br(H\ra X)$
from the event rate measurement and then divide by the total 
Higgs-strahlung or 
$WW$-fusion cross-section as obtained from corresponding studies.

\item
Select unbiased inclusive samples of Higgs decays, 
e.g., events in the recoil peak in the $\HZXll$ channel, and then 
determine from this sample the fractions of events corresponding to
a given decay mode $H\ra X$. 
Since in this direct method binomial
(or generally multi-nomial) statistics is employed, the errors
on the branching ratios are reduced by a factor of 
$\sqrt{1-\br(H\ra X)}$.

\end{itemize}

Examples of the first approach are given by a number 
of analyses examining the potential of the
linear collider for the measurement of 
the Higgs boson branching fractions into vector bosons 
and hadrons~\cite{battaglia_branchings,ee_hww,ee_hgg,ee_hgz}.
These analyses exploit either $WW$-fusion or the Higgs-strahlung
process with various possible $Z$ boson decays. 
The approach based on binomial statistics is described 
in Reference~\cite{brient_branchings}. Although relying only 
on Higgs-strahlung events with $\Zll$, this method yields 
errors very similar to the analyses based on the first approach,
exploiting a variety of $Z$ boson decays following $\EEHZ$.
Results on Higgs branching ratio measurements at $\sqrt{s}$ = 800 GeV 
and below are given in Table~\ref{tab:h_branchings}.

\begin{table}[htb!]
\begin{center}
\begin{tabular}{|c|c|c|c|c|c|c|}
\hline
$\mH$ (GeV) & 120 & 140 & 160 & 180 & 200 & 220 \\
\hline
Decay & \multicolumn{6}{|c|}{Relative precision (\%)} \\ 
\hline
$\bb$     & 2.4(a)/1.9(e) & 2.6(a) & 6.5(a) & 12.0(d) & 17.0(d) & 28.0(d) \\
$\cc$     & 8.3(a)/8.1(e) & 19.0(a)&        &         &         &         \\
$\tautau$ & 5.0(a)/7.1(e) & 8.0(a) &        &         &         &         \\
$gg$      & 5.5(a)/4.8(e) & 14.0(a)&        &         &         &         \\ 
$WW$      & 5.1(a)/3.6(e) & 2.5(a) & 2.1(a) &         &         &         \\
$ZZ$      &               &        & 16.9(a)&         &         &         \\
$\GG$     &23.0(b)/35.0(e)&        &        &         &         &         \\
$\Zg$     &               & 27.0(c)&        &         &         &         \\
\hline
\end{tabular}
\end{center}
\caption{
Precision on the Higgs boson branching ratio measurements
\cite{battaglia_branchings,ee_hww,ee_hgg,ee_hgz,brient_branchings}: 
(a) for 500 fb$^{-1}$ at 350 GeV; 
(b) for 500 fb$^{-1}$ at 500 GeV;
(c) for 1 ab$^{-1}$ at 500 GeV;
(d) for 1 ab$^{-1}$ at 800 GeV;
(e) as for (a) but using binomial statistics method.
\label{tab:h_branchings}
}
\end{table}

Recently, a dedicated analysis based on detailed 
flavor-tagging tools has been performed~\cite{ee_hqq} 
to evaluate the linear collider potential for the measurement
of hadronic decays of the light Higgs boson, $\mH$ = 120 GeV. 
The study is based on selection of an inclusive 
sample of hadronic Higgs decays using the Higgs-strahlung
process and exploiting all possible decay modes of the $Z$ boson. 
Each selected event is assigned a quantified
probability to contain $b$- or $c$-jets, 
referred to as $b$- and $c$-tag variables. The branching 
ratios $\Hbb$, $\Hcc$ and $\Hgg$ are determined
from the fit of a two-dimensional distribution of $b$-tag versus 
$c$- tag variables (\reffi{fig:bc_fit})
with three free normalization parameters quantifying the fractions of 
$\Hbb$, $\Hcc$ and $\Hgg$ events in the final 
selected samples. Results of this study are presented in 
Table~\ref{tab:hqq_branchings_org}. 

\begin{table}[htb!]
\begin{minipage}[c]{0.55\textwidth}
\begin{tabular}{|c|c|c|c|c|}
\hline
Decay  & $\Zll$ & $\Zvv$ & $\Zqq$ & Combined \\
\hline
       & \multicolumn{4}{|c|}{Relative precision (\%)} \\
\hline
$\Hbb$ &  3.0 &   2.1 &  1.5 &  1.1 \\
$\Hcc$ & 33.0 &  20.5 & 17.5 & 12.1 \\
$\Hgg$ & 18.5 &  12.3 & 14.4 &  8.3 \\
\hline
\end{tabular}
\end{minipage}
\begin{minipage}[c]{0.05\textwidth}
$\phantom{0}$
\end{minipage}
\begin{minipage}[c]{0.35\textwidth}
\caption{
Relative precision on the measurement of the 
hadronic branching ratios of Higgs for $\mH$ = 120 GeV.
Analysis is performed for $\sqrt{s}$ = 350 GeV, 
assuming an integrated luminosity of 500 fb$^{-1}$.
\label{tab:hqq_branchings_org}
}
\end{minipage}
\end{table}

\begin{figure}[htb!]
\begin{minipage}[c]{0.55\textwidth}
\psfig{figure=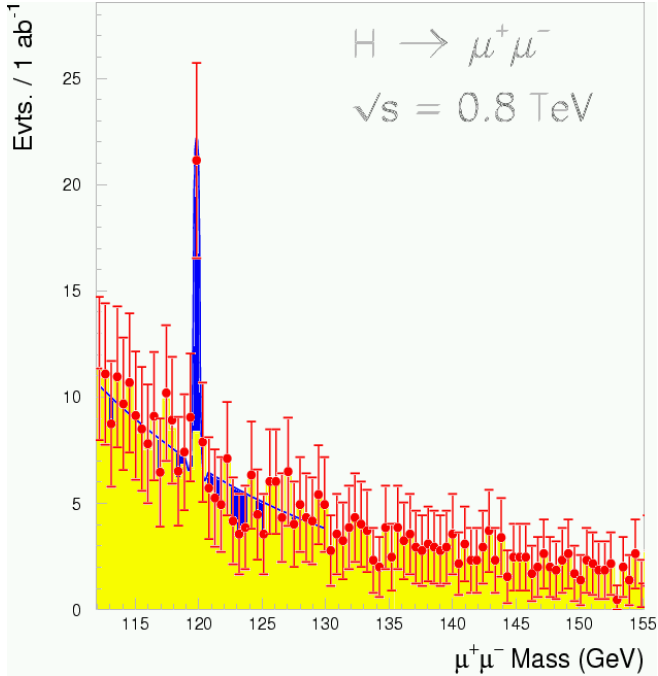,width=0.95\textwidth}
\end{minipage}
\begin{minipage}[c]{0.03\textwidth}
$\phantom{0}$
\end{minipage}
\begin{minipage}[c]{0.40\textwidth}
\caption{
Invariant mass of the two muons in the selected 
sample of $\Hmm$ decays following $WW$-fusion process. 
Higgs boson mass is 120 GeV. Center-of-mass energy 
is 800 GeV. An integrated luminosity of 1 ab$^{-1}$ is 
assumed.}
\label{fig:hmumu_800}
\end{minipage}
\end{figure}

One can improve the precision of branching ratio measurements
at higher $\EE$ collision energies. This is done taking advantage of 
the increase in the Higgs production rate with energy due to a 
$\log{\sqrt{s}}$ dependence of $WW$-fusion cross-section. Moreover, 
the specific luminosity is expected to increase with increasing 
center-of-mass energy.
Both these factors will enhance the signal statistics,
allowing not only to improve the precision on Higgs
branching ratios for the major decay modes, but also to measure   
rare Higgs decays, e.g., $\Hmm$~\cite{battaglia_hmm}. 
An example of this signal is shown in \reffi{fig:hmumu_800}.

\subsection{Top Yukawa Couplings of the Higgs Boson}

The top quark is exceptionally heavy as compared 
to the other quarks, with its mass being 
$m_t \sim v/\sqrt{2}$. 
The size of the top Yukawa coupling is nearly 1. 
The fact that the mass of the top quark is the same size 
as the scale of electroweak symmtry breaking 
may suggest that the physics of the top quark 
may be strongly related to the dynamics of the 
electroweak symmetry breaking. 
Therefore, the experimental determination of the top Yukawa coupling  
is expected to be a very sensitive 
probe of possible new physics that might be responsible
for generating the top quark mass.
For example, alternative models of electroweak 
symmetry breaking with new strong interactions, such as Technicolor 
and Topcolor models, substantially modify the top quark interaction 
with the Higgs sector and give rise to new signals that could be 
studied at the ILC.

If the Higgs boson mass is below 200 GeV, 
the top Yukawa coupling could be measured in associated 
$t \bar t H$ production, with an error of 
the order of 15\% at the LHC\cite{HcoupLHCSM} (with ``mild''
theory assumptions).
At the ILC\cite{Baer:1999ge} 
in the SM the production cross section becomes maximal 
around $\sqrt{s} = 700$ GeV for $m_h=120$ GeV.
The expected accuracy for the top Yukawa coupling in the SM is 4.2\%
at $\sqrt{s}=700$ GeV with $\int{\cal L} =500$ fb${}^{-1}$.
An error of 5.5\% can be achieved for $\sqrt{s} = 800 \gev$, 
$\int{\cal L} = 1000$~fb$^{-1}$ and $\mh = 120 \gev$~\cite{tdr}. 
However, recent studies show that for this Higgs mass and integrated
luminosity even at $\sqrt{s} = 500 \gev$ a measurement at the 10\%
level might be possible~\cite{tth500ILC}. 
For a combination of LHC and the 500~GeV~ILC, see
\refse{subsec:tthLHCILC}.


\subsection{Higgs Boson Self-Couplings}

Information about the structure of the Higgs potential 
will be obtained by measuring the triple Higgs boson coupling. 
Accurate information on this coupling is important to discriminate
models beyond the SM.
Even when the other Higgs boson couplings such as $hZZ$, $hWW$ 
and $hf\bar f$   
are in good agreement with the SM predictions, the Higgs
self-couplings can significantly deviate from the SM prediction due to
non-decoupling quantum effects of heavy particles. 
In the two Higgs doublet model, 
radiative corrections of O(100)\% on the $hhh$ coupling are  
possible if the additional heavy Higgs bosons 
show a non-decoupling property\cite{hhh-THDM}. 
Such a non-decoupling phenomenon is known to be required 
for a scenario of electroweak baryogenesis\cite{EWBG}. 
Therefore, the measurement of the triple Higgs boson coupling 
may be used to test such a model of the baryon asymmetry of the universe
\cite{hhh-ewbg}. 

\begin{figure}[htb!]
\begin{minipage}[c]{0.45\textwidth}
\psfig{figure=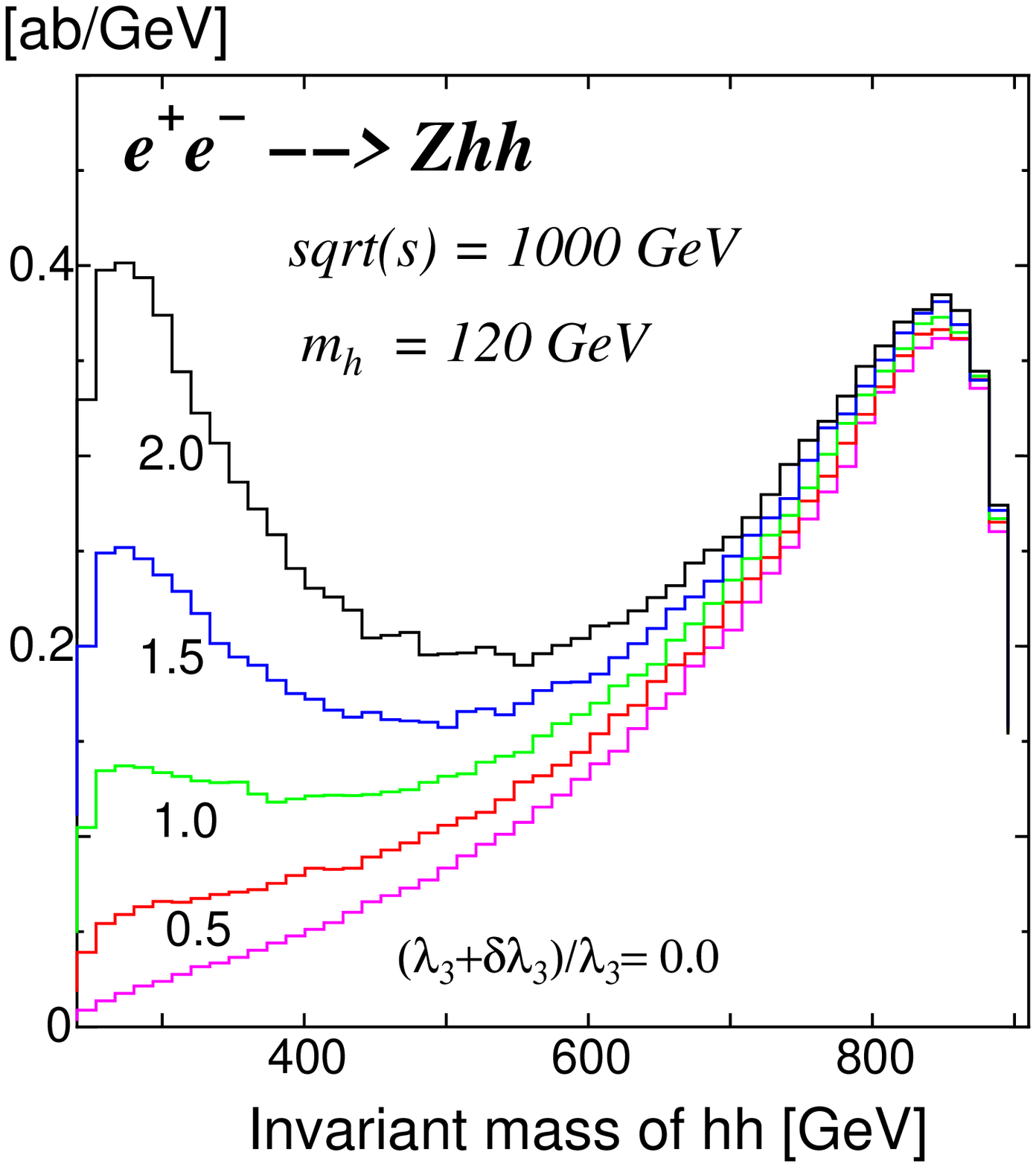,width=0.99\textwidth}
\end{minipage}
\begin{minipage}[c]{0.47\textwidth}
\psfig{figure=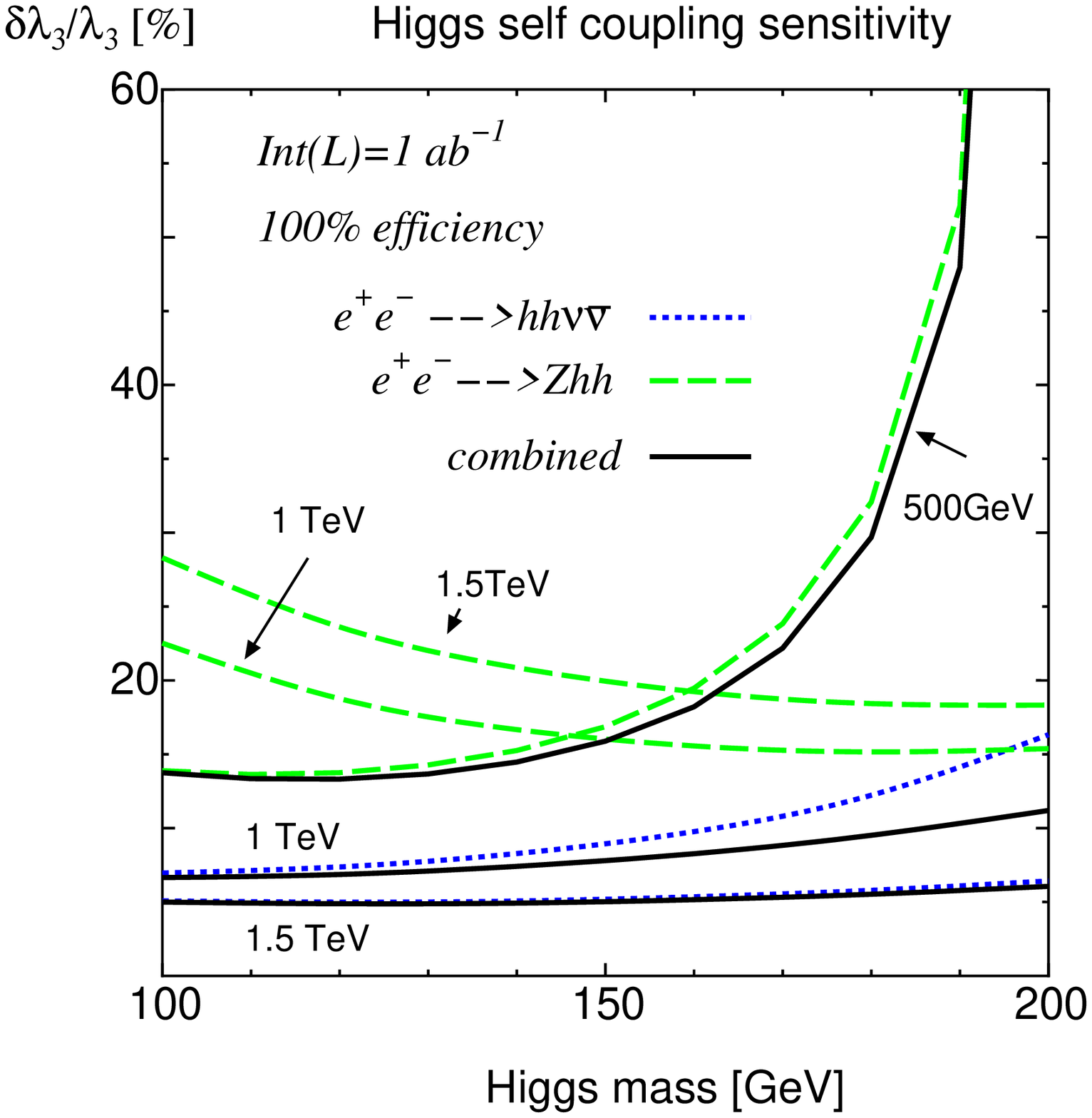,width=0.99\textwidth}
\end{minipage}
\caption{
(Left figure) The $hh$ invariant mass dependence 
of the $hhZ$ mode for several values of $\delta\lambda_3$.
(Right figure) The $\lambda_3$ measurement sensitivity; 
$hhZ$ (dashed line),
$hh\nu\bar{\nu}$ (dotted line) and  
combined results (solid line)
~\cite{hhh1}. \label{fig:self-coupling}
 }
\end{figure}

The $hhh$ coupling can be measured via the double Higgs boson 
production processes $e^+e^- \rightarrow Zhh$ and also 
$e^+e^- \rightarrow \nu\overline{\nu}hh$\cite{hhh}. 
In these processes, in addition to the diagram 
involving the $hhh$ coupling,
there are double Higgs radiation diagrams which depend only on 
the $hZZ$ or the $hWW$ coupling. 
Hence, the production cross section is not simply 
proportional to the square of the self-coupling constant.
Systematic studies on the self-coupling measurement 
have been performed for various values of $m_h$ and 
$\sqrt{s}$\cite{hhh1,hhh2,castanier01}.
As seen in Figure~\ref{fig:self-coupling} (left),   
the $hhZ$ mode is sensitive to the deviation $\delta\lambda_3$ 
of the $hhh$ coupling from the SM value 
in the low values of the invariant mass $M_{hh}$ of the $hh$ system
\cite{hhh2}. 
A cut on $M_{hh}$ is therefore useful to improve the sensitivity. 
For $\sqrt{s}\gsim$ 1 TeV, the sensitivity is further improved 
by using initial electron (and positron) polarization.  
The production cross section of the $WW$ fusion process 
can be increased by up to a factor of 2 (4) in principle with
electron (electron and positron) polarization.  

For $\sqrt{s}= 500 \gev$, $e^+e^- \rightarrow Zhh$ has much
larger cross section than 
the $e^+e^- \rightarrow \nu \bar{\nu}hh$ process.
For a higher energy, the production cross section of the
latter process is enhanced because of its t-channel nature,
whereas $e^+e^- \to Z h h$ decreases as $1/s$. 
For $\sqrt{s}\gsim$ 1 TeV, the $WW$ fusion process 
$e^+e^- \rightarrow \nu \bar{\nu}hh$ becomes important 
to determine the $hhh$ coupling.
The statistical errors in the $hhh$ measurement are 
shown in Figure~\ref{fig:self-coupling} (right)
as a function of the Higgs boson mass from $100$ to $200 \gev$, 
assuming an integrated luminosity of 1 ab$^{-1}$\cite{hhh2}.  
Dashed (dotted, sold) lines are the results for the $hhZ$ mode
($hh\nu\bar{\nu}$ mode, both modes combined). 
At $\sqrt{s}=500 \gev$, where the $hhZ$ mode is dominant,  
the $hhh$ coupling may be measured with about 20\% accuracy   
for relatively light Higgs bosons with $m_h~\lsim$ 150 GeV. 
For $\sqrt{s} \gsim 1$ TeV, where the $hh\nu\bar{\nu}$ mode is 
dominant, higher sensitivity ($\delta\lambda_3/\lambda_3~\lsim~10\%$) 
can be expected by using the invariant mass cut $M_{hh}<$ 600 GeV 
for the $hhZ$ mode and a 100\% polarized electron beam 
for the $hh\nu\bar{\nu}$ mode. 
%

The full calculation of the SM electroweak radiative 
corrections to the double Higgs-strahlung process $e^+e^- \to Zhh$  
has been presented in \citere{rad-eezhh1}.


\section{Higgs Boson Width Measurement}

At Higgs masses below approximately 2$\mW$, the detector 
resolution on the Higgs boson mass for the main decay channels,
$\Hbb$ and $\HWW$, is significantly larger than the 
natural width of the Higgs boson.
$\Gamma_H$. As a consequence, no direct measurement 
of the Higgs width is possible through the analysis of the reconstructed 
Higgs boson mass lineshape. However, using the relation
\begin{equation}
\Gamma_H = \frac{\Gamma(H\to X)}{\br(H\to X)}
\end{equation}
with $X = WW$ or $\ga\ga$, 
the width can be determined indirectly. 
For example, the partial width $\Gamma(H\to \ga\ga)$
can be measured with the $\ga\ga$ collider whereas
the branching fraction $\br(H\to \ga\ga)$ is 
accessible in $\EE$ collisions from the diphoton invariant mass
spectrum. The same procedure can be performed for the
$WW$-fusion process and the $H\to WW$ decay mode.
An accuracy in the determination 
of $\Gamma_H$ between 4\% and 15\% can be 
achieved~\cite{wwfusion} for $m_H$ up to 160 GeV.

\begin{figure}[htb!]
\begin{minipage}[c]{0.55\textwidth}
\psfig{figure=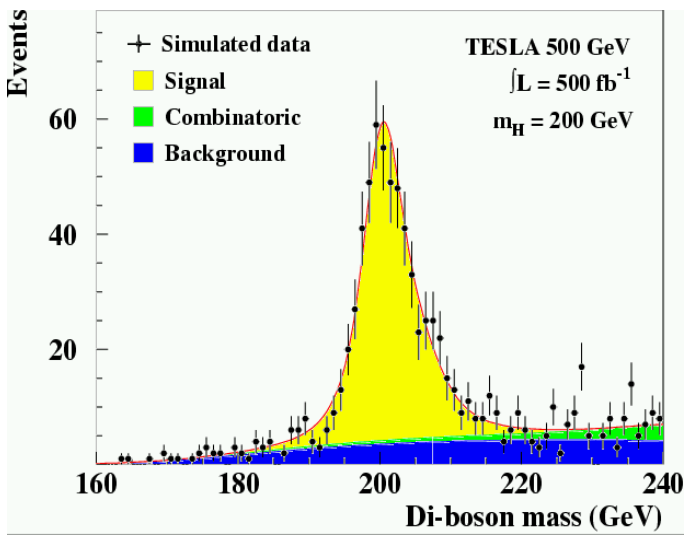,width=0.99\textwidth}
\end{minipage}
\begin{minipage}[c]{0.03\textwidth}
$\phantom{0}$
\end{minipage}
\begin{minipage}[c]{0.40\textwidth}
\caption{
Reconstructed Higgs boson mass lineshape in the 
$HZ\to WW\ell^+\ell^-$ and 
$HZ\to ZZ\ell^+\ell^-$ channels with subsequent 
hadronic decays of weak bosons for $\mH$ = 200 GeV. 
Selected sample corresponds to 500 fb$^{-1}$ of luminosity
collected at center-of-mass energy of 500 GeV. 
\label{fig:higgs_lineshape}
}
\end{minipage}
\end{figure}

The natural width of heavy Higgs boson with mass above $2\mZ$
becomes as large as few GeV and can be directly measured 
from the mass lineshape analysis~\cite{higgs_lineshape}. 
\reffi{fig:higgs_lineshape} presents the spectra
of reconstructed Higgs mass in the 
$HZ\to WW\ell^+\ell^-$ and 
$HZ\to ZZ\ell^+\ell^-$ channels with subsequent 
hadronic decays of weak bosons for $\mH$ = 200 GeV.
From the fit of the spectra with convolution of the detector
resolution functions and Breit-Wigner function, the Higgs width
can be determined with the relative precision of about 30\% 
for $\mH$ = 200$-$320 GeV.

\section{Spin and CP Quantum Numbers of Higgs Boson}
\label{sec:cp_higgs}
Full establishment of the Higgs mechanism implies 
measurements of Higgs spin and CP quantum numbers.
The SM Higgs boson (like the CP-even MSSM Higgs bosons) has spin and
CP-quantum numbers  
$J^{PC}$ = $0^{++}$. Information on 
CP-quantum numbers can be extracted by studying angular
dependences of the Higgs-strahlung process. The analysis for the SM
Higgs boson can also be applied to a SM-like MSSM Higgs boson.

\begin{figure}[htb!]
\begin{minipage}[c]{0.44\textwidth}
\psfig{figure=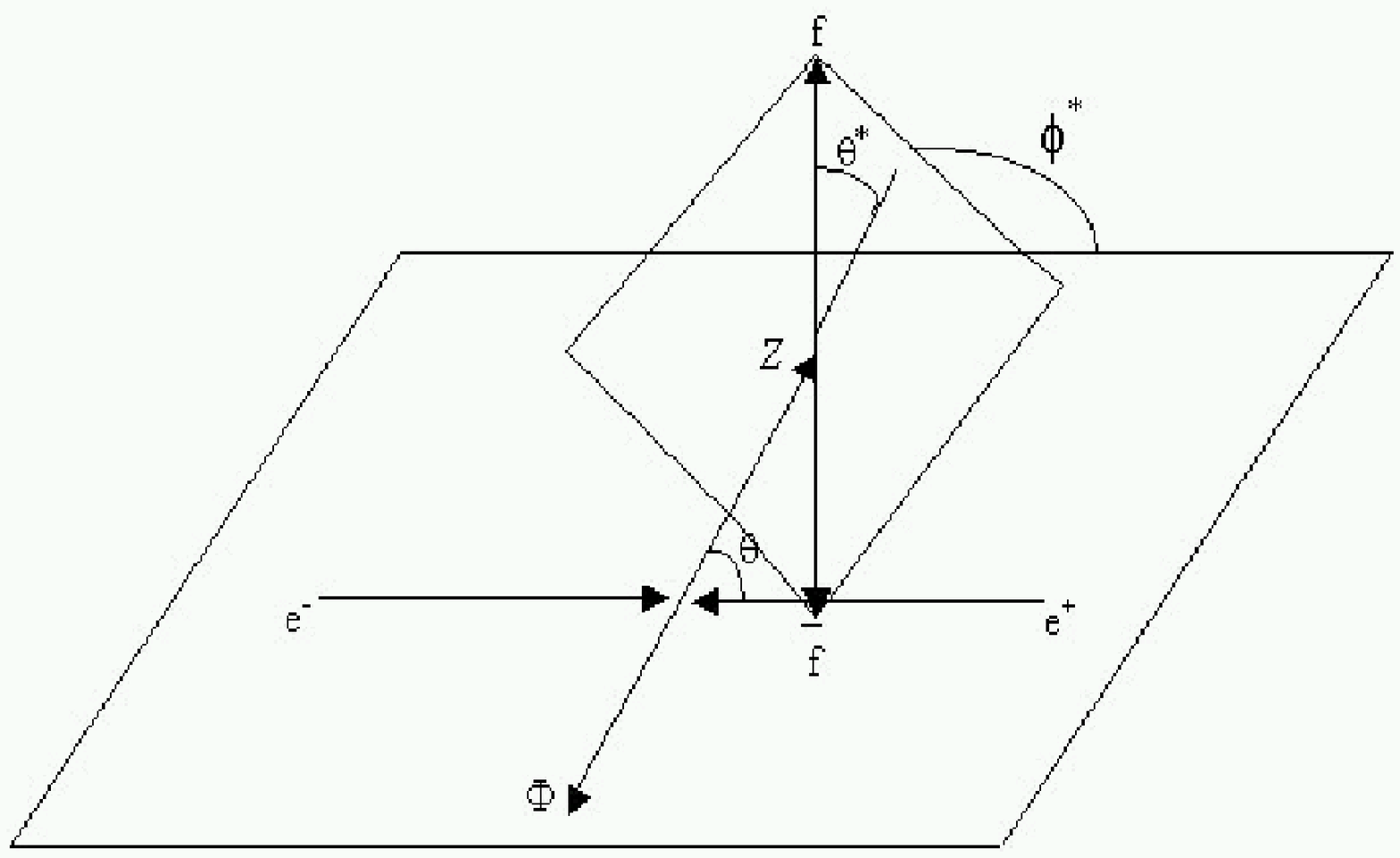,width=0.80\textwidth}
\end{minipage}
\begin{minipage}[c]{0.44\textwidth}
\psfig{figure=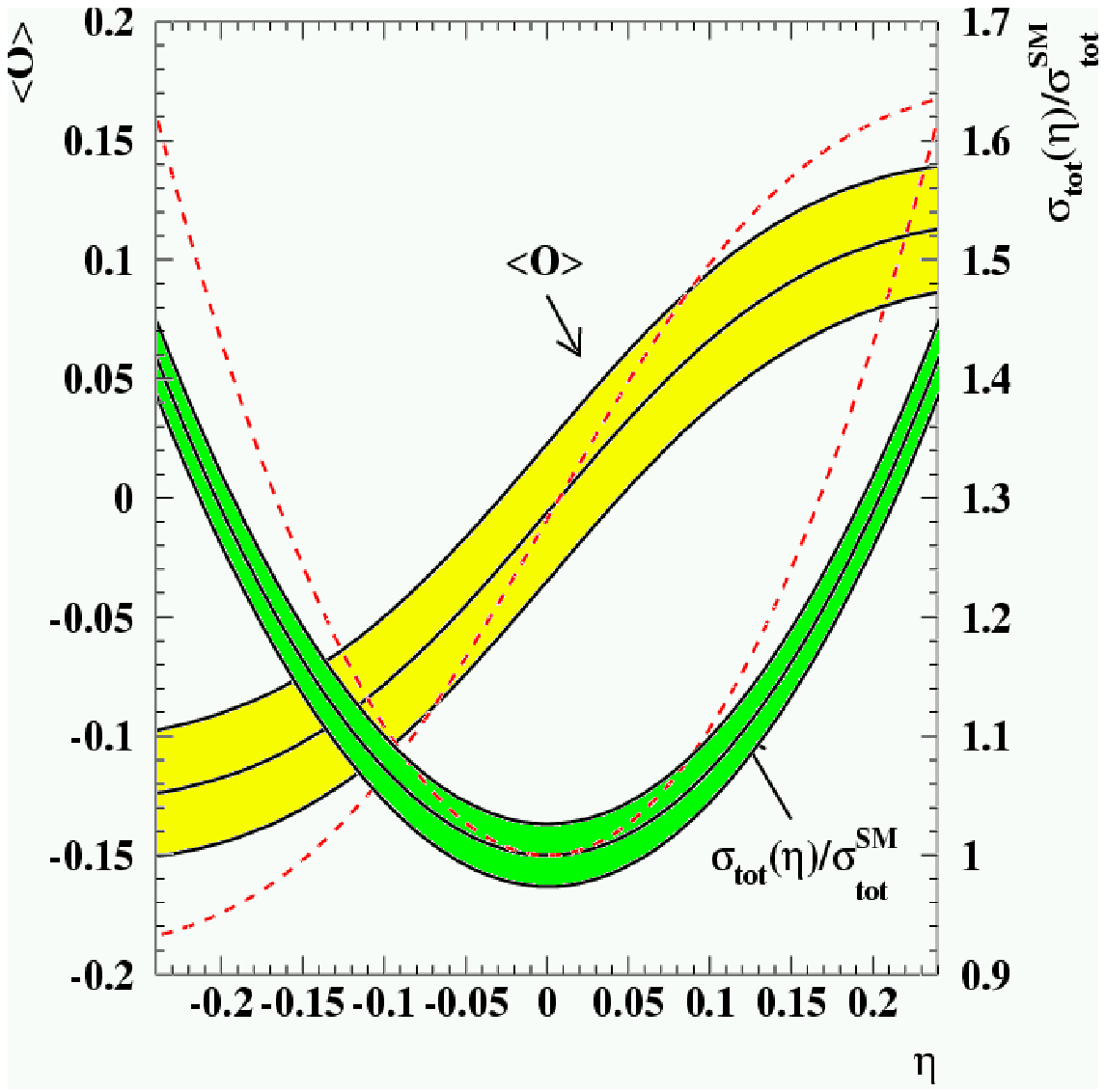,width=0.99\textwidth}
\end{minipage}
\begin{minipage}[c]{0.44\textwidth}
\caption{Definitions of the $Z$ production and decay angles 
$\theta$, $\theta^*$ and $\phi^*$ in the Higgs-strahlung
process.
\label{fig:angle_def}
}
\end{minipage}
\begin{minipage}[c]{0.06\textwidth}
$\phantom{0}$
\end{minipage}
\begin{minipage}[c]{0.44\textwidth}
\caption{
Gauge curve reflecting dependence of the mean value
of the optimal observable and the Higgs-strahlung cross-section
on the parameter $\eta$. Bands indicate
1$\sigma$ confidence level intervals as obtained 
for a data sample corresponding to 500 fb$^{-1}$ at $\sqrt{s}$ = 350 GeV.
\label{fig:cp_gauge} 
} 
\end{minipage}
\vspace{-1em}
\end{figure}

The following
quantities are sensitive to the CP quantum numbers of the Higgs:
the polar production angle of the Higgs boson, $\theta$;
the angle between the $Z$ momentum vector and the momentum vector 
of the fermion stemming from $Z$ decay, $\theta^*$; 
and the angle between the production and decay planes of the $Z$ boson,
$\phi^*$. The latter two angles are defined in the $Z$ boson rest frame.
The definitions of these angles are illustrated in 
\reffi{fig:angle_def}.

Studies presented in \citeres{cp_schumacher,cp_dova}
investigated the general case of a mixed scalar state $\Phi$ containing
both CP-even and CP-odd components. 
In the presence of a CP-odd admixture in the scalar field, the  
amplitude for the Higgs-strahlung process reads:
\BEQ
{\cal{M}}_{Z\Phi}={\cal{M}}_{ZH}+i\eta{\cal{M}}_{ZA},
\EEQ
where ${\cal{M}}_{ZH}$ is the CP-even SM-like s-wave 
amplitude, and ${\cal{M}}_{ZA}$ is a new CP-odd 
p-wave amplitude contributing with strength $\eta$ to the
Higgs-strahlung process. The presence of the latter modifies
the three-fold differential cross section 
$d^3\sigma/d(\cos\theta)d(\cos\theta^*)d(\cos\phi^*)$
with respect to the SM prediction. The parameter $\eta$ can be determined 
using the method of optimal observables~\cite{cp_schumacher}. 
Neglecting terms quadratic in $\eta$, the optimal 
observable is chosen to be 
\BEQ
{\cal{O}}=
\frac{2\re({\cal{M}}^*_{ZA}{\cal{M}}_{ZH})}{|{\cal{M}}_{ZH}|^2}.
\EEQ
The choice of the observable is very close to optimal for 
small $\eta$. At the next step, a ``gauge'' curve 
is produced, which quantifies the dependence of the mean
value of the optimal observable and its error on the parameter
$\eta$. This is done using reference distributions of 
the optimal observable corresponding to different values of $\eta$.  
Reference distributions are obtained for the selected samples in 
the $\HZXll$ channel. The parameter 
$\eta$ can then be read from this ``gauge'' curve including 
the confidence band.  The analyzing power
of this method can be enhanced by taking into account  
the dependence of the Higgs-strahlung cross section on $\eta$.
Technically, the parameter $\eta$ is calculated by minimizing the $\chi^2$,
\BEQ
\chi^2 = \frac{\big(\langle{\cal{O}}_{meas}\rangle-E(\langle{\cal{O}}\rangle)(\eta)\big)^2}
{\sigma_{exp}(\eta)}+
\frac{\big(N_{meas}-N_{exp}(\eta)\big)^2}{N_{exp}(\eta)},
\EEQ
where $\langle{\cal{O}}_{meas}\rangle$ and 
$N_{meas}$ are the measured mean value of the 
optimal observable and the number of observed signal events 
in the hypothetical data sample, $E(\langle{\cal{O}}\rangle)(\eta)$
and $N_{exp}(\eta)$ are the expected mean value 
of the optimal observable and expected number of signal events,
and $\sigma_{exp}(\eta)$ 
is the expected error on $E(\langle{\cal{O}}\rangle)(\eta)$.
$E(\langle{\cal{O}}\rangle)(\eta)$, $N_{exp}(\eta)$ and
$\sigma_{exp}(\eta)$ are determined from the ``gauge'' curves
shown in \reffi{fig:cp_gauge}.
For small values of $\eta$, $\eta$ $\leq$ 0.1, 
this parameter can be measured with a precision of about 0.03.
\begin{figure}[htb!]
\begin{minipage}[c]{0.49\textwidth}
\psfig{figure=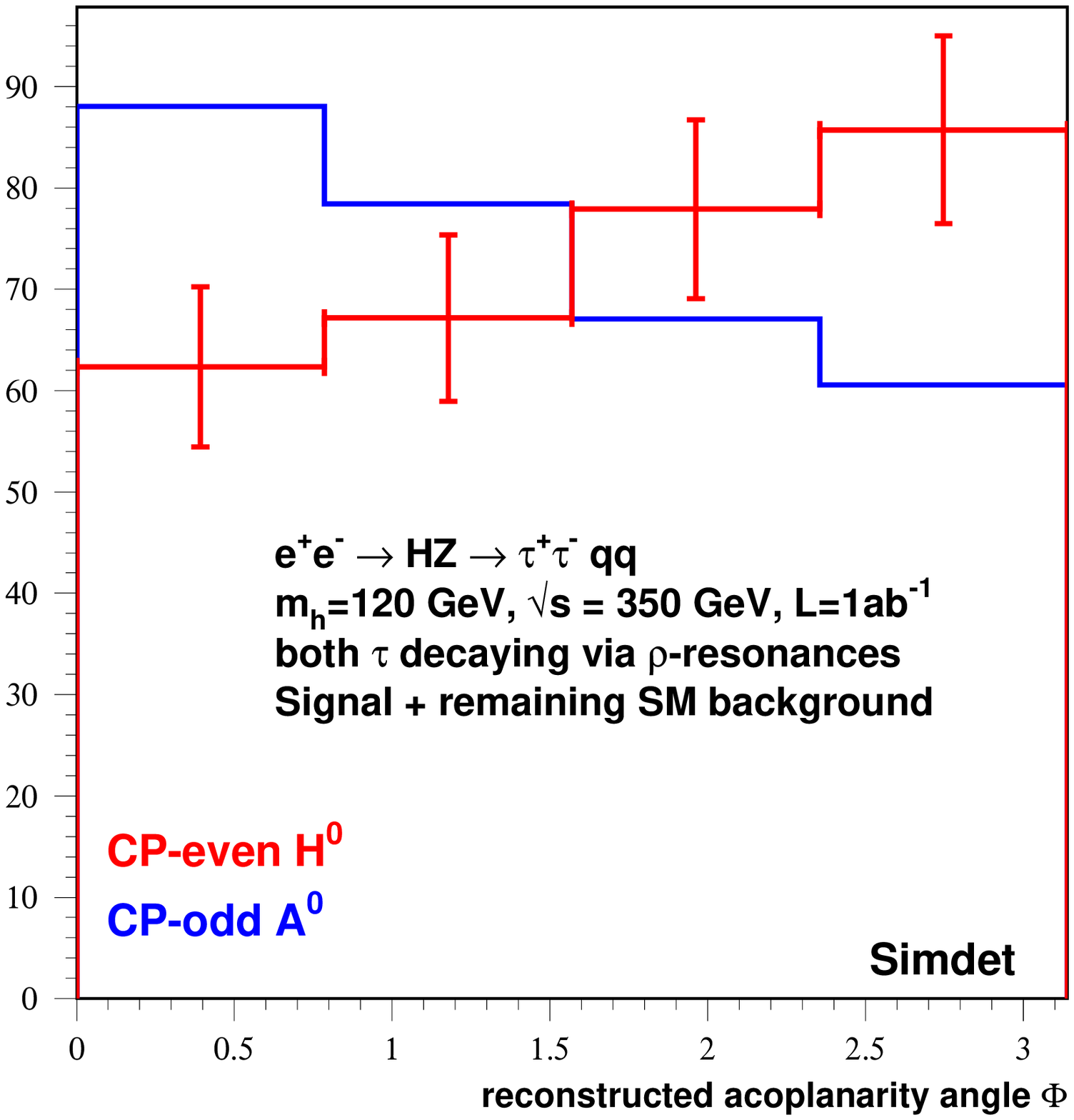,width=0.99\textwidth}
\end{minipage}
\begin{minipage}[c]{0.49\textwidth}
\psfig{figure=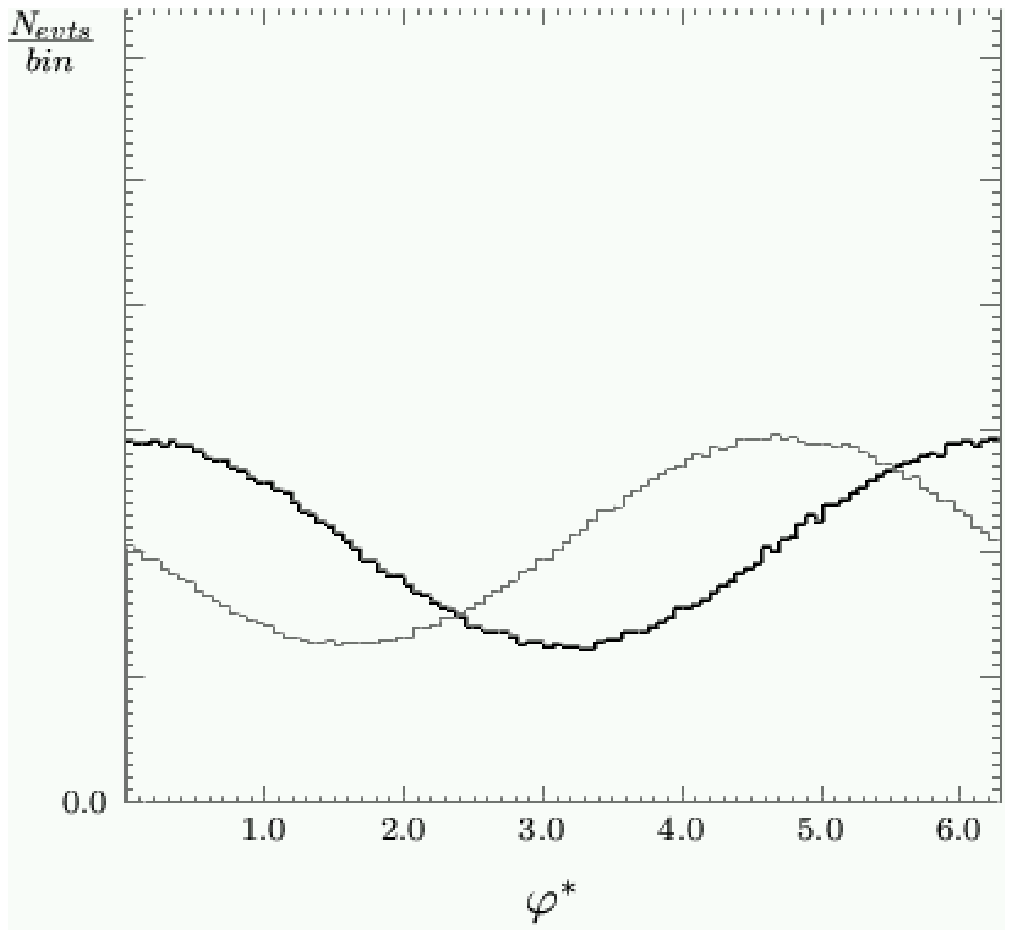,width=0.99\textwidth}
\end{minipage}
\begin{minipage}[c]{0.47\textwidth}
\caption{
The distribution of the acoplanarity angle $\Phi^*$ as defined for 
the $H\to \tau^+\tau^-\to \rho^+\bar{\nu_\tau}\rho^-\nu_\tau
\to \pi^+\pi^0\bar{\nu_\tau}\pi^-\pi^0\nu_\tau$ decay chain 
(see text) for the 
scalar state (error bars) and pseudoscalar state (histogram)
in the selected sample of $HZ\to \tau^+\tau^- q\bar{q}$ events.
Selected sample corresponds to 1 ab$^{-1}$ of 
data collected at $\sqrt{s}$ = 350 GeV. Higgs boson mass is 120 GeV.
\label{fig:cp_htt}
}
\end{minipage}
\begin{minipage}[c]{0.03\textwidth}
$\phantom{0}$
\end{minipage}
\begin{minipage}[c]{0.47\textwidth}
\caption{
The acoplanarity angle $\Phi^*$ in the 
$H\to \tau^+\tau^-\to \rho^+\bar{\nu_\tau}\rho^-\nu_\tau
\to \pi^+\pi^0\bar{\nu_\tau}\pi^-\pi^0\nu_\tau$ decay chain. 
Gaussian smearing of $\Phi^*$ and Higgs boson momenta are applied 
according to anticipated resolutions of the ILC detector. 
Thick line corresponds to a pure scalar state, while the thin line
to a mixed one. 
\label{fig:cp_htt_phase}
}
\end{minipage}
\end{figure}

An alternative method~\cite{cp_dova} 
consists of generating three-dimensional distributions
in $\cos\theta$, $\cos\theta^*$ and $\cos\phi^*$ for 
various values of $\eta$. These distributions are 
generated for the selected $\HZXll$ signal samples 
corresponding to 500 fb$^{-1}$ of data collected 
at $\sqrt{s}$ = 350 GeV. In the next step, the 
likelihood 
\BEQ
{\cal{L}} = 
\prod_{(\cos\theta)_i(\cos\theta)_j(\cos\phi^*)_k}
\frac{\mu_{ijk}^{N_{data}(i,j,k)} e^{-\mu_{ijk}}}{N_{data}(i,j,k)!}
\EEQ
is maximized, where $N_{data}(i,j,k)$ is the 
number of events of the hypothetical data sample and 
$\mu_{ijk}$ is the expected number in the $ijk$-th bin.
$\mu_{ijk}$ is calculated assuming a linear 
combination of the number of events of three Monte Carlo 
samples, corresponding to the production of scalar Higgs
(MC$(ZH)$), pseudoscalar Higgs (MC$(ZA)$)
and events for the interference term (MC$(IN)$):
\BEQ
\mu_{ijk} = \alpha {\rm MC}(ZH)_{ijk} + 
\beta {\rm MC}(IN)_{ijk} + \ga {\rm MC}(ZA)_{ijk},
\EEQ
where $\alpha$, $\ga$ and $\beta$ quantify contributions
from the production of the scalar state, pseudoscalar state
and their interference, respectively.
The likelihood is then maximized with respect to 
$\alpha$, $\beta$ and $\ga$. The quantity $\eta$ and its 
experimental error can then be determined by parameterizing 
the fraction of pseudoscalar component $\ga$ as a function of 
$\eta$. It has been shown that with this method, the
parameter $\eta$ can be measured with an accuracy of 
0.015 for $|\eta|$ $\leq$ 0.1.

The Higgs boson parity can be established also 
by analyzing the spin 
correlation effects in the Higgs boson decay into fermion 
pairs~\cite{hparity}.
The spin dependence of the decay probability is given by
\BEQ
\Gamma(H,A\to f\bar{f}) \sim 1 - s_z \bar{s}_z \pm s_\bot \bar{s}_\bot,
\EEQ
where $s$ and $\bar{s}$ denote spin vectors of fermion $f$ and 
anti-fermion $\bar{f}$ in their respective rest frames (with the 
$z$ axis oriented 
in the fermion flight direction). The positive sign in the transverse
spin correlation term holds for the scalar particle ($H$) and the
negative sign for the pseudoscalar ($A$). Although the dominant 
fermionic decay mode of the Higgs boson is into $b$ quark pairs,
it can hardly be utilized for measurement of the Higgs parity, as 
hadronization of quarks will inevitably dilute the spin information.
The most promising channel is the Higgs decay into tau leptons, with 
their subsequent decay into $\rho$ meson and neutrino~\cite{htautau}.
The angle $\Phi^*$ between the planes spanned by the momentum vectors of the 
products of the $\rho^+\to \pi^+\pi^0$ 
and $\rho^-\to \pi^-\pi^0$ 
decays in the $\rho^+\rho^-$ rest frame can be used to distinguish 
between the scalar and pseudoscalar states. Recently, a dedicated
study~\cite{htautau_exp} has been done which investigated 
the feasibility of measuring Higgs 
parity at ILC in the 
$H\to \tau^+\tau^-\to \rho^+\bar{\nu_\tau}\rho^-\nu_\tau
\to \pi^+\pi^0\bar{\nu_\tau}\pi^-\pi^0\nu_\tau$ channel. The
analysis is performed at the level of detailed 
detector simulation and includes accurate estimates of the 
SM background. Higgs boson decays
into tau leptons are selected using the Higgs-strahlung process with 
subsequent decays of the $Z$ boson into electron and muon pairs and quarks. 
Even in the presence of SM backgrounds, the acoplanarity angle $\Phi^*$ 
clearly discriminates between CP-even and CP-odd states,
as illustrated in \reffi{fig:cp_htt}. 

With 1 ab$^{-1}$ of 
data collected at $\sqrt{s} = 350 \gev$, scalar and pseudoscalar  
states can be distinguished at $\sim 4 \sigma$ level.
Furthermore, the method using the acoplanarity angle $\Phi^*$ 
can be applied in the general case of the mixed Higgs state, 
comprising both CP-even and CP-odd components.%
\footnote{
This
scenario requires redefinition of acoplanarity angle resulting 
in coverage of $[0,2\pi]$ range. For details see
Reference~\cite{cp_mixture}.
}%
~The presence of a CP-odd admixture
results in a phase shift in the distribution of 
the $\Phi^*$ angle as shown in \reffi{fig:cp_htt_phase}. 
From the measurement of 
this shift the mixing angle between scalar and pseudoscalar components
can be determined~\cite{cp_mixture} with an accuracy of about 0.2 rad 
if one assumes 1 ab$^{-1}$ of data collected at $\sqrt{s}$ = 350 GeV.

\begin{figure}[htb!]
\begin{center}
\psfig{figure=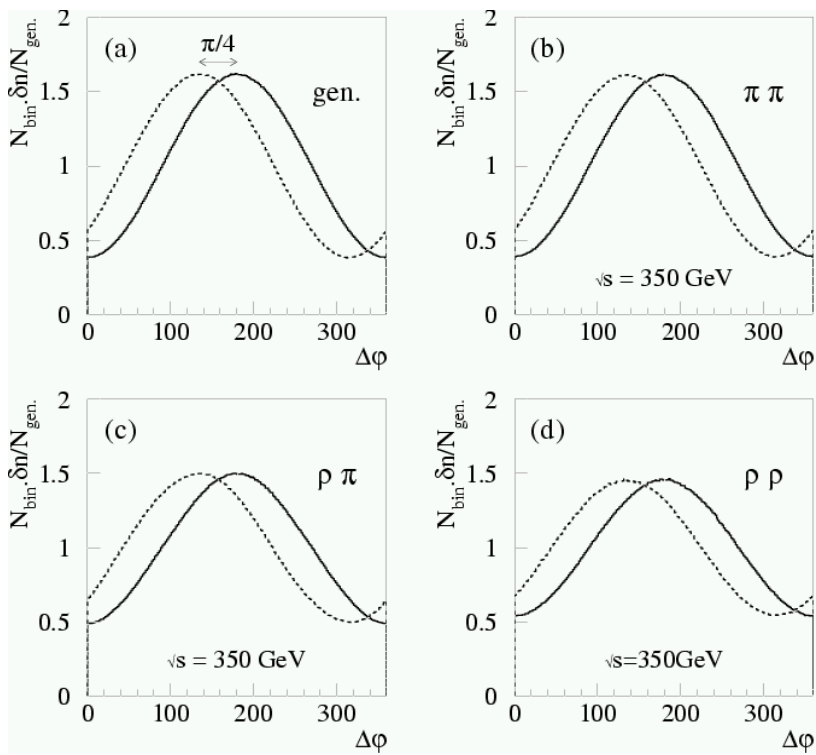,width=0.65\textwidth,
                                       height=0.60\textwidth}  
\end{center}
\caption{
The distributions of $\Delta\phi$: a) at the generation level; 
b), c) and d) after dedicated reconstruction~\cite{cp_rouge}
for the three channels $\pi\nu\pi\nu$,  $\pi\nu\rho\nu$ and $\rho\nu\rho\nu$ 
at center-of-mass energy of 350 GeV. Beamstrahlung effects only 
are taken into account. The histograms are normalized to the 
number of generated events and multiplied by the number of bins.
The full lines correspond to a pure scalar Higgs state ($\psi$ = 0), 
the dotted to a mixed state with $\psi$ = $\pi/8$. 
\label{fig:cphiggs_rouge_1}
}
\end{figure}

Recently, a new method to measure CP 
violation in a light Higgs boson decay from $\tau$-spin correlations
has been proposed~\cite{cp_rouge}. The method has been devised to
be insensitive to beamstrahlung and exploits the information encoded
in a unit vector $\vec{a}$, the $\tau$ polarization analyzer. 
The latter can be computed~\cite{tau_polariser} 
from the measured decay products for
the three hadronic decay modes, $\tau\ra \pi \nu, \rho\nu, a_1\nu$, 
which constitute 55\% of all decay channels of the $\tau$ 
lepton. For example, in the case of $\tau\ra \pi \nu$ the polarization 
analyzer is represented by the pion direction. For $\tau\ra \rho \nu$ and 
$\rho^\pm \ra \pi^\pm \pi^0$,
\begin{equation}
a^i = {\cal{N}}\Big(2(q\cdot p_\nu)q^i - (q\cdot q)p_\nu^i \Big),
\end{equation}  
where ${\cal{N}}$ is a normalization factor, $p_\nu = p_\tau-p_\rho$
is the neutrino momentum and $q = p_{\pi^{\pm}} - p_{\pi^0}$ is the 
difference between the momenta of the charged pion and neutral pion.
Once the vector $\vec{a}$ is reconstructed, the information on the 
CP phase in the light Higgs state can be accessed through the 
distribution of the azimuthal angle
\begin{equation}
dN/d(\Delta \phi) \sim 1 - \frac{\pi^2}{16}\cos(\Delta\phi - 2\psi), 
\label{eq:cp_analyser}
\end{equation}
where $\psi$ is the CP phase of the Higgs boson and 
$\Delta \phi$ is the difference in azimuthal angles of 
the $\vec{a}^\pm$ vectors  
in the $H\ra \tau^+\tau^-$ rest frame. The 
reconstruction of $\Delta \phi$ is possible 
in the Higgs-strahlung process. It requires information
on the visible decay products of the tau leptons,  
$Z$ boson momentum, primary vertex position and the Higgs boson mass.
The entire algorithm of reconstruction of $\Delta \phi$ is described 
in detail in Ref.~\cite{cp_rouge}. Visible decay products of tau 
leptons are directly measured using only tracking in the case of 
the $\tau \ra \pi \nu$ decay mode or using
calorimeter and tracking information in the case of 
$\tau\ra \rho \nu$. The momentum of the $Z$ boson can be determined from the
$Z\ra e^+e^-,\mu^+\mu^-,q\bar{q}$ decays. The primary vertex position in 
$r\phi$ plane is constrained by the beam spot, whereas the $z$ coordinate 
is determined from charged decay products of the $Z$ boson. Finally, 
it is assumed that the Higgs mass will be determined with high 
precision prior to this analysis. Distributions of $\Delta \phi$ 
when only beamstrahlung is taken into account are shown in Figure
\ref{fig:cphiggs_rouge_1}.
The sensitivity of the $\psi$ measurement
is expressed as 
\begin{equation}
S_{\psi}=1/\sigma_\psi\sqrt{N},
\end{equation}
where $\sigma_\psi$ is the error on $\psi$ expected 
from a maximum likelihood fit of the distribution for 
a sample of $N$ events. Defined in such a way, the sensitivity measures
the information per event on $\psi$ contained in the
distribution (\ref{eq:cp_analyser}).
Anticipated sensitivities in various decay channels for 
different center-of-mass energies are given in Table~\ref{tab:hp_sensitivity}.
\begin{table}[htb!]
\begin{minipage}[c]{0.45\textwidth}
\begin{center}
\begin{tabular}{cccc}
\hline
$\sqrt{s}$ (GeV) & \multicolumn{3}{c}{Sensitivity ($S_\psi$)} \\
\cline{2-4} 
 & $\pi\nu\pi\nu$ & $\pi\nu\rho\nu$ & $\rho\nu\rho\nu$ \\
\hline
230 & 0.92 & 0.88 & 0.83 \\
350 & 0.91 & 0.73 & 0.66 \\ 
500 & 0.88 & 0.64 & 0.55 \\
\hline
\end{tabular}
\end{center}
\end{minipage}
\begin{minipage}[c]{0.05\textwidth}
$\phantom{0}$
\end{minipage}
\begin{minipage}[c]{0.45\textwidth}
\caption{Sensitivity for the determination of 
CP phase $\psi$ of the Higgs boson for the 
three different channels $\pi\nu\pi\nu$, $\pi\nu\rho\nu$ 
and $\rho\nu\rho\nu$ at 
center-of-mass energies 230, 350 and 500 GeV.
\label{tab:hp_sensitivity}
}
\end{minipage}
\end{table}

\begin{figure}[htb!]
\begin{center}
\psfig{figure=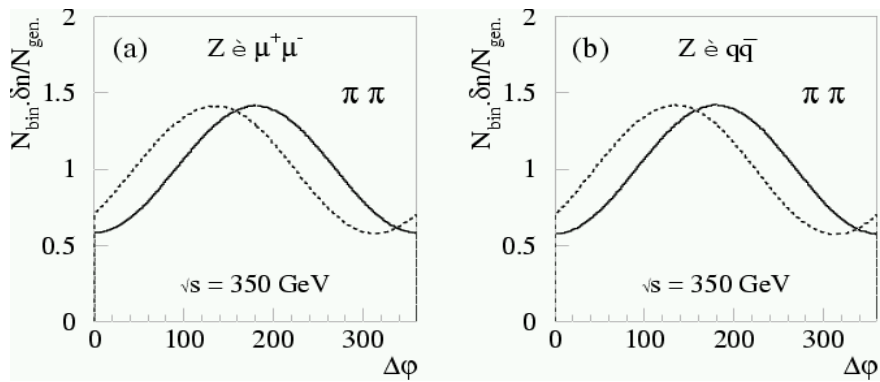,width=0.7\textwidth,
                                       height=0.30\textwidth}
\end{center}
\caption{
The distributions of $\Delta\phi$ in 
the $\pi\nu\pi\nu$ channel for the a) $Z\ra \mu^+\mu^-$ 
and b) $Z\ra q\bar{q}$ decays at center-of-mass energy 350 GeV.
All experimental effects are included.
The full lines correspond to a pure scalar Higgs state ($\psi$ = 0), 
the dotted to a mixed state with $\psi$ = $\pi/8$. 
\label{fig:cphiggs_rouge_2}
}
\end{figure}

To evaluate detector effects, a semi-realistic simulation of 
$H\ra \tau\tau \ra \pi\nu\pi\nu$ was performed. For the 
charged tracks an independent Gaussian smearing is performed 
on the five parameters: $\theta$, $\phi$, $1/p_T$ and the
two components of impact parameter resolution. Assumed widths 
of the Gaussians are:
\begin{displaymath}
\begin{array}{l}
\sigma (\theta) = \sigma(\phi) = 0.1 \ {\rm{mrad}} \\
\sigma (1/p_T) = 5\times 10^{-5} \ {\rm{GeV}}^{-1} \\
\sigma (r\phi) = \sigma (rz) = 
\Big(4.2 \oplus 4.0/(p\sin^{3/2}\theta)\Big) \ \mu{\rm{m}}. \\
\end{array}
\end{displaymath}
The jet energy resolution is assumed to be ${\rm{0.3}}/\sqrt{E_{jet}}$.
The reconstructed distributions of $\Delta\phi$ after 
inclusion of the detector effects are shown in 
Figure~\ref{fig:cphiggs_rouge_2} for the decays of $Z$ both into
$\mu^+\mu^-$ and $q\bar{q}$.  
The corresponding sensitivities are given in 
Table~\ref{tab:hp_sensitivity1}.
\begin{table}[htb!]
\begin{minipage}[c]{0.45\textwidth}
\begin{center}
\begin{tabular}{ccc}
\hline
$\sqrt{s}$ (GeV) & \multicolumn{2}{c}{Sensitivity ($S_\psi$)} \\
\cline{2-3} 
 & $Z\ra \mu^+\mu^-$ & $Z\ra q\bar{q}$ \\
\hline
230 & 0.69 & 0.71  \\
350 & 0.60 & 0.61  \\ 
500 & 0.58 & 0.58  \\
\hline
\end{tabular}
\end{center}
\end{minipage}
\begin{minipage}[c]{0.05\textwidth}
$\phantom{0}$
\end{minipage}
\begin{minipage}[c]{0.45\textwidth}
\caption{Sensitivity of the determination of 
CP phase $\psi$ of the Higgs boson for the 
$\pi\nu\pi\nu$ channel when all experimental 
effects are included. 
\label{tab:hp_sensitivity1}
}
\end{minipage}
\end{table}

The analyses of other decay modes, including the study of detector 
effects and $\pi^0$ reconstruction, have yet to be performed; nevertheless
it appears that a reasonable goal for the measurement of the phase   
$\psi$ should be to use all the above mentioned channels and reach 
a sensitivity better than 0.5, 
i.e., $\sigma_\psi < 0.6\pi/\sqrt{N_{\rm event}}$.

Finally, Higgs boson parity can determined in $\GG$ collisions 
of linearly polarized photon beams~\cite{hparity}. The matrix elements for 
the production of scalar and pseudoscalar Higgs bosons in 
$\GG$ collisions can be expressed as:
\BEQ
{\cal{M}}(\ga\ga\to H[0^{++}]) 
\sim \vec{\epsilon}_1\cdot\vec{\epsilon}_2,\hspace{5mm}
{\cal{M}}(\ga\ga\to A[0^{-+}]) 
\sim \vec{\epsilon}_1\times\vec{\epsilon}_2\cdot\vec{k}_\ga
\EEQ
where $\vec{\epsilon}_1$ and 
$\vec{\epsilon}_2$ denote polarization vectors of 
colliding photons and $\vec{k}_\ga$ is 
the momentum vector of one of the colliding photons.
Hence, the parity of the produced Higgs state can be 
extracted by measuring the dependence of the production rate on the relative
orientation of the polarization vectors of the linearly 
polarized photon beams. 

The measurement of Higgs spin can be performed by analyzing the 
energy dependence of the Higgs boson production 
cross section just above the kinematic threshold~\cite{hspin0,hspin}. 
For a spin zero particle
the rise of the cross section is expected to be 
$\sim \beta$, where $\beta$ is the velocity of the boson
in the center-of-mass system.  
\begin{figure}[htb!]
\begin{minipage}[c]{0.48\textwidth}
\psfig{figure=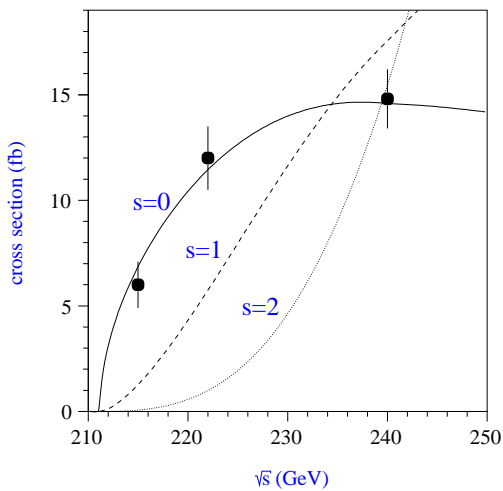,width=0.90\textwidth}
\end{minipage}
\begin{minipage}[c]{0.45\textwidth}
\caption{
The cross section of 
$\HZqqll$ just above the threshold assuming
$\mH$ = 120 GeV. The dots correspond to
a measurement and the curves are
predictions for several spins.
\label{fig:hspin}
}
\end{minipage}
\end{figure}
For a  spin-one particle the rise is $\sim \beta^3$
and for spin-two like $\sim \beta^5$.
With a very small luminosity of about
20 fb$^{-1}$ per energy point the scalar nature of 
the Higgs boson can be established and other spin 
hypotheses are disfavored, as shown in
\reffi{fig:hspin}.  It should be 
noted, however, that there are particular scenarios
for spin-1 and -2 particles which show a threshold
behavior similar in shape to the spin-0 case. This can be disentangled 
using angular information in addition.


\section{Heavy SM-like Higgs boson}
\label{sec:heavySMHiggs}

Most analyses for SM-like Higgs bosons have been performed for Higgs
masses below $\sim 150 \gev$. However, higher masses are still allowed
by the electroweak fits; see \refse{subsec:MHexp}. Alternatively, new
physics could compensate the effects of a heavier Higgs boson in the
electroweak precision observables~\cite{peskinwells,Choudhury:2002qb}. 
For heavy SM-like Higgs bosons, $\MH \gsim 150 \gev$, the main decay
channels are into $WW$ and $ZZ$. Most analysis performed for lower
masses can therefore not be taken over to this case.

\subsection{Heavy Higgs Production}

In addition to the treatment of different 
decay modes, a very heavy SM-like Higgs
demands special theoretical treatment, because its width becomes 
non-negligible.  For large $\MH$, the width of the Higgs is dominated by
decays into longitudinal gauge bosons, and grows as $\MH^3$,
\BEA
\Gamma \left(H \rightarrow V V \right) & \simeq & 
\frac{G_F \left( |Q_V| + 1 \right)}{\sqrt{2} 16 \pi} \MH^3
\left( 1 - \frac{4 M_V^2}{\MH^2} + 3 \frac{4 M_V^4}{\MH^4} \right)
\left( 1 - \frac{4 M_V^2}{\MH^2} \right)^{1/2} ,
\EEA
where $V = W,Z$, and $Q_V$ is the electric charge of $V$.
This behavior may be simply understood from the fact that the would-be 
Goldstone bosons are part of the Higgs doublet, and couple to the Higgs
proportionally to the Higgs quartic $\lambda = \MH / v$ (at tree level).
It is worth noting that the partial width into top quarks grows only as a
single power of $\MH$, which explains why the decay into tops never dominates.

\begin{figure}[htb!]
\begin{minipage}[c]{0.59\textwidth}
\psfig{figure=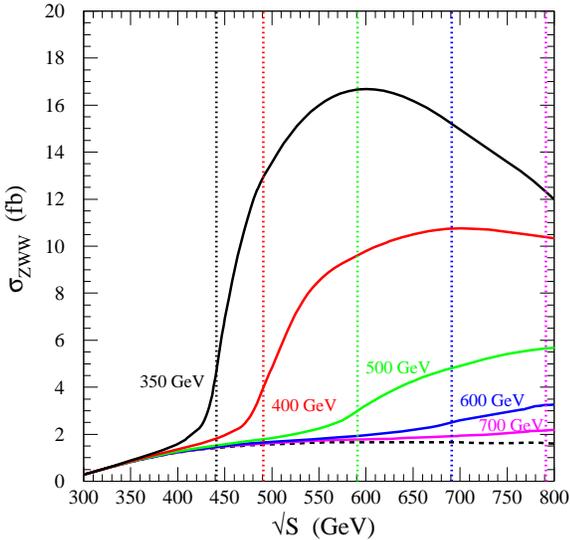,width=0.9\textwidth}
\end{minipage}
\begin{minipage}[c]{0.4\textwidth}
\caption{Cross sections for $Z W^+ W^-$ production as a function of $\sqrt{S}$
for a variety of Higgs masses.  The solid lines of each curve include the
appropriate SM Higgs width effects, whereas the dotted lines indicate the
nominal threshold for $HZ$ production.
\label{fig:xsecwidth}
}
\end{minipage}
\end{figure}

Once the width is large, the Higgs production signal and the SM background
processes can interfere in an important way.  As a result, the Higgs 
contribution in a given channel may be important below the putative threshold
energy for on-shell Higgs production, and the behavior of the signal slightly
above threshold may be modified in an important way by the width
\cite{Choudhury:2002qb}.  This is
illustrated in Figure~\ref{fig:xsecwidth} for the reaction 
$e^+ e^- \rightarrow W^+ W^- Z$, including the Higgs resonant production via
$e^+ e^- \rightarrow H Z$ followed by $H \rightarrow W^+ W^-$, as well as all
of the SM non-resonant production processes.  For heavy Higgses, typically
the associated production with a $Z$ boson, $e^+ e^- \rightarrow H Z$, and
the $W$-fusion process $e^+ e^- \rightarrow \nu_e \bar{\nu}_e H$ have
roughly similar rates.  The exception is when the Higgs is heavy enough that
on-shell $H$ and $Z$ cannot be produced, in which case the $W$-fusion rate
may dominate (see Figure~\ref{fig:xsec_sm}).

\subsection{Existing analyses}

\begin{figure}[htb!]
\begin{minipage}[c]{0.55\textwidth}
\epsfig{figure=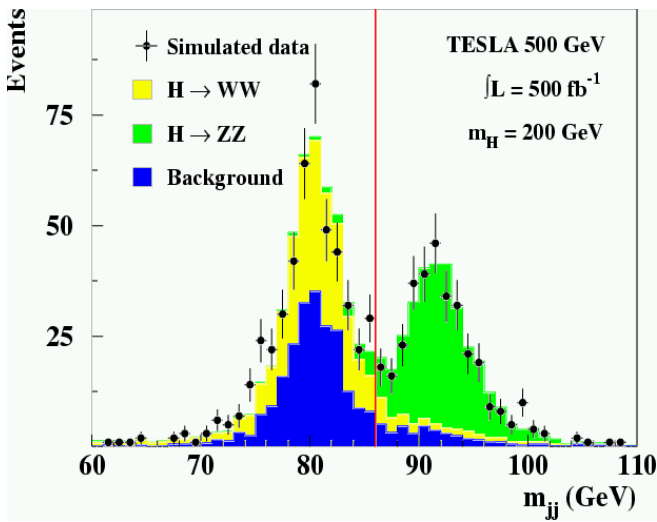,width=0.95\textwidth}
\end{minipage}
\begin{minipage}[c]{0.03\textwidth}
$\phantom{0}$
\end{minipage}
\begin{minipage}[c]{0.40\textwidth}
\caption{The distribution of the reconstructed vector boson mass 
in the sample of selected Higgs-strahlung events with
$\HWW$ or $\HZZ$ and $\Zll$. Distribution is obtained
with 5C kinematic fit as described in text.  
The $WW$ and $ZZ$ decays of the Higgs boson are clearly separated.
The Higgs mass is 200 GeV. The sample corresponds to 
integrated luminosity of 500 fb$^{-1}$ collected at $\sqrt{s}$ = 500 GeV.
 }
\label{fig:hhiggs_vv}
\end{minipage}
\end{figure}

A heavy SM-like Higgs boson decays predominantly into $WW$ and $ZZ$.
For a Higgs boson with mass $2\mZ < \mH < 2\mt$,
these final states constitute almost 100\% of all 
decay modes. This scenario has been investigated 
separately~\cite{higgs_lineshape}. In this study
Higgs events are selected exploiting the Higgs-strahlung
process with $\Zll$. Only hadronic decays of $W$ and $Z$ bosons
are considered. Separation of $WW$ and $ZZ$ decay modes 
is done on the basis of mass information. The hadronic system
assigned to the decay of the Higgs is clustered into four jets. 
For each possible jet-pairing a kinematical fit is performed 
imposing four-momentum conservation, taking into account the
center-of-mass energy ($\sqrt{s}$ = 500 GeV) 
and the measured momenta of the two 
leptons stemming from $Z$ decay. An additional constraint is imposed
forcing the invariant masses of the two dijet systems to be equal, 
leading to a five constraint (5C) fitting procedure. 
With this procedure, the $WW$ and $ZZ$ decay modes can be efficiently
separated by the mass obtained with the 5C kinematical fit. 
This is illustrated in \reffi{fig:hhiggs_vv}.

With clear separation of the $WW$ and $ZZ$ decay channels, the measurement
of the corresponding branching ratios of heavy Higgs can be performed. 
Results are given in Table~\ref{tab:hhiggs_vv}.

\begin{table}[htb!]
\begin{minipage}[c]{0.5\textwidth}
\begin{tabular}{|c|c|c|}
\hline
$\mH$ (GeV)  &  $\HWW$ & $\HZZ$ \\
             & \multicolumn{2}{|c|}{Relative precision (\%)} \\
\hline
200          &  3.5 & 9.9 \\
240          &  5.0 & 10.8 \\
280          &  7.7 & 16.2 \\
320          &  8.6 & 17.3 \\
\hline
\end{tabular}
\end{minipage}
\begin{minipage}[c]{0.03\textwidth}
$\phantom{0}$
\end{minipage}
\begin{minipage}[c]{0.4\textwidth}
\caption{
Precision on $\HWW$ and $\HZZ$ branching ratio measurements
for a heavy Higgs boson~\cite{higgs_lineshape}. 
Results are obtained assuming 500 fb$^{-1}$ of 
luminosity collected at $\sqrt{s}$ = 500 GeV.
\label{tab:hhiggs_vv}
}
\end{minipage}
\end{table}

\medskip
The prospect of measuring Higgs branching ratios with 
1 ab$^{-1}$ of $\EE$ collision data collected at 
1 TeV center-of-mass energy has been investigated in 
\citere{barklow_branchings}. The analysis is based 
on the measurement of topological cross sections 
$\sigma_{WW-{\rm fusion}}\br(H\ra X)$ with 
$X=b\bar{b},WW,gg,\ga\ga$ and 
makes use of results for the Higgs branching ratio measurements 
into $\bb$ and WW at 
lower center-of-mass energies. Electron and positron beam 
polarization of 80\% and 50\%, respectively, is assumed.
The study makes use of the following relations:
\BEA
\br(H\ra X) & = &
\Big(\sigma_{WW-{\rm fusion}} \br(H\to X)\Big)
\Big(\sigma_{WW-{\rm fusion}} \br(H\to WW)\Big)^{-1}
\br^*(H\to WW) \non \\
            & = &
\Big(\sigma_{WW-{\rm fusion}} \br(H\to X)\Big)
\Big(\sigma_{WW-{\rm fusion}}\br(H\to b\bar{b})\Big)^{-1}
\br^*(H\to b\bar{b}),
\EEA
where $\sigma_{WW{\rm -fusion}}\br(H\to X)$ is the 
topological cross section measured at 1 TeV and 
$\br^*(H\to WW)$ and $\br^*(H\to bb)$ are 
branching ratios measured at 350 GeV center-of-mass energies. 
Using these relations, a least-squares fit is performed to 
obtain measurement errors for $\br(H\to b\bar{b})$, 
$\br(H\to WW)$, $\br(H\to gg)$ and 
$\br(H\to \ga\ga)$. This procedure 
provide significant improvement (by a factor of two or more)
for decay modes with small branching fractions, such as
$\br(H\to b\bar{b})$ for 160 GeV $<$ $\mH$ $<$ 200 GeV,
$\br(H\to WW)$ for 120 GeV $<$ $\mH$ $<$ 140 GeV, and
$\br(H\to gg)$ and $\br(H\to \ga\ga)$ for 
all Higgs masses.
More details can be found in \citere{barklow_branchings}. 

\medskip
Another study~\cite{battaglia_hbb}, making use of the large number of
Higgs bosons produced in the $WW$~fusion channel, obtained the
following results. For 1 \iab of data at $\sqrt{s} = 800 \gev$ a 
$5 \si$ sensitivity to the bottom Yukawa coupling is achievable for
$\MH < 210 \gev$. A measurement of the branching ratio $\br(\Hbb)$ is
possible with the accuracies given in \refta{tab:hbb}.

\begin{table}[htb!]
\begin{minipage}[c]{0.5\textwidth}
\begin{tabular}{|c|c|}
\hline
$\mH$ (GeV)  &  $\Hbb$ \\
             & Relative precision (\%) \\
\hline
160          &   6.5 \\
180          &  12.0 \\
200          &  17.0 \\
220          &  28.0 \\
\hline
\end{tabular}
\end{minipage}
\begin{minipage}[c]{0.03\textwidth}
$\phantom{0}$
\end{minipage}
\begin{minipage}[c]{0.4\textwidth}
\caption{
Precision on $\Hbb$ branching ratio measurements
for a heavy Higgs boson~\cite{KlausiAmsterdam}. 
Results are obtained assuming 1 ab$^{-1}$ of 
luminosity collected at $\sqrt{s}$ = 800 GeV.
\label{tab:hbb}
}
\end{minipage}
\end{table}

\bigskip
Another analysis concerns the top Yukawa coupling measurement for
Higgs bosons heavier than $2 \, \mt$. In this case the top Yukawa 
coupling could be determined from the Higgs decay into $t\bar t$ 
pairs. This decay can be studied at the ILC using the Higgs resonant 
contribution to the process $W^+W^- \to t \bar t$. 
While this electroweak $t\bar t$ production process cannot be observed 
at the LHC due to the huge QCD background of $t \bar t$ production by
gluon fusion, it can be measured via a high energy 
$e^+e^-$ linear collider.
A complete simulation study of the $W^+W^- \to H \to t \bar t$ 
process from $e^+e^-$ collisions has been performed including 
realistic backgrounds and experimental effects\cite{Alcaraz:2000xr}.
The helicity cross sections for 
the $e^+e^- \to t\bar t \nu \bar \nu$ signal process, including 
ISR and beamstrahlung effects, are given in Table~\ref{tab:ttnn} 
for a collider energy of $\sqrt{s}=1 \tev$ and different 
Higgs boson masses. 
\begin{table}[htb!]
\begin{minipage}[c]{0.55\textwidth}
\begin{tabular}{ccccc} \hline
$m_H$ & $\sigma(LL) = \sigma(RR) $  & $\sigma(LR)$ 
 & $\sigma(RL)$ & Total \\\hline & & & & \\
100 & 0.11 & 0.22 & 0.21 & 0.65 \\
400 & 1.89 & 0.34 & 0.33 & 4.45 \\
600 & 0.89 & 0.22 & 0.21 & 2.21 \\
800 & 0.33 & 0.22 & 0.21 & 1.09\\[2mm]  \hline 
\end{tabular}
\end{minipage}
\begin{minipage}[c]{0.44\textwidth}
\caption[Tab1]{\label{tab:ttnn} Helicity cross sections (in fb) for the 
$e^+ e^- \to t \bar t \nu \bar \nu$ process, 
at $\sqrt s = 1$ TeV and for different Higgs boson 
masses (in GeV). $\sigma(\lambda,\lambda^\prime)$ denotes the cross section 
for production of a top quark with helicity
$\lambda$ and an antitop with helicity $\lambda^\prime$. CP invariance implies
$\sigma(RR)=\sigma(LL)$.}
\end{minipage}
\end{table}

\begin{figure}[htb!]
\begin{minipage}[c]{0.45\textwidth}
\psfig{figure=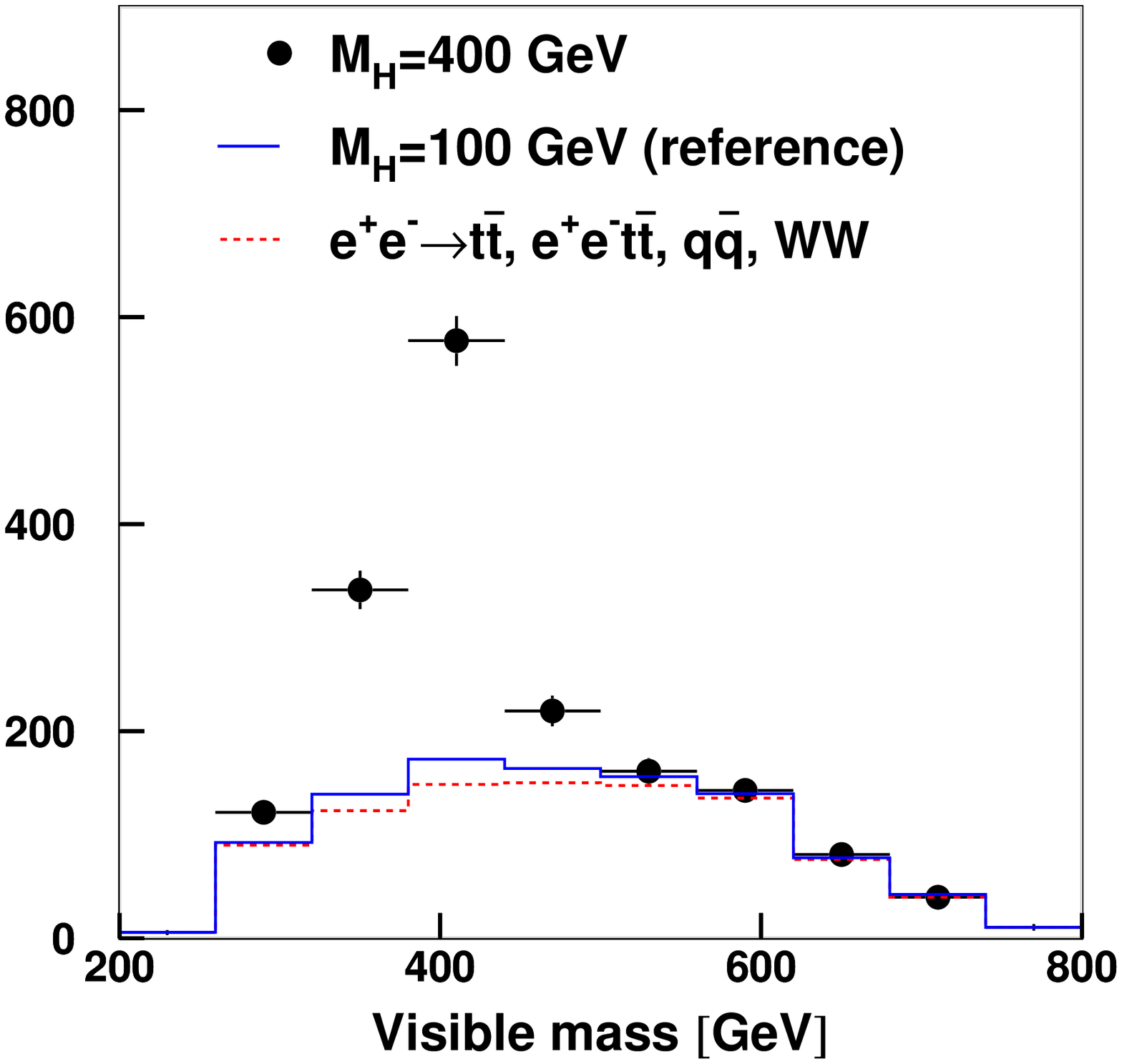,angle=0,width=1.1\textwidth}
\end{minipage}
\begin{minipage}[c]{0.47\textwidth}
\psfig{figure=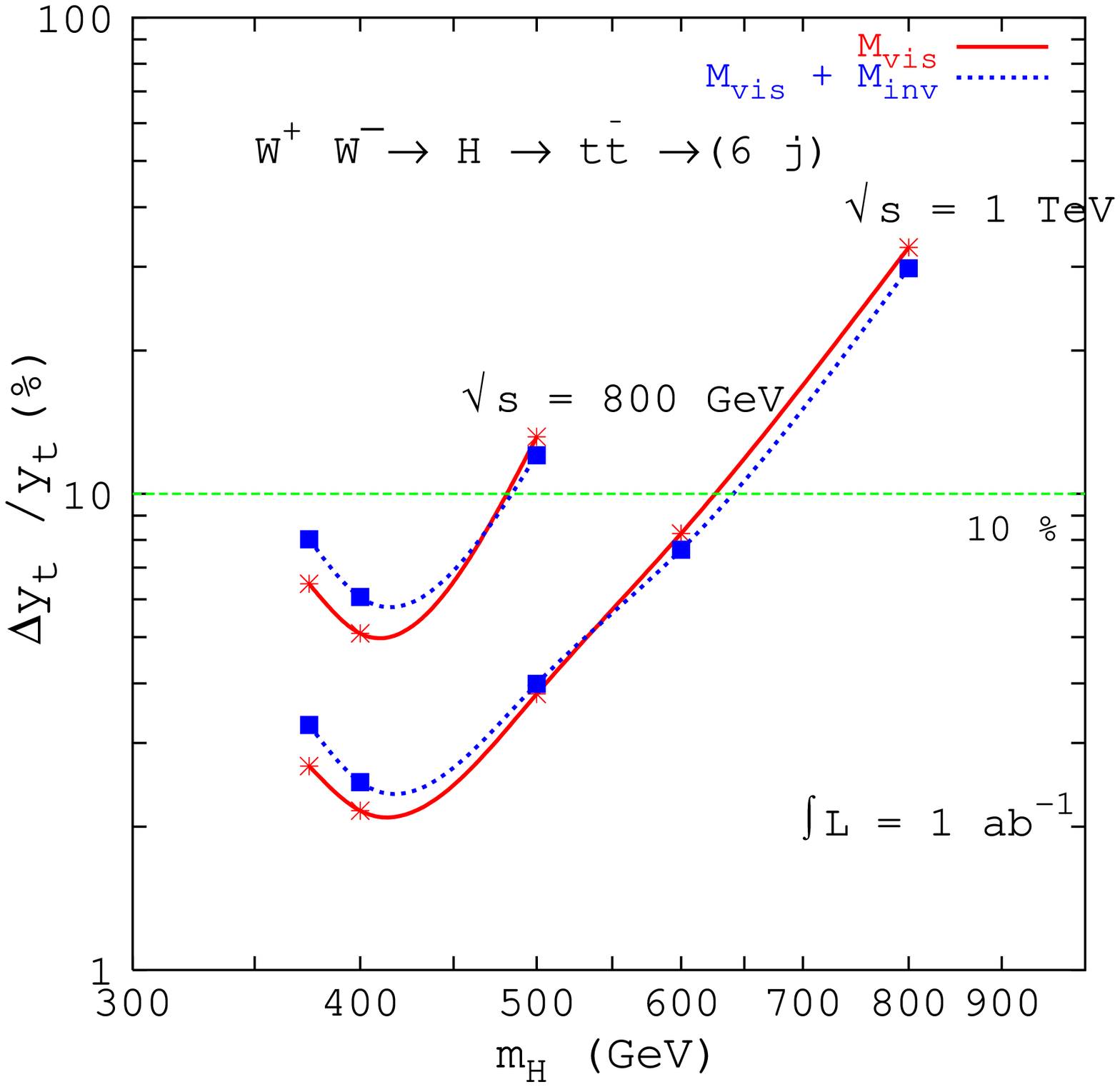,angle=-90,width=1.3\textwidth}
\end{minipage}
\caption{
[Left figure] Expected number of reconstructed 6-jet events 
as a function of the visible mass, at $\sqrt{s}=1$ TeV and with 
${\cal L}=1$ ab$^{-1}$ after all cuts, for a Higgs boson of 400 GeV
(dots) and for the backgrounds (dashed). The expectations (including 
background) for a Higgs of 100 GeV (solid) are also shown for 
comparison. 
[Right figure] Expected relative precision in the top Yukawa coupling 
mesurement as a function of the Higgs boson mass, from
6-jet events. 
\label{fig:topyukawa}}
\end{figure}

The relevant signal is a $t\bar t$ pair with missing transverse momentum
carried by the neutrinos or from beam electrons not observed
by the detector. The largest background is direct $t\bar t$ production. 
It can be suppressed by requiring missing mass greater than 
some minimal value, and further by choosing the jet association 
that best reconstructs the $t$ and $W$ masses. 
Another significant background is $e^+e^-t \bar t$, which 
can be reduced by requiring the missing transverse energy to be 
greater than some value such as 50 GeV as used in
Ref.~\cite{Alcaraz:2000xr}. 
The background from $\gamma t \bar t$ 
can be mostly eliminated with an appropriate cut 
on the $p_T^{t\bar t}$ of the $t \bar t$ pair\cite{Larios:1997ey}.  
The expected number of reconstructed 6-jet events as 
a function of the visible mass at $\sqrt{s}=1$ TeV 
with ${\cal L}=1$ ab$^{-1}$ is shown in Fig.~\ref{fig:topyukawa} (left), 
and the expected relative precision in the top Yukawa 
measurement is  shown in 
Fig.~\ref{fig:topyukawa} (right)\cite{Alcaraz:2000xr}. 
The QCD corrections of ${\cal O}(\alpha_s)$ to the $V_LV_L \to t \bar t$ 
scattering have been studied in Ref.~\cite{Godfrey:2004tj}, where 
$V_L$ represents the longitudinally polarized vector boson $V=W^\pm$ 
or $Z$. It has been found that corrections to the cross sections can 
be as large as 30\% and must be accounted for in any precision 
measurement of $VV \to t\bar t$.

\bigskip
The second question about heavier Higgs bosons is whether the Higgs
line-shape parameters (mass, decay width, Higgs-strahlung production cross
section) can be measured. A complete study of the mass range
200 GeV $< \mH <$ 320 GeV has been performed~\cite{LC-PHSM-2003-066}.
The final state $q\bar q q\bar q \ell^+\ell^-$ 
resulting from $HZ \to ZZZ$ and
from $HZ \to W^+W^- Z$ is selected. A kinematic fit is used to assign
the possible di-jet combinations to bosons ($W^+W^-$ or $ZZ$). The resulting
di-boson mass spectrum can be fitted by a Breit-Wigner distribution
convolved with a detector resolution function. A relative uncertainty
on the Higgs mass of 0.11--0.36\% is achievable from 500 \ifb at
500~GeV for Higgs masses between 200 and 320~GeV. The resolution on the total
width varies between 22 and 34\% for the same mass range. Finally, 
the total Higgs-strahlung cross-section can be measured with 3.5--6.3\%
precision. Under the assumption that only $\HWW$ and $\HZZ$
decays are relevant, their branching ratios can be extracted with
3.5--8.6\% and 9.9--17.3\%, respectively (see Table~\ref{tab:heavyh}).
The expected mass spectra for $\mH = 200  \gev$ and $\mH = 320  \gev$ are
shown in Fig.~\ref{fig:niels}. 

\begin{figure}[htb]
\centering
\epsfig{width=0.4\linewidth,file=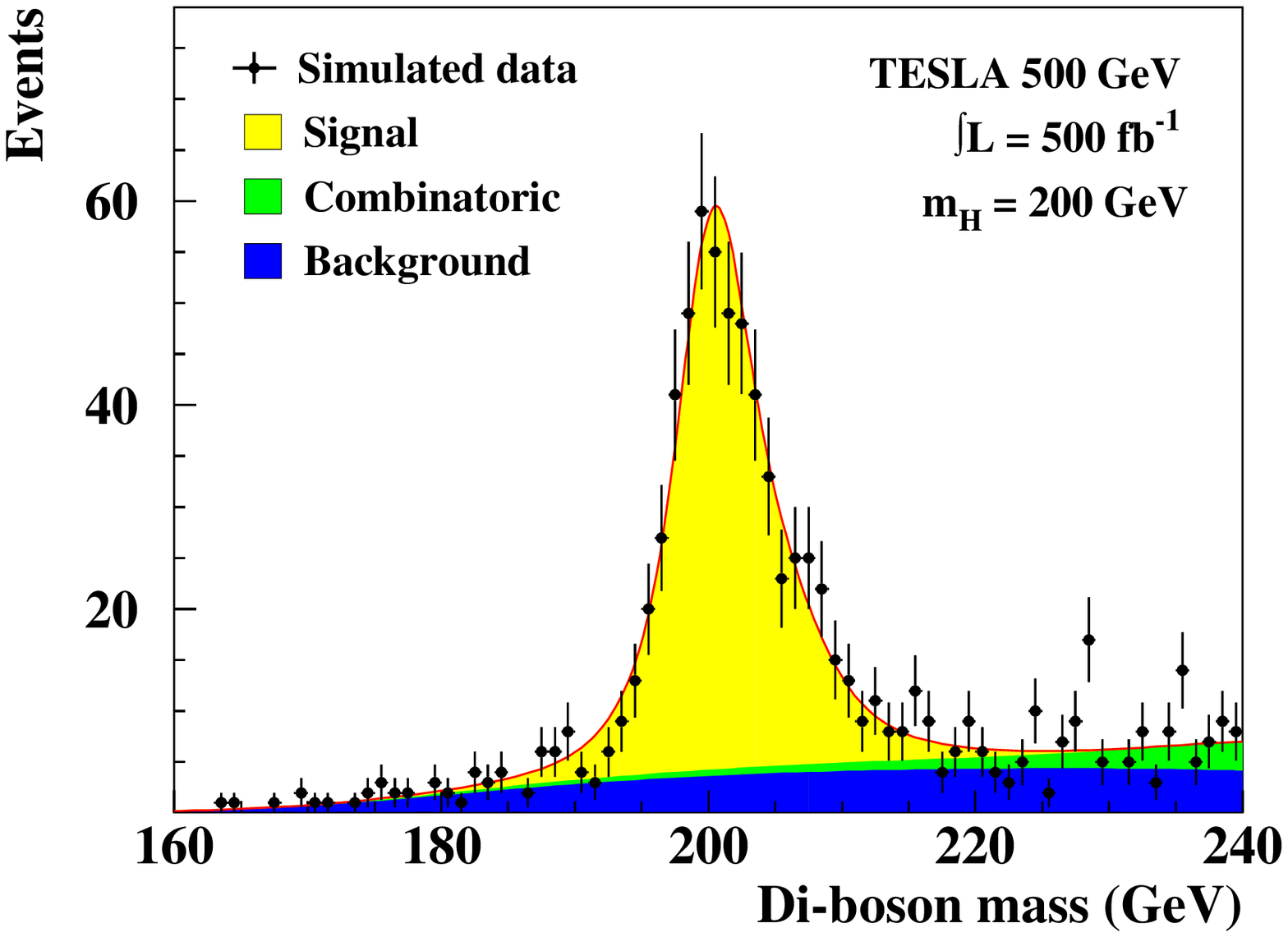}\hspace{1em}
\epsfig{width=0.4\linewidth,file=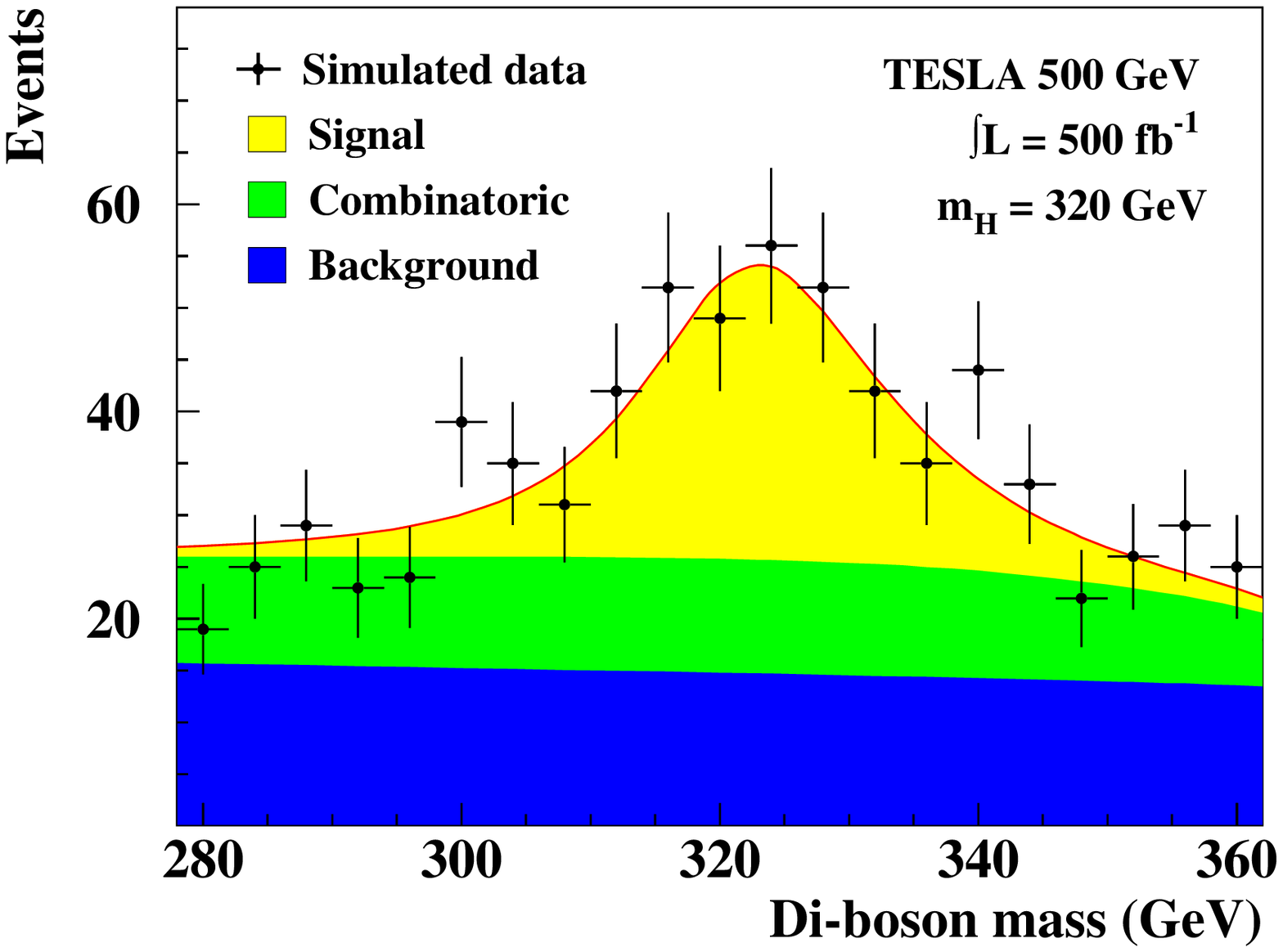}
\caption{Expected reconstructed Higgs boson mass spectra for
$\mH = 200  \gev$ and $\mH = 320  \gev$ from  500 \ifb at
500~GeV (from~\cite{LC-PHSM-2003-066}). }
\label{fig:niels}
\end{figure}

\begin{table}[htb!]
\begin{center}
\vspace{0.3cm}
\begin{tabular}{|c|c|c|c|}
\hline
$\mH$ (GeV) & $\De\si$ (\%) & $\De\mH$ (\%) & $\De\Ga_H$ (\%) 
                                                  \\ \hline
200 & 3.6 & 0.11& 34   \\ 
240 & 3.8 & 0.17& 27   \\
280 & 4.4 & 0.24& 23   \\
320 & 6.3 & 0.36& 26   \\ \hline

\end{tabular}
\caption{Expected precision on Higgs boson line-shape parameters 
for $200 < \mH < 320 \gev$ at the ILC with $\sqrt{s} = 500 \gev$. }
\label{tab:heavyh}
\end{center}
\end{table}


\subsection{Open questions}

In the field of heavy SM-like Higgs bosons many questions remain for
future studies. Besides extending the results quoted above to other
(possibly) higher Higgs mass values the following issues should be
addressed:

\begin{itemize}

\item
What are the chances to observe couplings of the heavy Higgs bosons
other than those already investigated? 

\item
What can be learned about the quantum numbers and the spin of heavy
(SM-like) Higgs bosons?

\item
Do these additional measurements pose new requirements for the detector?

\end{itemize}


\section{Heavy Neutral Higgs Bosons in the MSSM}

Besides the light CP-even Higgs boson in the MSSM there are two other
neutral Higgs bosons: the heavier CP-even $H$ and the CP-odd
$A$. For $\MA \gsim 150 \gev$ the two heavy Higgs bosons are very
similar in mass. The couplings of the Higgses to gauge bosons is
either strongly suppressed (for $HVV$) or even zero at the tree-level
(for $AVV$). Therefore the analyses performed for a heavy SM-like
Higgs boson as outlined in the previous section cannot be taken over
to the case of heavy MSSM Higgs bosons.

\medskip
Similar to the light CP-even Higgs boson, the
heavy neutral CP-even Higgs boson of MSSM can be
produced in $\EE$ collisions via Higgs-strahlung and 
weak boson fusion processes. Neutral Higgs bosons of MSSM 
can be also produced in pairs, $\EEhA$ and $\EEHA$, 
and singly through Yukawa processes, e.g.,
$e^+e^-\to h(A,H)b\bar{b}$; see \reffi{fig:heavy-higgs-contour}.
In the decoupling limit of the MSSM, the Higgs-strahlung and 
weak boson fusion processes involving $H$, as 
well as pair production process involving $h$, get 
strongly suppressed (see, however, \citere{eennH}). 
Hence, in this scenario the main production mechanisms of $H$ and $A$ bosons 
at an $\EE$ collider will be Yukawa processes and $\EEHA$. 
The pair production of heavy neutral Higgs bosons at an $\EE$ collider has been
studied in \citere{eeAHdesch}. The analysis 
is based on selection of $b\bar{b}b\bar{b}$ 
and $b\bar{b}\tau^+\tau^-$ final states, which are 
expected to be dominant channels for $\EEHA$ process
for a large part of MSSM parameter space. For the reconstruction of these 
final states, efficient $b$-tagging is important. The analysis 
employs the same flavor-tagging tools~\cite{bc_tagging} as used
in the study of the hadronic decays of the light Higgs boson~\cite{ee_hqq}.
The potential of the linear $\EE$ collider for detecting heavy neutral
Higgs bosons and measuring their properties has been evaluated
for various Higgs boson mass hypotheses. Although in large parts of
the MSSM parameter space $H$ and $A$ are almost 
degenerate in mass, scenarios with large mass splitting between
neutral Higgs bosons may occur. This feature is present in
CP-conserving as well as in CP-violating scenarios.

Studies have been performed for a linear collider
operated at center-of-mass energies of 500 and 800 GeV. 
%
\begin{figure}[htb!]
\begin{minipage}[c]{0.47\textwidth}
\psfig{figure=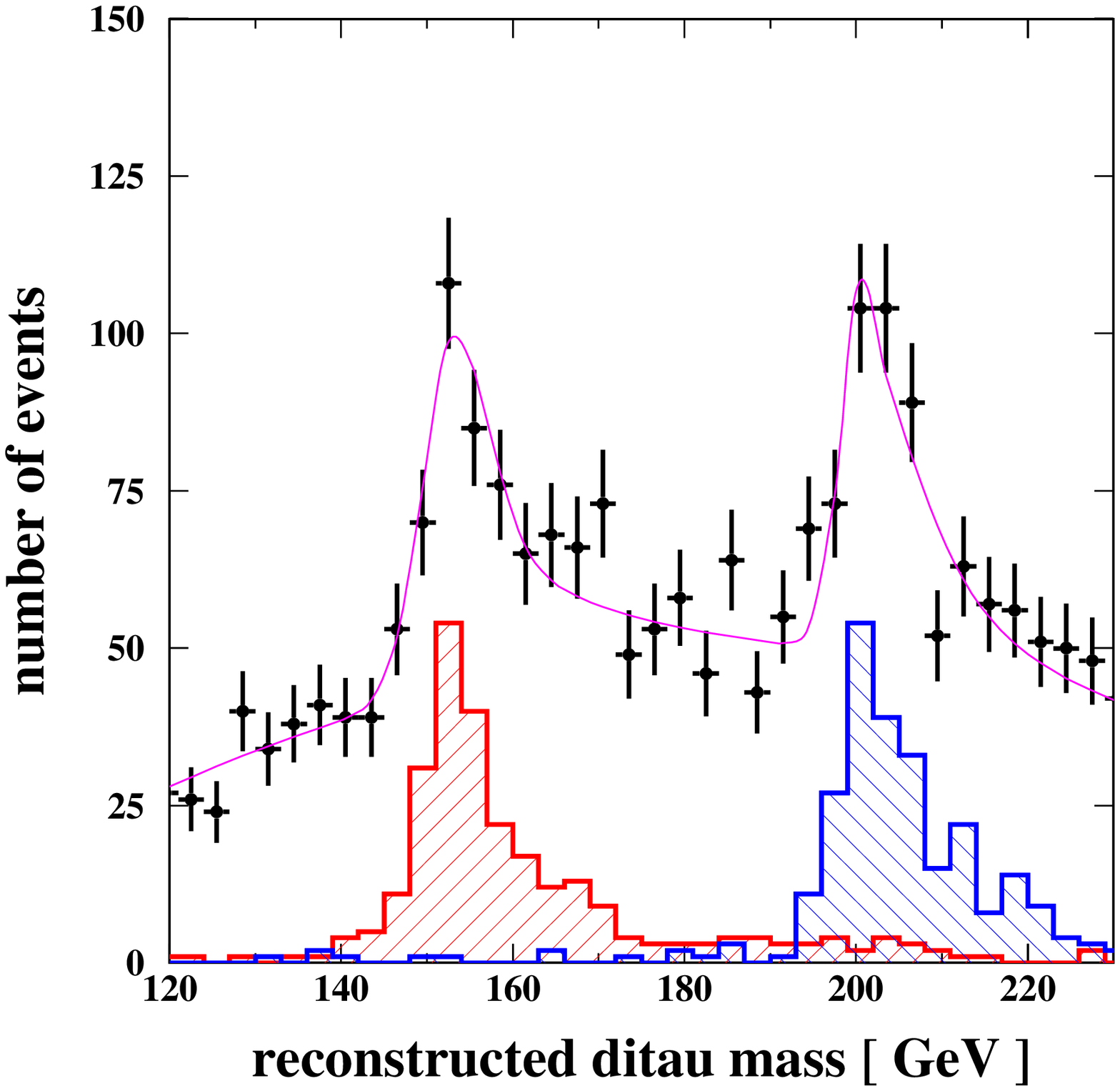,width=0.99\textwidth}
\end{minipage}
\begin{minipage}[c]{0.47\textwidth}
\psfig{figure=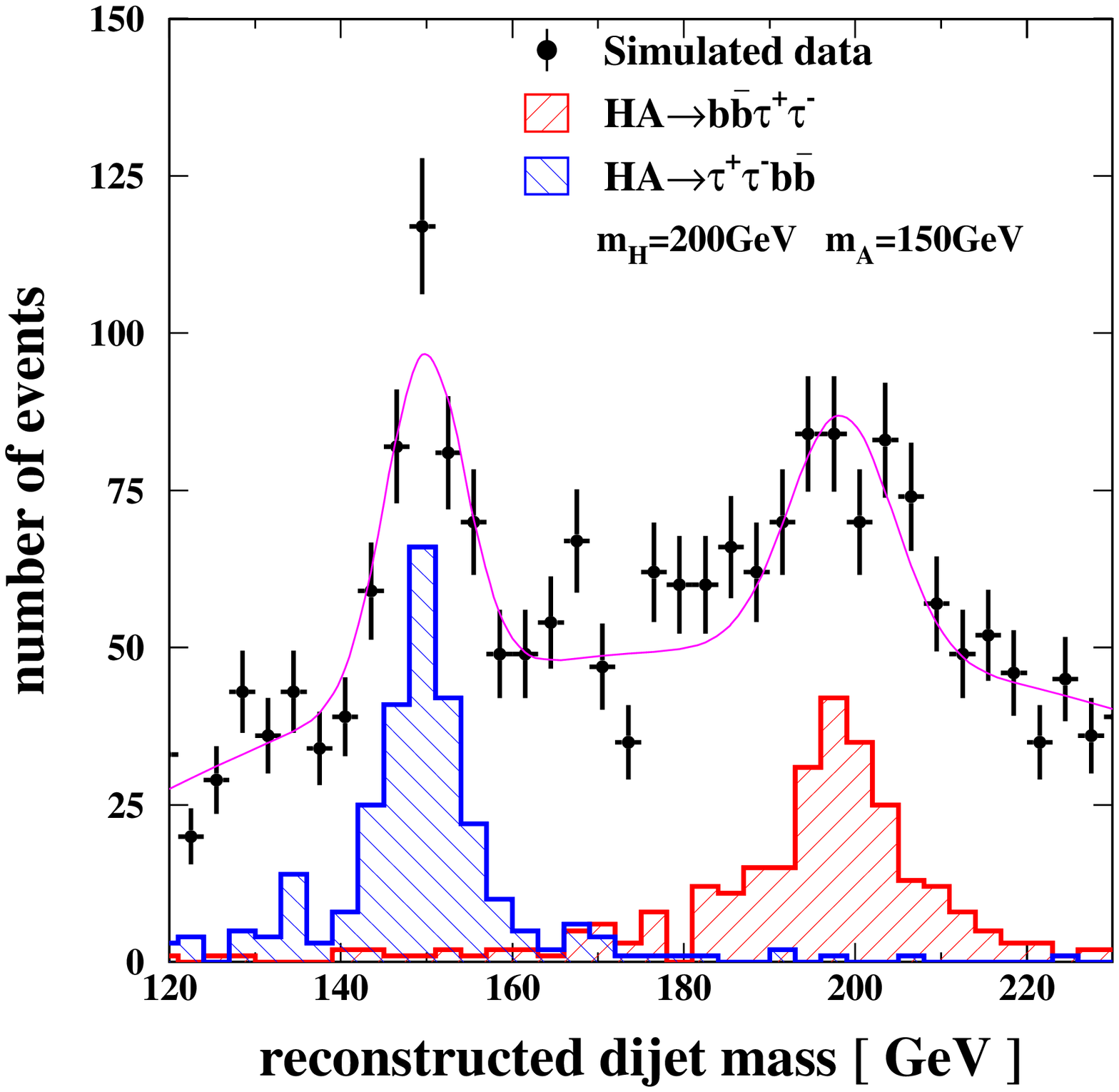,width=0.99\textwidth}
\end{minipage}
\caption{
Distributions of the reconstructed di-tau mass (left plot) and
di-jet mass (right plot) in the 
$HA\to b\bar{b}\tau^+\tau^-,\tau^+\tau^- b\bar{b}$ channels
for ($\mH$,$\mA$) = (200,150) GeV at $\sqrt{s}$ = 500 GeV. 
Assumed integrated luminosity is 500 fb$^{-1}$.
\label{fig:HA_bbtt}
}
\end{figure}
\begin{figure}[htb!]
\begin{minipage}[c]{0.47\textwidth}
\psfig{figure=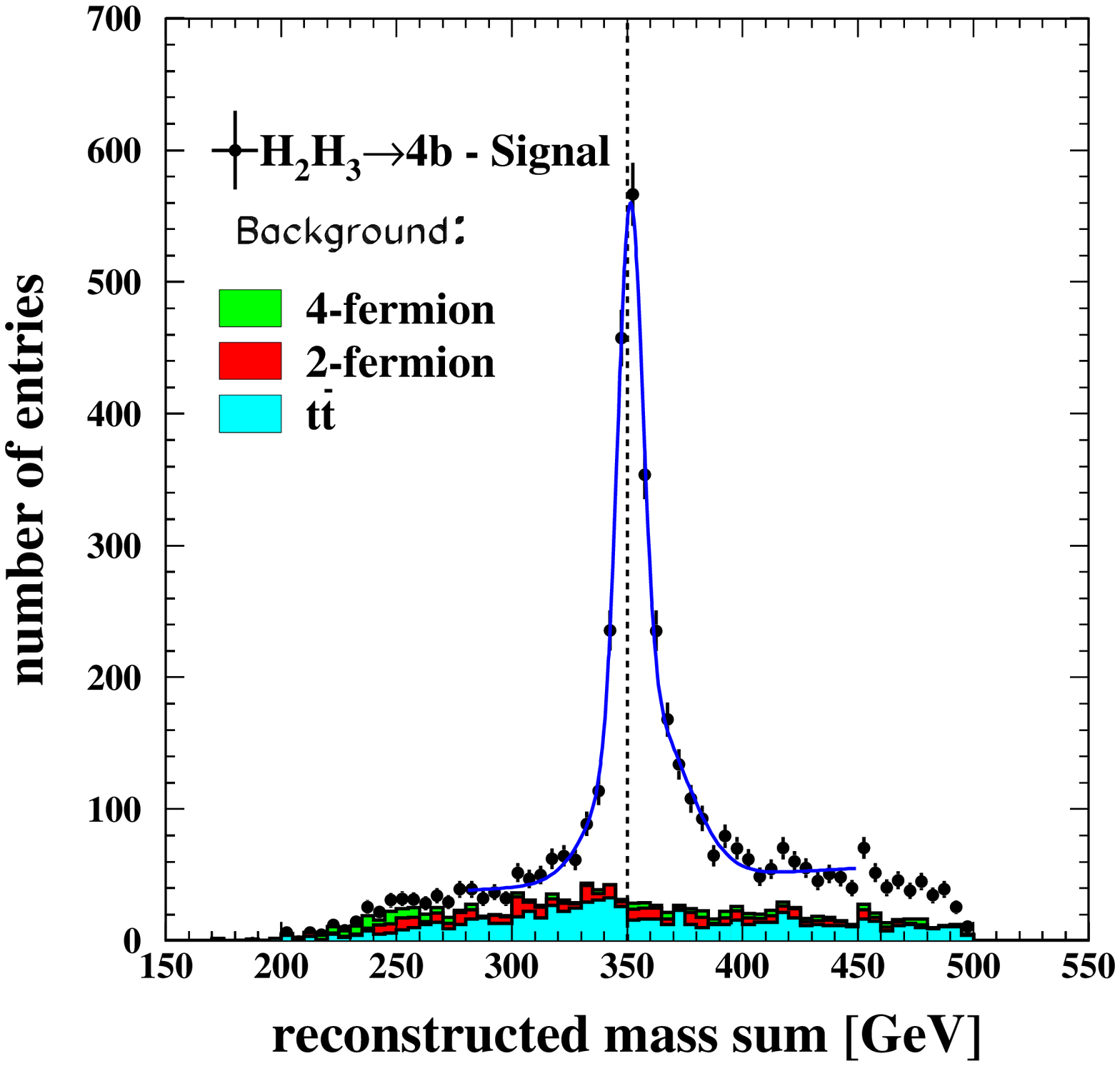,width=0.99\textwidth}
\end{minipage}
\begin{minipage}[c]{0.47\textwidth}
\psfig{figure=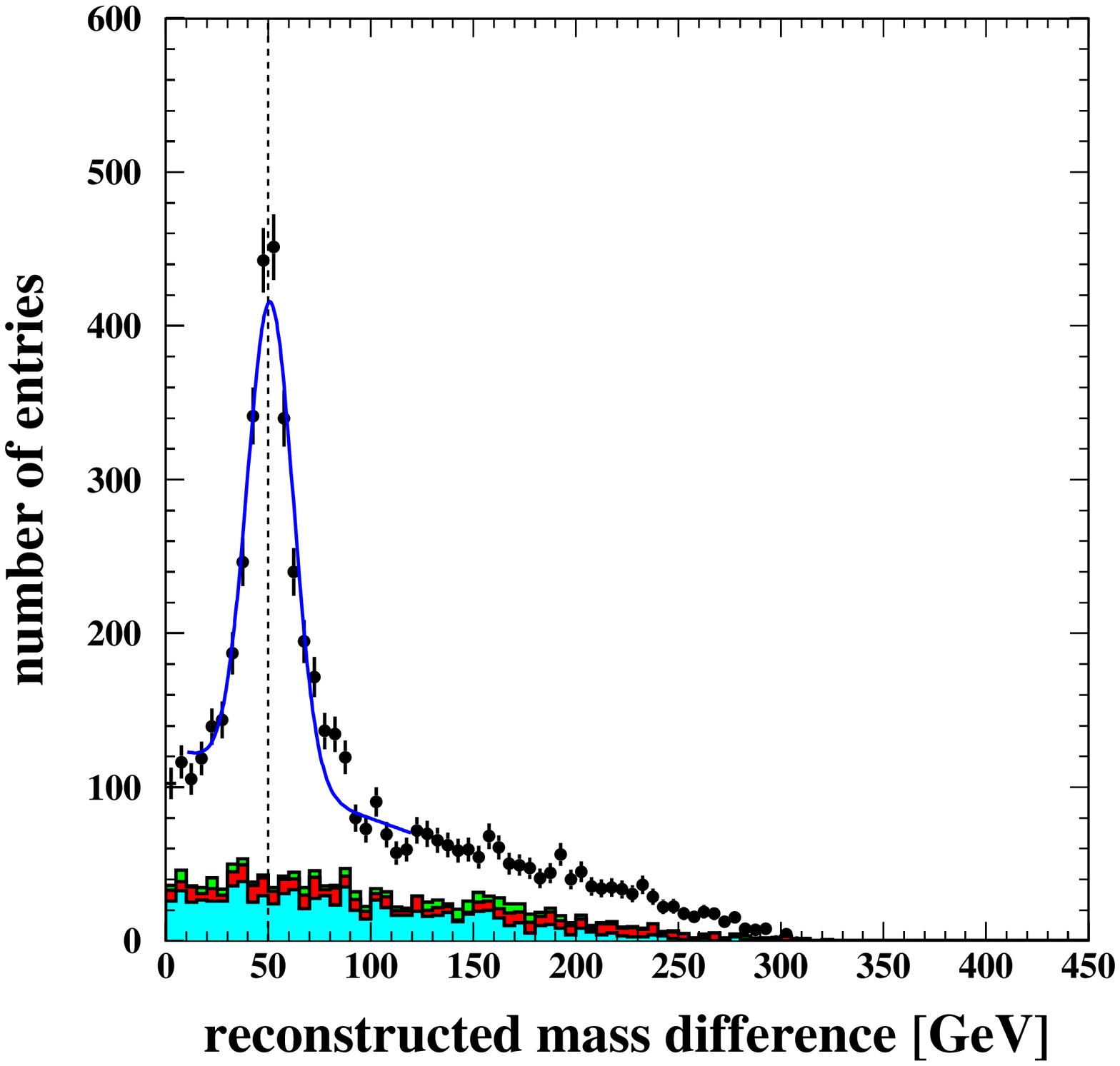,width=0.99\textwidth}
\end{minipage}
\caption{
Distributions of the reconstructed Higgs boson mass 
sum, $\mA+\mH$, (left plot) and
mass difference,  $|\mH-\mA|$, (right plot) in the 
$HA\to b\bar{b}b\bar{b}$ channel
for ($\mH$,$\mA$) = (200,150) GeV at $\sqrt{s}$ = 500 GeV. 
Assumed integrated luminosity is 500 fb$^{-1}$.
\label{fig:HA_bbbb}
}
\end{figure}
%
An integrated
luminosity of 500 fb$^{-1}$ is assumed for each energy. 
The branching fractions of the Higgs bosons into $b$-quarks 
and $\tau$-lepton 
pairs are set to their typical values in the MSSM: 
$\BrHbb$ = $\BrAbb$ = 90\%, $\BrHtt$ = $\BrAtt$ = 10\%.
Furthermore, the maximal allowed cross section for 
$\EEHA$ is assumed, corresponding to $|\sin(\alpha-\beta)|$ = 1.
These assumptions, however, do not restrict the generality of the study,
and results of the analysis can be applied for arbitrary 
values of Higgs branching fractions by appropriate rescaling the expected 
number of signal events in each analyzed channel. 
As an example \reffi{fig:HA_bbtt}
presents reconstructed ditau and dijet mass spectra
in the $b\bar{b}\tau^+\tau^-$ channel for the Higgs boson 
mass hypothesis ($\mH$,$\mA$) = (200,150) GeV. For the same mass 
hypothesis the distribution of the reconstructed Higgs boson
mass sum and mass 
difference in the $b\bar{b}b\bar{b}$ channel is presented 
in \reffi{fig:HA_bbbb}. 
Exploiting these two channels, the precision on
Higgs boson masses and production cross sections 
is evaluated as a function of the Higgs masses. 
Higgs boson masses can be
measured with accuracy ranging from 0.1 to 1 GeV 
for Higgs pair production far above and close to the 
kinematic limit, respectively.
The topological cross section 
$\sigma_{e^+e^-\to HA\to b\bar{b}b\bar{b}}$
can be determined with relative precision of 1.5$-$7\% 
and cross sections 
$\sigma_{e^+e^-\to HA\to b\bar{b}\tau^+\tau^-}$
and $\sigma_{e^+e^-\to HA\to \tau^+\tau^- b\bar{b}}$
with a precision of 4$-$30\%. Furthermore, Higgs boson widths
can be determined with relative accuracy ranging from 20 to 40\%,
depending on the Higgs masses. 
In the case of mass degenerate $H$ and $A$ bosons, 
the 5$\sigma$ discovery reach corresponds to a common 
Higgs boson mass of about 385 GeV in the 
$HA\to b\bar{b}b\bar{b}$ channel at 
$\sqrt{s}$ = 800 GeV, assuming an integrated luminosity 
of 500 fb$^{-1}$. The mass reach for heavy neutral
Higgs bosons can be extended by operating the linear collider at  
higher centre-of-mass energies. 
\reffi{fig:Higgs_SPS1a} shows as an example
the signal in the $HA\to b\bar{b}b\bar{b}$ channel
at $\sqrt{s}$ = 1 TeV
for the Higgs boson mass hypothesis ($\mH$,$\mA$)=(394.6,394.9) GeV, 
which corresponds to the SPS 1a benchmark point. 
With 1 ab$^{-1}$ of data collected at $\sqrt{s}$ = 1 TeV,  
$\mH$ and $\mA$ can be measured with an accuracy of 1.3 GeV, 
and the topological cross section with precision of 9\%~\cite{eeAHdesch}. 

\begin{figure}[htb!]
\begin{minipage}[c]{0.5\textwidth}
\psfig{figure=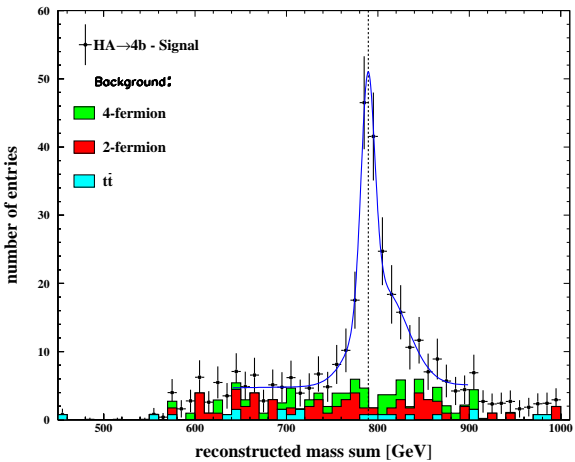,width=0.99\textwidth}
\end{minipage}
\begin{minipage}[c]{0.03\textwidth}
$\phantom{0}$
\end{minipage}
\begin{minipage}[c]{0.43\textwidth}
\caption{
Distribution of the reconstructed Higgs boson 
mass sum, $\mH+\mA$, for the SPS 1a benchmark point. 
Higgs boson mass hypothesis is ($\mH$,$\mA$)=(394.6,394.9) GeV.
Distribution corresponds to an integrated luminosity
of 1 ab$^{-1}$ collected at $\sqrt{s}$ = 1 TeV.
\label{fig:Higgs_SPS1a}
}
\end{minipage}
\end{figure}
%
\begin{figure}[htb!]
\begin{minipage}[c]{0.5\textwidth}
\psfig{figure=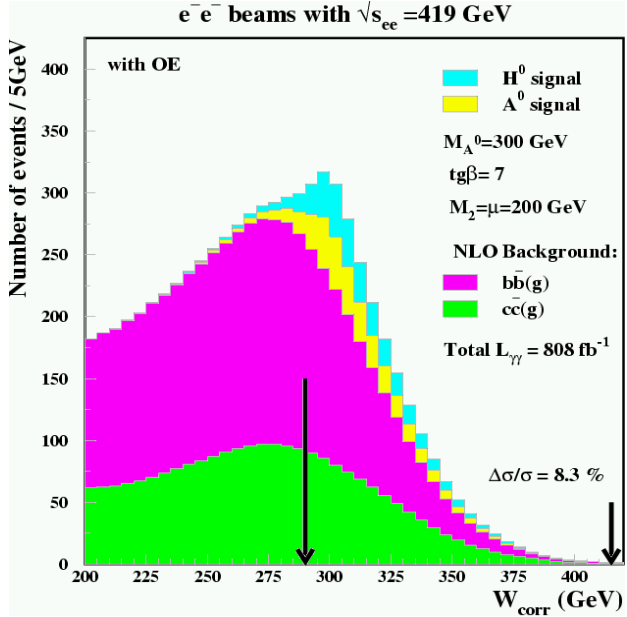,width=0.99\textwidth,
                             height=0.99\textwidth}
\end{minipage}
\begin{minipage}[c]{0.03\textwidth}
$\phantom{0}$
\end{minipage}
\begin{minipage}[c]{0.43\textwidth}
\caption{
Distribution of the invariant mass of the 
hadronic system in the 
$\ga\ga \to H,A\to b\bar{b}$ channel. 
Distribution corresponds to 808 fb$^{-1}$ of luminosity 
collected with photon collider operated at center-of-mass
energy of $\sqrt{s_{ee}}$ = 419 GeV~\cite{gg_hhiggs}.
Higgs boson masses are $\mH$ = $\mA$ = 300 GeV. 
\label{fig:gg_HA}
}
\end{minipage}
\end{figure}

Alternatively, heavy neutral Higgs bosons can be produced 
resonantly in $\GG$ collisions. The main advantage of the
photon collider with respect to the $\EE$ machine is its higher 
mass reach for heavy neutral Higgs particles. Neutral Higgs bosons
with masses up to 350 GeV can be detected with high statistical
significance at a photon collider operated at a modest 
center-of-mass energy of $\sqrt{s_{ee}}$ of 420 GeV~\cite{gg_hhiggs}.
More generally, Higgs bosons can be produced with masses up to 80\%
of the total $e^+e^-$ energy.
An example of the signal in the 
$\ga\ga \to H,A\to b\bar{b}$ 
channel is presented in \reffi{fig:gg_HA}.
With this process a $\ga$C can, e.g., cover the ``LHC wedge'' where
only a SM-like Higgs boson is within the reach of other colliders.

\bigskip
As was discussed above, in the decoupling limit of MSSM one of the
most promising channels to detect heavy neutral Higgs bosons will 
be associated Higgs boson production, $\EEHA$. Unfortunately, 
the mass reach for neutral Higgs bosons in this channel is limited:
the Higgs bosons can be produced only if $\mH+\mA$ $<$ $\sqrt{s}$.
$H$ and $A$ bosons can be also produced in Yukawa processes, such as
$\EEHAbb$. In this case the mass reach is extended almost up 
to collision energy. Although a number of theoretical 
studies have been performed which investigated the prospects of 
detecting heavy neutral Higgs
states at $\EE$ collider in Yukawa processes, no related
experimental analysis including realistic detector simulation 
and SM background estimates has been done so far. We hope that
this issue will be addressed by future studies.


\section{Charged Higgs Bosons}

The charged Higgs boson ($H^\pm$) is clear evidence 
of extended Higgs sectors beyond the SM. 
In the MSSM, the mass of the charged Higgs 
boson is related to that of the CP-odd Higgs boson by 
$m_{H^\pm}^{2}=\MA^2+\MW^2+\delta_{\rm loop}^{}$, 
so that $H^{\pm}$ and $A$ are almost degenerate in the large mass 
limit. By measuring this mass relation, the 
MSSM may be distinguished from the general two Higgs 
doublet model in which no such stringent relationship holds. 
While the derivative couplings 
$V H^+H^-$, $W^\pm H^\mp A$ and $W^\pm H^\mp H$
appear at tree level, there is no tree level 
$H^\pm W^\mp V$ ($V=\gamma$ or $Z$) coupling 
due to the $U(1)_{\rm em}$ gauge invariance and also 
the global $SU(2)_V$ symmetry 
in two Higgs doublet models including the MSSM.

\begin{figure}[htb!]
\begin{minipage}[c]{0.5\textwidth}
\psfig{figure=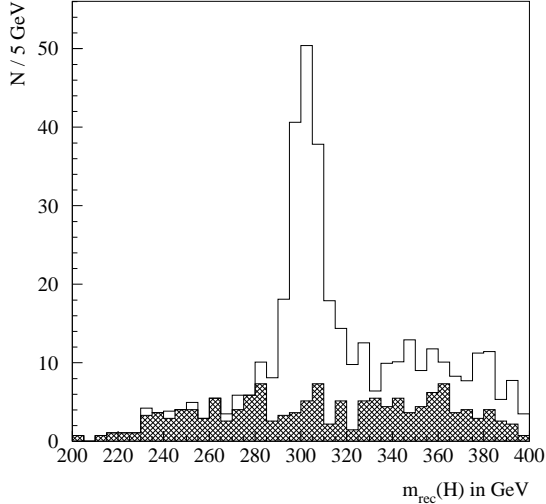,width=0.99\textwidth}
\end{minipage}
\begin{minipage}[c]{0.03\textwidth}
$\phantom{0}$
\end{minipage}
\begin{minipage}[c]{0.43\textwidth}
\caption{Fitted charged Higgs boson mass for $H^+H^- \to (t \bar b)(\bar t b)$
with $m_{H^\pm}^{}=300 \gev$. The histogram is normalized 
to an integrated luminosity of 1 ab$^{-1}$, with a 100\% 
branching ratio in the analyzed decay mode. The contribution
of the $tbtb$ background events is shown by the dark histogram. 
}
\label{fig:chiggs-1}
\end{minipage}
\end{figure}
%
At the ILC, a relatively light charged Higgs boson
is produced in pairs, $e^+e^- \to H^+H^-$, as long as 
the mass is smaller than one half of the collider energy, 
$m_{H^\pm} < \sqrt{s}/2$. 
A complete simulation of this process with 
the decay $H^+ \to t \bar b$ and $H^- \to b \bar t$ 
has been performed in Ref.~\cite{Battaglia:2001be}
for $\sqrt{s}=800 \gev$. 
The mass reconstruction is done by selecting 
the final state to be the four $b$ jets 
plus four non-$b$ tagged jets from the two $W$ bosons from top decay. 
The expected signal and background are shown in 
Fig.~\ref{fig:chiggs-1} for $m_{H^\pm}=300 \gev$, 
assuming the integrated luminosity to be $1$ ab$^{-1}$. 
The mass resolution is approximately 1.5\%.
A 5$\sigma$ level discovery will be possible 
for $m_{H^\pm}^{} < 350 \gev$ for $\sqrt{s} = 800 \gev$.

While the mass reach of pair production is limited to 
$\sqrt{s}/2$, the rare processes of single charged Higgs 
production may be useful for the search of heavier charged 
Higgs bosons. 
The dominant processes are 
$e^+e^- \to b \bar t H^+$, 
$e^+e^- \to \tau \bar \nu_\tau H^+$, 
$e^+e^- \to W^- H^+$ and $e^+e^- \to H^+e^-\bar\nu_e$ 
({\rm one-loop induced}). 
Their production rates have been calculated at leading order in 
\citeres{eeHW,eeenH,single_hpm_xsec,single_hpm_xsec2}.  
QCD corrections to $e^+e^- \to b \bar t H^+$ have been studied 
in \citere{single_hpm_xsec2}, and have turned 
out to be sizable. For more details about the progress in the cross
section calculations, see \refse{sec:2to3}. 
In Ref.~\cite{Moretti:2003cd}, the simulation 
for the process $e^+e^-\to t\bar b H^- + ~{\rm{c.c.}}
\to 4b +{\rm{j}}{\rm{j}}+ \ell + p_T^{\rm{miss}}$ 
($\ell=e,\mu$) has been performed for a charged Higgs 
boson mass above half of the collider energy, 
where the pair production channel $e^+e^-\to H^-H^+$ is no 
longer available. 
It has been shown that 
one can establish a statistically significant 
$H^\pm$ signal only over a rather limited mass region, 
of 20 GeV or so beyond $M_{H^\pm}\approx \sqrt s/2$, 
for very large or very small values of $\tan\beta$
and provided high $b$-tagging efficiency can be achieved; 
see \reffi{fig:chiggs-2}. 
%
\begin{figure}[htb!]
\begin{minipage}[c]{0.5\textwidth}
\epsfig{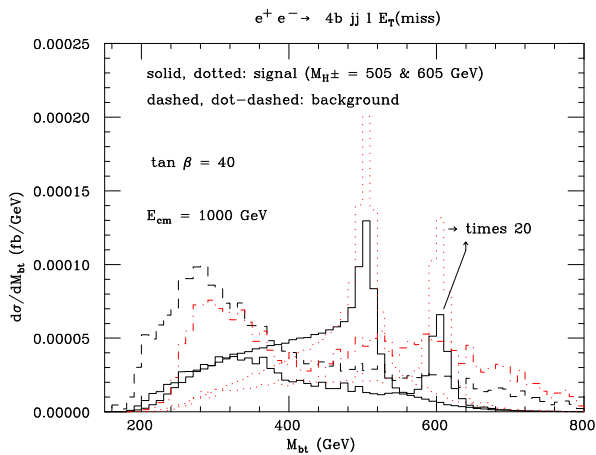}
\end{minipage}
\begin{minipage}[c]{0.03\textwidth}
$\phantom{0}$
\end{minipage}
\begin{minipage}[c]{0.43\textwidth}
\caption{(Solid and dashed lines) 
Differential distribution in the reconstructed Higgs mass 
from both $b$-jets not generated in top decays and the two 
top systems for the signal 
$e^+e^- \to b \bar t H^+ + t \bar b H^- \to t \bar t b \bar b$
and the background 
$e^+e^- \to t \bar t g^\ast \to t \bar t b \bar b$, 
yielding the signature,
after the
kinematic cuts, including 
all decay BRs. 
(Dotted and dot-dashed lines)
Same as above, but using only 
the $b$-jet with highest transverse momentum of the two 
not generated in top decays. See Ref.~\cite{Moretti:2003cd} for details.}
\label{fig:chiggs-2}
\end{minipage}
\end{figure}


\bigskip
The situation is different for charged Higgs masses smaller than
$\mt$. A precise measurement 
of $M_{H^{\pm}}$ in that range is a challenging task for any collider.  
In this case the charged Higgs decays dominantly to a tau lepton and neutrino 
($H^{\pm} \to \tau^{\pm} \nu$), and it is impossible to reconstruct directly 
the invariant mass of the di-tau final state.  However, due to the
polarization of the $\tau^{\pm}$ leptons, the energy of the $\tau^{\pm}$
decay products depends strongly on $M_{H^{\pm}}$, a feature that 
can be exploited to extract $M_{H^{\pm}}$ indirectly.  
The main background to a $H^{\pm}$ signal in the $\tau^{\pm}$ decay 
mode comes from the $W$-boson decays $W^\pm \to \tau^\pm \nu$. 
However, thanks to the different structure of the $H^{\pm}$ and 
$W^{\pm}$ electroweak interactions to $\tau^{\pm}$-leptons, 
the tau polarization is very different which will allow a separation
of $H^{\pm}\to\tau^{\pm}\nu$ and $W^{\pm}\to\tau^{\pm}\nu$
on a statistical basis. More specifically, the $\tau^{\pm}$ decay products
($\tau^+ \to \pi^+ \bar{\nu}$, or $\tau^+ \to \rho^+ \bar{\nu}$, etc.) 
have strikingly different topologies according as to whether they 
originate from a parent $W^{\pm}$ or $H^{\pm}$.

The key point is that $\tau^-$ leptons arising from 
$H^- \to \tau^- \nu$ decays are almost purely right-handed, 
in contrast to the left-handed $\tau^-$ leptons which arise from $W^-$ 
decays. This contrast follows from the helicity flip nature of 
the Yukawa couplings of the Higgs fields, and the helicity 
conserving nature of the gauge interactions.
The most dramatic difference is seen in the energy distribution for
the single pion channel ($H^+/W^+ \to \tau^+ \nu \to \pi^+ \nu \bar{\nu}$)
in the rest frame of the parent boson ($W^{\pm}$ or $H^{\pm}$)~\cite{BBCW}.
In practice, it is nearly impossible to reconstruct the rest frame of the
$W^+$ or $H^+$ bosons because the momentum of the neutrinos cannot be
measured. Instead, one can use the top quark pair production channel
with top quark decays ($t \to H^+ b$) and reconstruct the top quark
rest frame. 

The distributions for the energy of the single pion in the top rest
frame has the following form for $H^{\pm}$ and
$W^{\pm}$ cases, respectively~\cite{sfermion,sfermion2,sfermion3},
\begin{eqnarray}
\frac{1}{\Gamma}\frac{d\Gamma}{dy_{\pi}}=\frac{1}{x_{max}-x_{min}}
\left\{
\begin{array}{cc}
(1-P_{\tau})log\frac{x_{max}}{x_{min}} +
2P_{\tau}y_{\pi}(\frac{1}{x_{min}}-\frac{1}{x_{max}}),
&0<y_{\pi}<x_{min}\\
(1-P_{\tau})log\frac{x_{max}}{y_{\pi}} +
2P_{\tau}(1-\frac{y_{\pi}}{x_{max}}),
&x_{min}<y_{\pi}
\end{array}
\right.
\end{eqnarray}
where $y_{\pi}=\frac{2E_{\pi}^{top}}{M_{top}}$,
$x_{min}=\frac{2E_{\tau}^{min}}{M_{top}}$, 
$x_{max}=\frac{2E_{\tau}^{max}}{M_{top}}$, 
$E_{\tau}^{min}=\frac{M_{R}^2}{2M_{top}}$,
$E_{\tau}^{max}=\frac{M_{top}}{2}$.
For the $W$~boson, $P_{\tau} =-1$, and for the charged Higgs boson,
$P_{\tau} = 1$.

The energy distribution for a pion from $H^{\pm}$ decay has a maximum 
at the point 
$E(\pi^{\pm})={M_{H^{\pm}}}^2/(2M_t)$. 
This dependence allows one to extract $M_{H^{\pm}}$ from the shape of
the spectrum. 

The QCD corrected top-quark decay width 
(in terms of $q_{H^+} = m_{H^+}^2/m_t^2$) is given by~\cite{deltamb2},
\begin{eqnarray}
\Gamma_{QCD}^{imp}(t \to bH^+)=\frac{g^2}{64\pi M_W^2}m_t(1-q_{H^+})^2 \bar{m_b}^2(m_t^2) \tan^2\beta \times\nonumber\\
\left\{ 1+\frac{\alpha_S(m_t^2)}{\pi} 
\left[ 7-\frac{8\pi^2}{9}-2\log(1-q_{H^+})+
2(1-q_{H^+})\right.\right.\nonumber\\
\left.\left.+\left(\frac{4}{9}+\frac{2}{3}\log(1-q_{H^+})\right)(1-q_{H^+})^2 
\right] 
\right\} 
\end{eqnarray}
The $\Delta m_b$ corrections have been included using 
{\tt CPsuperH}~\cite{cpsh}. 
The numerical simulations have been performed assuming a center of mass
energy $\sqrt{s} = 500$~GeV and a total integrated luminosity
${\cal{L}} = 500$~fb$^{-1}$.
We have performed detailed computations and Monte Carlo simulations for
different sets of MSSM parameters, all leading to light charged
Higgs bosons~\cite{BBCW}. Here we present two scenarios,  
based on the mass parameters
$M_Q = M_U = M_D = 1$~TeV, $M_{\widetilde{g}} = M_2 = 1$~TeV,
$A_t = 500$, $\mu = \pm 500$~GeV, and $\tan\beta = 50$, which give
$M_{H^{\pm}} = 130$~GeV.  The branching ratio BR$(t\to H^\pm b)$
will be enhanced or suppressed depending on the sign of~$\mu$.
The branching ratios
for these examples are
\begin{center}
\begin{tabular}{crrl}
{\em (i)}   & $M_{H^\pm} = 130$~GeV, &
 \hspace*{10pt} $\mu < 0$ and $\tan\beta=50$: \hspace*{20pt}& 
  $BR(t\to H^+ b) = 0.24$ \cr
{\em (ii)}  & $M_{H^\pm} = 130$~GeV, &
 \hspace*{10pt} $\mu > 0$ and $\tan\beta=50$: \hspace*{20pt}& 
  $BR(t\to H^+ b) = 0.091$ \cr
\end{tabular}
\end{center}
The couplings have been implemented in CompHEP~\cite{CompHEP}, 
which has been used to compute the cross sections for signal and
background processes, including decays of top to $W^\pm$ and $H^\pm$
which subsequently decay to polarized $\tau^\pm$~leptons.
CompHEP was also used to generate events, and effects from initial
state radiation and Beamstrahlung were included. Polarized $\tau^\pm$ 
decays have been simulated using TAUOLA~\cite{tauola} interfaced to 
CompHEP. Hadronization and energy smearing in the final state are 
accounted for by means of PYTHIA~\cite{PYTHIA} using the
CompHEP-PYTHIA~\cite{interface} interface based on Les Houches 
Accord~\cite{Boos:2001cv}. Effects from final state radiation
have been implemented using the PHOTOS library~\cite{photos}.

A brief description of the fitting procedure is given in Ref~\cite{BBCW}.
We use the method of maximum likelihood to fit a spectrum created
from simulated signal and background events, and obtained the
following results for the cases described above:
\begin{center}
\begin{tabular}{rl}
{\em (i)}     & $M_{H^{\pm}}= 129.7 \pm 0.5$~GeV, \cr
{\em (ii)}    & $M_{H^{\pm}}= 129.4 \pm 0.9$~GeV,  \cr
\end{tabular}
\end{center}
Figure~\ref{fit} shows results of the simulations and fits.
The charged Higgs mass may be determined, in both cases,
with an uncertainty of the order of 1~GeV. 
This study is a theoretical
level analysis, and no systematics or detector effects have
been included.  
A full detector simulation will be necessary to test the 
robustness of this result.

\begin{figure}[htb!]
\centering
\includegraphics[width=75mm]{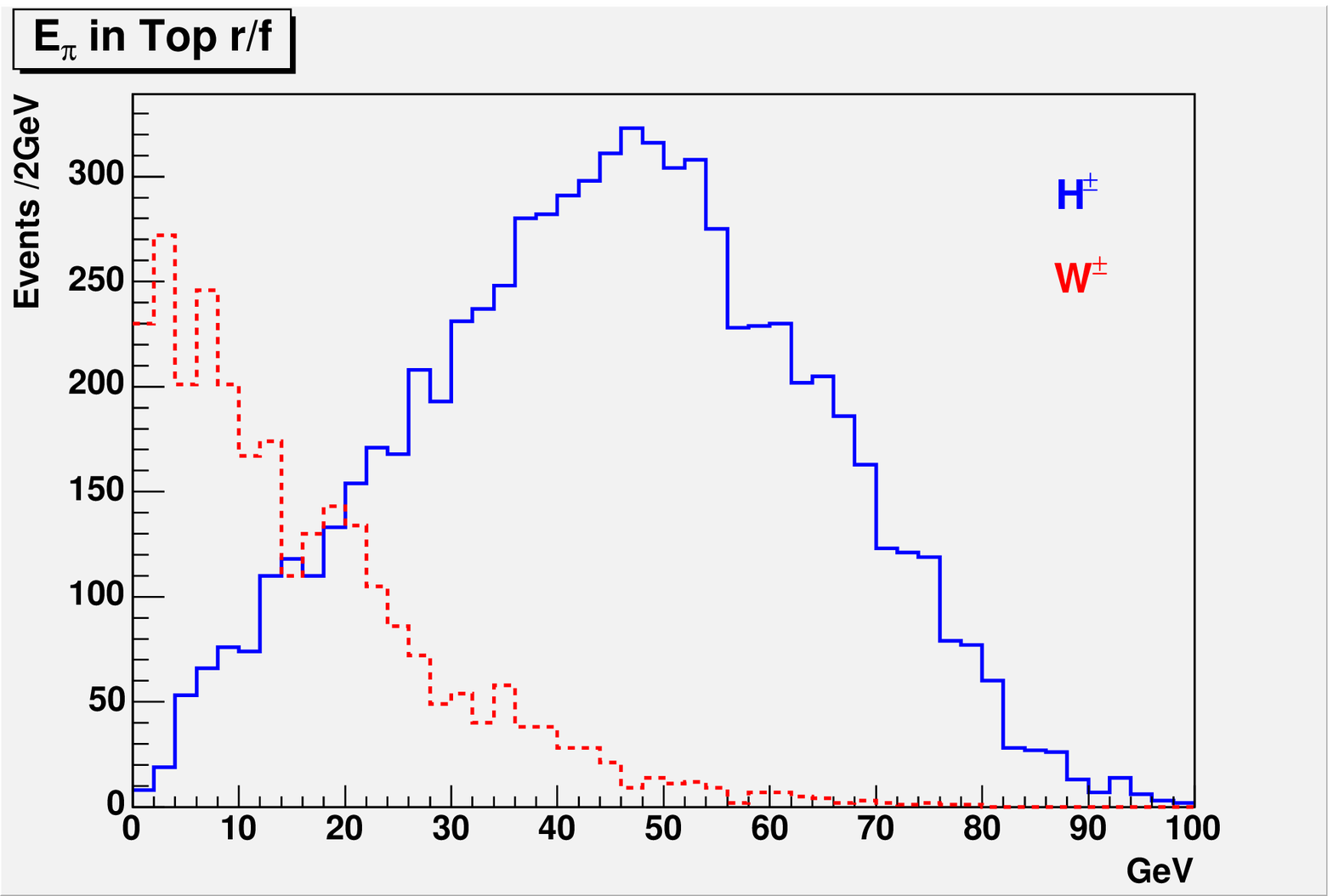}
\includegraphics[width=75mm]{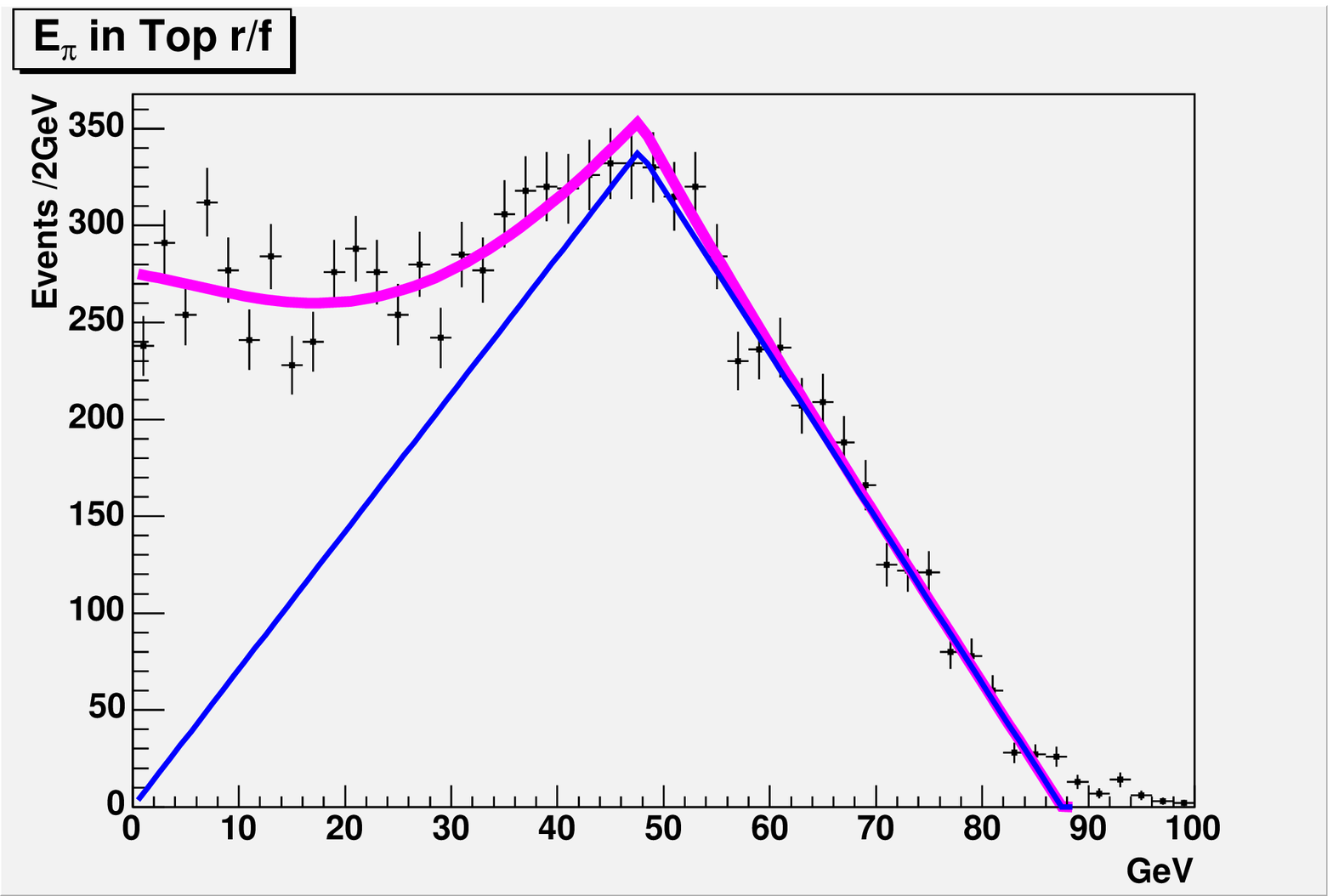}
\includegraphics[width=75mm]{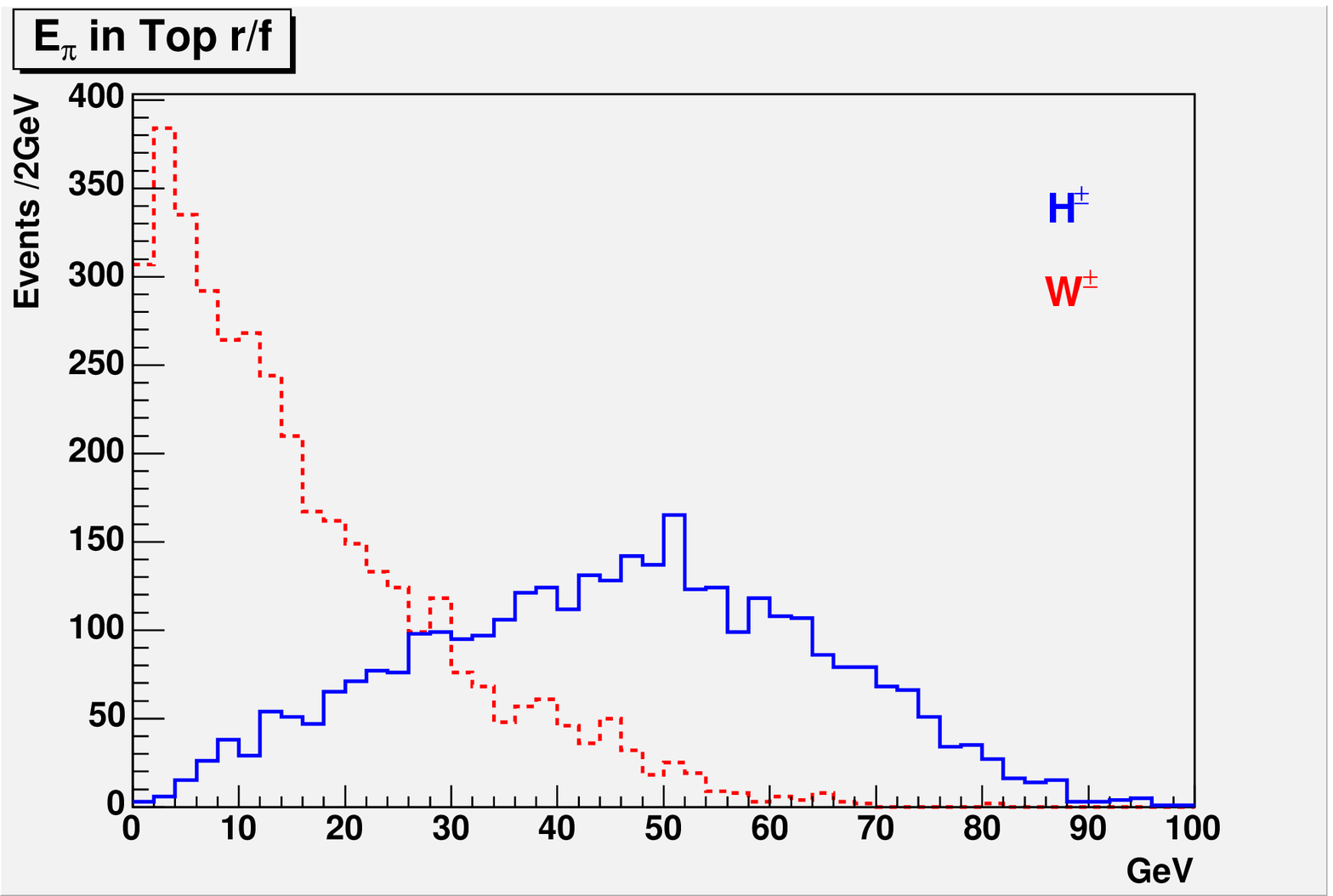}
\includegraphics[width=75mm]{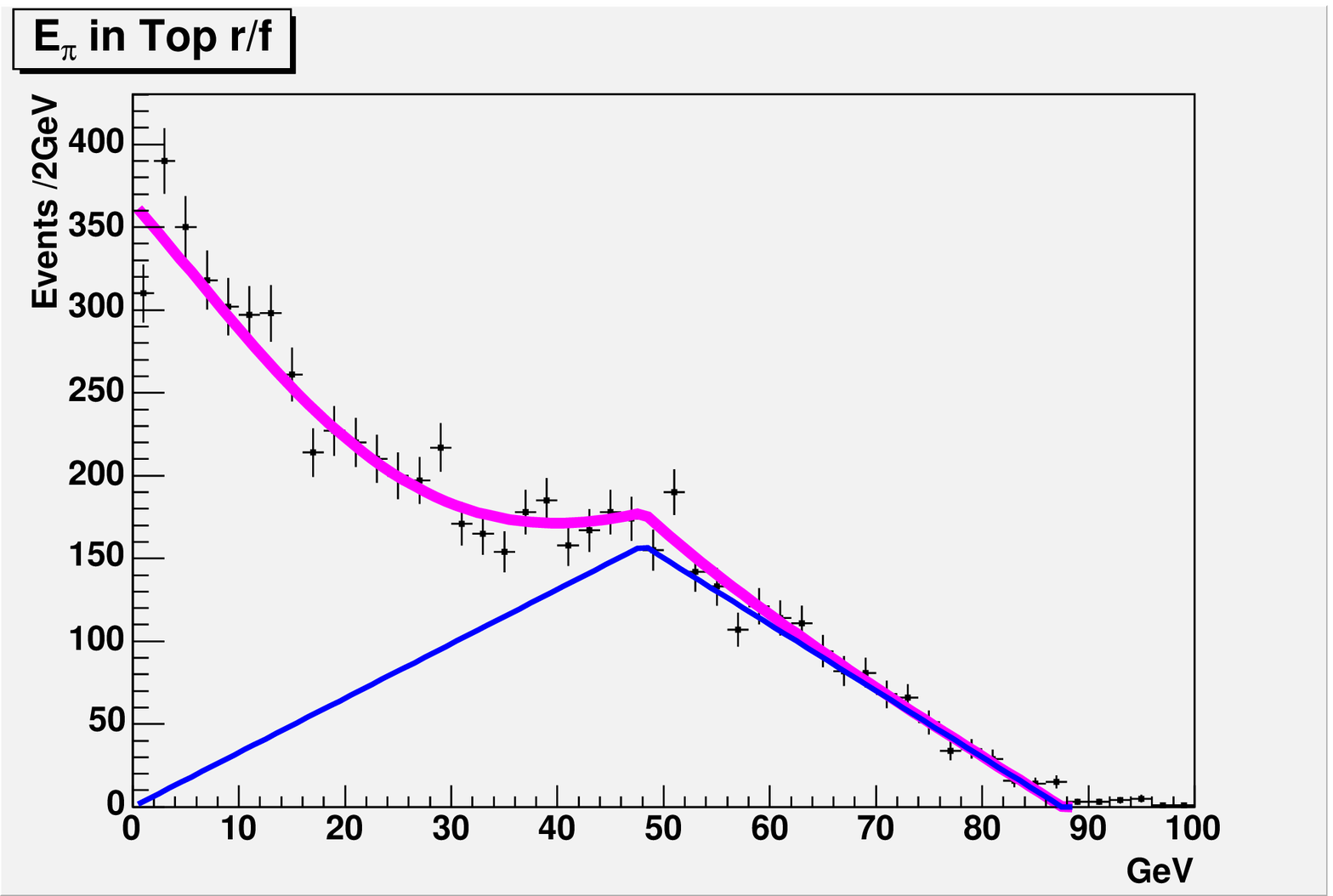}
\caption[.]{Generated $\pi^{\pm}$ energy spectra (left plots)
and the fit (right plots) for the two sets of MSSM parameters
described in the text.  \label{fit}}
\end{figure}


\section{Measuring $\tb$ in $\ga \ga$ Collisions}

The measurement of the mixing parameter $\tb$, 
one of the fundamental parameters in the Higgs sector of 
supersymmetric theories, is a difficult task, especially if $\tb$ is large.
In this section two methods are described to measure $\tb$ at the
$\ga$C. 

\smallskip
$\tau\tau$ fusion to Higgs bosons at a photon
collider~\cite{2a} can provide a valuable method for measuring
this parameter.  
The couplings in the limit of large $\tb$ (and moderate values of
$\MA$ in the case of the $h$) are given by (see e.g.\ \citere{gomez-bock})
\begin{eqnarray}
g_{\Phi\tau\tau}& =& \tb
{\rm ~~~~~~~~~~ for ~} \Phi=A   \nonumber\\
g_{\Phi\tau\tau}&\simeq& \tb
{\rm ~~~~~~~~~~ for ~} \Phi=h,H
\end{eqnarray}
[normalized to the Standard Model Higgs coupling
to a tau pair, $m_\tau/v$].
Thus the entire Higgs mass range up to the kinematical limit  
can be covered for large $\tb$ by this method.    

We consider the process
\begin{equation}
\ga \ga \to \tau\tau \; \tau\tau \to \tau \tau + h/H/A~.
\end{equation}
For large $\tb$ all
the Higgs bosons decay almost exclusively [80 to 90\%]
to a pair of $b$ quarks, and the final state consists of a pair of
$\tau$'s and a pair of resonant $b$ quark jets.
The main background channels 
are $\tau^+ \tau^-$ annihilation into a pair of $b$-quarks,
via $s$-channel $\ga/Z$ exchanges, naturally suppressed
as a higher-order electroweak process, and
diffractive $\ga \ga \to
(\tau^+ \tau^-) (b\bar{b})$ events, reduced kinematically 
by choosing proper cuts. The SUSY parameters have been chosen
according to the SPS~1b scenario~\cite{sps}, but $\MA$ has been varied.

\begin{figure}[htb!]
\begin{center}
\epsfig{figure=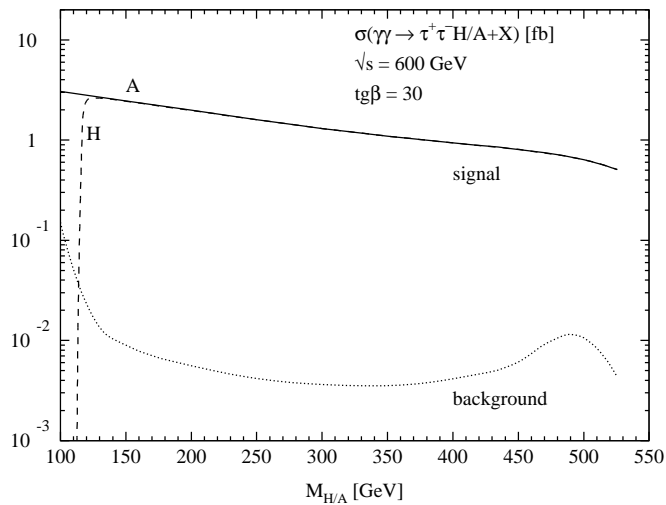,height=7.1cm}\\
\end{center}
\caption{The cross sections for the production of the $H/A$ 
  Higgs bosons in the $\tau\tau$ fusion process 
  at a $\ga\ga$ collider for
  $\tb=30$. The other parameters have been chosen according to the
  SPS~1b scenario (but with $\MA$ varied). 
 Also shown is the background cross section for 
  properly chosen experimental cuts.
  $\sqrt{s}$ denotes the $\ga\ga$ collider c.m.~energy, corresponding to
  approximately 80\% of the $e^\pm e^-$ linear collider energy.}
\label{fig:xsec}
\end{figure}

The cross
sections for the signals of $H$ and $A$ Higgs-boson production
in the $\tau\tau$ fusion process, together with the background 
processes, are presented in Fig.~\ref{fig:xsec}, including proper cuts.
The cross section for $\tau\tau$ fusion to the light Higgs boson $h$ 
is of the same size.

\begin{table}[htb!]
\begin{center}
\begin{tabular}{|c||c|cc||c|cccc|}
\hline\rule{0cm}{4mm}
 &\multicolumn{3}{|c||}{$E_{\ga\ga}=400 \gev$} 
 &\multicolumn{5}{|c|}{$E_{\ga\ga}=600 \gev$}\\
\hline
\rule{0cm}{5mm}$M_{\rm Higgs}$ & $A\oplus h$ &
\multicolumn{2}{|c||}{$A\oplus H$}& 
$A\oplus h$ &\multicolumn{4}{|c|}{$A\oplus H$}\\ 
 \phantom{i}[GeV]  
&  100  &  200  &  300  & 100   &  200  &  300  &  400  &  500  \\
\hline
\rule{0cm}{5mm}$\tb$ & \scriptsize{I}   & \scriptsize{II}   & 
              \scriptsize{III} &
              \scriptsize{IV}  & \scriptsize{V}    & \scriptsize{VI}  &
              \scriptsize{VII} & \scriptsize{VIII}     \\
\hline\hline
\rule{0cm}{5mm}10  & 8.4\% & 10.7\% & 13.9\% & 8.0\% & 9.0\% & 
11.2\% & 13.2\% & 16.5\% \\
30  & 2.6\% & 3.5\% & 4.6\% & 2.4\% & 3.0\% & 3.7\% & 4.4\% & 5.3\% \\
50  & 1.5\% & 2.1\% & 2.7\% & 1.5\% & 1.8\% & 2.2\% & 2.6\% & 3.2\% \\
\hline
\end{tabular}
\caption{Relative errors $\Delta\tb/\tb$ on $\tb$ 
in measurements for $\tb=$ 10, 30 and 50, based on: 
combined $A\oplus h$ [I,IV] and $A\oplus H$ [II,III,V--VIII] production, 
assuming [$E_{\ga\ga}= 400 \gev$, ${\cal L}= 100$~fb$^{-1}$]
and  [$E_{\ga\ga}= 600 \gev$, 
${\cal L}= 200$~fb$^{-1}$]. Proper cuts and efficiencies are applied 
on the final--state $\tau$'s and $b$ jets.}
\label{tab:stat-error}
\end{center} 
\end{table}

The statistical accuracy with which large $\tb$ values can be measured in
$\tau\tau$ fusion to $h/H/A$ Higgs bosons is exemplified for three 
$\tb$ values, $\tb=10$, 30 and 50, in Table~\ref{tab:stat-error}. 
Results for scalar $H$ production are 
identical to pseudoscalar $A$ in the mass range above 120 GeV. 
The two channels $h$ and 
$A$, and $H$ and $A$ are combined in the overlapping mass ranges in which
the respective two states cannot be discriminated. 
In Table~\ref{tab:stat-error}
the relative errors 
$\Delta\tb/\tb$ are presented.
Since in the region of interest 
the $\tau\tau$ fusion cross sections are proportional to $\tan^2\beta$ and 
the background is small, the absolute errors $\Delta\tb$ are nearly 
independent of $\tb$, varying between 
\begin{eqnarray}
\Delta\tb \simeq 0.9 {\;\;\;\rm and \;\;\;} 1.3          \label{deltat}
\end{eqnarray}
for Higgs mass values away from the kinematical limits.

This analysis compares favorably well with the
corresponding $b$-quark fusion process at the LHC \cite{R6}.  Moreover, the 
method can be applied readily for a large range of Higgs mass values and thus 
is competitive with complementary methods in the $e^+e^-$ mode of a linear 
collider \cite{R5}.

\bigskip
An alternative method to measure $\tan\beta$ is provided by the
associated $tH^\pm$  
production in $\gamma\gamma$ collisions~\cite{Doncheski:2003te}.  
The subprocess $b\gamma \to H^- t$ utilizes the $b$-quark content of the 
photon.  $\tan\beta$ enters through the $tbH^\pm$ vertex,
so that the amplitude 
squared for the subprocess $b\gamma\to t H^-$ is given by: 
\begin{equation}
\sum |M(b\gamma \to t H^\pm) |^2  \propto
\sqrt{2}G_F \pi\alpha (m_b^2 \tan^2\beta + m_t^2/\tan^2\beta).
\end{equation}
Fig.~\ref{fig:godfrey-gaga} shows the 
cross section as a function of $\tan\beta$ with the measurement 
precision superimposed.  
It has been found that  $\gamma\gamma\to tH^\pm +X$ can 
be used to make a good determination of $\tan\beta$ for most of the parameter 
space with the exception of the region around $\tan\beta\simeq 7$
where the cross section is at a point of inflection.  This measurement 
provides an additional constraint on $\tan\beta$ which 
complements other processes.   

\begin{figure}[htb!]
\begin{minipage}[c]{0.5\textwidth}
\epsfig{file=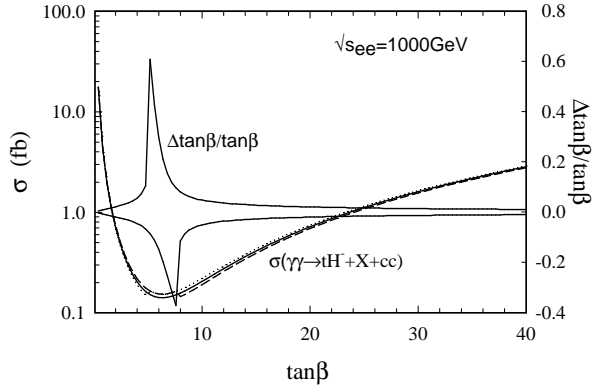,width=0.95\linewidth,clip=}\\
\end{minipage}
\begin{minipage}[c]{0.03\textwidth}
$\phantom{0}$
\end{minipage}
\begin{minipage}[c]{0.43\textwidth}
\caption{
$\sigma(\gamma\gamma\to t H^- +X)$ vs. $\tan\beta$ for the 
backscattered laser case and the sensitivities to $\tan\beta$ based 
only on statistical errors (solid lines) for $\sqrt{s}_{ee}=1$~TeV and 
$M_H = 200$~GeV.  For the cross sections, the solid 
line represents the expected cross section at the nominal value of 
$\tan\beta$, while the dashed (dotted) line represents the expected cross 
section at $\tan\beta - \Delta\tan\beta$ ($\tan\beta + \Delta\tan\beta$).
From~\citere{Doncheski:2003te}.} 
\label{fig:godfrey-gaga}
\end{minipage}
\end{figure}



\section{Measurements in Exotic Scenarios}

An interesting possibility is that the Higgs decays into particles not
present in the Standard Model.  In particular, if there are
exotic light objects which did not obtain their mass directly from the Higgs
mechanism, their couplings to the Higgs may be large despite their small
masses, and they may dominate over the standard Higgs decay modes.  Such a
scenario depends very strongly on how these new objects manifest themselves
inside the detector, and can play a very significant role in how well both the
LHC and the ILC can determine the Higgs properties.  In fact, the tiny
expectation for the SM Higgs width (provided $\MH \lsim 150 \gev$) implies
that the Higgs is particularly sensitive to new decay modes, even if they are
not very strongly coupled.

\subsection{Higgs decaying into unflavored jets}

The first such scenario arises when the Higgs decays primarily into light
hadrons, without any particular bottom or charm content.  This actually occurs
in some limits of the minimal supersymmetric standard model
\cite{Dobrescu:2000jt}.  If, for example,
the pseudo-scalar Higgs $A$ is light enough, the dominant decay mode of the
SM-like Higgs may be $H \rightarrow AA$.  If the $A$ mass is less than $2 m_b$,
and $\tan \beta$ is larger than 10 or so, the dominant $A$ decay will be into
strange quarks, with separation too small to be resolved as separate jets.
Thus, the decay $H \rightarrow AA \rightarrow s \bar{s} s \bar{s}$ results
in two hard jets of energy roughly $\MH /2$ in the Higgs rest 
frame\footnote{This situation is also likely to arise in the NMSSM for regions
of parameter space with small fine-tuning \cite{Dermisek:2005ar}.}.

A second possibility is that the Higgs decays into down-type squarks
\cite{Berger:2002vs}.  The
most likely candidate is the scalar bottom $\tilde{b}$, which could further
decay into a pair of anti-quarks 
through a baryon-number violating interaction of
the type $UDD$.  Again, if the mass of the $\tilde{b}$ is relatively small,
the two quarks may not be sufficiently separated to be resolved as individual
jets, and the Higgs decays effectively into two jets.  Such a light
bottom squark was postulated in \cite{Berger:2000mp} 
to address an excess in the
open bottom quark production cross section at run I of the Tevatron.
The MSSM
Higgs width into a pair of scalar bottoms can be compared with the SM width
into $b \bar{b}$ as,
\BEA
\frac{\Gamma_{\tilde b}}{\Gamma_b}
= \frac{(\mu \tan \beta)^2}{2 m_h^2} \sin^2 2\theta_b
\left( 1 - 4 \frac{m_{\tilde b}^2}{m_h^2} \right)^{\frac{1}{2}}
\EEA
where $\mu$ is the higgsino mass parameter, $\tan \beta$ is the ratio of
Higgs VEVs, and $\theta_b$ is the sbottom mixing angle.  The decay into
$\tilde{b}$ dominates whenever $\mu \tan \beta / m_h$ is large, and the mixing
angle is not too small and the $\tilde{b}$ mass not too large.

\begin{figure}[htb]
\begin{minipage}[c]{0.55\textwidth}
\epsfysize=3.0in \epsffile{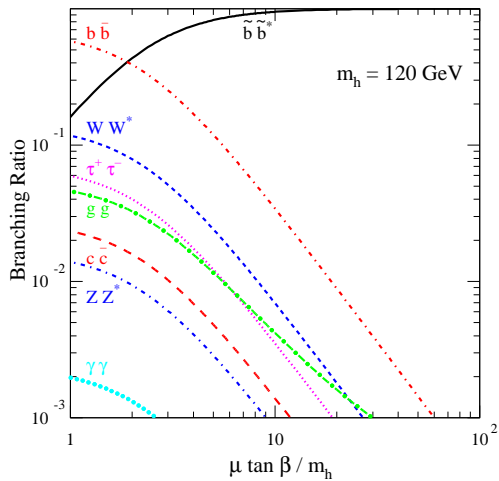}
\end{minipage}
\begin{minipage}[c]{0.45\textwidth}
\caption{Branching fractions for various Higgs boson 
decay channels as a function of the ratio $\mu \tan \beta/m_h$, 
with $m_h = 120 \gev$.  The sbottom mass is fixed to 
$m_{\tilde{b}} = 5 \gev$.} 
\label{fig:hjj_br}
\end{minipage}
\end{figure}

However the new decay mode arises, this scenario can be modelled by assuming
that the Higgs receives a new contribution to its width into jets 
(and assuming the width into other SM particles remains SM-like).  For
definiteness, we discuss the case with the light $\tilde{b}$, but the results
are largely insensitive to this choice.  As the width into the new decay mode
becomes large, it eventually overwhelms the SM decay modes, driving their
branching ratios to be small.  Figure~\ref{fig:hjj_br} 
shows the branching ratios
for all of the important decay modes as a function of $\mu \tan \beta / m_h$.
One should note that the light $\tilde{b}$ {\em have} been included in the 
loop-generated $H$-$g$-$g$ and $H$-$\gamma$-$\gamma$ couplings, and thus there
is some model-dependence in those curves.  As can be seen from the figure,
once $\mu \tan \beta / m_h$ is larger than about 10, the decay into $\tilde{b}$
completely dominates, with the canonical SM decay into $b \bar{b}$ having
a branching ratio of less than $10\%$.

The primary feature of the new
decay mode is that, by reducing the decay into SM particles while leaving
the Higgs production cross sections unchanged, it reduces the statistics
available to measure the Higgs interactions with SM particles and thus
limits the precision with which they may be measured.  At the same time,
the decay into jets is experimentally very challenging, and can prevent
the LHC from discovering the Higgs by depressing the decay into observable
SM particles below LHC sensitivities and opening a new decay channel which
cannot be seen at the LHC because of huge QCD jet backgrounds.
A careful analysis shows that if the width into 
the new hadronic state is more than a few times the width into
bottom quarks, the LHC completely
loses the ability to detect the light CP-even Higgs \cite{Berger:2002vs}.  
This could lead one to the perplexing situation in which
supersymmetry and perhaps one or more of the MSSM heavy Higgses are discovered,
but the Higgs actually responsible for electroweak symmetry breaking is too
difficult to extract from the large hadronic backgrounds.  

The ILC will
easily detect such a Higgs, because its primary search mode is not sensitive
to the Higgs decay mode (see \refse{sec:hstrahl}).  At the same time, the high
sensitivity expected at the ILC will help allow it to make important
measurements of the  Higgs.  As the decay width into jets increases, the
statistics for SM decays decreases, and some measurements are degraded.
However, since the decay into jets can in general
be extracted from the background,
measurements of the couplings associated with Higgs production remain 
accessible and precise.  Figure~\ref{fig:hjj_err}
plots the projected  uncertainties in the Higgs couplings
as a function of the ratio of widths 
$\Gamma (H \to jj) / \Gamma (H \to b \bar{b})$.  Note that
$H \to jj$ includes the SM decay into bottom quarks as well as the
new contribution into unflavored jets.

\begin{figure}[htb]
\begin{minipage}[c]{0.5\textwidth}
\epsfysize=3.0in \epsffile{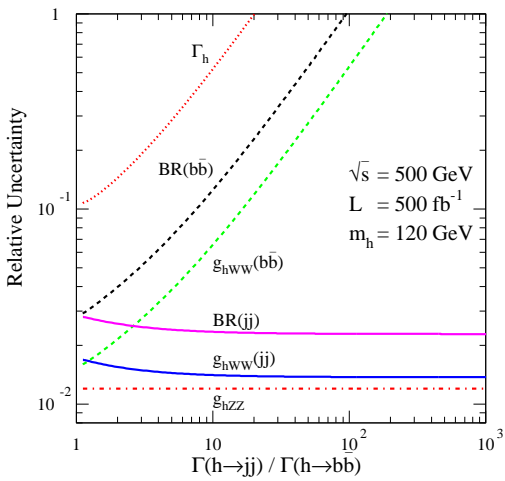}
\end{minipage}
\begin{minipage}[c]{0.45\textwidth}
\caption{Expected accuracy in the measurements of 
the $b \overline{b}$ and jet-jet 
branching fractions, the $hZZ$ and $hWW$ coupling 
strengths, and the total width of the Higgs boson, as a function of the 
ratio of the jet-jet and the $b \overline {b}$ widths, assuming the Higgs 
boson couplings to $b \overline{b}$, $ZZ$, and $WW^*$ are standard.} 
\label{fig:hjj_err}
\end{minipage}
\end{figure}

\subsection{Invisibly decaying Higgs}
\label{subsec:invisHiggs}

The Higgs might also decay predominantly into a neutral exotic particle, such
as the lightest neutralino of a supersymmetric theory.  In such a case,
the Higgs becomes effectively invisible, because the decay products do not
interact with the detector material.
However, even in this case a precise mass measurement is possible with
the recoil method down to the level of about 
$\sim 100 \mev$~\cite{tdr}. Concerning the other (now rare) decay
channels, the existing studies apply with appropriately scaled signal
rates; see \refses{sec:HiggsCouplings},
\ref{sec:heavySMHiggs}.


\chapter{Theoretical Developments}
\label{sec:theory}

The ILC does not only offer an even greater physics potential, 
but in turn represents a 
great challenge for theorists to understand phenomena at 
the experimentally achievable level of precision.

At energies exceeding the reach of LEP2, many new processes will
be accessible, such as top-quark pair production, Higgs production,
or reactions with new-physics
particles, as predicted, e.g., by SUSY models. To exploit the 
potential of the ILC, predictions for such reactions should be based
on full transition matrix elements and improved by radiative corrections
as much as possible. 
The higher level of accuracy at the ILC requires a
corresponding level of precision in the theoretical predictions.

In the following we focus directly on Higgs physics, but there
are other important areas (related to EWSB) where higher-order
corrections are indispensable. This comprises
precision calculations for $\MW$ (i.e., for $\mu$ decay);
precision observables on the $Z$ resonance (e.g., for the effective
leptonic weak mixing angle, $\sweff$);
radiative corrections to $2 \to 3, 4, \ldots$ processes in general;
single $W$ production;
etc.


\section{Radiative corrections to Higgs masses and couplings in the MSSM}
\label{sec:mhMSSM}

Within the MSSM the masses of the $\cp$-even neutral Higgs bosons as
well as the masses of the charged Higgs bosons are
calculable in terms of the other MSSM parameters. 
The status of the available corrections to the masses and mixing
angles in the MSSM Higgs sector (with real parameters) 
can be summarized as follows. For the
one-loop part, the complete result within the MSSM is 
known~\cite{ERZ,mhiggsf1lA,mhiggsf1lB,mhiggsf1lC}. The by far dominant
one-loop contribution is the \order{\alt} term due to top and stop 
loops ($\alt \equiv h_t^2 / (4 \pi)$, $h_t$ being the 
superpotential top coupling).
Concerning the two-loop
effects, their computation is quite advanced and has now reached a
stage such that all the presumably dominant
contributions are known. They include the strong corrections, usually
indicated as \order{\alt\als}, and Yukawa corrections, \order{\alt^2},
to the dominant one-loop \order{\alt} term, as well as the strong
corrections to the bottom/sbottom one-loop \order{\alb} term ($\alb
\equiv h_b^2 / (4\pi)$), i.e., the \order{\alb\als} contribution. The
latter can be relevant for large values of $\tb$. Presently, the
\order{\alt\als}~\cite{mhiggsEP1b,mhiggsletter,mhiggslong,mhiggsEP0,mhiggsEP1},
\order{\alt^2}~\cite{mhiggsEP1b,mhiggsEP3,mhiggsEP2} and the
\order{\alb\als}~\cite{mhiggsEP4,mhiggsFD2} contributions to the self-energies
are known for vanishing external momenta.  In the (s)bottom
corrections the all-order resummation of the $\tb$-enhanced terms,
\order{\alb(\als\tb)^n}, is also performed \cite{deltamb1,deltamb2}.
Recently the \order{\alt\alb} and \order{\alb^2} corrections
became available~\cite{mhiggsEP4b}. 
Finally a full two-loop effective potential 
calculation (including even the momentum dependence for the leading
pieces) has been published~\cite{mhiggsEP5}. However, no computer code
is publicly available.

In the case of the MSSM with complex parameters (cMSSM) the higher
order corrections have so far been restricted, after the first more
general investigations~\cite{mhiggsCPXgen}, to evaluations in the effective
potential (EP)
approach~\cite{mhiggsCPXEP,mhiggsCPXEP1} and to the renormalization
group (RG) improved \onel\ EP
method~\cite{mhiggsCPXRG1,mhiggsCPXRG2}. These results have been
restricted to the corrections coming from the (s)fermion sector and
some leading logarithmic corrections from the gaugino sector.
A more
complete one-loop calculation has been attempted in \citere{mhiggsCPXsn}.
More recently the leading \onel\ corrections have also been evaluated
in the Feynman-diagrammatic (FD) method, using the on-shell renormalization
scheme~\cite{mhiggsCPXFD1}. The full one-loop result can be found in
\citere{mhiggsCPXFDproc,habilSH}.  

The upper limit is estimated to be 
$\mh \lsim 135 \gev$~\cite{mhiggslong,mhiggsAEC}. 
The remaining theoretical uncertainty on the lightest $\cp$-even Higgs
boson mass has been estimated to be below 
$\sim 3 \gev$~\cite{mhiggsAEC,PomssmRep}. 
The above calculations have been implemented into public codes. 
The program {\tt FeynHiggs}~\cite{mhiggslong,mhiggsAEC,feynhiggs,feynhiggs2}
is based on the results obtained in the FD
approach, it includes all available corrections. The code
{\tt CPsuperH}~\cite{cpsh} is based on the RG
improved effective potential approach.
For the MSSM with real parameters the two codes can differ by up to
$\sim 4 \gev$ for the light $\cp$-even Higgs boson mass due to
subleading higher-order corrections that are included only in 
{\tt FeynHiggs}.


\section{Single Higgs production at one-loop in $2 \to 3$ processes}
\label{sec:2to3}

Recently some one-loop calculations of electroweak radiative corrections
have been presented for $2\to3$ processes that are interesting 
at the ILC: 
$e^+e^-\to\nu\bar\nu H$ \cite{Belanger:2002ik,Denner:2003yg} and
$e^+e^-\to t\bar t H$ \cite{You:2003zq,Belanger:2003nm,eetth}.
The results of \citeres{Belanger:2002ik,Belanger:2003nm}
were obtained with the
{\sc Grace-Loop} \cite{Fujimoto:zx} system (see below).
In \citeres{Denner:2003yg,You:2003zq,eetth} the 
technique \cite{Denner:2002ii} 
for treating tensor 5-point integrals was employed. 
While \citeres{Belanger:2002ik,You:2003zq,Belanger:2003nm} make
use of the slicing approach for treating soft-photon emission,
the results of \citeres{Denner:2003yg,eetth} have been obtained by
dipole subtraction and checked by phase-space slicing
for soft and collinear bremsstrahlung.

In $e^+e^-$ annihilation there are two main production mechanisms
for the SM Higgs boson. In the Higgs-strahlung process,
$e^+e^-\to Z H$, a virtual $ Z$ boson decays into a $ Z$ boson
and a Higgs boson. The corresponding cross section rises sharply at
threshold ($\sqrt{s}\gsim\MZ+\MH$)
to a maximum at a few tens of GeV above $\MZ+\MH$ and then
falls off as $s^{-1}$, where $\sqrt{s}$ is the CM
energy of the $e^+e^-$ system. In the $W$-boson-fusion process,
$e^+e^-\to\nu_e\bar\nu_e H$, the incoming $e^+$ and $e^-$ each emit
a virtual $W$~boson which fuse into a Higgs boson. The cross section of
the $W$-boson-fusion process grows as $\ln s$ and thus is the dominant
production mechanism for $\sqrt{s}\gg\MH$. 
A complete calculation of the \order{\al} electroweak corrections to 
$e^+e^- \to \nu \bar\nu H$ in the SM has been performed in 
\citeres{Belanger:2002ik,Denner:2003yg}.
The results of \citeres{Belanger:2002ik,Denner:2003yg} are in good
agreement, see, e.g.,~\cite{DidiAmsterdam}.
The agreement of the correction is within 0.2\% or better with respect to
the lowest-order cross sections.

Sample results taken from \citere{Denner:2003yg} are shown in
\reffi{fig:eennHcorr} for two Higgs boson masses, $\MH = 115, 150 \gev$.
The higher-order corrections are of the order of a few percent if the
tree-level result is expressed in terms of
$\gf$~\cite{Denner:2003yg}. 
\begin{figure}
\centerline{\includegraphics[width=.5\textwidth]{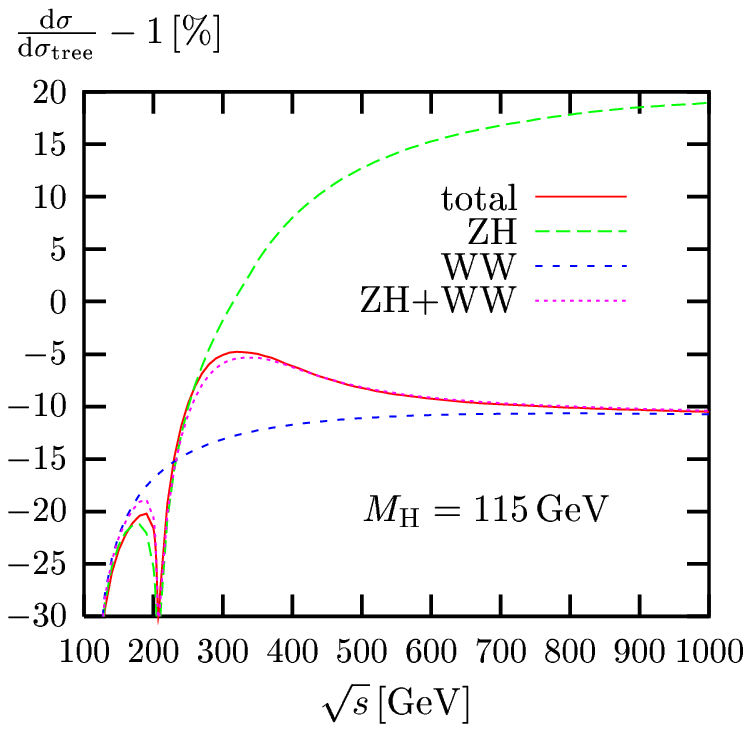}
\includegraphics[width=.5\textwidth]{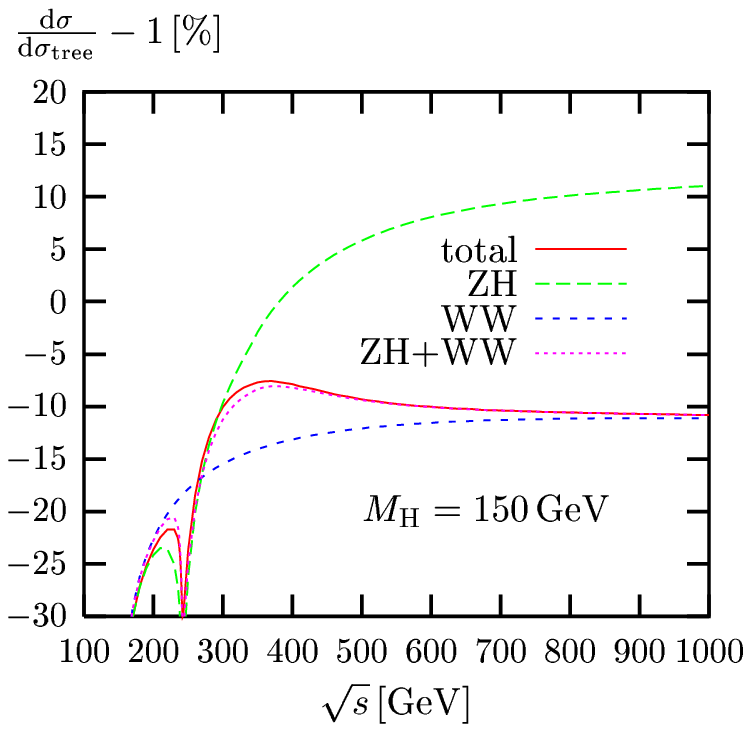}}
\caption{Relative electroweak corrections to the complete process
$e^+e^- \to \nu\bar\nu H$ 
and to the contributions from $ZH$-production and
$WW$-fusion channels for $\MH = 115 \gev$ and 
$\MH = 150 \gev$~\cite{Denner:2003yg}.
}
\label{fig:eennHcorr}
\end{figure}%

\begin{figure}[htb!]
\begin{center}
\mbox{\includegraphics[width=8cm,height=8.5cm]{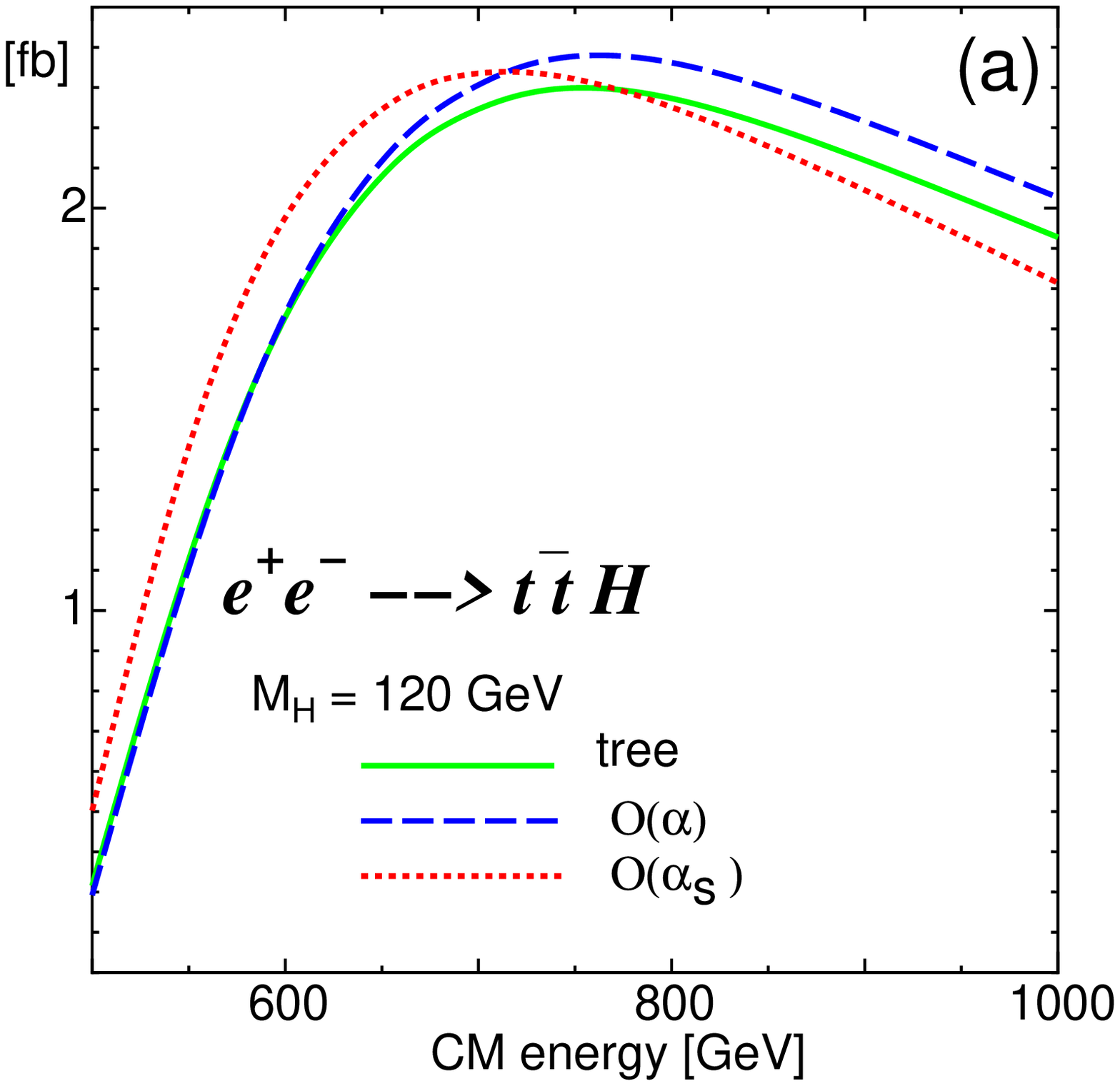}
\includegraphics[width=8cm,height=8.5cm]{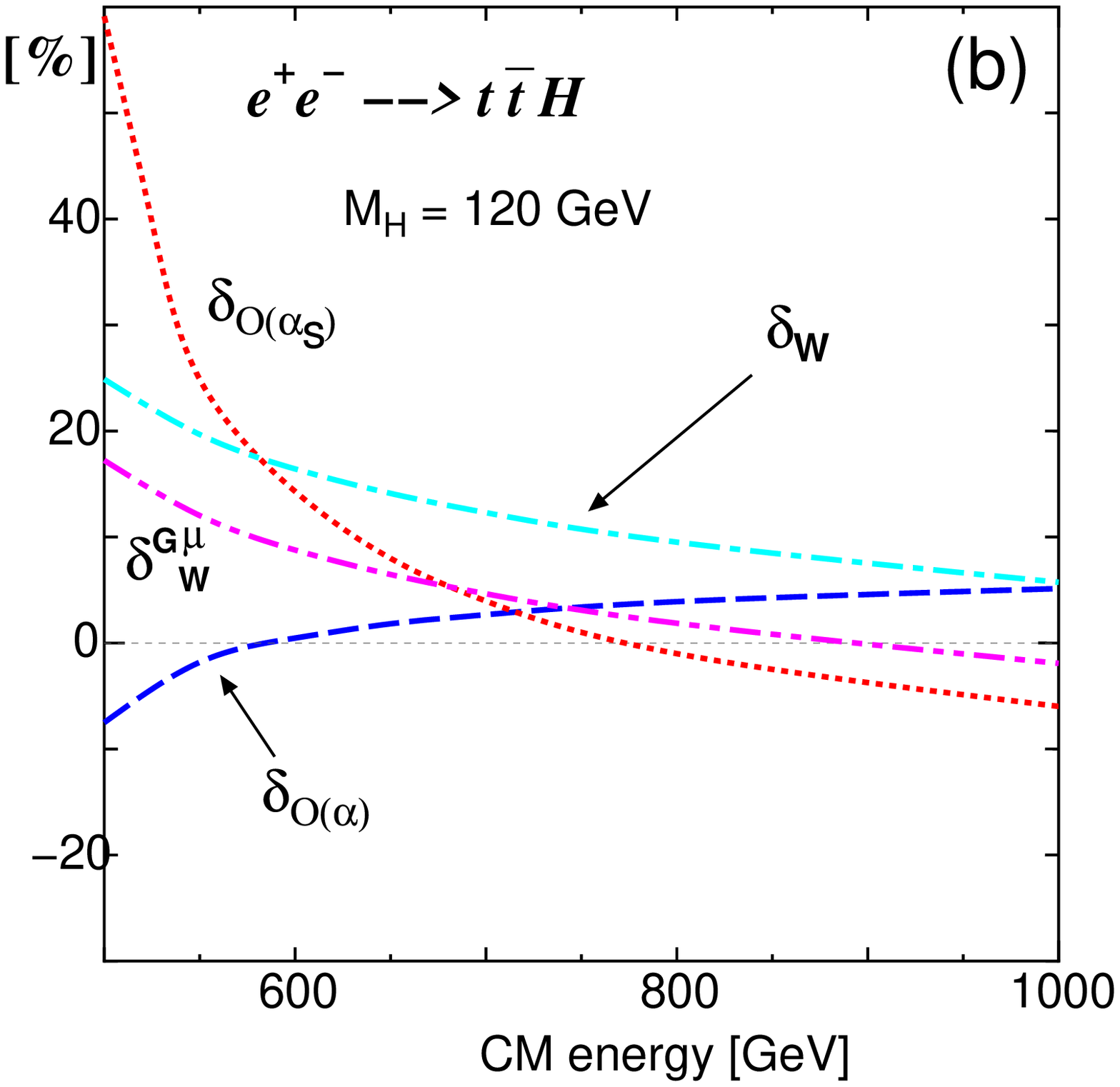}}
\caption{ (a) Total cross section as a function of the CMS
energy for $M_H=120$ GeV. Shown are the results for the total cross sections
for the tree level, full \order{\al} and \order{\als}
level in (a). The relative corrections are shown in
(b). Solid lines are tree level, dashed lines are the full \order{\al}
and dotted lines are the \order{\als}
corrections. In addition, the
genuine weak correction $\delta_W$ and the relative correction
$\delta_W^{G_\mu}$ in the $G_\mu$ scheme are presented. For more
details see \citere{Belanger:2003nm}.}
\label{fig:eettH}
\end{center}
\end{figure}

The Yukawa coupling of the top quark could be measured at a future
ILC with high energy and luminosity at the level of $\sim5\%$ 
\cite{tdr} by analyzing the process 
$e^+e^- \to t \bar t H$.
A thorough prediction for this process, thus, has to control
QCD~\cite{eetthQCD} (possibly SUSY QCD~\cite{eetthSUSYQCD}) 
and electroweak corrections~\cite{Belanger:2003nm,eetth}. 
The results of \citeres{Belanger:2003nm,eetth} have been compared with
other calculations, see e.g.\ \citere{DidiAmsterdam}.
Agreement within $\sim 0.1\%$ is found. This also holds for other energies
and Higgs-boson masses.
The results of the previous calculation \cite{You:2003zq}
roughly agree with the ones of \citeres{Belanger:2003nm,eetth} at
intermediate values of $\sqrt{s}$ and $\MH$, but are at variance
at high energies (TeV range) and close to threshold (large $\MH$).

Sample results taken from \citere{Belanger:2003nm} are shown in
\reffi{fig:eettH} for $\MH = 120 \gev$.
The higher-order corrections are of the order of 10 percent if the
tree-level result is expressed in terms of
$\gf$~\cite{Denner:2003yg}. More higher-order corrections possibly
have to be evaluated in order to reach the required ILC accuracy.

\bigskip
We now turn to the corresponding predictions within the MSSM
At the one-loop level, first
the contributions of fermion and sfermion loops in the MSSM
to $e^+e^- \to \nu \bar \nu h$
have been evaluated in \citere{eennH,eennHWiener}. The corrections
to the light Higgs boson production have been found to be at the
percent level~\cite{eennH}. 

Since at the tree level and in the decoupling limit the heavy neutral MSSM
Higgs bosons decouple from the $Z$, the mass reach for their discovery at
the ILC is limited to approximately $\sqrt{s} / 2$ from the pair production
process. It has been investigated how single production
mechanisms could extend the mass reach of the ILC. In particular,
the $WW$-fusion process $\ee\to\nu_e\bar{\nu_e} H$ 
has been investigated~\cite{eennH}. 
Its tree level cross-section is proportional to
$\cos(\beta-\alpha)$. Depending
on the SUSY parameters, radiative corrections might increase the cross-section
for $\ee\to\nu_e\bar{\nu_e} H$, possibly allowing discovery beyond the 
pair production kinematic limit for certain choices of the MSSM parameters.
Using left-polarized electron beams and right-polarized positron beams
the cross-section can further be enhanced.
A particular scenario where this is the case has been chosen 
in~\cite{eennH} ($\msusy = $ 350 GeV, $\mu = 1000 $ GeV,
$M_2 = 200 $ GeV and large stop mixing). Cross-section contours for this
scenario are shown in Fig.~\ref{fig:heavymssm}. 

\begin{figure}[htb]
\centering
\epsfig{width=12cm,height=6cm,file=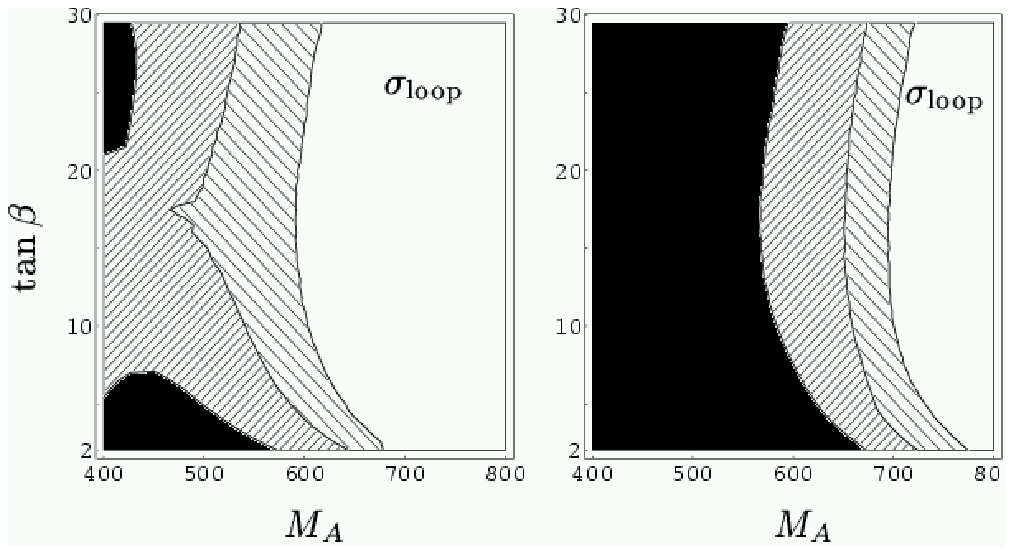}
\caption{Cross-section contours for $\ee\to H \nu\bar{\nu}$
for a particular MSSM scenario (see text) 
in the $\mA-\tb$ plane for $\sqrt{s} = 1 $ TeV.
The different shadings correspond to:
white: $\sigma \le 0.01$ \ifb, 
light shaded: 0.01 \ifb $\le \sigma \le$ 0.02 \ifb, 
dark shaded: 0.02 \ifb $\le \sigma \le$ 0.05 \ifb, 
black: $\sigma \ge$ 0.05 \ifb
(from~\cite{eennH}). The left figure is for unpolarized beams, the
right figure for an electron (positron) polarization of 0.8 (0.6).}
\label{fig:heavymssm}
\end{figure}

In a CP violating scenario the three neutral Higgs bosons,
$H_1$, $H_2$, $H_3$, 
are mixtures of the CP even and CP odd Higgs fields. Consequently, they
all couple to the $Z$ boson and to each other. These
couplings may be very different from those of the CP conserving case.
In the CP violating scenario
the Higgs-strahlung processes $e^+e^- \to H_i Z$ ($i=1,2,3$) and 
pair production processes
$e^+e^- \to H_i H_j$ ($i\neq j$) may all occur, with widely varying
cross-sections.


\section{Higgs production at one-loop in $2 \to 2$ processes}

The most promising channels for the production of the $\cp$-even
neutral MSSM Higgs bosons in the first phase of an ILC
are the Higgs-strahlung processes~\cite{hprod},
\BEQ 
e^+e^- \to Z\,H_i ~, 
\label{eetohZ}
\end{equation}
($H_{1,2} = h,H$) and the associated production of a scalar and a
pseudoscalar Higgs boson,  
\BEQ
e^+e^- \to A\,H_i~.  
\label{eetohA}
\end{equation} 
The full one-loop corrections~\cite{eehZhA1L} and the leading two-loop
corrections~\cite{eehZhA2L} are available. They have been combined to
obtain the currently most accurate results for these production cross
sections. The results have been implemented into the code 
{\tt FeynHiggsXS}~\cite{eehZhA2L}. 

\begin{figure}[htb!]
\begin{center}
\begin{tabular}{p{0.48\linewidth}p{0.48\linewidth}}
\mbox{\epsfig{file=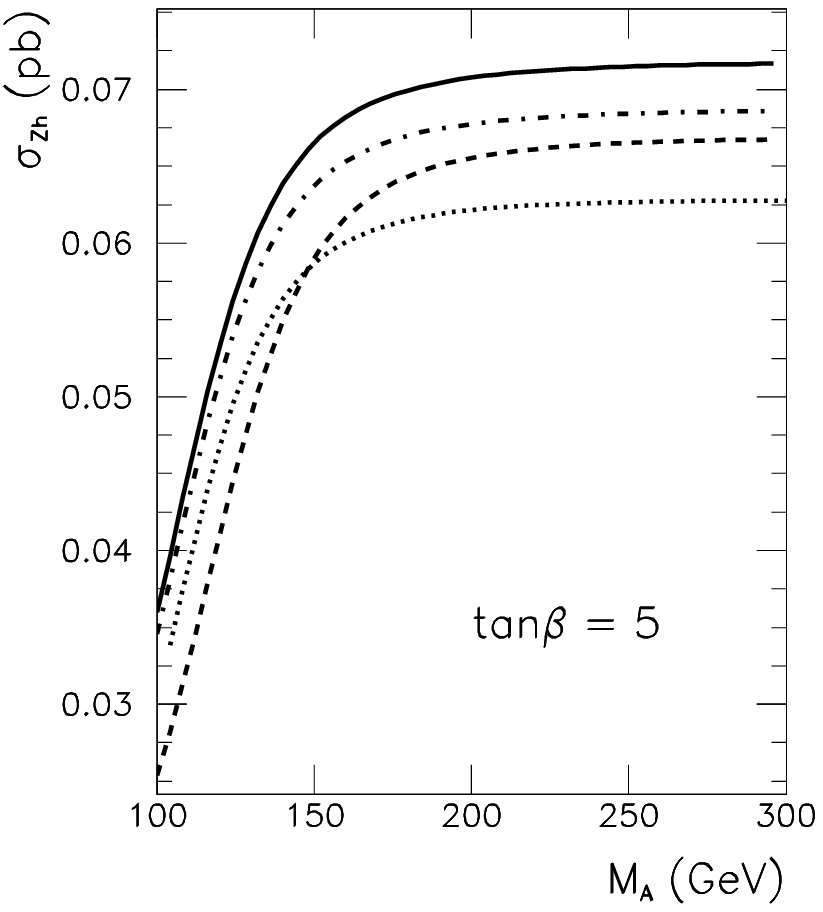,width=\linewidth,height=0.9\linewidth}}&
\mbox{\epsfig{file=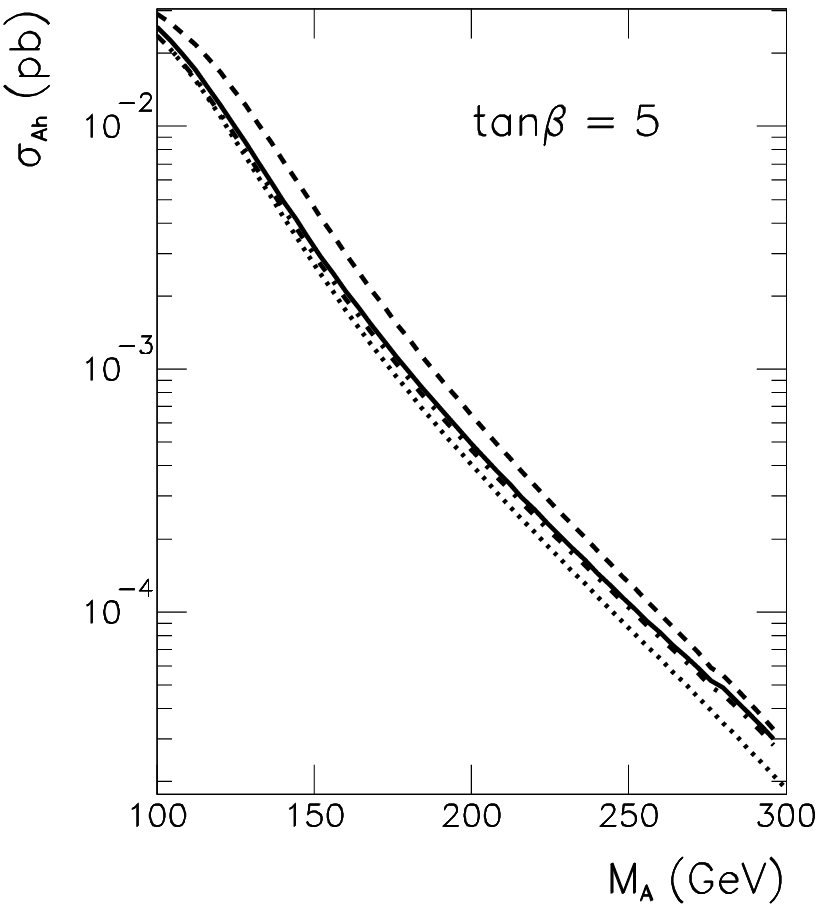,width=\linewidth,height=0.9\linewidth}}\\
\mbox{\epsfig{file=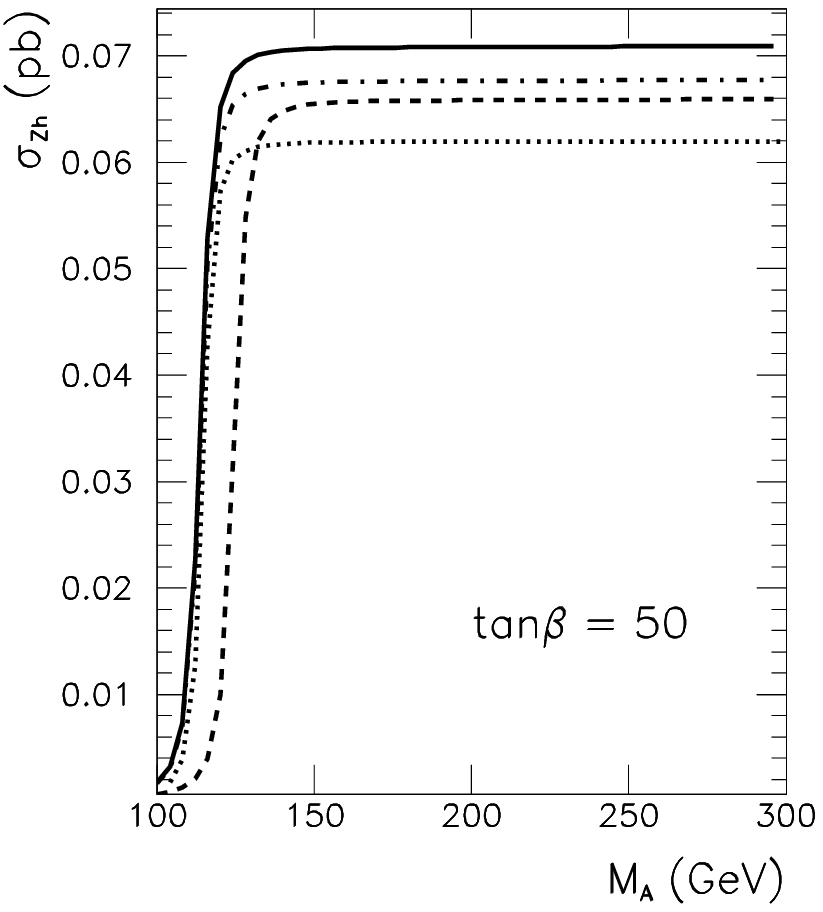,width=\linewidth,height=0.9\linewidth}}&
\mbox{\epsfig{file=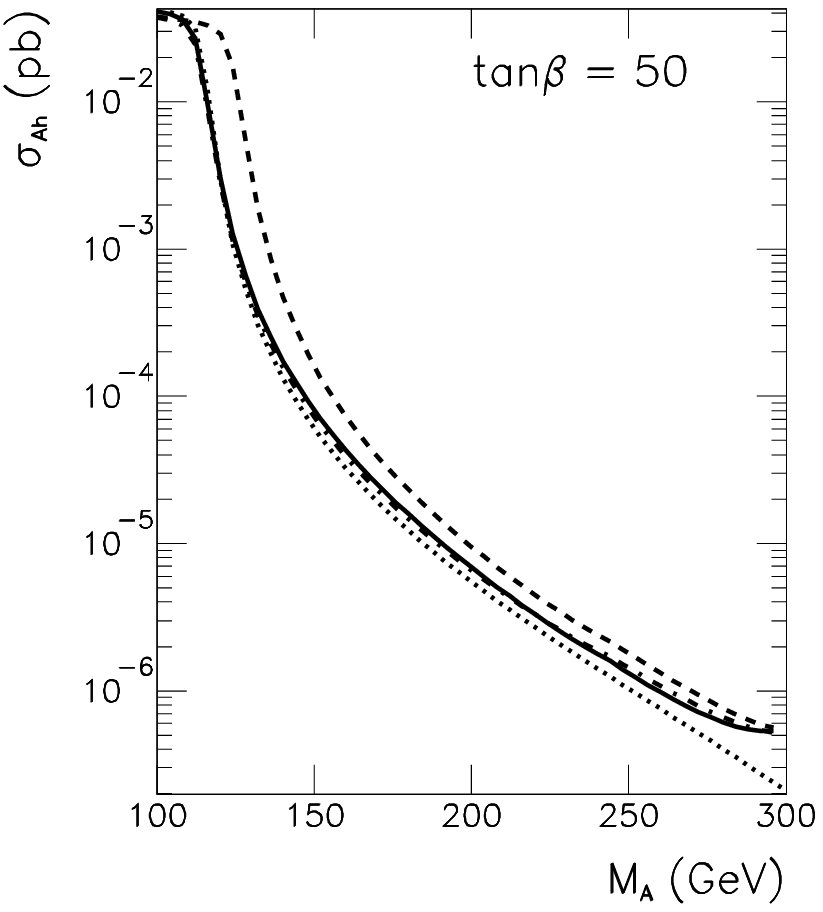,width=\linewidth,height=0.9\linewidth}}\\
\end{tabular}
\vskip -5mm 
\caption{$\sigma_{Zh}$ and $\sigma_{Ah}$ as a function of
  $\MA$ at $\sqrt{s} = 500 \gev$, shown for $\tb = 5$ and $\tb = 50$,
  in the no-mixing scenario (see text), as evaluated with 
  {\tt FeynHiggsXS}.  The solid
  (dot-dashed) line represents the two-loop result including
  (excluding) box contributions, the dotted line shows an effective
  coupling approximation and the dashed line shows the one-loop result.}
\label{fig:sigma_MA}
\end{center}
\end{figure}

In \reffi{fig:sigma_MA} the dependence of $\sigma_{Zh}$ and
$\sigma_{Ah}$ on $\MA$ are shown in the ``no-mixing'' benchmark
scenario~\cite{benchmark} ($\msusy = 2 \tev$, $\Xt = 0$, $\At = \Ab$, 
$\mu = 200 \gev$, $M_2 = 200 \gev$, $\mgl = 0.8 \, \msusy$) for  
$\tb = 5$ and $\tb=50$. The solid
(dot-dashed) line represents the two-loop result including
(excluding) box contributions, the dotted line shows an effective
coupling approximation and the dashed line shows the one-loop result. 
For $\sigma_{Zh}$, the $A$ boson decouples
quickly; the dependence on $M_A$ becomes very weak for
$\MA \gsim 250 \gev$, when $\sigma_{Zh}$ is already practically
constant. In the same limit,
$\sigma_{Ah}$ goes quickly to zero due to suppression of the effective
$ZhA$ coupling, which is $\sim\cos(\aeff - \be)$; 
also the kinematical suppression
plays a role, but this becomes significant only for sufficiently large
$\MA$, $\MA>350 \gev$.  For large $\tb$ the decoupling of $M_A$ is
even more rapid. The differences between the FD two-loop result and
the effective coupling approximation 
for the Higgs-strahlung cross section tend also to a constant, but
they increase with $\MA$ for the associated production. The latter can
be explained by the growing relative importance of 3- and 4-point
vertex function contributions compared to the strongly suppressed
Born-like diagrams. As can be seen from~\reffi{fig:sigma_MA}, for
$\tb=50$ and $\MA\geq 300 \gev$ the two-loop result is almost an order of
magnitude larger than the result of the effective coupling approximation, and 
starts to saturate. This can be
attributed to the fact that the (non-decoupling) vertex and box
contributions begin to dominate the cross section value. However, such
a situation occurs only for very small $\sigma_{Ah}$ values,
$\sigma_{Ah} \approx 10^{-3}$~fb, below the expected experimental ILC
sensitivities.

Comparing the one-loop and two-loop results an uncertainty larger than
$\sim 5\%$ can be attributed to the theory evaluation of the
production cross sections. 
Another type of possibly large corrections in the MSSM 
are the so-called Sudakov logs (see \citere{sudakov} and references
therein). They appear in the form of $\log(q^2/\msusy^2)$ (where $q$
is the momentum transfer)  in the production cross sections of SUSY
particles at $e^+e^-$ colliders.
In order to reach the required level of accuracy of 1-2\% further
two-loop corrections as well as the corresponding Sudakov logs will
have to be calculated.


\section{Double Higgs production processes at one-loop order}

Full electroweak radiative corrections to the double 
Higgs-strahlung $e^+e^- \to ZHH$ has been calculated 
in Refs.~\cite{rad-eezhh2,rad-eezhh1}. 
The two results are in good agreement, within 0.1\%, 
for $\sqrt{s} \lsim 800$ GeV for $m_{H}=115,150,200$ GeV.
In Ref.~\cite{rad-eezhh1}, 
the computation is performed with the help of
{\tt GRACE-loop}.

It has been shown that the QED corrections are large especially 
around the threshold, so that a proper resummation of the initial
state radiation needs to be performed in this case. 
At energies where the cross section can be substantial, however,  
the corrections are modest, especially for a Higgs boson mass 
of $M_H \simeq 120$ GeV. The genuine weak corrections
near the peak of the cross sections are also not large,  
not exceeding $\sim 5\%$, and therefore are  
below the expected experimental precision. 
As discussed in Ref.~\cite{hhh1,hhh2}, the invariant mass $M_{HH}$ 
distribution can be useful to extract the triple Higgs boson coupling.
It is found that the genuine weak corrections, 
contrary to the QED corrections,
hardly affect the shape of the distribution at least for energies
where this distribution is to be exploited\cite{rad-eezhh1}: 
see Figure~\ref{fig:eezhh2} 
and also Figure~\ref{fig:self-coupling} (left).
Therefore, an anomalous triple Higgs boson coupling could still be
distinguished, if large enough, in this distribution provided that 
a proper inclusion of the initial QED corrections is allowed 
in the experimental simulation.
\begin{figure}[htb!]
\begin{minipage}[c]{0.55\textwidth}
\psfig{figure=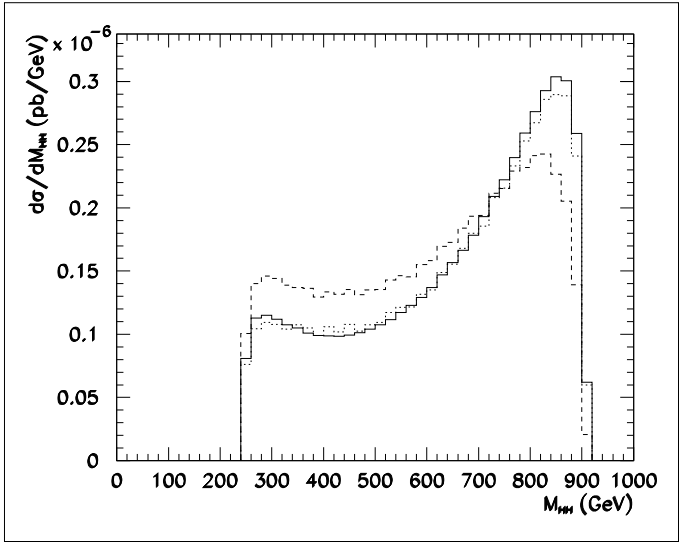,width=0.99\textwidth}
\end{minipage}
\begin{minipage}[c]{0.03\textwidth}
$\phantom{0}$
\end{minipage}
\begin{minipage}[c]{0.40\textwidth}
\caption{$d\sigma/d M_{HH}$ for
$M_H=120$ GeV at $\sqrt{s}=1$ TeV. We show the tree-level (full
curve), the effect of including only the genuine weak corrections
(dotted curve) and the effect of including the full ${\cal O}(\alpha)$
corrections (dashed curve)\cite{rad-eezhh1}.}
\label{fig:eezhh2}
\end{minipage}
\end{figure}

Recently, the corrections to the double Higgs production  
were also calculated in Ref.~\cite{grace-eenunuhh}.


\newcommand{\imag}{{\rm Im}}
\newcommand{\real}{{\rm Re}}

\section{Heavy MSSM Higgs bosons with CP violation at the $\ga\ga$ collider}

The 2-doublet Higgs sector of the MSSM
is automatically CP conserving at the tree
level. However, CP violation can be induced by radiative
corrections transmitting CP violating phases from the soft SUSY
breaking Lagrangian to the Higgs system, in particular the
relative phases between the higgsino mass parameter $\mu$ and the
trilinear sfermion-Higgs parameters $A_f$.
The Higgs system in 2-doublet models is generally described by a
$3 \times 3$ complex mass matrix, composed of a real dissipative
part and an imaginary absorptive part. The real dissipative part
includes the parameters of the Higgs potential extended by loop
corrections. The 
imaginary part includes the corresponding absorptive contributions
of the loops, i.e., the widths in the diagonal elements.

CP asymmetries are naturally enhanced \cite{r9,r9b} in the decoupling
regime where $M_A \simeq M_H$. Near mass degeneracy of the
scalar and pseudoscalar states in CP violating theories allows for 
frequent mutual transitions which induce large CP-odd mixing effects.
In this limit the CP violation effects are located in the heavy Higgs
sector and the light Higgs field can be ignored%
\footnote{
In general, however, it would be desirable to include the full 
$3 \times 3$ mixing (with complex entries).}. The mass matrix
of the heavy Higgs sector can be reduced to a $2 \times 2$ form:
\begin{eqnarray}
{\mathcal M}^2_{HA}
  = \left(\begin{array}{cc}
   M^2_H - i M_H \Gamma_H  & \Delta^2_{HA}  \\[1mm]
    \Delta^2_{HA}        & M^2_A - i M_A \Gamma_A
    \end{array}\right)
\label{eq:complex_mass_matrix_squared}
\end{eqnarray}
The CP violation effects are encoded in $\Delta^2_{HA}$ connecting
the scalar $H$ and the pseudoscalar $A$ fields through dissipative
and absorptive contributions. Diagonalizing this mass matrix leads
to the CP-mixed mass eigenstates $H_2$ and $H_3$. The rotation is
described by the complex mixing parameter
\begin{equation}
X = \frac{1}{2} \tan 2\theta = \frac{\Delta^2_{HA}} {M^2_H - M^2_A
- i [M_H\Gamma_H - M_A\Gamma_A]}
\end{equation}
The shift in the masses and widths and the CP coupling of the
states are quantitatively accounted for by the mixing parameter
$X$:
\begin{eqnarray}
&& \hskip -0.7cm \left[M^2_{H_3}\!-\!i M_{H_3} \Gamma_{H_3}\right]
  \!\mp\!\left[M^2_{H_2}\!-\!i M_{H_2} \Gamma_{H_2}\right]
 \!=\!\left\{\left[M^2_A\!-\!i M_A \Gamma_A\right]
         \!\mp \!\left[M^2_H\!-\!i M_H \Gamma_H\right]\right\}
           \left\{ \hskip -0.2cm \begin{array} {l}
           { }\\[-6mm]
           \times \sqrt{1+4X^2} \\[3mm]
           \times \hskip 0.3cm 1
           \end{array}\right.  \nonumber\\[-0.61cm]
&&
\end{eqnarray}
For a typical SUSY scenario,
\begin{eqnarray}
\msusy = 0.5 \tev \approx \MA,\quad |A_t|= 1.0\tev,\quad
|\mu|= 1.0\tev,\quad \phi_\mu=0\,; \quad \tan\beta=5
\label{eq:parameter_set}
\end{eqnarray}
the mixing parameter $X$ and the shifts of masses and widths are
illustrated in \reffis{fig:fig1}(a), (b) and (c). $\phi_A$ denotes the
CP-violating phase of $\At$. 

\begin{figure}[htb!]
\centering
\includegraphics[width=4.3cm,height=4.5cm,clip=true]{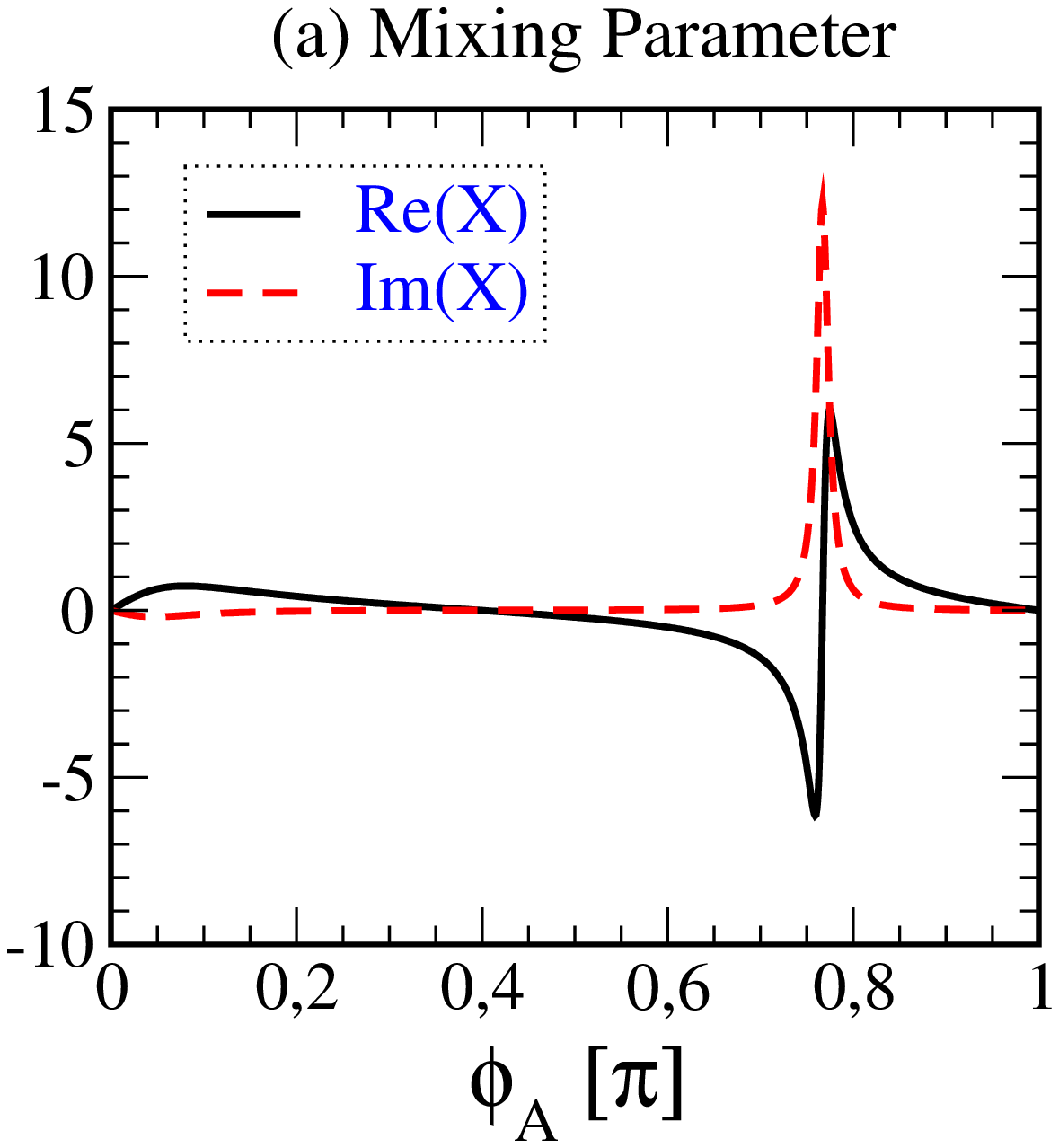}~~
\includegraphics[width=9.6cm,height=4.5cm]{eps/shift_mssm.eps}
\vskip -0.4cm
\caption{The dependence on the phase of $\At$, $\phi_A$, 
of (a) the mixing parameter $X$ and
     of the shifts of (b) masses and (c) widths for the parameter set
     (\ref{eq:parameter_set}); in (b,c) the mass and width differences
     without mixing  are shown by the broken lines.
    $\Re{\rm e}/\Im{\rm m} X(2\pi-\phi_A)
     =+\Re{\rm e}/\!\!-\!\Im{\rm m} X(\phi_A)$ for angles above $\pi$
(taken from \citere{r9}).}
\label{fig:fig1}
\end{figure}

Several asymmetries can be defined which reflect the presence of
CP violating mixing effects either indirectly or directly. Direct
CP violation can be studied in a classical way by measuring
asymmetries of inclusive cross sections between left- and
right-polarized photons \cite{r9}. At a photon collider, the
circular polarization of laser photons is transferred completely to
the high energy photons in Compton back-scattering of laser light
near the maximum of the photon energy \cite{tdr,r2,r3}.

The CP violating asymmetry is defined by
\begin{equation}
A_{\rm hel} = \frac{\sigma_{++} - \sigma_{--}} {\sigma_{++} +
\sigma_{--}}~,
\end{equation}
where the subscripts denote the helicities of the two colliding photons.
On top of the $H_2, H_3$ Higgs states the asymmetries are
parameterized by the complex mixing parameter,
\begin{eqnarray}
 A_{\rm hel} [H_2] \, \simeq \, {\cal A}_{\rm hel} [H_3]
     \, \simeq\, \frac{2\, \imag (\cos\theta \sin\theta^\ast)}
                             {|\cos\theta|^2+|\sin\theta|^2}
\end{eqnarray}
in the asymptotic decoupling limit. Even though corrections due to
non-asymptotic Higgs masses are still quite significant for the
SUSY scenario introduced above, remarkably large asymmetries are
predicted nevertheless in $\gamma \gamma$ fusion, shown in
Fig.~\ref{fig:fig2} as a function of the CP violating phase of the trilinear
stop-Higgs coupling $A_t$.

%
\begin{figure}[htb!]
\begin{minipage}[c]{0.55\textwidth}
\includegraphics[width=7cm,height=5cm]{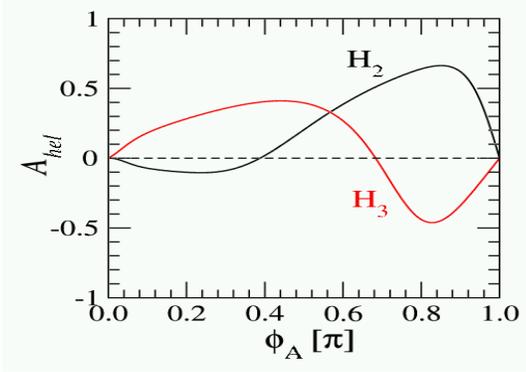}
\end{minipage}
\begin{minipage}[c]{0.03\textwidth}
$\phantom{0}$
\end{minipage}
\begin{minipage}[c]{0.40\textwidth}
\caption{The $\phi_A$ dependence of the CP--violating asymmetry
         ${\cal A}_{\rm hel}$ on top of the $H_2$, $H_3$ Higgs states,
         respectively.}
\label{fig:fig2}
\end{minipage}
\end{figure}
%

Thus a photon collider provides an instrument that allows us to
perform a unique classical experiment on CP violation in the Higgs
sector of supersymmetric theories.


\section{Charged Higgs Bosons}

Charged Higgs bosons can be pair-produced at the ILC via 
$e^+e^- \to H^+ H^-$
if $\mHpm < \sqrt{s} / 2$\cite{komamiya}. 
A complete simulation of this process for
the decay $H^+ \to t\bar{b}$ has been performed for $\sqrt{s} = 800 $ GeV,
1 ab$^{-1}$, and $\mHpm = 300 $ GeV~\cite{Battaglia:2001be}.
The mass resolution is approximately 1.5\%. A 5$\sigma$ discovery will
be possible for $\mHpm < 350 $ GeV.

Since in pair production the mass reach for charged Higgs bosons is 
limited to $\sqrt{s}/2$, also the rare processes of single charged Higgs
production may be considered. The dominant processes for single charged Higgs
production are $\ee\to b\bar{t} H^+, \ee\to \tau^- \bar{\nu_\tau} H^+$,
and $\ee\to W^- H^+$. Their cross-sections have been calculated at leading
order in~\citeres{single_hpm_xsec,single_hpm_xsec2}. Although the
$H^\pm W^\mp$  production is a $2 \to 2$ process\cite{eeHW}, 
the cross section is suppressed because of the 
absence of the tree level $H^\pm W^\mp Z$ coupling 
in multi Higgs doublet models.
QCD corrections to $\ee\to b\bar{t} H^+$
have recently become available~\cite{single_hpm_xsec2} and are sizable.
In general, parameter regions for which the production cross-section
exceeds 0.1~fb are rather small
for charged Higgs masses beyond the pair production threshold.
Cross-section contours for $\sqrt{s} = 500 $ GeV and 800 GeV are shown in
Fig.~\ref{fig:singlecharged}.

\begin{figure}[htb]
\centering
\epsfig{width=0.4\linewidth,file=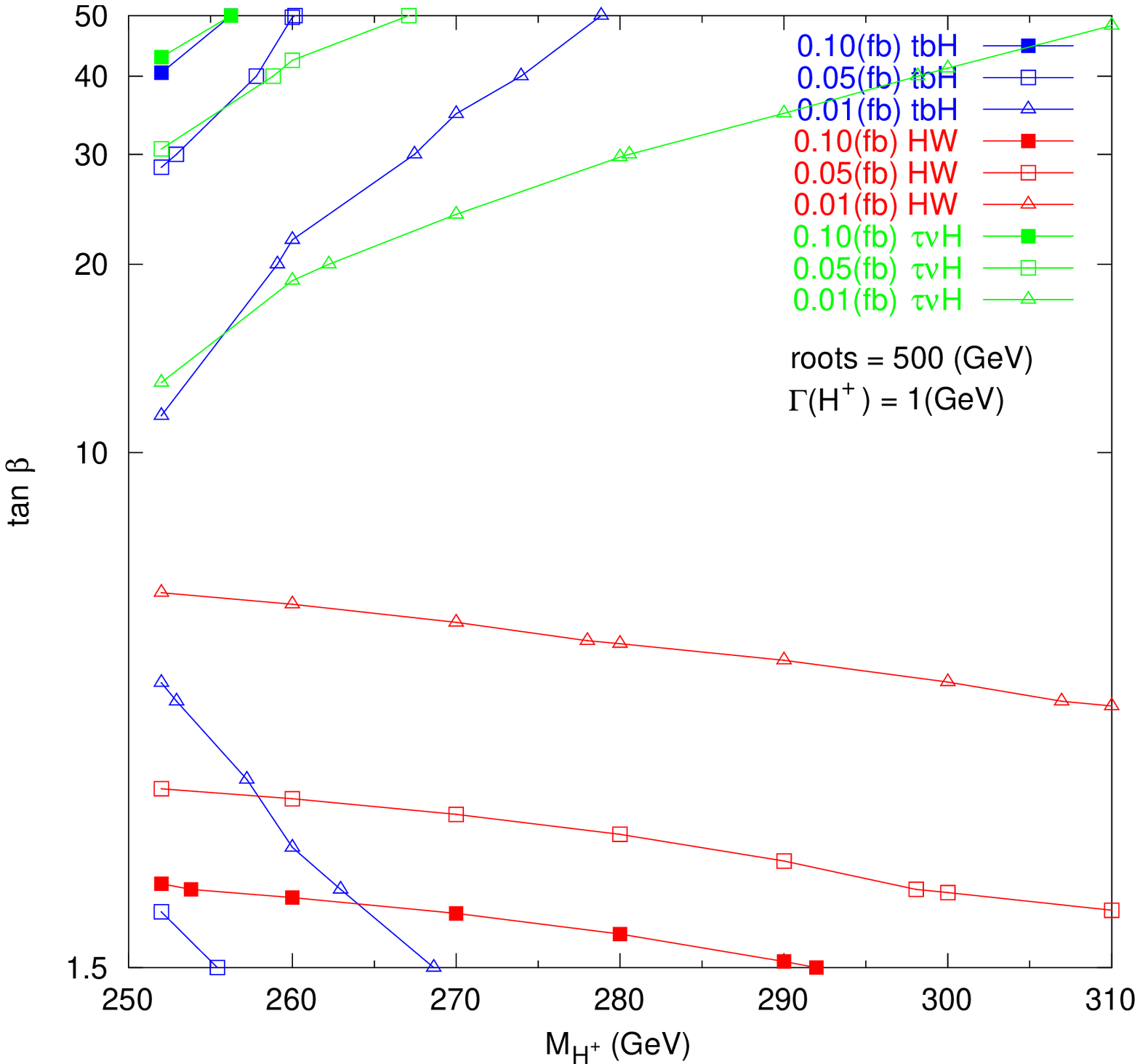} 
\epsfig{width=0.4\linewidth,file=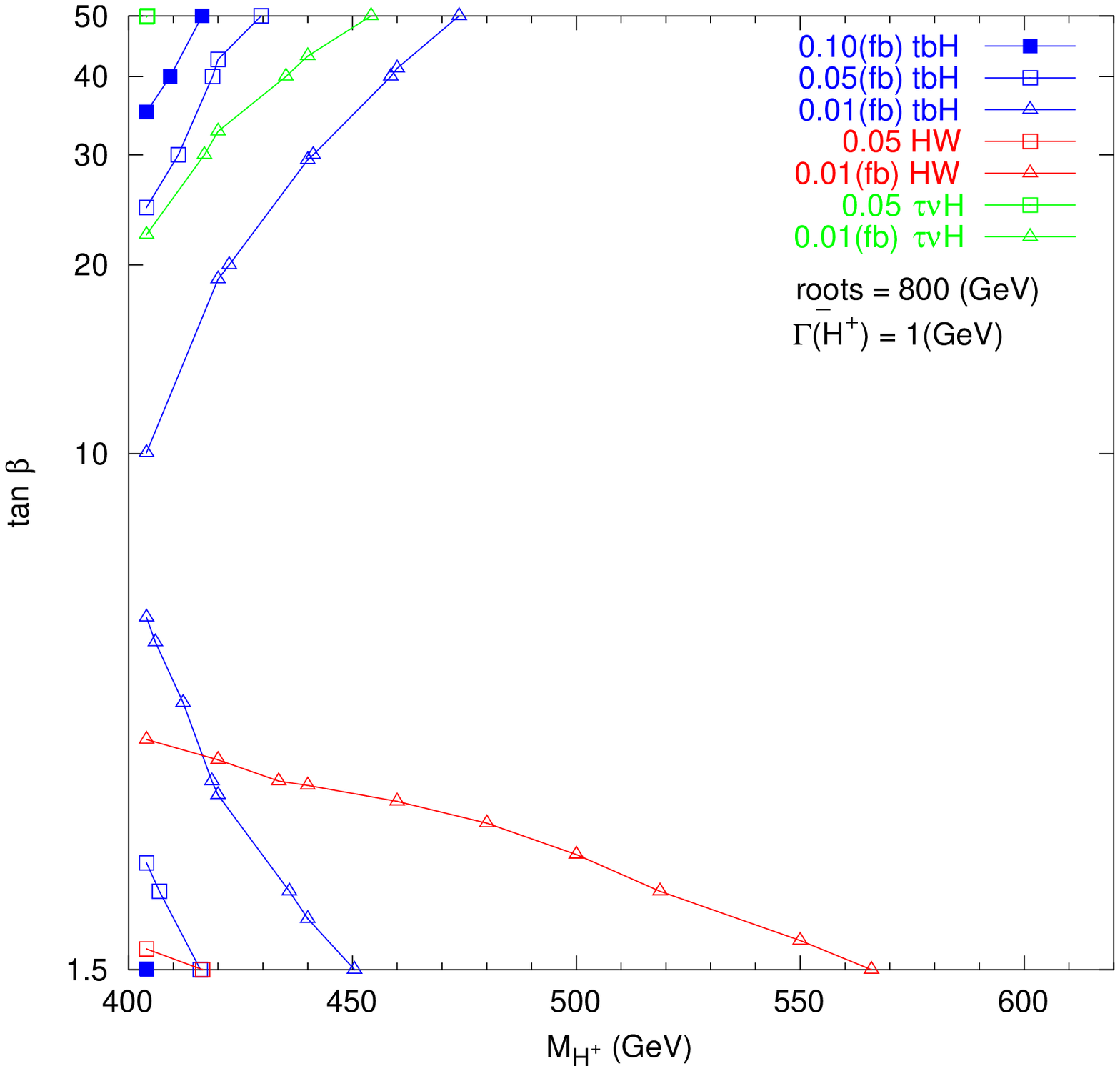}
\caption{Cross-section contours for the processes
$e^+e^- \to b \bar t H^+$ (blue/dark), 
$e^+e^- \to \tau^- \bar \nu_\tau H^+$ (green/light grey),
and $e^+ e^- \to W^- H^+$ (red/medium grey) 
at $\sqrt{s} = 500 $ GeV (left) and at 800 GeV
(right)~\cite{KlausiAmsterdam}. }
\label{fig:singlecharged}
\end{figure}

At a $\gamma\gamma$ collider, production of 
a heavy charged Higgs boson is typically one order of magnitude 
larger than the production of the same final state 
at an $e^+e^-$ collider with the same collision energy due to the 
t-channel nature. 
For $\sqrt{s}_{\gamma\gamma} > 2 m_{H^\pm}^{}$, 
pair production is dominant similarly to the electron-positron 
collision\cite{komamiya,eeHpHm}.  When the mass of the charged Higgs boson is 
larger than the threshold of pair production, single production 
processes become important. 
Cross sections of various single production processes 
have been calculated in Ref.~\cite{gg-mk}.
Moreover, a polarized $\gamma\gamma$ collider can determine the
chirality of the Yukawa couplings of fermions with the charged Higgs boson
via single charged Higgs boson production and, thus, discriminate
models of new physics\cite{single-charged-photon}.


\def\anti{\bar}
\newcommand{\mupmum}{{\mu^+\mu^-}}
\newcommand{\zstar}{{Z^\star}}
\newcommand{\wstar}{{W^\star}}
\newcommand{\stopone}{{\tilde t_1}}
\newcommand{\stoptwo}{{\tilde t_2}}
\newcommand{\mstop}{{m_{\stop}}}
\newcommand{\msquark}{{m_{\tilde q}}}
\newcommand{\mstopone}{{m_{\stopone}}}
\newcommand{\mstoptwo}{{m_{\stoptwo}}}
\newcommand{\mstopmean}{{\sqrt{\mstopone \mstoptwo}}}
\newcommand{\sbot}{{\tilde b}}
\newcommand{\sbotone}{{\tilde b_1}}
\newcommand{\sbottwo}{{\tilde b_2}}
\newcommand{\msbot}{{m_{\sbot}}}
\newcommand{\msbotone}{{m_{\sbotone}}}
\newcommand{\msbottwo}{{m_{\sbottwo}}}
\newcommand{\lam}{{\lambda}}
\newcommand{\kap}{{\kappa}}
\newcommand{\akap}{{A_\kappa}}
\newcommand{\alam}{{A_\lambda}}
\newcommand{\gam}{{\gamma}}
\newcommand{\vev}[1]{{\langle #1 \rangle}}
\newcommand{\what}{{\widehat}}
\newcommand{\tildeil}{{\widetilde}}
\newcommand{\hu}{{h_u}}
\newcommand{\hd}{{h_d}}
\newcommand{\hi}{{h_1}}
\newcommand{\hii}{{h_2}}
\newcommand{\ai}{{a_1}}
\newcommand{\aii}{{a_2}}
\newcommand{\mhi}{{m_{\hi}}}
\newcommand{\mhii}{{m_{\hii}}}
\newcommand{\mai}{{m_{\ai}}}
\newcommand{\maii}{{m_{\aii}}}
\newcommand{\mueff}{{\mu_{\rm eff}}}
\newcommand{\noi}{{\noindent}}
\newcommand{\mhusq}{{m_{H_u}^2}}
\newcommand{\mhdsq}{{m_{H_d}^2}}
\newcommand{\mssq}{{m_S^2}}
\newcommand{\mqsq}{{m_Q^2}}
\newcommand{\musq}{{m_U^2}}
\newcommand{\mdsq}{{m_D^2}}
\newcommand{\mlsq}{{m_L^2}}
\newcommand{\mesq}{{m_E^2}}
\newcommand{\mz}{{m_Z}}
\newcommand{\wpm}{{W^{\pm}}}
\newcommand{\hpm}{{H^{\pm}}}
\newcommand{\mhm}{{m_{\hm}}}
\newcommand{\call}{{{\cal L}}}
\newcommand{\calm}{{{\cal M}}}
\def\wtil{\widetilde}
\def\what{\widehat}
\newcommand{\tauptaum}{{\tau^+\tau^-}}
\newcommand{\mbb}{{m_{b\anti b}}}
\newcommand{\bm}{{\boldmath}}
\newcommand{\bfbm}{{\bf\boldmath}}
\newcommand{\glsp}{{$\wtil g$-LSP}}
\newcommand{\ifmath}[1]{{\relax\ifmmode #1\else $#1$\fi}}
\newcommand{\half}{{\ifmath{{\textstyle{1 \over 2}}}}}
\newcommand{\threehalf}{{\ifmath{{\textstyle{3 \over 2}}}}}
\newcommand{\quarter}{{\ifmath{{\textstyle{1 \over 4}}}}}
\newcommand{\sixth}{{\ifmath{{\textstyle{1 \over 6}}}}}
\newcommand{\third}{{\ifmath{{\textstyle{1 \over 3}}}}}
\newcommand{\twothirds}{{{\textstyle{2 \over 3}}}}
\newcommand{\fivethirds}{{{\textstyle{5 \over 3}}}}
\newcommand{\fourth}{{\ifmath{{\textstyle{1\over 4}}}}}
\newcommand{\mplanck}{{M_{\rm P}}}
\newcommand{\mpl}{{\mplanck}}
\newcommand{\sur}{{{\tilde u_R}}}
\newcommand{\msur}{{{m_{\sur}}}}
\newcommand{\stl}{{{\tilde t_L}}}
\newcommand{\str}{{{\tilde t_R}}}
\newcommand{\sbl}{{{\tilde b_L}}}
\newcommand{\sbr}{{{\tilde b_R}}}
\newcommand{\sqbar}{{\ov{\sq}}}
\newcommand{\slep}{{\tilde \ell}}
\newcommand{\slepbar}{{\ov{\slep}}}
\newcommand{\mslep}{{m_{\slep}}}
\newcommand{\slepl}{{\tilde \ell_L}}
\newcommand{\mslepl}{{m_{\slepl}}}
\newcommand{\slepr}{{\tilde \ell_R}}
\newcommand{\mslepr}{{m_{\slepr}}}

\newcommand{\sel}{{\tilde e}}
\newcommand{\selbar}{{\ov{\sel}}}
\newcommand{\msel}{{m_{\sel}}}
\newcommand{\sell}{{\tilde e_L}}
\newcommand{\msell}{{m_{\sell}}}
\newcommand{\selr}{{\tilde e_R}}
\newcommand{\mselr}{{m_{\selr}}}

\newcommand{\cptwo}{{\tilde \chi^+_2}}
\newcommand{\cmtwo}{{\tilde \chi^-_2}}
\newcommand{\cpmtwo}{{\tilde \chi^{\pm}_2}}
\newcommand{\mcptwo}{{m_{\cptwo}}}
\newcommand{\mcpmtwo}{{m_{\cpmtwo}}}
\newcommand{\stautwo}{{\tilde \tau_2}}
\newcommand{\mstautwo}{{m_{\stauone}}}

\newcommand{\To}{{\Rightarrow}}
\newcommand{\lra}{{\leftrightarrow}}
\newcommand{\msusyslash}{{m_{\susyslash}}}
\newcommand{\susy}{{{\rm SUSY}}}
\newcommand{\hsm}{{h_{\rm SM}}}
\newcommand{\mhsm}{{m_{\hsm}}}
\newcommand{\hl}{{h}}
\newcommand{\hh}{{H}}
\newcommand{\ha}{{A}}
\newcommand{\hp}{{H^+}}
\newcommand{\hm}{{H^-}}
\renewcommand{\hpm}{{H^{\pm}}}
\newcommand{\mhl}{{m_{\hl}}}
\newcommand{\mhh}{{m_{\hh}}}
\newcommand{\mha}{{m_{\ha}}}
\newcommand{\mhp}{{m_{\hp}}}
\newcommand{\cotb}{{\cot\beta}}
\renewcommand{\mz}{{M_Z}}
\newcommand{\mw}{{M_W}}
\newcommand{\mgut}{{M_U}}
\newcommand{\mx}{{M_X}}
\newcommand{\mstring}{{M_S}}
\renewcommand{\wp}{{W^+}}
\newcommand{\wm}{{W^-}}
\renewcommand{\wpm}{{W^{\pm}}}
\newcommand{\wmp}{{W^{\mp}}}
\newcommand{\chitil}{{\tilde\chi}}

\newcommand{\cnone}{{\tilde\chi^0_1}}
\newcommand{\cnonestar}{{\tilde\chi_1^{0\star}}}
\newcommand{\cntwo}{{\tilde\chi^0_2}}
\newcommand{\cnthree}{{\tilde\chi^0_3}}
\newcommand{\cnfour}{{\tilde\chi^0_4}}
\newcommand{\snu}{{\tilde\nu}}
\newcommand{\snul}{{\tilde\nu_L}}
\newcommand{\msnul}{{m_{\snul}}}

\newcommand{\snue}{{\tilde\nu_e}}
\newcommand{\snuel}{{\tilde\nu_{e\,L}}}
\newcommand{\msnuel}{{m_{\snul}}}

\newcommand{\snubar}{{\ov{\snu}}}
\newcommand{\msnu}{{m_{\snu}}}
\newcommand{\mcnone}{{m_{\cnone}}}
\newcommand{\mcntwo}{{m_{\cntwo}}}
\newcommand{\mcnthree}{{m_{\cnthree}}}
\newcommand{\mcnfour}{{m_{\cnfour}}}
\newcommand{\wt}{{\tilde}}
\newcommand{\wh}{{\widehat}}
\newcommand{\cpone}{{\wt \chi^+_1}}
\newcommand{\cmone}{{\wt \chi^-_1}}
\newcommand{\cpmone}{{\wt \chi^{\pm}_1}}
\newcommand{\mcpone}{{m_{\cpone}}}
\newcommand{\mcpmone}{{m_{\cpmone}}}

\newcommand{\staur}{{\wt \tau_R}}
\newcommand{\staul}{{\wt \tau_L}}
\newcommand{\stau}{{\wt \tau}}
\newcommand{\mstaur}{{m_{\staur}}}
\newcommand{\stauone}{{\wt \tau_1}}
\newcommand{\mstauone}{{m_{\stauone}}}
\newcommand{\emem}{{e^-e^-}}
\newcommand{\gamhsm}{{\Gamma_{\hsm}^{\rm tot}}}
\newcommand{\gamhl}{{\Gamma_{\hl}^{\rm tot}}}
\newcommand{\gamha}{{\Gamma_{\ha}^{\rm tot}}}
\newcommand{\gamhh}{{\Gamma_{\hh}^{\rm tot}}}

\newcommand{\fbi}{{~{\rm fb}^{-1}}}
\newcommand{\fb}{{~{\rm fb}}}
\newcommand{\abi}{{~{\rm ab}^{-1}}}
\newcommand{\ab}{{~{\rm ab}}}
\newcommand{\pbi}{{~{\rm pb}^{-1}}}
\newcommand{\pb}{{~{\rm pb}}}
\newcommand{\h}{{h}}
\newcommand{\ma}{{m_{\ha}}}
\renewcommand{\hi}{{\h_1}}
\renewcommand{\hii}{{\h_2}}
\newcommand{\hiii}{{\h_3}}
\renewcommand{\mhi}{{m_{\hi}}}
\renewcommand{\mhii}{{m_{\hii}}}
\newcommand{\mhiii}{{m_{\hiii}}}
\newcommand{\beq}{\begin{equation}}   
\newcommand{\eeq}{\end{equation}}
\newcommand{\bea}{\begin{eqnarray}}   
\newcommand{\eea}{\end{eqnarray}}
\newcommand{\baa}{\begin{array}}      
\newcommand{\eaa}{\end{array}}
\newcommand{\bit}{\begin{itemize}}    
\newcommand{\eit}{\end{itemize}}
\newcommand{\bce}{\begin{center}}     
\newcommand{\ece}{\end{center}}
\newcommand{\beqa}{{\begin{eqnarray}}}
\newcommand{\eeqa}{{\end{eqnarray}}}
\newcommand{\bed}{{\begin{description}}}
\newcommand{\eed}{{\end{description}}}
\newcommand{\calo}{{{\mathcal O}}}


\section{NMSSM}
\label{sec:NMSSM}

\subsection{Overview}
\label{sec:nmssmintro}

One attractive supersymmetric model is the Next to
Minimal Supersymmetric Standard Model (NMSSM)~\cite{allg,King:1995vk} which
extends the MSSM by the introduction of just one singlet superfield,
$\what S$, in addition to the Higgs up- and down-type superfields
$\what H_u$ and $\what H_d$. [Hatted (unhatted) capital letters denote
superfields (scalar superfield components).]  When the scalar
component of $\what S$ acquires a TeV scale vacuum expectation value
(a very natural result in the context of the model), the
superpotential term $\lam \what S \what H_u \what H_d$ generates an
effective $\mu\what H_u \what H_d$ interaction for the Higgs doublet
superfields with $\mu\equiv\mueff=\lam \vev{S}\equiv \lam s$.  Such a
term is essential for acceptable phenomenology. No other SUSY model
generates this crucial component of the superpotential in as natural a
fashion.  
At the same time, the NMSSM preserves all the most attractive
features of the MSSM.  First, since the NMSSM introduces only an additional
singlet superfield, the $SU(3)$, $SU(2)_L$ and $U(1)$
coupling constants continue to unify at the unification scale, $\mgut\sim
{\rm few}\times 10^{16}\gev$, just as in the MSSM.
Further, radiative electroweak symmetry breaking remains
a natural possibility within the NMSSM.
The electroweak symmetry breaking generates non-zero vevs for 
all the Higgs fields, $\langle H_u \rangle = h_u$, 
$\langle H_d \rangle = h_d$ and $\langle S \rangle = s$.
A possible cosmological domain wall problem \cite{abel1} can be avoided
by introducing suitable non-renormalizable operators \cite{abel2} that
do not generate dangerously large singlet tadpole diagrams
\cite{tadp}. Alternatively, it has been argued
that the domain walls are themselves the source of dark energy
\cite{domainwalldarkenergy}. 

The trilinear term in the Higgs potential reads
\beq \label{1.2}
\lambda A_{\lambda} S H_u H_d + \frac{\kappa} {3} A_\kappa S^3 \,. 
\eeq
Aside from $\lam$, $\kap$, $\mueff$, $\akap$ and $\alam$,
the final Higgs sector parameter is $\tb = h_u/h_d$. 
Thus, as compared to the three 
independent parameters needed in the 
MSSM context (often chosen as $\mu$, $\tan \beta$ and $\mha$), the
Higgs sector of the NMSSM is described by the six parameters
\beq \label{6param}
\lambda\ , \ \kappa\ , \ A_{\lambda} \ , \ A_{\kappa}, \ \tan \beta\ ,
\ \mu_\mathrm{eff}\ .
  \eeq
A possible sign convention~\cite{nmhdecay} is such that
$\lambda$ and $\tan\beta$ are positive, while $\kappa$,
$A_\lambda$, $A_{\kappa}$ and $\mu_{\mathrm{eff}}$ should be allowed
to have either sign. 

The particle content of the NMSSM differs from the MSSM by the
addition of one CP-even and one CP-odd state in the neutral Higgs
sector (assuming CP conservation), and one additional neutralino.
Altogether, in the Higgs sector we have the CP-even
Higgs bosons $\hi,\hii,\hiii$, the CP-odd Higgs bosons $\ai,\aii$
and the usual charged Higgs pair $\hpm$.  The five neutralinos
($\wtil\chi^0_{1,2,3,4,5}$)
are typically mixtures of the gauginos, higgsinos and the singlino.
In some limits, the singlino can be a fairly pure state
and it can either be the heaviest or the lightest of the neutralinos.


\subsection{The Higgs boson mass spectrum}
\label{sec:NMSSMHiggsspectrum}

The mass of the lightest CP-even Higgs boson at the tree-level is given
by
\BEQ
m_{h, {\rm tree}}^2 \le \MZ^2 \KL \cos^2 2\be 
     + \frac{2\,\la^2}{g^2 + g'^2} \sin^2 2\be \KR~,
\EEQ
where $\la$ should be below $\la \lsim 0.7$ to remain perturbative up
to the GUT scale. 
The evaluation of higher-order corrections to Higgs boson masses and
mixing angles is less advanced than in the MSSM. No full one-loop
calculation is available so far although
there has been much
work~\cite{radcor1,radcor2,yeg,higrad2,higsec1,LC-TH-2003-034}. 
 Most recently the leading two-loop
Yukawa corrections have become available~\cite{NMSSMmhYuk2}.
All available corrections have been implemented into the code 
{\tt NMHDECAY}~\cite{nmhdecay}. No genuine higher-order corrections to
Higgs boson production and decay are available. However, in many cases
(especially if the new singlet mass scale is large) 
the results from the MSSM can be taken over to the NMSSM case.

An important issue for NMSSM Higgs phenomenology is the
mass and nature of the lightest CP-even and CP-odd Higgs bosons.
In particular, if the $\ai$ is  very light or  even 
just moderately light there
are dramatic modifications in the phenomenology of Higgs discovery
at both the LHC and
ILC~\cite{Gunion:1996fb,Dobrescu:2000jt,Ellwanger:2001iw,Ellwanger:2003jt,LC-TH-2003-034,Ellwanger:2004gz,nmhdecay,higsec3,Ellwanger:2005uu}.
A light $\ai$ is natural in the context of the model, see
\citere{Dobrescu:2000jt} for a discussion. 
A very light $\ai$ is experimentally not ruled out, especially if it
has a large singlet component. In some regions of the NMSSM
parameter space, one can also get the lightest CP-even state, $\hi$,
to be very light as well.  

Another interesting feature of the NMSSM is
that the standard measure of fine-tuning 
\beq F={\rm Max}_p F_p \equiv {\rm
  Max}_p\left|{d\log \mz\over d\log p}\right|\,, 
\label{eq:finetuning}
\eeq 
where the parameters $p$ comprise all GUT-scale soft-SUSY-breaking
parameters, can be very low for Higgs boson mass values above the LEP
limit~\cite{finetuning,bast,King:1995vk,Dermisek:2005ar}.


\subsection{LEP Higgs searches}

Within the NMSSM it is also possible to have a Higgs boson
with a mass below the LEP limits~\cite{Dermisek:2005ar}.
It escapes LEP constraints by virtue of having unusual decay modes for
which LEP limits are weaker. In particular, parameters for which
$\hi\to \ai\ai$ decays are dominant over $\hi\to b\anti b$ decays must
be carefully considered%
\footnote{
This is also possible in the MSSM~\cite{haa}; corresponding holes from
the LEP Higgs searches in the MSSM parameter planes exist~\cite{newleplimits}.
}%
.  Particular care is required if there is a
mixture of $\hi\to\ai\ai$ and $\hi\to b\anti b$ decays, and the
dominant $\ai$ decay is to $b\anti b$. 
In this situation, the single channel $Z2b$ and $Z4b$ rate limits from LEP
are less sensitive than the combination of both channels, since in the
$Z2b$ interpretation the $Z4b$ final state is ignored, and vice versa,
which leads to a loss in sensitivity compared with the full combination.

This has particular relevance to the NMSSM scenarios with the
very lowest fine-tuning, $F<10$, as examined in
\citere{Dermisek:2005gg}. Two types of scenarios emerge for $F<10$.
In both, $\br(\hi\to b\anti b)\sim 0.1\div0.2$ and
$\br(\hi\to \ai\ai)\sim 0.85\div0.75$. In scenarios of type I (II), 
$\mai>2\mb$ ($\mai<2\mb$) and $\br(\ai\to
b\anti b)\sim 0.92$ (0). More details can be found in
\citeres{Dermisek:2005gg,newleplimits}. 
It has been found that the  
only $F<10$ scenarios consistent with the full LEP Higgs Working Group (LHWG)
analysis are of type II. Type I scenarios are typically excluded at about the
$98\div 99\%$ CL after data from all experiments are combined. 

A very important feature of the type-II $F<10$ scenarios is that
a significant fraction of them can easily explain the well-known excess
in the $Zh\to Zb\anti b$ final state in the vicinity of $\mh\sim
100\gev$.  This is illustrated in Fig.~\ref{zbbplot} 
\cite{Dermisek:2005gg}, where all $F<10$ NMSSM parameter
choices (from a lengthy scan) with $\mai<2\mb$ are shown to predict
$\mhi\sim 96\div 105\gev$ 
along with a $C_{\rm eff}=[g_{ZZh}^2/g_{ZZ\hsm}^2]\br(h\to
  b \anti b)$ that would explain the observed excess with respect
to the expected limit.  
The statistical significance of this excess is of order $2\sigma$.
The properties of the four most ideal points for describing the excess
are given in Table~\ref{fourpoints}.  Note that for point 3,
$\mai<2\mtau$ implying that $\ai\to$ quarks and gluons.
%
A general feature with $F < 10$ is that they predict the Higgs boson
properties needed to describe the LEP $Zh\to Z+b$'s excess.
Another possibility, occuring at higher $F$~values, is that the
$\hi$ has a $ZZ\hi$ coupling that is reduced in strength
relative to the SM $ZZ$-Higgs coupling.  
This can give rise to other scenarios with light Higgs bosons that are 
consistent with the LHWG limits on the $Z+b$'s final state.

\begin{figure}[htb!]
\begin{minipage}[c]{0.55\textwidth}
\centerline{\includegraphics[width=2.5in,angle=90]
                            {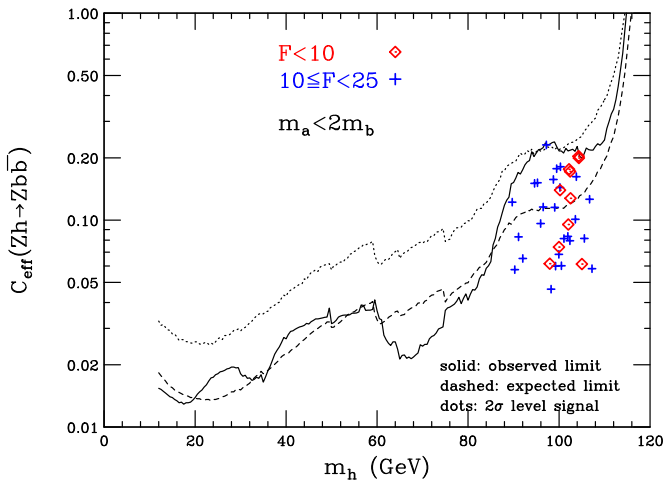}}
\end{minipage}
\begin{minipage}[c]{0.03\textwidth}
$\phantom{0}$
\end{minipage}
\begin{minipage}[c]{0.40\textwidth}
\caption{Expected and observed 95\% CL limits on
  $C_{\rm eff}=[g_{ZZh}^2/g_{ZZ\hsm}^2]\br(h\to b\anti b)$ from
  \citere{lep_higgs} are shown vs. $\mh$.
  Also plotted are the predictions for 
  NMSSM parameter choices in our scan that give fine-tuning measure $F<25$
and  $\mai<2\mb$ with fixed $\tanb=10$ and
gaugino masses of $M_{1,2,3}(\mz)=100,200,300\gev$. \cite{Dermisek:2005gg}}
\label{zbbplot}
\end{minipage}
\end{figure}

\begin{table}[htb!]
\begin{center}
\begin{tabular}{|c|c|c|c|c|c|}
\hline
$\mhi$ & $\mai$ & \multicolumn{3}{c|} {Branching Ratios} & 
$N_{\rm SD}^{\rm max}$ \cr
(GeV) & (GeV) & $\hi\to b\anti b$  & $\hi\to \ai\ai$ &
$\ai\to\tauptaum$  & $300\fbi$ \cr 
\hline
\hline
102.4 & 9.0 & 0.173 & 0.793 &0.875 & 3.6 \cr
102.5 & 5.4 & 0.128 & 0.848 &0.938 & 2.4 \cr
100.2 & 3.1 & 0.141 & 0.832 &0.0 & 2.5 \cr
102.2 & 3.6 & 0.177 & 0.789 &0.814 & 3.3 \cr
\hline
\end{tabular}
\end{center}
\vspace{-0.5em}
\caption{Some properties of the $\hi$ and $\ai$ 
for four points with $F<10$ and $\mai<2\mb$
from a $\tanb=10$, $M_{1,2,3}(\mz)=100,200,300\gev$ NMSSM scan.
 $N_{\rm SD}^{\rm max}(300\fbi)$ is the statistical
significance of the best 
``standard'' LHC Higgs detection channel after accumulating integrated
luminosity of $L=300\fbi$. }
\label{fourpoints}
\end{table}


\subsection{ILC measurements}

In this section we analyze the implications for the ILC
of an NMSSM scenario in which there is a Higgs with fairly SM-like
couplings, but 
that decays to two lighter Higgs bosons, emphasizing the most probable
case of $h \to aa$ and $\mh$ near $100\gev$. 

To set the stage for the ILC,
an important question is the extent to which the type of $h\to aa$
Higgs scenario (whether NMSSM or other) described here can be explored
at the Tevatron and the LHC.
This has been examined in the case of the NMSSM in
\cite{Gunion:1996fb,Ellwanger:2001iw,Ellwanger:2003jt,Ellwanger:2004gz,Ellwanger:2005uu},
with the conclusion that 
observation of any of the NMSSM Higgs bosons may be very difficult at
hadron colliders if the decay mode $\hi\to\ai\ai$ is dominant.
For example, the largest statistical significance in any of the
``standard'' search channels at the LHC, given in the last column of
Table~\ref{fourpoints} for  the four sample type-II scenario points,
is below $5\sigma$ even after $L=300\fbi$ of integrated luminosity is
accumulated. 
In this case, if any signals are observed at the Tevatron or LHC, it
seems very possible that they will be quite tenuous, perhaps difficult
to interpret, and possibly of quite low statistical significance.
In such a circumstance, the ILC becomes essential. It is the ideal
machine for detecting a $\mh\sim 100\gev$ Higgs boson
with unusual decays.  One simply employs the search channel
of inclusive $e^+e^-\to Z^*\to Z\h\to \ell^+\ell^- X$
in which $\mx$ is computed from the incoming and outgoing lepton
information.  An unmistakable peak in $\mx$ is predicted, and
it is completely independent of the $\h$ decay mode.
Detection of $\gam\gam\to\h\to aa$ is also predicted to be completely
straightforward (regardless of the dominant $a$ decay mode) 
at the   $\gam\gam$ collider facility that will
probably accompany the ILC~\cite{Gunion:2004si}. 
Typical signal and background plots are shown in Ref.~\cite{Gunion:2004si}
assuming that the $\gam\gam$ collider is operated at an
electron beam energy of $75\gev$ and with laser and electron
polarizations chosen to produce the general purpose broad spectrum
in $E_{\gam\gam}$.


It should be noted that much of the discussion above regarding Higgs
discovery is quite generic.  Whether the $a$ is truly the NMSSM
CP-odd $\ai$ or just a lighter Higgs boson into which
the SM-like $h$ pair-decays, hadron collider detection of the $h$
in its $h\to aa$ decay mode will be very challenging --- only
the ILC can currently guarantee its discovery.


\subsection{Open questions}

As mentioned above, the NMSSM studies are in general less advanced
than the corresponding MSSM studies. The following open questions
should be answered so that the ILC results can be fully exploited in
the context of the NMSSM:

\begin{itemize}

\item
Calculations for the Higgs masses, mixing angles and couplings have to
be refined to reach the required level of accuracy. This task is
similar to the MSSM case, where, however, already more precise
calculations exist. Also the renormalization in the NMSSM Higgs sector
is less advanced than in the MSSM. More precise calculations are also
needed for the NMSSM Higgs boson production cross sections and
branching ratios. 

\item
In the important low fine-tuning case that $h_1 \to a_1 a_1$ with
$\ai\to \tauptaum$ 
is dominant, excellent
$\tau$-tagging capabilities at the ILC are necessary. 
Further experimental studies are needed. 
In a more general case where $\hi\to \ai\ai$ and $\ai\to b\anti b$, 
excellent $b$-tagging capabilities would be crucial. This includes the
distinction between $b$ and $\bar b$ in order to reduce the number
of $b$-jet combinations in multi-$b$ final states.

\item As in the MSSM, also in the NMSSM the lightest neutralino is a
  good candidate for cold dark matter, see
  e.g.~\cite{Gunion:2005rw,Belanger:2005kh} and references therein. It
  is possible that Higgs physics plays an important role here if the
  LSPs annihilate via a Higgs boson to SM particles. More precise
  calculations for the Higgs boson couplings are necessary to reach
  the required precision. More experimental efforts have to be
  undertaken to extract the corresponding Higgs boson masses and
  widths with the required accuracies. This is discussed further in
  \refse{sec:cosmo}.  There, the difficult scenario of annihilation of
  the lightest neutralino via a rather light singlet-like CP-odd Higgs
  boson (as possible in the low-$F$ NMSSM scenarios) is discussed.

\item
There are other, non-Higgs-sector-related issues. These comprise the
question of baryogenesis within the NMSSM, the investigation of the
LSP quantities at the ILC, and the determination of neutralino sector
observables in order to reconstruct all corresponding fundamental
parameters. 

\end{itemize}


\section{Gauge Extensions of the MSSM}

Another way to increase the mass of the lightest SUSY Higgs boson can
arise from 
theories with extended gauge interactions.  If the Higgs is charged under
one or more of the new gauge groups, there are additional $D$-term 
contributions 
to its quartic term in the Higgs potential, and thus the light CP even
Higgs may be considerably heavier 
than the tree-level MSSM expectation.  Early efforts in this direction
included additional $U(1)$ gauge groups \cite{Haber:1986gz}.  
These are effective,
but the possibility of large Higgs masses is limited if one requires
perturbativity up to the grand unification scale.  More recently,
non-Abelian gauge groups which are asymptotically free have appeared
which allow for large Higgs masses, but whose interactions are free from
Landau poles to arbitrarily high scales \cite{Batra:2003nj,Martinez:2004rh}.

A simple example is provided by supersymmetric topflavor, which has gauge
structure $SU(3)_C \times SU(2)_1 \times SU(2)_2  \times U(1)_Y$.  The
MSSM Higgses and the third family are charged under $SU(2)_1$ whereas the
first two generations are charged under $SU(2)_2$.  Precision EW constraints
require the masses of the extra gauge bosons to be greater than a few TeV.
The Higgs mass at tree level may be expressed as,
\BEA
m_{h^0}^2 & \leq & \frac{1}{2} \left( g^2 \Delta + g_Y^2 \right) v^2
\cos^2 \beta
\EEA
where $g$ and  $g_Y$ are the usual MSSM $SU(2)_L$ and $U(1)_Y$ gauge couplings,
and $\Delta$ parameterizes the new contribution from the extended $SU(2)$
structure,
\BEA
\Delta & \equiv & \frac{1 + \frac{2 m^2}{u^2} \frac{1}{g_2^2}}
{1 + \frac{2 m^2}{u^2} \frac{1}{g_1^2 + g_2^2}}
\EEA
where $m^2$ is a soft mass for the Higgs which breaks 
$SU(2)_1 \times SU(2)_2 \rightarrow SU(2)_L$, $u$ is the breaking scale, and
$g_1$ and $g_2$ are the underlying $SU(2)$ couplings, satisfying
$1 / g_1^2 + 1 / g_2^2 = 1 / g_L^2$.  Tree level Higgs masses on the order
of 350 GeV are relatively easy to obtain in this model, which clearly evades
the LEP bounds~\cite{lep_higgs}. This motivates studies of heavier
(SM-like) Higgs bosons, see \refse{sec:heavySMHiggs}. 

The model contains the extra $Z^\prime$ and $W^\prime$s from the broken
extended gauge symmetry.  These couple more strongly to the third family
than to the first two and the $W^\prime$ is expected to be discovered at the
LHC if this model is realized in nature \cite{Sullivan:2002jt}.  
The ILC could further potentially
pin down the model and its parameters by indirectly measuring the
properties of the $Z^\prime$, and studying the neutralino and chargino
sectors, which now include the partners of these new gauge bosons as well as
the usual MSSM components.  Detailed studies of the branching ratios,
production cross sections, etc., would be interesting to pursue.

The presence of extra asymptotically free gauge interactions may also
result in an interesting modification of the NMSSM.  The new strong gauge
interactions may drive the coupling $\lambda$ small at high energies, 
effectively removing the perturbative unification bound on the value of
$\lambda$ (see \refse{sec:NMSSMHiggsspectrum}), and realizing an
otherwise theoretically unmotivated region of 
parameter space \cite{Batra:2004vc}.  A sample Higgs spectrum (at one loop) is
presented in Figure~\ref{fig:genmssm}, 
and shows several interesting features including
a rather heavy set of CP even fields, and a charged scalar as the lightest of
all of the Higgses.

\begin{figure}[htb]
\centering
\epsfig{width=0.4\linewidth,file=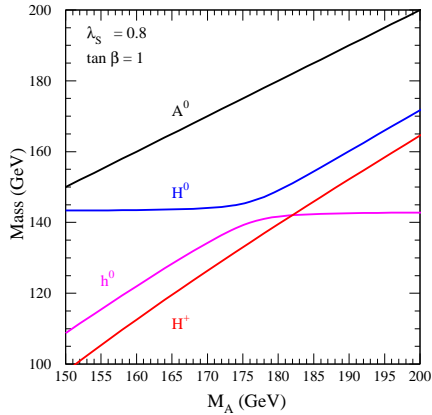} 
\caption{Higgs spectrum in an NMSSM-like model with large $\lambda$
\cite{Batra:2004vc}.}
\label{fig:genmssm}
\end{figure}


\section{Fat Higgs Models}

A final possibility is to interpret the strongly coupled NMSSM as composite
theory which confines at the strong coupling 
scale \cite{Harnik:2003rs,Chang:2004db,Delgado:2005fq}.
The original Fat Higgs \cite{Harnik:2003rs} 
introduces an additional $SU(2)$ interaction
and a number of {\em preons} charged under it.  At an intermediate scale,
$\Lambda$, the $SU(2)$ theory confines, forming a 
singlet superfield $S$ and Higgs-like
superfields $H$ and $\bar{H}$ which play the role of the two MSSM Higgses.
There is a dynamical super-potential induced of the form,
\BEA
W_{dyn} & = & \lambda S \left( H \bar{H} - v_0^2 \right)
\EEA
where naive dimensional analysis predicts $\lambda \sim 4 \pi$ at the
confinement scale, and $v_0^2 \sim m \Lambda / 4 \pi$ 
is related to the mass of the pair of underlying
preon fields ($m$) which bind into $S$.  The mass parameter $m$ is ultimately
what determines the electroweak scale.

Notably different from the NMSSM, the contribution to the Higgs quartic
$\lambda$ is much larger because the effective cut-off is much lower than
the unification scale\footnote{Unification of the MSSM gauge couplings remains
possible (and calculable, despite the strong dynamics) provided some addtional
matter is included to arrange for the unification.}.  In addition, the
electroweak symmetry is broken even in the supersymmetric limit.  Higgs
spectra similar to those expected from the gauge extensions with large
$\lambda$ result, with similar phenomenology quite distinct from the MSSM
and usual NMSSM; see Figure~\ref{fig:genmssm} for an example.

Variations of the original Fat Higgs either restrict the composite sector
to include only the gauge singlet \cite{Chang:2004db} or generate the top quark
from the strong dynamics as well as the Higgses \cite{Delgado:2005fq}, 
motivating the
fact that the top mass is much larger than any of the other fermions by
postulting that the large top Yukawa interaction is a residual of the force
that bound a set of preons into both top and Higgs.


\renewcommand{\Re}{{\rm Re}~}
\renewcommand{\Im}{{\rm Im}~}
\newcommand{\Lama}{\Lambda_1}
\newcommand{\Lamb}{\Lambda_2}
\newcommand{\Lamc}{\Lambda_3}
\newcommand{\Lamd}{\Lambda_4}
\newcommand{\Lame}{\Lambda_5}
\newcommand{\Lamf}{\Lambda_6}
\newcommand{\Lamg}{\Lambda_7}
\newcommand{\go}{G^0}
\newcommand{\mhpm}{m_{\hpm}}
\newcommand{\cosb}{\cos\beta}
\newcommand{\sinb}{\sin\beta}
\newcommand{\kpuo}{\kappa^{U,0}}
\newcommand{\rhuo}{\rho^{U,0}}
\newcommand{\kpdo}{\kappa^{D,0}}
\newcommand{\rhdo}{\rho^{D,0}}
\newcommand{\kpu}{\kappa^U}
\newcommand{\rhu}{\rho^U}
\newcommand{\kpd}{\kappa^D}
\newcommand{\qlo}{Q^0_L}
\newcommand{\uro}{U^0_R}
\newcommand{\dro}{D^0_R}
\newcommand{\ql}{Q_L}
\newcommand{\ur}{U_R}
\newcommand{\mud}{M_U}
\newcommand{\mdd}{M_D}

\section{The Significance of $\tan\beta$ in 
  Two-Higgs-Doublet Models}

There are a number of discussions in this report about the 
measurement of the parameter $\tan\beta$ at the ILC.  
A precise theoretical understanding of the significance of this
parameter is important. Recent progress in this direction is
reviewed.


\subsection{$\tan\beta$ as an (un)physical parameter}

In the two-Higgs-doublet model (2HDM), the Higgs sector of the
Standard Model is extended to include
two complex hypercharge-one Higgs
doublets~\cite{higgs_hunters,branco}.  
After the breaking of electroweak symmetry (preserving the
electromagnetic U(1)$_{\rm EM}$ symmetry group), the 
Higgs fields' vacuum expectation values (vevs) 
can be chosen to take the following form:
\beq
\langle \Phi_1 \rangle={1\over\sqrt{2}} \left(
\begin{array}{c} 0\\ v_1\end{array}\right), \qquad \langle
\Phi_2\rangle=
{1\over\sqrt{2}}\left(\begin{array}{c}0\\ v_2 e^{i\xi}
\end{array}\right)\,,\label{potmin}
\eeq
where $0\leq\xi<2\pi$ and the $v_i$ are real and positive and
normalized such that
$v^2\equiv v_1^2+v_2^2=4m_W^2/g^2=(246~{\rm GeV})^2$.
In addition, the ratio of the magnitudes of the two vevs is denoted by
$\tan\beta\equiv v_2/v_1$, where $0\leq\beta\leq\pi/2$.  

The relative phase $\xi$ is clearly not a physical parameter of the
model, since one is free to rephase the $\Phi_2$ field to absorb this
phase.  Perhaps more surprisingly, $\tan\beta$ is also not generally a
physical parameter.  Since the two Higgs fields $\Phi_1$ and
$\Phi_2$ possess identical gauge quantum numbers, one is free to
redefine the scalar fields via an arbitrary U(2) 
transformation~\cite{Ginzburg:2004vp,davidson},
$\Phi_i\to U_{ij}\Phi_j$.  Different choices of 
the unitary matrix $U$ correspond to
different basis choices for the scalar fields.  Physical predictions
of the model must, of course, be basis independent.

If $\tan\beta$ is not a physical parameter, what is the meaning of
$\tan\beta$ that appears so often in this report?  In fact,
$\tan\beta$ can be promoted to a physical parameter by 
imposing some restrictions on possible terms that appear in the 2HDM
Lagrangian.  For example, by imposing
supersymmetry on the model, the two Higgs fields $\Phi_1$ and $\Phi_2$
are distinguishable in a particular basis, and this basis
provides a physical definition of $\tan\beta$.  
However, if
phenomena consistent with a two-Higgs-doublet sector are observed, it
may not be immediately obvious what theoretical 
constraints govern the underlying
2HDM.  Thus, in order to carry out a truly model-independent study of
2HDM phenomena, one cannot employ $\tan\beta$ as defined below
\refeq{potmin}.  

Here we consider the most general 2HDM. 
One natural basis choice that exists is the so-called 
Higgs basis~\cite{Donoghue:1978cj,type2,silva,lavoura,lavoura2,davidson}, 
consisting of the scalar fields $\Phi_a$ and $\Phi_b$, in
which only one of the neutral Higgs fields (which by convention will
be denoted by $\Phi_a^0$)
has a non-zero vacuum
expectation value.  This basis choice is not unique, since one is
always free to independently rephase the individual Higgs fields.  
One linear combination of these two rephasing transformations (corresponding
to $\Phi_a\to e^{i\psi}\Phi_a$, $\Phi_b\to e^{i\psi}\Phi_b$) is
the global hypercharge transformation
U(1)$_{\rm Y}$.  This transformation does
not modify any of the scalar
potential parameters.  However, we shall employ it simply to remove
the phase from the vacuum expectation value of $\Phi_a^0$.
In this convention, the Higgs basis is parameterized by one angle
$\chi$:
\BEA
\Phi_a&=&\Phi_1\cosb+e^{-i\xi}\,\Phi_2\sinb\,, \label{higgsbasis1}\\
e^{i\chi}\,\Phi_b&=&-e^{i\xi}\,\Phi_1\sinb+\Phi_2\cosb\,.\label{higgsbasis2}
\EEA
One then obtains
\beq \label{abbasis}
\Phi_a=\left(\begin{array}{c}
G^+ \\ {1\over\sqrt{2}}\left(v+\varphi_1^0+iG^0\right)\end{array}
\right)\,,\qquad
e^{i\chi}\,\Phi_b=\left(\begin{array}{c}
H^+ \\ {1\over\sqrt{2}}\left(\varphi_2^0+i A\right)\end{array}
\right)\,.
\eeq
Indeed only $\Phi_a^0$ develops a
vacuum expectation value, which by convention is chosen to be real and
positive. 
Of the original eight scalar degrees of freedom, three Goldstone
bosons ($G^\pm$ and~$G^0$) are absorbed (``eaten'') by the $W^\pm$ and
$Z$.  In \refeq{abbasis},
$\varphi_1^0$, $\varphi_2^0$ are CP-even neutral Higgs fields,
$ A$ is a CP-odd neutral Higgs field, and $H^\pm$ is the physical
charged Higgs boson pair.
If CP is conserved,
then $ A$ is the physical CP-odd Higgs scalar and $\varphi_1^0$
and $\varphi_2^0$ mix to produce two physical neutral CP-even Higgs
states $ h$ and $ H$.  If the Higgs sector is CP-violating, then
$\varphi_1^0$, $\varphi_2^0$, and $ A$ all mix to produce three
physical neutral Higgs states of mixed CP properties.  

In the Higgs basis, the scalar potential takes the following form:
\BEA
\mathcal{V}&=& M_{11}^2\Phi_a^\dagger\Phi_a+M_{22}^2\Phi_b^\dagger\Phi_b
-[M_{12}^2\,e^{i\chi}\Phi_a^\dagger\Phi_b+{\rm h.c.}]\nonumber\\*[6pt]
&& +\edz\Lambda_1(\Phi_a^\dagger\Phi_a)^2
+\edz\Lambda_2(\Phi_b^\dagger\Phi_b)^2
+\Lambda_3(\Phi_a^\dagger\Phi_a)(\Phi_b^\dagger\Phi_b)
+\Lambda_4(\Phi_a^\dagger\Phi_b)(\Phi_b^\dagger\Phi_a)
\nonumber\\*[6pt]
&& +\left\{\edz\Lambda_5\,e^{2i\chi}(\Phi_a^\dagger\Phi_b)^2
+\big[\Lambda_6\,e^{i\chi}(\Phi_a^\dagger\Phi_a)
+\Lambda_7\,e^{i\chi}(\Phi_b^\dagger\Phi_b)\big]
\Phi_a^\dagger\Phi_b+{\rm h.c.}\right\}, \label{pothiggs}
\EEA
$M_{11}^2$ and $M_{12}^2$ are fixed by the
scalar potential minimum conditions, and $M_{22}^2$ is
directly related to the physical charged Higgs boson squared-mass:
\beq \label{chhiggsmass}
\mhpm^2=M_{22}^2+\edz v^2\Lamc\,.
\eeq

The conditions for CP-invariance of the Higgs sector are very simple.
The 2HDM is CP-conserving if and only if the Higgs basis scalar
potential parameters satisfy~\cite{lavoura,davidson} 
\beq \label{cpconds}
\Im(\Lame^*\Lamf^2)=\Im(\Lame^*\Lamg^2)=\Im(\Lamf^*\Lamg)=0\,.
\eeq
If \refeq{cpconds} is
satisfied, then there exists a choice of $\chi$ such that all the
Higgs basis scalar potential parameters are simultaneously
real~\cite{Branco:2005em,cpbasis,Ivanov:2005hg}. 
%
By working in the Higgs basis, it is clear that $\tan\beta$ never appears.
Thus, the couplings of the Higgs bosons to the vector bosons
and the Higgs self-couplings can depend only on the scalar
potential parameters (in the Higgs basis), 
the gauge couplings and the mixing angles 
that determine the neutral Higgs boson mass eigenstates in terms of
$\varphi_1^0$, $\varphi_2^0$ and $A$.  

In practice, the parameter $\tan\beta$ is most often 
associated with the Higgs-fermion Yukawa
couplings.  Thus, we now focus on the coupling of the 2HDM to three
generations of quarks and leptons.
The Higgs couplings to fermions are model dependent.
The most general structure for the 2HDM Higgs-quark Yukawa couplings,
often referred to as the type-III
model~\cite{typeiii}, 
is given in the Higgs basis by (with an analogous form for the
Higgs-lepton Yukawa couplings):
\beq \label{ymodeliii}
\!\!\!\!\!\!\!\!
-\mathcal{L}_Y=\overline \qlo 
\widetilde\Phi_a \kpuo  \uro +\overline Q_L^0 \Phi_a \kpdo \dro
+ \overline \qlo \widetilde\Phi_b \rhuo \uro +\overline \qlo  \Phi_b \rhdo \dro
+{\rm h.c.}\,,
\eeq
where $\widetilde\Phi\equiv
i\sigma_2 \Phi^*$,
$\qlo $ is the weak isospin quark doublet,
and $\uro$, $\dro$ are weak isospin quark singlets.
$\kappa^{Q,0}$ and $\rho^{Q,0}$ ($Q=U$, $D$)
are Yukawa coupling matrices in quark
flavor space.

The fermion mass eigenstates are related to the interaction eigenstates
by the following unitary transformations:
\beq
U_L=V_L^U U_L^0\,,\qquad\quad U_R=V_R^U U_R^0\,,\qquad\quad
 D_L=V_L^D D_L^0\,,\qquad\quad D_R=V_R^D D_R^0\,,
\eeq
and the Cabibbo-Kobayashi-Maskawa matrix is defined as $K\equiv V_L^U
V_L^{D\,\dagger}$.  It is also convenient to define ``rotated''
linear combinations of the Yukawa coupling matrices:
\beq \label{rotyuks}
\kpu\equiv V_L^U \kpuo V_R^{U\,\dagger}\,,\qquad
\rhu\equiv V_L^U \rhuo V_R^{U\,\dagger}\,,\qquad
\kpd\equiv V_L^D \kpdo V_R^{D\,\dagger}\,,\qquad
\rho^D\equiv V_L^D \rhdo V_R^{D\,\dagger}\,.
\eeq
The quark mass terms are identified by  replacing the scalar
fields with their vacuum expectation values.
The unitary
matrices $V_L^U$, $V_L^D$, $V_R^U$ and $V_R^D$ are chosen so that
$\kappa^D$ and $\kappa^U$ are diagonal with real non-negative entries.
These quantities are proportional to the \textit{diagonal} 
quark mass matrices:
\beq \label{qmasses}
M_D=\frac{v}{\sqrt{2}}\kappa^D\,,\qquad\qquad
M_U=\frac{v}{\sqrt{2}}\kappa^U\,.
\eeq
However, the matrices
$\rho^D$ and $\rho^U$ are \textit{complex}
non-diagonal matrices, which are not constrained in a general model.

The Higgs-fermion interaction, expressed in terms of quark mass
eigenstates, takes the following form~\cite{branco,davidson,inprep}:
\BEA \label{modeliiihqq}
 && \hspace{-0.4in} -\mathcal{L}_Y=
\frac{1}{v}\,\overline D\mdd D\varphi_1^0+\frac{1}{\sqrt{2}}\overline{D}
(\rho^D P_R+{\rho^D}^\dagger P_L)D\varphi_2^0
+\frac{i}{\sqrt{2}}\overline D(\rho^D P_R-{\rho^D}^\dagger P_L)D A
+\frac{i}{v}\overline D\mdd\gamma_{5}D\go \nonumber \\[4pt]
&& +\frac{1}{v}\,\overline U\mud U\varphi_1^0+\frac{1}{\sqrt{2}}
\overline{U}(\rhu P_R+{\rhu}^\dagger P_L)U\varphi_2^0
-\frac{i}{\sqrt{2}}\overline 
U(\rhu P_R-{\rhu}^\dagger P_L)U A
-\frac{i}{v}\overline U\mdd\gamma_{5}U\go \nonumber \\[4pt]
&&+\left\{\overline U\left[K\rho^D P_R-{\rhu}^\dagger KP_L\right] DH^+
+\frac{\sqrt{2}}{v}\,\overline U\left[K\mdd P_R-\mud KP_L\right] DG^+ 
+{\rm h.c.}\right\}\,,
\EEA
where $P_{R,L}=\edz(1\pm\gamma_{5})$.
The physical couplings of the quarks to the 
Higgs fields are then obtained by expressing
$\varphi_1^0$, $\varphi_2^0$ and $A$ in terms of the neutral
Higgs mass eigenstates.
Once again, the parameter $\tan\beta$ does not appear.  

The most general 2HDM is phenomenologically untenable
over most of its parameter space.
In particular, since $\rho^D$ and $\rho^U$ are
non-diagonal, \refeq{modeliiihqq} [even when expressed in terms of the
neutral Higgs mass eigenstates] exhibits tree-level Higgs-mediated
flavor changing neutral currents (FCNCs).  
One way to avoid this problem is to go the 
region of parameter space where $M_{22}^2$ is very
large (or equivalently $m_{H^\pm}\ll m_Z$).  In this 
\textit{decoupling limit}, the second doublet $\Phi_2$ is very
heavy and decouples from the effective low-energy theory.
To a very good approximation, $\varphi_1^0$
is a CP-even Higgs mass eigenstate with couplings
nearly indistinguishable from those of the Standard Model Higgs 
boson~\cite{ghdecoupling,inprep2}.
In particular, tree-level Higgs-mediated FCNCs are suppressed, and
the corresponding 2HDM is phenomenologically viable (at the expense of
the fine-tuning necessary to guarantee that the second Higgs doublet
is very heavy).

The more common procedure for avoiding FCNCs is to impose a symmetry
on the theory in order to restrict the form of the Higgs-fermion
interactions in a particular basis~\cite{Weinberg}.  
This procedure then establishes
the \textit{symmetry basis}.  If the scalar
fields of the symmetry basis are denoted by $\Phi_1$ and $\Phi_2$,
whose relation to the Higgs basis is given by 
\refeqs{higgsbasis1} -- (\ref{higgsbasis2}), then the parameter $\tan\beta$ is
promoted to a physical parameter of the theory.  For example, a
discrete symmetry can be imposed such that 
$\Phi_2$ decouples from the fermions in the symmetry basis.  
This yields the so-called
type-I Higgs-fermion couplings~\cite{type1,hallwise}.  
Another discrete symmetry can be
found in which down-type right-handed fermions couple exclusively to
$\Phi_1$ and up-type right-handed fermions couple exclusively to
$\Phi_2$.  This  yields the so-called
type-II Higgs-fermion couplings~\cite{type2,hallwise}. 
As a final example, the MSSM also exhibits 
type-II Higgs-fermion couplings~\cite{susyhiggs}.

For simplicity, consider a \textit{one-generation} type-II 
CP-conserving model.  It is straightforward to show that the Higgs
basis Yukawa couplings satisfy the following equation~\cite{davidson}:
\beq  \label{consistcond}
\kappa^U\kappa^D+\rho^U\rho^D=0\,.
\eeq
Moreover, one can compute $\tan\beta$ (where $\beta$ is the relative
orientation of the Higgs basis and the symmetry basis):
\beq  \label{tanbdefs}
\tanb=\frac{-\rho^D}{\kappa^D}=\frac{\kappa^U}{\rho^U}\,,
\eeq
where $\kappa^Q=\sqrt{2}M_Q/v$.  Thus, it follows that:
\beq
\rho^D=-\frac{\sqrt{2}m_d}{v}\tanb\,,\qquad \rho^U=
\frac{\sqrt{2}m_u}{v}\cotb\,.
\eeq
Inserting this result into 
\refeq{modeliiihqq}, and expressing $\varphi_1^0$ and $\varphi_2^0$
in terms of the CP-even Higgs mass eigenstates $h$ and $H$
yields the well-known
Feynman rules for the type-II Higgs-quark interactions.


\subsection{Interpretation of ILC measurements}

How should one interpret precision measurements of 2HDM Yukawa
couplings?  A priori, one does not know whether a symmetry basis exists.  
Thus, one should attempt to determine experimentally 
the $\rho^U$ and $\rho^D$ coupling
matrices. Testing relations such as \refeq{consistcond} could reveal the
presence of a symmetry basis, in which case one can determine $\tanb$ from
\refeq{tanbdefs}.  However, one could discover that the equality of
\refeq{tanbdefs} does not hold.
For example, in a one-generation type-III
CP-conserving model, a more general procedure would consist of defining
three $\tanb$-like parameters:
\beq \label{othertanb}
\tanb_d\equiv \frac{-\rho^D}{\kappa^D}\,,\qquad
\tanb_u\equiv  \frac{\kappa^U}{\rho^U}\,,
\eeq
and a third parameter $\tanb_e\equiv -\rho^E/\kappa^E$
corresponding to the Higgs-lepton interaction.  The equality of these
three $\tanb$-like parameters would imply the existence of a symmetry
basis (and an underlying type-II Higgs-Yukawa coupling).  Deviations
from these equalities would imply a more complex realization of
the electroweak symmetry breaking dynamics.
 
We can illustrate this procedure in a very simple model approximation.
In the MSSM at large $\tanb$ (and supersymmetric masses significantly
larger than $m_Z$), the effective Lagrangian that describes the
coupling of the Higgs bosons to the third generation quarks is given
by:
\beq \label{leff}
-\mathcal{L}_{\rm eff}=
h_b(\overline q_L\Phi_1) b_R
+ h_t(\overline q_L \widetilde\Phi_2) 
t_R +\Delta h_b\,(\overline q_L \Phi_2) b_R
+{\rm h.c.}\,,
\eeq
where $\overline q_L\equiv (\overline u_L\,,\,\overline d_L)$.
The term proportional to $\Delta h_b$ is generated at one-loop
due to supersymmetry-breaking effects.  
The tree-level relation between $m_b$ and $h_b$ is
modified~\cite{deltamb1,deltamb2}: 
\beq
m_b=\frac{h_b v}{\sqrt{2}}\,\cosb(1+\Delta_b)\,,
\eeq
where $\Delta_b\equiv (\Delta h_b/h_b)\tanb$.  That is, $\Delta_b$ is
$\tanb$-enhanced, and governs the leading one-loop correction
to the physical Higgs couplings to third generation quarks.
In typical models at large $\tanb$, $\Delta_b$ can be
of order 0.1 or larger and of either sign.  An explicit
expression for $\Delta_b$ is given in
\citeres{Carena:2002es,deltamb1,deltamb2}. 
In the approximation scheme above, 
$\kappa^U\simeq h_t\sinb$, $\rho^U\simeq h_t\cosb$, and~\cite{davidson}
\beq
\kappa^D\simeq h_b\cosb(1+\Delta_b)\,,\qquad\quad
\rho^D \simeq -h_b\sinb\left(1-\frac{\Delta_b}{\tan^2\beta}\right)\simeq
-h_b\sinb\,.
\eeq
It follows that:
\beq
\tan\beta_d\,\simeq\,\frac{\tanb}{1+\Delta_b}\,,\qquad\quad
\tan\beta_u\simeq\tan\beta\,.
\eeq
Thus, supersymmetry-breaking loop-effects can yield observable
differences between $\tan\beta$-like parameters that probe the
underlying nature of supersymmetry-breaking.


\section{Lepton Flavor Violation in the Higgs boson decay at ILC}

Lepton flavor violation (LFV) in charged leptons 
directly indicates new physics beyond the SM.  
In the models such as based on SUSY, in addition to
the gauge boson mediated LFV process, the LFV Yukawa 
couplings are naturally induced by 
slepton mixing~\cite{lfvyukawa}.  
The direct search for such LFV Yukawa couplings 
would be possible by measuring the Higgs decay 
process $h, (H, A) \to \tau^\pm \mu^\mp$ at collider 
experiments\cite{lfv-hdecay0,lfv-hdecay,lfv-hdecay2}. 
Such a possibility has been examined at the LHC in Ref.~\cite{assamagan}.
In Ref.~\cite{ota}, this possibility has been discussed 
and the feasibility of this process at a ILC via the 
Higgs-strahlung process has been studied   
under the constraints from current data on LFV rare tau 
decay processes.
A possibility of a search for the LFV Yukawa interaction 
via the deep inelastic scattering process $e^- N \to \tau^- X$ 
at the fixed target option of ILC has also been studied\cite{kuno}.


\section{Dimension-Six Higgs Operators}

\subsection{The SM as a low energy effective theory of 
new physics models}

Under the constraint from the current precision data,  
the SM with a light Higgs boson would be an effective 
theory to the cutoff scale which could be considered 
to be as high as the Planck scale without contradicting 
theoretical arguments such as vacuum stability and triviality. 
On the other hand, such scenario of the SM with very high 
cutoff scale has been considered to be unnatural, because 
due to quantum corrections the Higgs boson mass is quadratically 
sensitive to the cutoff (new physics) scale ($\sim \Lambda^2$).
Demanding relatively small (less than 90\%) hierarchy, 
the scale of the new physics naturally would be expected 
to be of the order of 1-10 TeV. 
The effect of new physics would then enter into 
the low energy theory as higher dimensional 
operators.
If a light Higgs boson is discovered at the LHC but no additional
particles are seen at the LHC or the ILC, it is important to
search for small deviations of the Higgs boson potential from the
SM predictions to probe new physics scales.
If the reason for such small deviations is beyond-the-SM 
physics at large scales $\Lambda$, the effective operator approach
can be chosen to parameterize the low-energy behavior of such models.

The effective Lagrangian of the SM with the cutoff $\Lambda$ can be 
described as 
\begin{eqnarray}
  {\cal L}_{\rm eff} = {\cal L}_{\rm SM}^{} 
  + \sum_n \frac{f_n}{\Lambda^2} {\cal O}_n 
  + O(1/\Lambda^{n+2}),  
\end{eqnarray}
where ${\cal O}_n$ are dimension six operators, and 
$f_n$'s are dimensionless ``anomalous couplings''. 
It is expected that $f_n$'s are of order of unity at the 
cutoff scale $\Lambda$.
There are eleven SM gauge invariant bosonic operators 
in dimension six (dim-6) terms\cite{hagiwara93}, 
and eight of them are relevant to the Higgs boson field:
\begin{eqnarray}
  {\cal O}_{BW} &=& \Phi^\dagger B_{\mu\nu} W^{\mu\nu} \Phi,
   \label{o-BW} \\
  {\cal O}_{\Phi,3} &=& (D_\mu\Phi)^\dagger \Phi^\dagger\Phi (D^\mu\Phi),
   \label{o-P3}\\ 
  {\cal O}_{WW} &=& \Phi^\dagger W_{\mu\nu} W^{\mu\nu} \Phi, 
   \label{o-WW}\\ 
  {\cal O}_{BB} &=& \Phi^\dagger B_{\mu\nu} B^{\mu\nu} \Phi, 
   \label{o-BB}\\ 
  {\cal O}_{W} &=& (D_\mu\Phi)^\dagger W^{\mu\nu} (D_\nu\Phi), 
   \label{o-W}\\ 
  {\cal O}_{B} &=& (D_\mu\Phi)^\dagger B^{\mu\nu} (D_\nu\Phi), 
   \label{o-B}\\ 
  {\cal O}_{\Phi,1} &=& \frac{1}{2} \partial^\mu(\Phi^\dagger\Phi)
                                    \partial_\mu(\Phi^\dagger\Phi),
   \label{o-P1}\\ 
  {\cal O}_{\Phi,2} &=& \frac{1}{3} (\Phi^\dagger\Phi)^3.
   \label{o-P2} 
\end{eqnarray}
These operators are separated to the following three categories. \\

\noindent
[Class I] (${\cal O}_{BW}^{}$, ${\cal O}_{\Phi,3}^{}$) \\
These operators contribute to the gauge boson two point functions 
(${\cal O}_{BW}^{} \Rightarrow S, 
  {\cal O}_{\Phi,3}^{} \Rightarrow T$), 
so that they receive tight constraints from 
the current precision data:
their coefficients are bounded at the 95\% C.L. as \cite{zhang03}
\begin{eqnarray}
 -0.07 < \frac{f_{BW}^{}}{(\Lambda/{\rm TeV})^2} < 0.04, 
 -0.02 < \frac{f_{\Phi,3}^{}}{(\Lambda/{\rm TeV})^2} < 0.02. 
\end{eqnarray}\\

\noindent
[Class II] 
(${\cal O}_{WW}^{}$, ${\cal O}_{BB}^{}$, ${\cal O}_{W}^{}$, 
${\cal O}_{B}^{}$) \\
These four operators lead to interactions of the 
Higgs boson with gauge bosons such as 
\begin{eqnarray}
 {\cal L}_{\rm eff}^H &=& 
  g_{H\gamma\gamma}^{} H A_{\mu}A^{\nu} + 
  g_{HZ\gamma}^{(1)}  A_{\mu\nu}Z^\mu \partial^{\nu}H + 
  g_{HZ\gamma}^{(2)} H A_{\mu\nu}Z^{\mu\nu}+ 
  g_{HZZ}^{(1)}  Z_{\mu\nu}Z^\mu \partial^{\nu}H \nonumber\\&& + 
  g_{HZZ}^{(2)} H Z_{\mu}Z^{\mu\nu}+ 
  g_{HWW}^{(1)} (W_{\mu\nu}^+W^{-\mu} \partial^{\nu}H+h.c.) + 
  g_{HWW}^{(2)} H W_{\mu\nu}^+W^{-\mu\nu}.
\end{eqnarray}
Combinations of the coefficients of 
${\cal O}_{WW}^{}$, ${\cal O}_{BB}^{}$, ${\cal O}_{W}^{}$ and 
${\cal O}_{B}^{}$ contribute to 
these $V$-$H$ couplings $g_{HVV'}^{}$.
Theoretical bounds from partial wave unitarity on these 
coefficients are rather weak.  
By using the precision data for the LEP II Higgs search results 
and one-loop corrections, these coefficients are  
constrained not so strongly as 
\begin{eqnarray}
&& -4 < \frac{f_{B}^{}}{(\Lambda/{\rm TeV})^2} < 2, \hspace{6mm}
 -6 < \frac{f_{W}^{}}{(\Lambda/{\rm TeV})^2} < 5, \nonumber \\
&& -17 < \frac{f_{BB}^{}}{(\Lambda/{\rm TeV})^2} < 20, \hspace{6mm}
 -5 < \frac{f_{WW}^{}}{(\Lambda/{\rm TeV})^2} < 6. 
\end{eqnarray}
The measurement of the triple gauge boson couplings can also 
contribute to constrain these anomalous couplings, but the present 
results give weaker bounds. 
They are expected to be constrained more precisely at future 
collider experiments; for example, via 
$WW \to WW$\cite{zhang03} at the LHC, via
$e^+e^- \to W^+W^-\gamma$ ($ZZ\gamma$) at 
the ILC\cite{gonzalez-garcia99}, 
and $\gamma\gamma \to ZZ$ at a $\gamma\gamma$ collider\cite{taoprep}.  
The effects of the anomalous couplings 
of this class modify the production and decay 
of the Higgs boson, so that they can be well tested 
via these processes 
at the LHC and the ILC after the Higgs boson is detected. 
\\

\noindent
[Class III] (${\cal O}_{\Phi,1}^{}$, ${\cal O}_{\Phi,2}^{}$) \\
The last two are genuine dim-6 interactions of the Higgs 
boson, which are not severely constrained by the current 
precision electroweak data: see next subsection.


\subsection{Genuine Dimension Six Higgs Operators}
\label{sec:dim6}

Recently, phenomenological consequences of 
the genuine dim-6 Higgs interactions 
have been studied~\cite{LC-TH-2003-035}. 
The operators ${\cal O}_{\Phi,1}^{}$ and ${\cal O}_{\Phi,2}^{}$ 
lead to a Lagrangian
\begin{equation}
    {\mathcal{L}}' = \sum_i^2 {a_i \over v^2}{\mathcal{O}}_{\Phi,i},
    \label{lp}
\end{equation}
where $a_i=f_{\Phi,i}^{}\, v^2/\Lambda^2$. 
This is the consistent formulation respecting the SM gauge symmetry.
These operators contribute to the $HVV$ and $HHVV$ interactions 
and the Higgs boson self-interactions $H^3$ and $H^4$.
In terms of the canonically normalized Higgs field 
and the physical Higgs boson mass, the interactions 
are expressed by 
\begin{eqnarray}
 {\cal L}_{HVV}^{}&=& 
  \left(M_W^2 W_\mu^+W^{-\mu}+\frac{1}{2}M_Z^2 Z_\mu Z^\mu\right)
   \left((1-\frac{a_1}{2})\frac{2H}{v}+(1-a_1)\frac{H^2}{v^2}\right), \\
 {\cal L}_{H^3}^{}&=&  - \frac{m_H^2}{2v}
  \left((1-\frac{a_1}{2}+\frac{2a_2}{3}\frac{v^2}{m_H^2})H^3 
  - \frac{2a_1 H \partial_\mu H\partial^\mu H}{m_H^2}\right),\\
 {\cal L}_{H^4}^{}&=&  - \frac{m_H^2}{8v}
  \left((1-a_1+4a_2\frac{v^2}{m_H^2})H^3 
  - \frac{4a_1 H \partial_\mu H\partial^\mu H}{m_H^2}\right).
\end{eqnarray}
The measurement of $HVV$ vertices can test $a_1$ while 
by measuring the Higgs self-coupling constants $a_2$ 
(as well as $a_1$) can be probed.     
For $H^3$ and $H^4$ interactions, the effect of $a_2$ 
not only changes the Higgs potential, 
but also gives kinetic corrections 
which are enhanced at high energies. 

In~\cite{LC-TH-2003-035}, it has been shown that the parameter $a_1$ 
can be well measured at the 
ILC to an accuracy of 0.005 (0.003) corresponding 
to a scale $\Lambda \approx 4 $ TeV, from 1 \iab of data 
at 500 (800) GeV through the measurement of the production 
cross-sections from Higgs-strahlung 
and $WW/ZZ$-fusion for $\mH = 120 $ GeV. 
The parameter $a_2$ modifies the form of the Higgs potential 
and thus the Higgs pair production cross-section. 
At the LHC the measurement of the $HH$ production cross section 
is difficult because of the small rate with huge backgrounds\cite{baur03}.
At the ILC with the same integrated luminosity (1 ab$^{-1}$), 
for $\mH = 120 $ GeV, $a_2$ can be measured to 0.13 (0.07) 
at 500 (800) GeV corresponding to a scale $\Lambda \approx $ 
1 TeV\cite{castanier01}; see Fig.~\ref{fig:dim6}.

\begin{figure}[htb!]
\centering
\epsfig{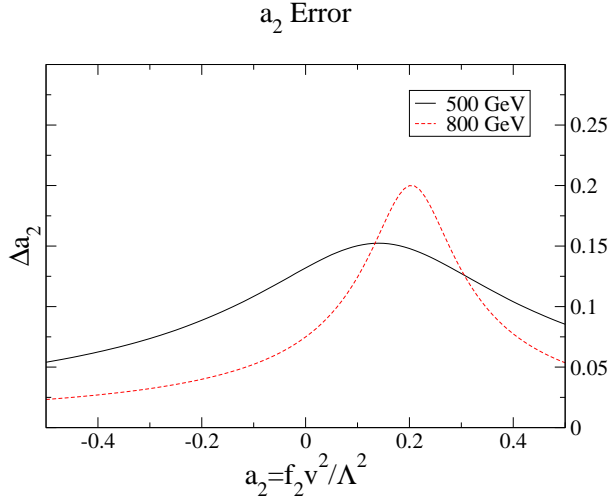}
\caption{Combined statistical accuracy on $a_2$ with an 
integrated luminosity of 1 ab$^{-1}$ for $\sqrt{s}=500$ GeV 
and 800 GeV for $m_H^{}=120$ GeV, 
using the Higgs-strahlung and the $WW$-fusion channels.}
\label{fig:dim6}
\end{figure}


\subsection{Open questions}

In order to exhaust the possibilities of the physics of dimension six
operators, the following items should be considered in future studies.

\begin{itemize}

\item
Fermionic operators should be included in 
the analysis.

\item
It is also important to calculate the anomalous 
coupling in various new physics scenarios. 
By classifying models in terms of anomalous couplings, 
a systematic search for new physics beyond the SM 
is possible with the future precision 
data at the ILC, even when only a light Higgs 
boson is found at the LHC.

\end{itemize}


\section{Little Higgs}
\label{sec:LittleHiggs}

Recently, a new class of models for electroweak symmetry breaking has
been proposed, the ``Little Higgs'' models~\cite{LH-reviews}.  Similar to
supersymmetry as a model-building principle, the Little-Higgs
mechanism stabilizes the mass of scalar particles against its
quadratic cut-off sensitivity.  The Higgs boson appears as a member of
the Goldstone boson multiplet pertaining to a global symmetry that is
spontaneously broken at some scale $\Lambda \sim \mathcal{O}(10
 \tev)$.  Naive dimensional analysis connects the cutoff $\Lambda$ of
such models with the decay constant of these bosons $F$ by $F \sim
\Lambda/4\pi$.  This scale $F$ is supposed to be above the electroweak
scale $v=246 \gev$, presumably in the $\tev$-region.

Being Goldstone bosons, the scalars in Little-Higgs models have
neither a mass nor a potential as long as the global symmetry is
exact.  However, the SM gauge and Yukawa interactions of the Higgs
explicitly break the global symmetry, thereby radiatively generating a
potential for the Higgs by the Coleman-Weinberg mechanism.  This idea
dates back to the 1970s~\cite{Georgi:1974yw}, but at that time no
mechanism was known that suppressed the one-loop corrections to the
Higgs self-energy without fine-tuning among the couplings.

This drawback is cured by the mechanism of collective symmetry
breaking, originally observed in the context of deconstructed extra
dimensions~\cite{Arkani-Hamed:2001ca}.  The structure of gauge and
Yukawa interactions can be arranged such that if any one of their
couplings is set to zero, the emerging global symmetry makes the Higgs
an exact Goldstone boson.  Only if all couplings are different from
zero, a potential is generated.  If the global symmetry is broken in
this particular way, one-loop contributions to the Higgs potential
depend only logarithmically on the cut-off, while a quadratic
divergence appears only at two-loop order.  This allows for a
three-scale hierarchy without fine-tuning: $v \sim F/4\pi \sim
\Lambda/(4\pi)^2$.

To fully implement collective symmetry breaking in all sectors and
thus cancel all one-loop quadratic cutoff sensitivity, a complete set
of new gauge bosons, scalars and fermions is needed.  Just like the
MSSM extends the SM by new weakly-interacting particles with a
characteristic soft-SUSY-breaking mass scale, Little-Higgs models thus
lead to an extended weakly-interacting effective theory, where most of
the new particles are expected with masses of order $F\gsim
1 \tev$.

In order to successfully implement collective symmetry breaking,
either the global symmetry representation must be reducible or the
gauge group must be non-simple.  Consequently, there are two basic
lines of model building.  Particularly economic models emerge from a
simple global symmetry group with an irreducible representation, such
as the ``Littlest Higgs'' with its $SU(5)\to SO(5)$ breaking
chain~\cite{Arkani-Hamed:2002qy}.  The non-simple gauge structure of
such models generically introduces $W'$ and $\gamma'/Z'$ recurrences
of the SM vector bosons (for additional U(1) factors there always is
the freedom not to gauge it, see the discussion at the end of this
subsection).  By contrast, simple-(gauge-)group models 
tend to have a proliferation of scalar states in multiple
representations and typically lead to a two-Higgs-doublet model at low
energies, while the extra gauge bosons corresponding to nondiagonal
generators of the extended EW gauge group are nontrivial to
detect~\cite{Kaplan:2003uc}.  Finally, the ``moose models'' that were 
found first~\cite{Arkani-Hamed:2002qx} establish a relation to
higher-dimensional gauge theories~\cite{Contino:2003ve} and feature
factorized symmetries both in the scalar and vector sectors.

Collective symmetry breaking in the fermion sector requires the
introduction of extra fermions, most notably heavy top-like states
that implement a see-saw mechanism for the top mass in conjunction
with the cancellation of fermion loops in the Higgs potential.  In
simple-group models, the irreducibility of the gauge group calls for
extra fermion partners in all three generations.

By construction, Little-Higgs models involve tree-level mixing between
the SM fields and their heavy partners.  This leads to shifts in
electroweak precision observables of the order $v^2/F^2$ which are at
variance with data if $F$ is below the $\tev$ range.  Detailed
analyses~\cite{Csaki:2002qg} generically constrain the model
parameters to $F \gsim 2-3 \tev$.  An elegant way of removing some
of the dangerous contributions is the implementation of a custodial
$SU(2)$ symmetry~\cite{Chang:2003un}.  Incidentally, in some models
the problem is also ameliorated if the Higgs boson is rather heavy
while the hierarchy $v\ll F\ll\Lambda$ is kept~\cite{Kilian:2003xt}.
The mixing of states is eliminated by the implementation of a discrete
symmetry, $T$-parity~\cite{Birkedal-Hansen:2003mp}.  Analogous to
$R$-parity in the MSSM and $KK$-parity in extra-dimension models, this
leads to a one-loop suppression of shifts in precision observables and
makes the lightest $T$-odd particle a suitable dark-matter candidate.
Finally, embedding the little-Higgs mechanism in a supersymmetric
model~\cite{Birkedal:2004xi} can nicely remove the fine-tuning
problems that are present both in the MSSM and in non-supersymmetric
models.

If the Little-Higgs mechanism is realized in nature, we need
experimental means to verify this fact and to identify the particular
model.  If possible, we should observe the new particles, the symmetry
structure in the couplings that is responsible for stabilizing the
Higgs potential, and the non-linearity of the representation that is
characteristic for Goldstone scalars at low energies.  This can be
achieved by a combined analysis of LHC and ILC
data~\cite{Csaki:2002qg,Han:2003gf,LHatPC}, while flavor data will play a
minor role~\cite{Huo:2003vd}.  The prospects for uncovering the
ultimate UV completion of the model~\cite{Birkedal:2004xi,Katz:2003sn}
remain uncertain.

The LHC provides us with a considerable detection reach for new vector
bosons.  If found, their properties (masses, mixing angles and
branching ratios) can be determined with considerable accuracy.
Simultaneously, new vector bosons can be studied by their indirect
effect on contact interactions at the ILC, where the search limit
extends into the $10 \tev$ range.  New quarks can also be searched
for at the LHC, although with less sensitivity, and they affect
top-quark precision observables that can be measured at the ILC.

For the scalar sector of Little Higgs models we have to consider a
wide range of possibilities, essentially arbitrary combinations of
singlets, doublets, triplets, and more.  While such particles may be
nontrivial to isolate in LHC data, at least the lighter part of the
Higgs spectrum can be covered by searches at the ILC.  Precision
studies of vector-boson and Higgs anomalous couplings can even uncover
traces of the nonlinear Goldstone-boson representation and thus
establish the main ingredient of the Little-Higgs paradigm.

Finally, a rather generic feature of Little-Higgs phenomenology
originates from the enlarged global symmetry.  This symmetry
necessarily involves spontaneously broken $U(1)$ subgroups.
Consequently, the spectrum contains either $Z'$ bosons (gauged
$U(1)$)~\cite{Csaki:2002qg}, or light pseudoscalar particles
otherwise~\cite{Kilian:2004pp}.  The latter can be studied in
diphotons at the LHC, in top-quark production at the ILC, and as
resonances in $\gamma\gamma$ fusion at a photon collider.


\section{Models with Higgs Triplets}

Some of the ``Little Higgs'' models discussed in \refse{sec:LittleHiggs}
postulate the existence of an SU(2)$_L$ triplet of scalar fields in 
addition to the conventional doublet. 
Unfortunately, the physical states associated with this triplet
are expected to be too heavy for direct observation 
at a TeV linear collider. Nonetheless, ``Little Higgs'' models do not
exhaust all possible models predicting triplets of scalar fields. 
These may occur, for example, in the models with extra spatial 
dimensions~\cite{HTriplet_XD} (see also \refse{sec:ExtraDimHiggs}) and
in other scenarios.  
Even the Standard Model Higgs sector can be extended to 
contain both doublet and triplet fields
without contradicting existing experimental data~\cite{HTriplet_SM}.
Models with Higgs triplets predict an enriched spectrum of physical
states, including not only neutral and singly charged particles, 
but also doubly charged states. These can be directly 
produced and detected at the linear $\EE$ collider. One shouldn't
also exclude the possibility to detect a doubly charged Higgs boson,
$H^{--}$, as a resonance 
at an $e^-e^-$ collider, though this will require 
an anomalously strong $Hee$ coupling.  The existence of 
scenarios in which the $H^{--}$ state can be 
produced in $e^-e^-$ collisions with sufficiently high rate enabling
its detection would emphasize the importance of a future 
ILC facility with several collision options.
Clearly, rich phenomenology of the models with Higgs triplets invites
investigation of the ILC potential for probing such a models.


\section{Higgs Bosons and Extra Dimensions}
\label{sec:ExtraDimHiggs}

In recent years, models with extra spatial dimensions have become very popular.
They primarily address the electroweak hierarchy problem either by
diluting gravity in a large extra dimensional volume that the SM
interactions do 
not feel (as in the large or ADD extra dimensions~\cite{add}) or
by explaining the apparent difference in scales by a warped metric 
in a small extra dimension (as in the warped or RS models~\cite{rs}).
Large extra dimensions necessarily have the Standard Model confined to a
three dimensional sub-manifold, or brane, whereas RS exists in several
versions which primarily differ as to how much of the SM is in the bulk.

The `classic' signatures involve the effects of the Kaluza-Klein (KK) tower of
gravitons.  In the large extra dimension case, the size of the extra dimensions
is expected to be large enough that the KK states are extremely tightly spaced
and thus appear as a continuum of objects, each one of which interacts only
with ordinary gravitational strength.  Relevant impact on collider observables
occurs because there are so many states that they collectively have a large
effect.  In the case of warped extra dimensions the states have masses with
spacing on the order of TeV.  However, each is much more strongly coupled
than the ordinary graviton, and so individually they can have a large
impact on processes at colliders.  The upshot is that both theories can
induce deviations of SM processes 
like $e^+e^-\to f \bar f$ and $e^+e^-\to W^+W^-$ from the virtual exchange 
of KK modes \cite{addhewett,rshewett}, or
their real emission together with SM fermions or gauge bosons~\cite{gravemis}.
These modes have been studied with experimental models of
detectors, see e.g.\ \citere{tdr}. 

More recently, also the impact of extra dimensions on the Higgs boson
phenomenology has been studied. In the ADD scenario, three effects have been
analyzed (for another analysis, see \citere{Hall:1999fe}): 

\begin{enumerate}
\item
A modification of the quasi-resonant $W^+W^- \to H$ production process
through interference of the SM amplitude
with the imaginary part of the graviton/graviscalar KK exchange 
amplitude~\cite{LC-TH-2003-011}. In order to
yield a significant modification, a large total Higgs width is needed 
(i.e., large $\mH$), which implies on the other hand a large center-of-mass
energy. While the graviscalar contribution only modifies the normalization
of the cross-section 
(by few percent for $\sqrt{s} = $ 1 TeV, $\mH = $ 500 GeV and 2 extra 
dimensions at a fundamental Planck scale of 1 TeV),
a significant change of the angular distribution is expected from the
spin-2 graviton exchange. 

\begin{figure}[htb!]
\begin{center}
\epsfxsize=10cm   
\epsfbox{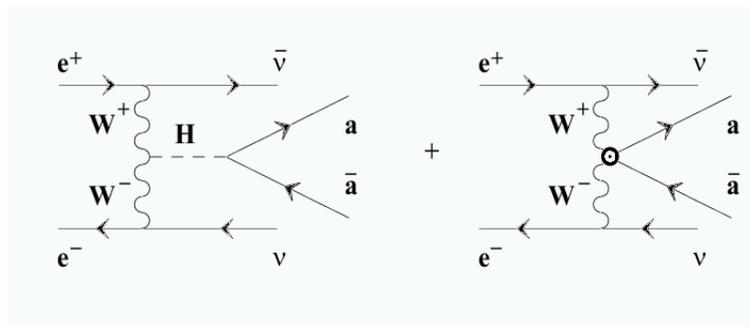}
\caption{
Effective Feynman diagrams from Equation (\ref{eff}) corresponding to 
single Higgs production with the subsequent Higgs decay into $a \bar a$, where
$a$ is a heavy particle, such as $W$, $Z$ or $t$ \cite{Datta:2003cf}.}
\label{Hbrane2}
\end{center}
\end{figure}

\item
A modification of the process $e^+e^-\to H H Z$ and the existence of
the process $e^+e^-\to H H\gamma$ which is absent at 
tree level in the SM~\cite{Deshpande:2003sy}. For a 1 TeV ILC and 
$\mH = $ 120 GeV, a sizable correction to $e^+e^-\to H H Z$ both in
normalization and angular distribution is expected for fundamental Planck
scale up to a few TeV. Furthermore, the cross-section for $e^+e^-\to H H\ga$
exceeds 0.1 fb for a fundamental Planck scale below approximately 2 TeV.
In~\cite{Deshpande:2003sy}, expected 5$\sigma $ discovery limits on
the fundamental Planck scale of 880--1560 (1640--2850) GeV have been derived
at $\sqrt{s} = $ 500 (1000) GeV for 6--3 extra dimensions.

\begin{figure}[htb!]
\begin{center}
\epsfxsize=8cm   
\epsfbox{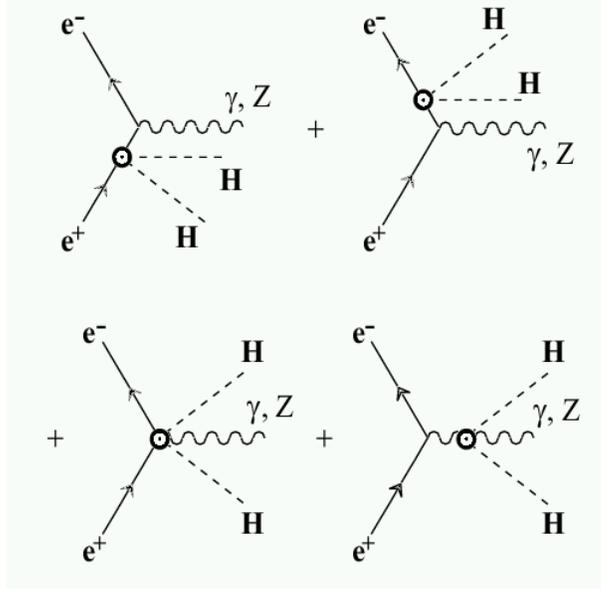}
\caption {
Feynman diagrams which contribute to the processes $e^+e^- \rightarrow ZHH$ or
$e^+e^- \rightarrow \gamma HH$ \cite{Deshpande:2003sy}. In this case,
a five particle effective vertex generated by the virtual graviton
exchange or branon radiative corrections has to be taken into account,
which is also provided by the Lagrangian (\ref{eff}).} 
\label{Hbrane3}
\end{center}
\end{figure}

\item
 Direct Higgs pair production $e^+e^- \rightarrow HH$
 \cite{Rizzo:1999qv,Delerue:2004tc}.  
The signature of this channel is two pairs of $b$-jets with an invariant mass 
close to the Higgs mass. The main SM backgrounds 
are $e^+e^- \rightarrow W^+W^-,\, ZZ,\, ZH$ and $b\bar bb\bar b$.  
The differential cross section gives a characteristic signal of the spin 
2 interaction. For an ILC operating at 1 TeV and a fundamental Planck 
scale of 2 TeV, the cross section is of the 
same order as the SM $e^+e^- \rightarrow ZH$ for $m_H=120$ GeV. 
Indeed, 700 Higgs pair events 
can be generated with an integrated luminosity of 500 fb$^{-1}$. 

\end{enumerate}

\begin{figure}[htb!]
\begin{center}
\epsfxsize=5cm   
\epsfbox{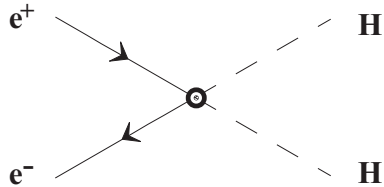}
\caption{
Feynman diagrams associated with the Higgs pair production 
$e^+e^- \rightarrow HH$ \cite{Delerue:2004tc}. The effective 
four particle vertex is provided by the exchange of virtual KK-graviton
in the rigid ADD and by a branon loop for flexible branes (See Equation (\ref{eff})).}
\label{Hbrane1}
\end{center}
\vspace{2em}
\end{figure}


\subsection{Radion Effects}

Extra dimensional models contain moduli which describe the size
and shape of the extra dimensions.   
These scalar fields, usually called the radion, can be
understood as a zero mass component of the higher dimensional metric, i.e.,
$g_{xy}$ where $x$ and $y$ are two of the compactified directions.  In the case
of a five dimensional model, there is a single scalar field which parameterizes
the size of the extra dimension.
In order to stabilize the size of the extra dimensions, these fields must
acquire masses, and the specifics of the stabilization mechanism will in
general play some role in determining the properties of the radion
\cite{radions}.
Generically, they are
expected to couple to SM particles through the trace of the energy-momentum
tensor, i.e., very similar to the SM Higgs boson, up to the trace anomaly 
of QCD.
The lightest radion might in fact be lighter than the lightest graviton
excitation and thus could be the discovery channel for the model. 

In flat space, the radion is not usually relevant for colliders (see
however \citere{Battaglia:2004js}).  The 
couplings are too small and not compensated by a large number of fields.
They may nonetheless play an important role in cosmology and could even
make up a fraction of the dark matter \cite{Kolb:2003mm}.
In warped backgrounds, the radion is much more relevant, because it
picks up the same enhancement of its couplings as the graviton KK modes.
Thus, the influence on the Higgs sector could be quite dramatic
through mixing of the radion with the Higgs.
This leads to mass eigenstates which are a mixture of
both, and thus a strong modification of both the Higgs boson and 
radion properties, in particular their
couplings to gauge bosons and fermions. For a review of the radion 
phenomenology, see e.g.~\cite{radionrev}. Analyses of radion physics
at the LHC and the ILC can be found in \citere{lhcilc}. The radion
sector is governed 
by 3 parameters: the strength of the radion-matter interactions described
by an energy scale $\Lambda_\phi$, the mass of physical radion $m_\phi$,
and the radion-Higgs mixing parameter $\xi$. In Fig.~\ref{fig:radionprop},
the effective couplings squared of the Higgs boson and the radion (relative
to those of a SM Higgs boson) are shown
for the choice $\Lambda_\phi = 5 $ TeV, and three values of the radion mass
(20, 55, 200 GeV) as a function of $\xi$. Large deviations of the Higgs
couplings from their SM values are expected if there is large radion-Higgs 
mixing present. The radion itself has couplings which are reduced by a factor
$v/\Lambda_\phi$ with respect to those of a SM Higgs in the case of no mixing,
which requires high luminosity for direct discovery. The sensitivity of the
trilinear Higgs coupling to radion admixtures 
has been studied as well in~\cite{radionrev}.

\begin{figure}[htb!]
\centering
\epsfig{height=0.4\linewidth,file=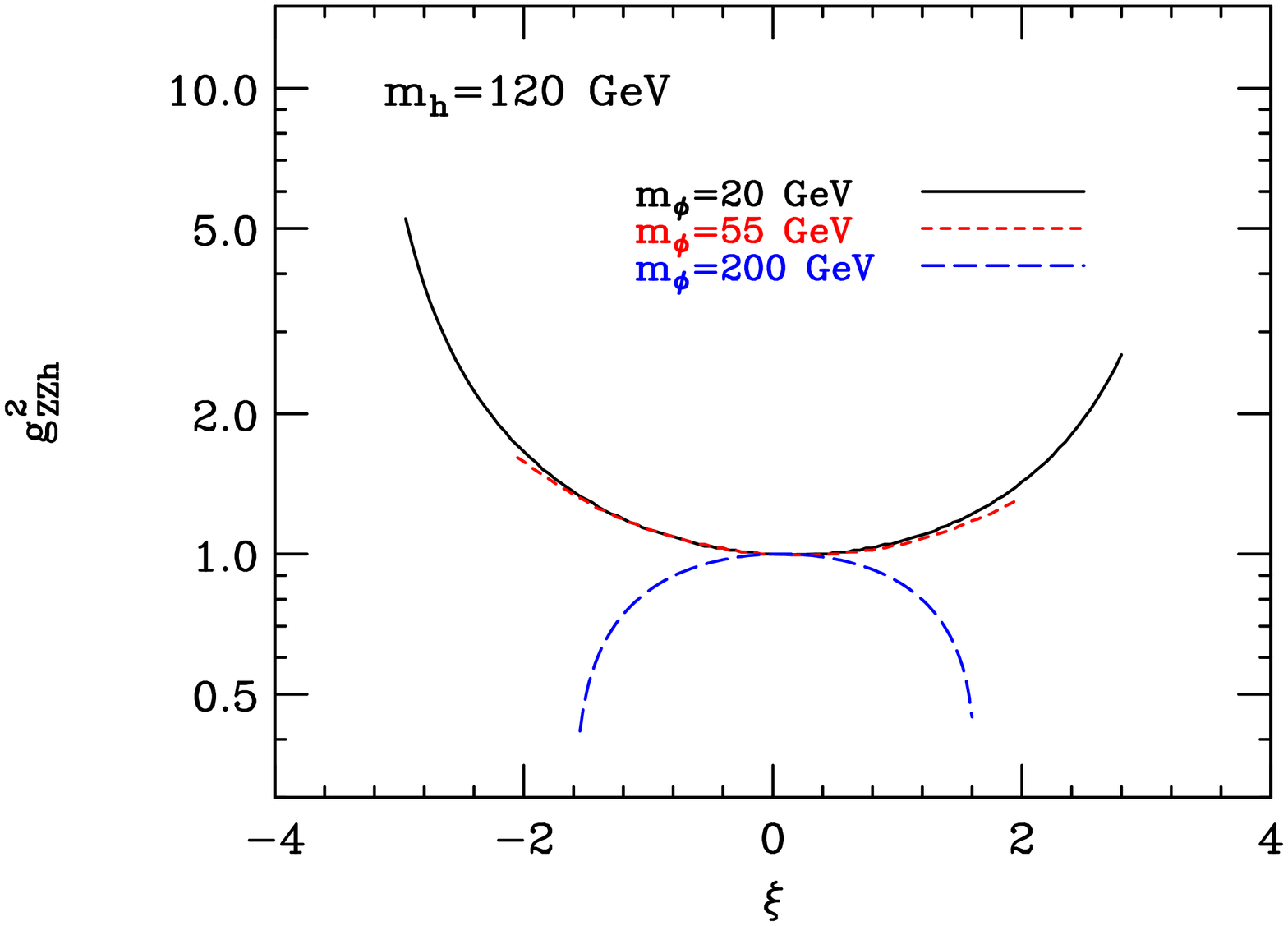,angle=90}
\epsfig{height=0.4\linewidth,file=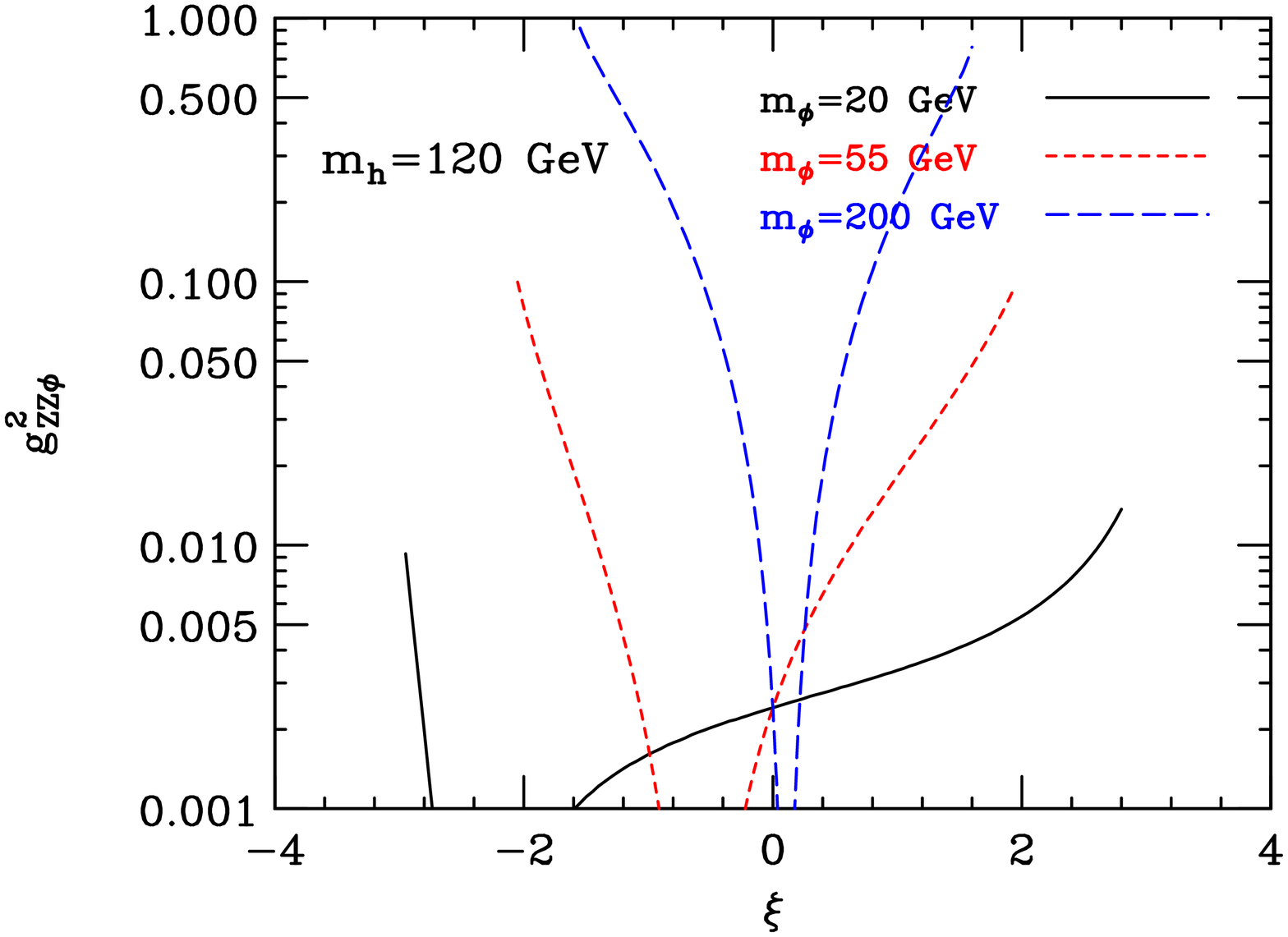,angle=90}
\caption{Effective coupling of the Higgs boson (left) and the radion (right)
to Z boson (from \cite{radionrev}).}
\label{fig:radionprop}
\end{figure}


\subsection{Branon Effects}

The previous discussion has assumed that the branes are perfectly rigid. 
However, it is expected that the dynamics responsible for generating the brane
will in fact leave it with a finite tension which will allow it to 
oscillate. The brane fluctuations are manifest
as new pseudo-scalar particles, which are called branons \cite{DoMa}. 
For appropriate choices of parameters, 
branons are natural dark matter candidates, being stable and 
weakly coupled to Standard Model fields \cite{CDM}. The 
collider signature for branons is missing energy; in particular, 
the mono-jet and single photon signals for hadron colliders \cite{Hadro} 
and the single $Z$ and single photon channels for $e^+e^-$ colliders
\cite{ACDM}. This last signature has been studied experimentally by
LEP~\cite{L3}, finding the most constraining bound for the brane 
tension scale: $f>180$ GeV, for light branons.  These collider 
searches are complementary to astrophysical and cosmological 
ones \cite{CDML} and they need to be completed with branon 
radiative correction analyses. Indeed, these quantum corrections 
are very similar to the virtual KK-graviton effects in the 
rigid ADD model since the two effects predict the same effective 
Lagrangian \cite{rad}:
\begin{eqnarray}
{\mathcal \triangle}{\cal L}_{\rm eff}= W_1 T_{\mu\nu}T^{\mu\nu}+W_2 T_\mu^\mu
T_\nu^\nu  \label{eff}\,,
\end{eqnarray}
where $W_1$ and $W_2$ are two constants which depends on the 
renormalization procedure.  This means that branons lead to analogous 
four body interactions and Higgs boson phenomenology.

\chapter{Analysis Tools}
\label{sec:tools}

In this section we give a brief overview about the state-of-the-art
tools that are 
available for the evaluation of Higgs boson observables. The first
subsection summarizes the computer codes that have been developed for
the calculation of Higgs boson mass spectra and branching
fractions. The second subsection briefly describes event generators
available for Higgs boson physics. All codes are listed in
\refta{tab:codes}, together with their corresponding web address
(see also the Les Houches Web Repository for BSM Tools
\cite{bsmrepository}).

\begin{table}[htb!]
\begin{center}
\begin{tabular}{ll}
\hline\hline
{\tt HDECAY} & \verb+people.web.psi.ch/spira/hdecay/+ \\
{\tt FeynHiggs} & \verb+www.feynhiggs.de/+ \\
{\tt CPsuperH} & \verb+www.hep.man.ac.uk/u/jslee/CPsuperH.html+ \\
{\tt NMHDECAY} & \verb+www.th.u-psud.fr/NMHDECAY/nmhdecay.html+ \\
\hline
{\tt HERWIG} & \verb+hepwww.rl.ac.uk/theory/seymour/herwig/+ \\
{\tt ISAJET} & \verb+www.phy.bnl.gov/~isajet/+ \\
{\tt MADGRAPH} & \verb+madgraph.hep.uiuc.edu/+ \\
{\tt PYTHIA} & \verb+www.thep.lu.se/~torbjorn/Pythia.html+ \\
{\tt SHERPA} & \verb+www.physik.tu-dresden.de/~krauss/hep/+ \\
{\tt WHIZARD} & \verb+www-ttp.physik.uni-karlsruhe.de/whizard/+ \\
\hline\hline
\end{tabular}
\end{center}
\caption{Web links for Higgs-related tools.}
\label{tab:codes}
\end{table}


\section{Higgs spectra and branching fractions}

\subsection{Standard Model Higgs}

The Standard Model Higgs decay branching fractions and total width
are computed by the packages {\tt HDECAY}~\cite{Djouadi:1997yw} 
and {\tt FeynHiggs}~\cite{feynhiggs,feynhiggs1.2,feynhiggs2}.
We outline here the known radiative corrections and sources of 
theoretical uncertainty in the SM Higgs branching ratios.  We focus
here on {\tt HDECAY}:
for some of the decays to $b \bar b$, $c \bar c$, $s \bar s$,
we believe that {\tt HDECAY} is slightly better than
{\tt FeynHiggs} in terms of containing SM 
higher order radiative corrections to the decay widths;
however, the other decays are treated to comparable accuracy.

The radiative corrections to Higgs decays to fermion and boson pairs
have been reviewed in Ref.~\cite{Spirareview}; we give here a brief
sketch of the known corrections and refer to Ref.~\cite{Spirareview}
for references to the original calculations.
The full QCD corrections to the Higgs decay to $q \bar q$ are known
up to three loops neglecting the quark mass in the kinematics; at the 
two loop level they are known for massive final-state quarks in the 
leading-$N_F$ approximation.
The electroweak corrections to the Higgs decay to quark
or lepton pairs are known at one-loop; in addition, the QCD corrections 
to the leading top-mass-enhanced electroweak correction term are known up
to three loops, to order $G_Fm_t^2 \alpha_s^2$.
All of these corrections to the Higgs partial widths to fermions
are included in a consistent way in the program 
{\tt HDECAY}~\cite{Djouadi:1997yw}.

For the Higgs masses below the $WW$ threshold, $M_H \lesssim 160$~GeV,
decays into off-shell gauge bosons ($WW$, $ZZ$) are important and 
affect the total Higgs width, thus feeding in to the Higgs branching fractions.
Off-shell decays to $W$~and $Z$~bosons are taken into account. 
One-loop electroweak corrections to Higgs decays to $WW$
and $ZZ$ are known, together with the QCD corrections
to the leading $\mathcal{O}(G_F m_t^2)$ result up to three loops.
These corrections to $\Gamma_{W,Z}$ amount to less 
than about 5\% in the intermediate Higgs mass range \cite{Spirareview}
(translating to less than roughly 2\% in ${\rm BR}(H \to b \bar b)$
for $M_H \simeq 120$ GeV) and have been neglected in {\tt HDECAY}, although
their inclusion would seem straightforward.

The Higgs decay into gluon pairs, $H \to gg$, is of order 
$\mathcal{O}(G_F \alpha_s^2)$
at the leading one-loop order and therefore suffers a large scale dependence.
In the SM, the state of the art is the $\mathcal{O}(G_F \alpha_s^4)$ 
contribution in the large $m_t$ limit computed in 
Ref.~\cite{Chetyrkin:1997iv}.  
This correction is of order 20\% for a light
Higgs, increasing the total Higgs hadronic width by only about 1\%,
and the remaining scale uncertainty from varying the 
renormalization scale between $M_H/2$ and $2 M_H$ is roughly $\pm 5\%$.  
This correction appears not to be included in {\tt HDECAY}, although 
its inclusion would be trivial.

The SM $H \to \gamma\gamma$ decay partial width receives QCD corrections, 
which of course only affect the quark loop diagrams.  Because the 
external particles in the $\gamma\gamma H$ vertex are color neutral, 
the virtual QCD corrections are finite by themselves.  Since no real
radiation diagrams contribute, the QCD corrections to $H \to \gamma\gamma$
are equivalent to those to the inverse process $\gamma\gamma \to H$.  
This is in contrast to the QCD corrections to the $ggH$ vertex.
The QCD corrections to $\Gamma_{\gamma}$ in the SM 
are known analytically at the 
two-loop [$\mathcal{O}(G_F \alpha_s)$] order \cite{ggH2loopQCD} 
and as a power expansion up to third order in $M_H/m_t$ at three-loop 
[$\mathcal{O}(G_F \alpha_s^2)$] order \cite{ggH3loopQCD}.  They are small
for Higgs masses $M_H < 2 m_t$; the $\mathcal{O}(\alpha_s)$ corrections
are only of order 2\% for $M_H < 2 M_W$, and the $\mathcal{O}(\alpha_s^2)$
corrections are negligible, demonstrating that the QCD corrections are 
well under control.
The SM $H \to \gamma\gamma$ decay partial width also receives electroweak
radiative corrections.  The electroweak corrections are much more 
difficult to compute than the QCD corrections and a full two-loop calculation
does not yet exist.  The electroweak correction due to two-loop diagrams
containing light fermion loops and $W$ or $Z$ bosons 
(with the Higgs boson coupled to the $W$ or $Z$ boson, because the light 
fermion Yukawa couplings are neglected) was computed recently in 
Ref.~\cite{ggHEWlightf} and contributes between $-1\%$ and $-2\%$ for 
$M_H \lesssim 140$ GeV.
The leading $\mathcal{O}(G_F m_t^2)$ electroweak correction due to 
top-mass-enhanced two-loop diagrams containing third-generation 
quarks was also computed recently in Ref.~\cite{ggHEWGFmt2} as an expansion
to fourth order in the ratio $M_H^2/(2M_W)^2$.\footnote{The 
$\mathcal{O}(G_F m_t^2)$ electroweak correction was also considered in
Ref.~\cite{ggHEWGFmt2old}, whose results disagree with that of 
Ref.~\cite{ggHEWGFmt2}.  The source of this disagreement is addressed
in Ref.~\cite{ggHEWGFmt2}.}  The expansion appears to
be under good control for $M_H \lesssim 140$ GeV, where this correction 
contributes about $-2.5\%$ almost independent of $M_H$.  The 
leading $\mathcal{O}(G_F M_H^2)$ correction was computed
in Ref.~\cite{ggHEWGFMH2} for large $M_H$; however, this limit is
not useful for a light Higgs boson.
We conclude that the electroweak radiative corrections to 
$H \to \gamma\gamma$ appear to be under control at the 1--2\% level.

The Higgs branching ratios in the SM also have a parametric 
uncertainty due to the experimental uncertainties in the 
SM input parameters.  The largest sources of parametric uncertainty
are the bottom and charm quark masses and (to a lesser extent) the 
strong coupling $\alpha_s$, which 
contributes via the QCD corrections to the $H q \bar q$ couplings.  
These parametric uncertainties were evaluated 
in, e.g., Refs.~\cite{Djouadi:1995gt,MSSMHiggs}.  
Ref.~\cite{MSSMHiggs} found a parametric uncertainty in 
${\rm BR}(H \to b \bar b)$ of about 1.4\% for $M_H = 120$ GeV, using 
the standard $\alpha_s = 0.1185 \pm 0.0020$ \cite{alphas} and 
a somewhat optimistic $m_b(m_b) = 4.17 \pm 0.05$ GeV 
($\overline{\rm MS}$) \cite{mbmb}.
The parametric uncertainty in ${\rm BR}(H \to b \bar b)$ is suppressed due to 
the fact that $\Gamma_b$ makes up about 2/3 of the Higgs total
width at $M_H = 120$ GeV, leading to a partial cancellation of the 
uncertainty in the branching ratio; the uncertainty instead feeds into
the other Higgs branching ratios to $W^+W^-$, $\tau\tau$, etc.
We also expect the parametric uncertainty in ${\rm BR}(H \to b \bar b)$
to be somewhat larger at higher Higgs masses, where $\Gamma_b$ no longer 
dominates the total width.

The best measurements of $\alpha_s$ come from LEP-I and II; the Tevatron
and LHC are unlikely to improve on this.  The ILC is expected to improve the 
precision of $\alpha_s$ by about a factor of two \cite{gigaz,ILCalphaS}.
The bottom quark mass is extracted from heavy quarkonium spectroscopy 
and $B$ meson decays with a precision limited by theoretical uncertainty.
There are prospects to improve the bottom quark mass extraction through
better perturbative and lattice calculations \cite{bottommass}
and more precise measurements of the upsilon meson properties from CLEO
\cite{CLEO}.


\subsection{MSSM Higgs}

\subsubsection{CP-conserving MSSM}

The three major codes used to compute the Higgs spectrum and branching 
fractions in the CP-conserving MSSM are {\tt HDECAY}~\cite{Djouadi:1997yw},
{\tt FeynHiggs}~\cite{feynhiggs,feynhiggs2}, and
{\tt CPsuperH}~\cite{cpsh}.  

{\tt HDECAY} computes the MSSM Higgs spectrum and mixing angles 
with the user's choice of packages: 
{\tt Subh} \cite{Carena:1995bx} and {\tt Subhpole} \cite{Carena:1995wu} 
(now subsumed into 
{\tt CPsuperH} \cite{cpsh,mhiggsCPXRG2,Carena:2001fw})
as well as Haber et al.~\cite{Haber:1996fp}, which use a 
renormalization-group-improved effective field theory approach to compute
Higgs masses and mixing angles; and 
{\tt FeynHiggsFast}%
\footnote{
{\tt FeynHiggsFast}, though still available, has been superseded by
{\tt FeynHiggs2.2}.
}%
\cite{feynhiggs},
which contains the two-loop diagrammatic computation 
(the most recent version uses {\tt FeynHiggsFast} by default).  
A reconciliation between the two-loop diagrammatic computation
of the mass of the lightest CP-even Higgs boson and the effective field theory
computation was performed in \citere{bse}.
The code contains, in addition to the SM modes, 
Higgs decays to SUSY particles and SUSY contributions to loop-induced decays.
In the case of Higgs decays to SM particles through tree-level couplings,
the MSSM rates are computed by scaling the SM couplings by the appropriate
Higgs mixing angles.

{\tt FeynHiggs}, in addition to computing the MSSM Higgs mass spectrum and 
mixing angles using the on-shell Feynman-diagrammatic approach at the 
two-loop level, also computes the Higgs decay branching fractions and
the Higgs boson production cross sections at the Tevatron and the LHC
for all relevant channels (in a simplified parametric approach).
In some
of the SUSY decays, {\tt FeynHiggs} includes corrections that are not in 
{\tt HDECAY}, e.g., the $\Delta_b$ corrections in the charged Higgs decay to
$t\bar b$.  Details and references were given in Sec.~\ref{sec:mhMSSM}.

{\tt CPsuperH} contains a computation of the MSSM Higgs boson pole masses
and mixing angles based on the renormalization-group-improved effective
potential computation in the MSSM~\cite{Carena:2001fw}. Since the case
with real parameters is only a special case of the complex MSSM, more
details about the two codes will be
discussed further in the next subsection.


\subsubsection{CP-violating MSSM}

The two major codes used to compute the Higgs spectrum and branching
fractions in the MSSM with explicit CP violation are 
{\tt FeynHiggs} \cite{feynhiggs,feynhiggs2,mhiggsCPXFD1,habilSH} and 
{\tt CPsuperH} \cite{cpsh}. 

The program 
{\tt FeynHiggs}~\cite{feynhiggs,feynhiggs1.2,feynhiggs2} 
is based on the results obtained in the Feynman-diagrammatic (FD)
approach~\cite{mhiggsletter,mhiggslong,mhiggsAEC,mhiggsEP4b}.
It computes the MSSM Higgs spectrum and mixing angles, taking into
account the full phase dependence at the one-loop and a partial phase
dependence at the two-loop level.
The code 
{\tt CPsuperH}~\cite{cpsh} is based on the renormalization group (RG)
improved effective potential
approach~\cite{mhiggsRG1a,mhiggsRG1,Carena:1995wu,bse}. 
For the MSSM with real parameters the two codes can differ by up to
$\sim 4 \gev$ for the light CP-even Higgs boson mass, mostly due to
formally subleading two-loop corrections that are included only in 
{\tt FeynHiggs}. For the MSSM with complex parameters the phase
dependence at the two-loop level is included in a more advanced
way~\cite{mhiggsCPXRG,mhiggsCPXRG2} in {\tt CPsuperH}, but, on the other hand, 
{\tt CPsuperH} does not contain all the subleading one-loop contributions
that are included~\cite{mhiggsCPXFD1,mhiggsCPXFDproc,habilSH} in 
{\tt FeynHiggs}. 
In the case of real parameters, the remaining theoretical uncertainty on 
the light CP-even Higgs boson mass has been estimated to be below 
$\sim 3 \gev$~\cite{mhiggsAEC,PomssmRep}, if {\tt FeynHiggs} is used.
In the case of complex parameters, no detailed estimate for the additional
theory uncertainty due to the complex phases has been performed.
However, the intrinsic uncertainties should be larger than for real
parameters by at most a factor of two.


\subsection{Beyond the MSSM}

\subsubsection{The NMSSM}

The Next-to-Minimal Supersymmetric Standard Model (NMSSM) contains
a Higgs sector enlarged by a singlet chiral superfield compared to
the MSSM.
The Higgs masses, couplings and decay widths in this model are computed
by the Fortran code {\tt NMHDECAY} \cite{nmhdecay,Ellwanger:2005dv}.
The computation of the Higgs spectrum includes leading electroweak 
corrections and leading two-loop terms.  The computation 
of the decay widths is carried out as in {\tt HDECAY} \cite{Djouadi:1997yw} 
(temporarily without three-body decays); Higgs-to-Higgs decays are included
and can be important.
Each point in parameter space can be checked against low-energy observables, 
as well as negative Higgs boson searches at LEP and negative 
sparticle searches at LEP and the Tevatron, including unconventional 
channels relevant for the NMSSM. Version 2.0 \cite{Ellwanger:2005dv} 
allows to compute the dark matter relic density via a link to a NMSSM 
version of the {\tt MicrOmegas} code \cite{Belanger:2004yn}.

\subsubsection{Other extended models}

There are many extended Higgs models in the literature for which no
public codes exist.  The Higgs couplings have been studied
in detail using private codes for some of these models, as follows:
\begin{itemize}

\item Randall-Sundrum model: mixed radion and 
Higgs \cite{Hewett:2002nk,Dominici:2002jv},

\item Randall-Sundrum model: Higgs with a profile in the warped extra 
dimension \cite{Lillie:2005pt},

\item Universal extra dimensions and Kaluza-Klein
  effects~\cite{Petriello:2002uu}, 

\item Littlest Higgs 
model \cite{Han:2003gf,LHatPC,Logan:2004th,Gonzalez-Sprinberg:2004bb}.

\end{itemize}


\section{Event generators}

Here we give brief descriptions of Monte Carlo event generators 
that produce Higgs events for the ILC.  Many of these rest upon the 
spectrum and branching fraction calculators like {\tt FeynHiggs}
and/or {\tt HDECAY} discussed in the previous sections.%
%
\footnote{
The depth of the descriptions does not reflect the importance of the
respective generator, but the information provided to the authors.
}%

\begin{itemize}

\item {\tt HERWIG}

{\tt HERWIG}~\cite{Marchesini:1991ch,Corcella:1999qn,Corcella:2000bw,Corcella:2001pi,Corcella:2001wc}  
is a general-purpose Monte Carlo (MC) event generator for high-energy
collider processes within the SM and MSSM \cite{Moretti:2002eu,Moretti:2002hu}.
The {\tt HERWIG} source codes and related information 
can be found in \cite{HWweb}.

Higgs production and decay is available within the SM and MSSM.
While the SM implementation is built-in, the MSSM generation
relies on input information provided by other programs.
As the {\tt HERWIG}  generator does 
not contain any  built-in models for SUSY-breaking scenarios,
in all cases the general MSSM particle spectrum
and decay tables must be provided just like those for any other
object, with the following caveats:
(i) (coloured) SUSY particles do not radiate (which is reasonable if their
decay lifetimes are much shorter than the QCD confinement scale); 
(ii) CP-violating SUSY phases are not included.
A package, {\tt ISAWIG}, has been created to work with 
{\tt ISAJET} \cite{Paige:2003mg}
to produce a file containing the SUSY (and Higgs) particle masses, lifetimes,
couplings and mixing 
parameters. This package takes  the outputs of the {\tt ISAJET}
MSSM programs and produces a data file in a format that can be read into
{\tt HERWIG} for the subsequent process generation. 
Of particular relevance to Higgs production and decay is also the interface
to the {\tt HDECAY} program \cite{IWweb}. 
As of 2005, also a link to {\tt FeynHiggs} is available. 
However, the user can
produce her/his own file provided that the correct format is used; see
the corresponding web page, where
examples of input files can be found (the SUSY benchmark 
points recommended in Ref.~\cite{sps} are also uploaded). Finally, 
the implementation
of the SUSY Les Houches accord of Ref.~\cite{Skands:2003cj} will soon be
available.

The SM and MSSM Higgs production and decay processes of relevance for the
physics of an ILC \cite{Moretti:2002rj} currently implemented in 
{\tt HERWIG} are
are associated with the process numbers ${\tt IPROC}=300,400,900,1000$ 
and 1100 (i.e., Higgs-strahlung, vector-boson fusion, pair production and
associated production with heavy fermions). Also the decay of SUSY
particles to Higgs bosons is considered~\cite{Moretti:2002eu,Moretti:2002hu}.

\item {\tt ISAJET 7.73}

The event generator {\tt Isajet}~\cite{Paige:2003mg} can be used to generate
SM Higgs boson production via the 
process $e^+e^-\to ZH_{SM}$ using the
\verb|E+E-| reaction type, and stipulating the final state to be
\verb|'Z0'| and \verb|'HIGGS'|.
The Higgs mass is input using the \verb|HMASS| keyword.
The $H_{SM}$ decay modes are calculated internally at tree level.
The beam polarization can be adjusted using the \verb|EPOL| keyword,
which allows input of $P_L(e^-)$ and/or $P_L(e^+)$. 
The events can be generated with initial state photon
bremsstrahlung and/or beamstrahlung using the \verb|EEBREM|
or \verb|EEBEAM| keywords. Input of the beamstrahlung parameter
$\Upsilon$ and bunch length $\sigma_z$ is needed.
Then an effective electron and positron distribution function
is calculated, 
and the hard scattering events are convolved with the $e^\pm$ PDFs, 
to give a variable CM energy for the hard scattering.
Fox-Wolfram QCD showers and independent hadronization are used
to convert final state quarks and gluons into jets of hadrons.

{\tt Isajet} can also generate Higgs production in the MSSM. 
One may either run 
with weak scale MSSM input parameters, or with high scale parameters
in supergravity models, GMSB models or AMSB models. The full SUSY and Higgs 
mass spectrum is calculated using 2-loop RGEs to obtain an iterative
solution, and full one-loop radiative corrections are included. The 
various Higgs masses are computed using the RG improved one-loop effective 
potential evaluated at an optimized scale to account for leading two-loop 
terms. 
Agreement for the Higgs boson masses with {\tt FeynHiggs} is in the 
few GeV range, typically.
The production reactions include $e^+e^-\to Zh$, $ZH$, $Ah$, 
$AH$ and $H^+H^-$ (along with the various SM and MSSM $2\to 2$ processes). 
The sparticle mass spectrum and decay table can be output separately
using the \verb|isasugra| or \verb|isasusy| subprograms.
Leading QCD corrections are included for Higgs decay to heavy quarks.
It should be noted that the various MSSM Higgs bosons can be produced
at large rates 
via sparticle cascade decays. Also, the {\tt Isajet} decay table includes
Higgs boson decays to sparticles, which can sometimes modify and even 
improve prospects for MSSM Higgs boson discovery.

As with SM Higgs production, beam polarization and brems/beamstrahlung
effects can be included in the event generation. 
While spin correlation effects are neglected, full decay
matrix elements are used for three-body sparticle decays.
Decays to left and right tau lepton helicity states are computed, to give
on average the correct energy distributions from decays to tau leptons.

\item {\tt MadGraph}

{\tt MadGraph} \cite{Stelzer:1994ta} is an automated tree-level matrix
element generator based on helicity amplitudes as implemented in 
the {\tt HELAS} library \cite{Murayama:1992gi}.
The associated event generator {\tt MadEvent} \cite{Maltoni:2002qb} can then
be used to generate unweighted events.  For showering/hadronization,
interfaces to {\tt PYTHIA} and {\tt HERWIG} are available.

Higgs production in both the SM and MSSM \cite{SUSYMadGraph} 
is handled at tree-level.  
In the MSSM, the Higgs spectrum and mixing angles are supplied from a
SUSY Les Houches Accord (SLHA) \cite{Skands:2003cj} input file, which
can be generated with the user's choice of spectrum calculator.
Higgs decay widths in both the SM and MSSM are taken from {\tt HDECAY}, 
which is integrated into the {\tt HELAS} library.  

\item {\tt PYTHIA}

\newcommand{\tsc}[1]{\texttt{#1}}
\newcommand{\pT}{{p_\perp}}

\tsc{PYTHIA} \cite{Sjostrand:2000wi,PYTHIA}
is a general-purpose event generator for hadronic events
in $e^+e^-$, $eh$, and $hh$ collisions (where $h$ is any hadron
or photon). For a recent brief introduction to this and many other
codes, see \cite{Dobbs:2004qw}. The current version is always
available from the \tsc{PYTHIA} web page with update notes and 
a number of useful examples.

\texttt{PYTHIA} contains a large
 number of Higgs boson processes, both for production (in many cases with
 polarisation effects included, see \cite{PYTHIA}, Section
 8.8) and decay. 
The partial widths of the Higgs bosons are very strongly dependent on 
the mass, and are calculated
as a function of the actual Higgs mass, i.e., not just
at the nominal mass. Leading order expressions are used, with
running fermion masses and, in some cases, overall $K$-factors,
to absorb the leading radiative corrections.
Other higher-order effects are not included.
Since the Higgs is a spin-0 particle, it decays isotropically. In decay
processes such as $h^0 \to W^+ W^- / Z^0 Z^0 \to 4$ fermions, angular 
correlations are included \cite{Barger:1993wt}. In decays to 2 jets, the parton
 shower off the final state is merged with the 3-jet matrix elements
 for hard jet emission \cite{Norrbin:2000uu}, thus incorporating part
of the NLO correction.  

The relevant physics
 scenarios included in \texttt{PYTHIA} can be categorised as follows:
\begin{itemize}
\item Light Standard Model Higgs (narrow-width approximation): In $e^+e^-$ annihilation, the main processes 
are $e^+e^-\to h^0Z^0$, 
usually dominant close to threshold, 
and vector boson ($Z^0 Z^0$ and $W^+ W^-$) fusion which is important at
high energies. The vector boson fusion process is a full $2\to 3$ body calculation.
The loop-induced $\gamma\gamma$ fusion process may
also be of interest, in particular when the effects of
beamstrahlung photons and backscattered photons are included. 
For details, 
see \cite{PYTHIA}, Section 8.5.1. 
\item Heavy Standard Model Higgs: for details,
see \cite{PYTHIA}, Section 8.5.2. 
\item SUSY/2HDM: the Higgs sector in Supersymmetric models
 is a special case of the general 2-Higgs Doublet Model included 
in \texttt{PYTHIA}. The internal \texttt{PYTHIA} routines for fixing
 the MSSM Higgs sector rely on the effective potential approach of
 \cite{Carena:1995bx,Carena:1995wu}, 
including the important radiative corrections
at large $\tan\beta$. 
Other Higgs boson spectrum calculations can be
 interfaced, either via the SUSY Les Houches Accord
 \cite{Skands:2003cj,Allanach:2004ub}, or via the run-time interfaces 
to both \texttt{Isasusy} \cite{Baer:1993ae} and 
\texttt{FeynHiggs}
 \cite{feynhiggs,feynhiggs2}. Recently, an interface to NMSSM models via
 the Les Houches Accords \cite{Boos:2001cv,Skands:2003cj}
has also been implemented \cite{nmssmgen}.
The additional production processes possible in a 2HDM are included,
such as Higgs boson pair production.  In production and decay, however,
the effects of virtual superpartners are {\it not} included.
For details, see \cite{PYTHIA}, Sections 8.5.3--8.5.5 (2HDM)
and 8.7 (SUSY). 
\item Left--Right symmetry / Higgs Triplets: the particle content is
  expanded by right-handed $Z_R^0$ and $W_R^{\pm}$ and right-handed
  neutrinos. The Higgs fields are in a triplet representation, leading
  to doubly-charged Higgs particles,  one set for each of the two 
  SU(2) groups. Also the number of  
neutral and singly-charged Higgs states is increased relative to the 
Standard Model. \texttt{PYTHIA} implements the scenario of
  \cite{Huitu:1996su,Barenboim:1996pt}. The main decay modes implemented are
$H_L^{++} \to W_L^+ W_L^+, \ell_i^+ \ell_j^+$ ($i, j$ generation 
             indices)  and
$H_R^{++} \to W_R^+ W_R^+, \ell_i^+ \ell_j^+$. For details, see \cite{PYTHIA}, Section 8.6.3
\item Technicolor: while not a model with a fundamental
  Higgs boson, Technicolor still belongs to the category of 
  Higgs-related topics. Techni-rho production is possible using
  the Heavy Standard Model Higgs processes or through
  a vector dominance mechanism \cite{Lane:2002sm}; see \cite{PYTHIA},
  Section 8.6.7. 
\item Little Higgs: Standard \texttt{PYTHIA} does not explicitly 
  include a Little
  Higgs model, but important features can be handled using 
  the options for 4th generation fermions
  and possibly $Z'/W'$ bosons.  For example, single heavy top quark $T$
  production and decay can be simulated using the $t'$.
  See
  \cite{PYTHIA}, Sections 8.6.1 and 8.6.2. 
\end{itemize}

\item {\tt SHERPA}

{\tt SHERPA} \cite{Gleisberg:2003xi} is a full event generator which is 
capable of describing events in $e^+e^-$, $pp$, $p\bar p$, and $\gamma\gamma$
collisions; for the latter the photon spectra produced through 
laser-backscattering are
parametrised according to {\tt CompAZ}. For the description of the 
perturbative aspects of such events, it includes a fully automated matrix 
element generator for the calculation of cross sections at the tree-level 
({\tt AMEGIC++} \cite{Krauss:2001iv}), a module for the simulation of 
subsequent soft QCD radiation through the parton shower ({\tt APACIC++} 
\cite{Krauss:2005re}) and a merging of both according
to the prescription of \cite{Catani:2001cc,Krauss:2002up,Schalicke:2005nv}.

All production and decay processes in the SM and the MSSM are treated at 
tree-level, including processes involving Higgs bosons. In the case of the 
MSSM, the particle spectra and mixing angles are taken from a SUSY Les 
Houches Accord (SLHA) \cite{Skands:2003cj} input file. Additional interfaces 
to {\tt IsaSUSY}, {\tt HDECAY} and {\tt FeynHiggs} are available.

\item {\tt WHIZARD} 

{\tt WHIZARD}~\cite{Whizard} is a program system designed for the
efficient calculation of multi-particle scattering cross sections and
simulated event samples.  Tree-level matrix elements are generated
automatically for arbitrary partonic processes by calling one of the
external programs {\tt O'Mega}~\cite{Omega}, {\tt MADGRAPH} or 
{\tt CompHEP}~\cite{CompHEP}.  Matrix 
elements obtained by alternative methods (e.g., including loop
corrections) may be interfaced as well. The program is able to calculate
numerically stable signal and background cross sections and generate
unweighted event samples with reasonable efficiency for processes with up
to six final-state particles. 

Polarization is treated exactly for both the initial and final states. 
Final-state quark or lepton flavors can be summed over automatically where
needed.  For ILC physics, beamstrahlung ({\tt CIRCE}) and initial-state
radiation spectra are included for electrons and photons.  Exact color
flow information is maintained, which is used when
fragmenting and hadronizing events with the built-in {\tt PYTHIA}
interface. The events can be written to file in STDHEP or ASCII format. 

Currently, {\tt WHIZARD} supports the Standard Model and the MSSM,
optionally with anomalous couplings, but model extensions or completely
different models can be added.  Higgs production and decay couplings, and
spectra in the case of the MSSM, are adopted from the corresponding matrix
element generator.

\end{itemize}

\chapter{Interplay between LHC and ILC in the Higgs Sector}
\label{sec:LHCILC}

If a Higgs-like state is discovered at the LHC and the ILC,
independent of the realization of the Higgs mechanism, important
synergistic effects arise from the interplay of LHC and
ILC~\cite{lhcilc}. 

\section{Overview}

If a state resembling a Higgs boson is detected, it is crucial to
experimentally test its nature as a Higgs boson. To this 
end the couplings of the new state to as many particles as possible must 
be precisely determined, which requires observation of the candidate
Higgs boson in several different production and decay channels.
Furthermore the spin and the other CP-properties of the new state need to 
be measured, and it must be clarified whether there is more than one Higgs 
state. The LHC will be able to address some of these questions, but in
order to make further progress a comprehensive program of precision
Higgs measurements at the ILC will be necessary. 
While the ILC will
provide a wealth of precise experimental information on a light Higgs
boson, the LHC may be able to detect heavy Higgs bosons which lie
outside the kinematic reach of the ILC (it is also possible, however,
that the ILC will detect a heavy Higgs boson that is not experimentally
observable at the LHC due to overwhelming backgrounds). 
Even in the case where only one
scalar state is accessible at both colliders, important synergistic effects 
arise from the
interplay of LHC and ILC. This has been demonstrated for the example 
of the Yukawa coupling of the Higgs boson to a pair of top quarks, see
below. 

The LHC and the ILC can successfully work together in determining the CP
properties of the Higgs bosons. In an extended Higgs sector with
CP violation there is a non-trivial mixing between all neutral Higgs states.
Different measurements at the LHC and
the ILC (in both the electron--positron and the photon--photon collider
options) have sensitivity to different coupling parameters. In the
decoupling limit, the lightest Higgs boson is an almost pure CP-even
state, while the heavier Higgs states may contain large CP-even and
CP-odd components.
Also in this case, high-precision measurements of the properties of
the light Higgs boson at the ILC may reveal small deviations from the SM
case, while the heavy Higgs bosons might only be accessible at the LHC.
In many scenarios, for instance the MSSM, CP-violating effects are
induced via loop corrections. The CP properties therefore depend on the
particle spectrum. The interplay of precision measurements in the Higgs
sector from the ILC and information on the SUSY spectrum
from the LHC can therefore be important for revealing the CP structure.
As an example, if CP-violating effects in the Higgs sector in a SUSY
scenario are established at the ILC, one would expect CP-violating
couplings in the scalar top and bottom sector. The experimental
strategy at the LHC could therefore focus on the CP properties of
scalar tops and bottoms. 

While the most studied Higgs boson models are the SM and the MSSM, more
exotic realisations of the Higgs sector cannot be ruled out. Thus, it is
important to explore the extent to which search strategies need to be
altered in such a case.

A possible scenario giving rise to non-standard properties of the Higgs
sector is the presence of large extra dimensions, motivated for instance 
by a ``fine-tuning'' and ``little hierarchy'' problem of the MSSM, see
\refse{sec:ExtraDimHiggs}. 
A popular class of such models comprise those
in which some or all of the SM particles live on 3-branes in the extra
dimensions. Such models inevitably require the existence of a radion
(the quantum degree of freedom associated with fluctuations of the distance
between the 3-branes or the size of the extra dimension(s)).
The radion has the same quantum numbers as the Higgs boson, 
and in general the two will mix.  Since the radion has couplings that
are very different from those of the SM Higgs boson, the two physical
eigenstates will have  unusual properties corresponding to
a mixture of the Higgs and radion
properties; the prospects for detecting them
at the LHC and ILC must be carefully analysed. 
In cases of this kind the ILC can observe both the Higgs boson and the
radion, and covers most of the parameter space where detection of either
state at the LHC is difficult.

A case where the LHC detects a heavy (500~GeV--1~TeV) SM-like Higgs
boson rather than a light CP-even Higgs boson, as apparently needed to
satisfy precision electroweak constraints, can also occur in a general
two-Higgs-doublet model. The source of the extra contributions mimicking
the effect of a light Higgs boson in the electroweak precision tests
may remain obscure in this case. The significant improvement in the
accuracy of the electroweak precision observables obtainable at the ILC
running in the GigaZ mode and at the $WW$ threshold will be crucial to
narrow down the possible scenarios.

Another challenging SUSY scenario is the MSSM with an extra singlet
(NMSSM, see \refse{sec:NMSSM}), especially if the extra singlet
dominates the lightest Higgs.
While such a state has reasonably large production cross sections at the
LHC, it would be difficult to detect as it mainly decays hadronically. 
Such a state could be discovered at the ILC. From the measurement of its
properties, the masses of the heavier Higgs bosons could be predicted,
guiding in this way the searches at the LHC. For a very heavy 
singlet-dominated Higgs state, on the other hand, 
the kinematic reach of the LHC
will be crucial in order to verify that a non-minimal Higgs sector is
realised. Thus, input from both the LHC and the ILC will be needed in
order to provide complete coverage of the NMSSM  parameter space.

If no light Higgs boson exists, quasi-elastic scattering processes of
$W$ and $Z$ bosons at high energies provide a direct probe of the dynamics
of electroweak symmetry breaking. The amplitudes can be measured in 
6-fermion processes both at the LHC and the ILC. The two colliders are 
sensitive to different scattering channels and yield complementary 
information.  
The combination of LHC and ILC data will considerably increase the LHC
resolving power. In the low-energy range it will be possible to
measure anomalous triple gauge couplings down to the natural value of
$1/16\pi^2$. The high-energy region where resonances may appear can be
accessed at the LHC only. The ILC, on the other hand, has an indirect
sensitivity to the effects of heavy resonances even in excess of the
direct search reach of the LHC. Detailed measurements of cross
sections and angular distributions at the ILC will be crucial for
making full use of the LHC data.


\section{Higgs coupling determination}

\subsection{Combined LHC/ILC analysis}
\label{subsec:tthLHCILC}

At the LHC, without any theory assumption, no coupling measurement of
a Higgs boson can
be performed. With ``mild'' theory assumptions one might hope for a $t
\bar t H$ coupling measurement at the level of
12-19\%~\cite{HcoupLHCSM}. At the ILC, on the other hand, 
the $t\bar t H$ coupling can be measured only at high
energies, $\sqrt{s} \gsim 800 \gev$. However, even in the first stage
of the ILC, the $\sqrt{s} = 500 \gev$ data can be combined with the
LHC data on Higgs branching ratios to obtain information about the 
$t \bar t H$ coupling (see \citere{deschischumi} for a first
evaluation). Using information from all Higgs production and 
decay modes at the LHC~\cite{HcoupLHCSM} and from the ILC, i.e.,

\begin{itemize}
\item
$\MH$,

\item
$\si_{\rm tot}(e^+ e^- \to HZ)$,

\item
$\si_{\rm tot}(e^+ e^- \to HZ) \times \br(H \to X)$
\quad ($X = b \bar b, \; \tau^+\tau^-, \; gg, \; WW^*$), and

\item
$\si_{\rm tot}(e^+ e^- \to \nu \bar \nu H) \times \br(\Hbb)$,
\end{itemize}
a determination of $g_{t \bar t H}$ down to the level of 11-14\% can
be performed without any additional assumptions on the underlying
theory (for $\mH \lsim 200 \gev$, and assuming SM-like BR
measurements)~\cite{talkSHslac}. 

\smallskip
In a similar way the coupling of the Higgs boson to photons can be
constrained in a better way than with the LHC or the ILC results
alone. In the combined analysis of LHC and ILC data~\cite{talkSHslac}
(with the ILC running at $\sqrt{s} = 500 \gev$ and without the $\ga\ga$
collider option) the $\ga\ga H$ coupling can be determined to the level of
7-8\% for $\mH \lsim 200 \gev$ (and assuming SM-like BR measurements).


\subsection{Model-Independent Analysis for the Higgs Boson Couplings}

Once the (lightest) Higgs boson is found at the LHC, its mass, production 
cross section, and decay branching ratios will be measured as precisely as
possible at the LHC and the ILC in order to determine the Higgs boson
couplings to gauge bosons and fermions. The predictions for these
couplings differ among various models beyond the SM. 
By determining these couplings accurately, these models may be 
distinguished, even when no further new particle is found.

These Higgs boson couplings may be written in a 
model-independent way as\cite{jlc,kiyoura}, 
\begin{eqnarray*}
  {\cal L} =   x \frac{m_b}{v}   h\overline{b}b
               + y (   \frac{m_t}v h\overline{t}t
                     + \frac{m_c}v h\overline{c}c)
             + z \frac{m_\tau}v  h\overline{\tau}\tau
               + u (g m_W^{} h W_\mu W^\mu 
                   + \frac{g_Z^{}}{2} m_Z^{} h Z_\mu Z^\mu),
\end{eqnarray*}
where the four parameters $x$, $y$, $z$, and $u$
represent the multiplicative factors in the Higgs-boson coupling
constants with down-type quarks, up-type quarks, charged-leptons and
massive gauge bosons, respectively. 
The SM corresponds to $x$ = $y$ = $z$ = $u$ = $1$.
This expression is valid at lowest order 
in the SM, MSSM, NMSSM and
multi Higgs doublet models 
without tree-level flavor mixing.
The expected accuracy of the parameter determination has been
evaluated with respect to the SM case.
It has been shown that the $u$ and $x$ parameters are determined to the 
few-percent level, and $y$ and $z$ are constrained to
less than 10\%.
From the correlation of the four parameters
determined at the ILC, it may also be possible to distinguish various
models.
For example, the Type-I two-Higgs-doublet model, which predicts the relation
$x$=$y$, and the MSSM lie in different regions of the $x$-$y$ space.
For a large $\tan\beta$ value, the allowed range of
the $x$-$z$ space can deviate from
the $x$=$z$ line for the MSSM because of the SUSY corrections to the
$hb\overline{b}$ vertex\cite{epsb,deltamb2}.
%


\section{What if Only a Light Higgs and Nothing Else is Found at the LHC?}

If (only) a state resembling a Higgs boson is detected, it is crucial to
experimentally test its nature as a Higgs boson. 
The couplings of the supposed Higgs boson to as many particles as
possible must measured; see also the previous subsection.
Furthermore the spin and the other CP-properties of the new state need to 
be measured. 
The LHC will be able to address some of these questions, but in
order to make further progress a comprehensive program of precision
Higgs measurements at the ILC will be necessary. The significance of the
precision Higgs program is particularly evident from the fact that many
extended Higgs theories over a wide part of their parameter space have a
lightest Higgs scalar with nearly identical properties to those of the
SM Higgs boson. In this decoupling limit additional states of
the Higgs sector are heavy and may be difficult to detect both at the LHC and
ILC.


\subsection{Indirect Determination of the Heavy Higgs Boson Masses in
  the MSSM}

In the MSSM, information on the heavy Higgs boson masses can be  
extracted from the experimental information on branching ratios 
of the lightest Higgs boson~\cite{deschi} (see also \citere{MA0} for an
earlier study).

The analysis has been performed in an SPS~1a--based scenario~\cite{sps},
where $M_A$ has been kept as a free parameter, but so heavy that 
the LHC only detects one light Higgs boson. For the
parameters of the SPS~1a scenario this corresponds to the region
$M_A \gsim 400$~GeV. 

The precise measurements of Higgs branching ratios at the ILC together
with accurate determinations of (parts of) the SUSY spectrum at the LHC
and the ILC (see \citere{lhcilc}) will allow in this
case to obtain indirect information on $M_A$ (for a discussion of
indirect constraints on $M_A$ from electroweak precision observables,
see Ref.~\cite{gigaz}). 
When investigating the sensitivity to $M_A$ it is crucial to take into
account realistic experimental errors on the other SUSY parameters that
enter the prediction of the Higgs branching ratios. 
In \citere{deschi} all the SUSY parameters have been varied 
according to error estimates for the measurements at
LHC and ILC in this scenario. The sbottom masses and the gluino mass
can be obtained from mass reconstructions at the LHC with ILC input;
see Ref.~\cite{lhcilc}; the precisions have been assumed to be
$\Delta m_{\tilde g} = \pm 8$~GeV and 
$\Delta m_{\tilde b_{1,2}} \approx \pm 7.5$~GeV.
The lighter stop (which in the SPS~1a scenario has a mass 
of about 400~GeV, see Ref.~\cite{sps}) will be accessible 
at the ILC, leading to an accuracy
of about $\Delta m_{\tilde t_1} = \pm 2$~GeV. The 
the stop mixing angle, $\theta_{\tilde t}$, can also be determined
with high accuracy~\cite{deschi}.
For $\tan\beta$ an uncertainty of $\Delta \tan\beta = 10\%$ has been
used (this accuracy can be 
expected from measurements at the ILC in the gaugino sector for the SPS~1a 
value of $\tan\beta = 10$ \cite{Desch:2003vw}). Concerning the top
quark mass an error of $\Delta m_t = \pm 0.1$~GeV has been assumed
from the ILC, so that 
the parametric uncertainties on the $m_h$ predictions become negligible.
$\mh$ is measured at the ILC with high accuracy, but a theory error from
unknown higher-order corrections of $\pm 0.5$~GeV has been
included~\cite{mhiggsAEC}. 

The analysis is based on the comparison of the theoretical
prediction~\cite{hff,PomssmRep} for 
the ratio of branching ratios 
\begin{equation}
r \equiv \frac{\left[{\rm BR}(h \to b \bar b)/
                     {\rm BR}(h \to WW^*)\right]_{\rm MSSM}}
              {\left[{\rm BR}(h \to b \bar b)/
                     {\rm BR}(h \to WW^*)\right]_{\rm SM~~~}\,}
\label{eq:sec221_r}
\end{equation}
with its prospective experimental measurement. 
Even though the experimental error on the ratio of the two BR's is
larger than that of the individual BR's, the quantity $r$
has a stronger sensitivity
to $M_A$ than any single branching ratio.

Assuming a certain precision of $r$, indirect bounds on $\MA$ can be
obtained. 
For the experimental accuracy of $r$ two different values have been
considered: a 4\% accuracy resulting from a first phase of ILC
running with $\sqrt{s} \lsim 500$~GeV~\cite{tdr,nlc,jlc,talkbrient},
and a 1.5\% accuracy which can be achieved from ILC running at 
$\sqrt{s} \approx 1$~TeV~\cite{barklow_branchings}.
Fig.~\ref{fig:BlackHole} shows that a 4\% accuracy on $r$ allows 
to establish an indirect upper bound on $M_A$ 
for $M_A$ values up to $M_A \lsim 800$~GeV (corresponding
to an $r$~measurement of $r \gsim 1.1$).
With an accuracy
of 1.5\%, on the other hand, a precision on $\Delta M_A / M_A$ of 
approximately 20\% (30\%) can be achieved for $M_A = $ 600 (800) GeV.
The indirect sensitivity extends to even higher values of $M_A$.
The comparison with the idealized situation where all SUSY parameters
(except $M_A$) were precisely known 
illustrates the importance of taking 
into account the parametric errors as well as the
theory errors from unknown higher-order corrections. Detailed
experimental information on the SUSY spectrum and a precision
measurement of $m_t$ are clearly indispensable for exploiting the
experimental precision on $r$.

\begin{figure}[htb!]
\begin{center}
\epsfig{figure=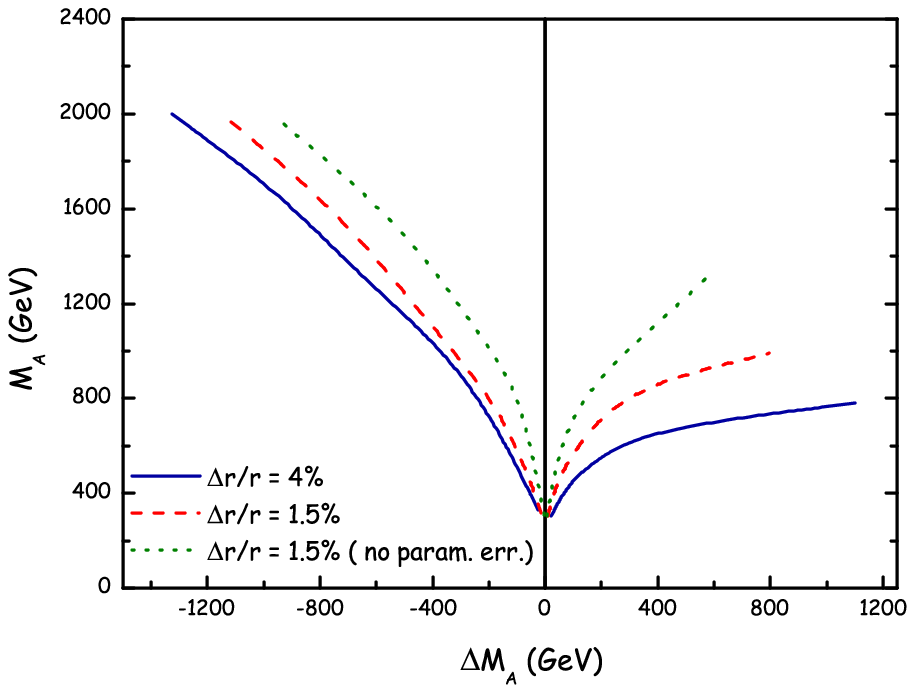, width=12cm}
\caption{The 1$\sigma$ bound on $M_A$, $\Delta M_A$,
versus $M_A$ obtained from a
comparison of the precision measurement of $r$ (see text) at the ILC 
with the MSSM prediction. The results for $\Delta M_A$ are shown
for a 4\% accuracy of $r$ (full line) and a 1.5\% accuracy of $r$
(dashed line). The parametric uncertainties in the
prediction of $r$ resulting from LHC/ILC measurement errors on
$\tan\beta, m_{\tilde{b}_{1,2}}, m_{\tilde{t}_1}, m_{\tilde{g}},
m_h $, and $m_t$ are taken into account. Also shown is the accuracy on
$M_A$ which would be obtained if these uncertainties were neglected 
(dotted line).
}
\label{fig:BlackHole}
\end{center}
\end{figure}

\subsection{Distinction between the MSSM and the NMSSM}

\begin{figure}[htb!]
\begin{minipage}[c]{0.50\textwidth}
\epsfig{file=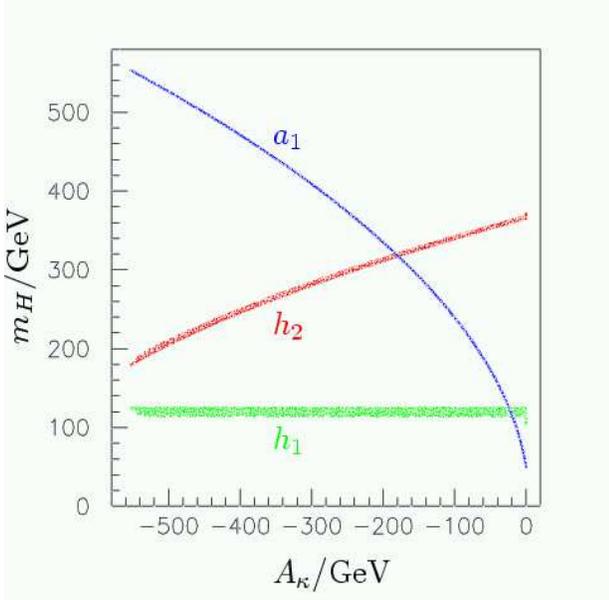}
\end{minipage}
\begin{minipage}[c]{0.03\textwidth}
$\phantom{0}$
\end{minipage}
\begin{minipage}[c]{0.45\textwidth}
\caption{\label{nmssmhiggsmass}
  The masses of the two light scalar Higgs bosons, $m_{h_1}$,
  $m_{h_2}$, and of the lightest pseudoscalar Higgs boson $m_{a_1}$ as
  function of the trilinear Higgs coupling $A_{\kappa}$ in the NMSSM
  for $\mu_\mathrm{eff}=457.5$ GeV, $\tan\beta=10$, $\lambda=0.5$ and 
  $\kappa = 0.2$, while $A_\lambda$ and $A_\kappa$ have been scanned
  over all possible values.  
  $h_1$ is MSSM-like whereas $h_2$ and $a_1$ are singlet-dominated
  Higgs particles.}
\end{minipage}
\end{figure}

It is possible that the MSSM and the NMSSM result in similar
low-energy phenomenology. Then neither the LHC nor the ILC alone 
can identify one
or the other model.
Especially the Higgs sector does not allow the identification of the NMSSM
\cite{LC-TH-2003-034} if scalar and/or pseudoscalar Higgs bosons with
dominant singlet character escape detection due to their suppressed
couplings to the other MSSM particles. This is illustrated in one
specific example. The scenario is fixed by $\mu_\mathrm{eff}=457.5 \gev$, 
$\tan\beta=10$, $\lambda=0.5$ and $\kappa = 0.2$. 
A scan  over the remaining parameters 
in the Higgs sector, the trilinear couplings
$A_\lambda$ and $A_\kappa$, has been performed using
{\tt NMHDECAY}~\cite{nmhdecay}. The parameter
points which survive the theoretical and experimental constraints are
in the region $2740~\textrm{GeV} < A_\lambda < 5465$~GeV and
$-553~\textrm{GeV} < A_\kappa < 0$, see \reffi{nmssmhiggsmass}.
In this parameter region the lightest scalar Higgs boson $h_1$ is
always MSSM-like whereas the second lightest scalar $h_2$ and the
lightest pseudoscalar $a_1$ are singlet-dominated.
For $-450~\textrm{GeV} \lsim A_\kappa \lsim -100$~GeV 
$h_2$ and $a_1$ have very pure singlet character and masses 
$\gsim 200$~GeV, which excludes their appearance in
chargino/neutralino decays in this scenario. 
All other heavier Higgs bosons have masses of ${\cal O}(A_\la)$, hence
they cannot be produced at the LHC or the ILC.
In this case a combined LHC/ILC analysis in the neutralino sector,
as carried out in \citere{MSSMvsNMSSM}, is necessary. It 
can find the phenomenological differences
between the two models and thus reveal also the nature of the Higgs sector.

For $A_\kappa \lsim -450$~GeV the smaller mass of the $h_2$ and a stronger
mixing between the singlet and MSSM-like states in $h_1$ and $h_2$
might allow a discrimination in the Higgs sector while for
$A_\kappa \gsim -100$~GeV the existence of a light pseudoscalar $a_1$ may
give the first hints of the NMSSM \cite{NMSSMh1a1a1}. See
\refse{sec:NMSSM} for further details.

\chapter{Cosmological connections}
\label{sec:cosmo}

\section{Cold dark matter and MSSM Higgs physics}
\label{sec:mssmcosmo}

A large fraction of the total amount of matter in the universe appears
to be cold dark matter (CDM). The density of CDM in the universe is
tightly constrained by WMAP and other astrophysical and cosmological
data~\cite{WMAP}. Within the MSSM (or other supersymmetric extensions
of the SM) it is well known that the lightest supersymmetric particle
(LSP) is an excellent candidate for cold dark matter~\cite{EHNOS}, 
with a density that falls naturally within the
range~\cite{WMAP},
\BEQ
0.094 < \Omega_{\rm CDM} h^2 < 0.129~.
\label{cdmdensity}
\EEQ
Assuming that the cold dark matter is composed predominantly of the
lightest neutralino, a value of $\Omega_{\rm CDM} h^2 \sim 1$ can
easily be obtained. However, in order to exactly match
\refeq{cdmdensity} a mechanism in the early universe is needed that
efficiently reduces the CDM density.

One possible mechanism is the rapid Higgs pole annihilation, 
\BEQ
\tilde\chi^0_1 \, \tilde\chi^0_1 \to A \to b \bar b~,
\label{rapidannihilation}
\EEQ
involving the pseudo-scalar MSSM Higgs boson. This mechanism is very
efficient for $\MA \approx 2\,m_{\tilde\chi^0_1}$. 
The role of the ILC is to determine with high accuracy the masses and
couplings of the involved particles in the process in
\refeq{rapidannihilation}. The corresponding theoretical determination
of $\Omega_{\rm CDM}$ then has to match the experimental value of
\refeq{cdmdensity}. 

At leading order the annihilation cross section is given by
\BEQ
\si_{\tilde\chi \tilde\chi \to A \to {\rm all}} \approx
\frac{2 \, y_{A\tilde\chi\tilde\chi}^2 \, \Ga_A}{\MA} \;
\frac{s_{\tilde\chi\tilde\chi}}
     {(s_{\tilde\chi\tilde\chi} - \MA^2)^2 + \Ga_A^2 \MA^2}, 
\label{eq:chichiAXS}
\EEQ
where $s_{\tilde\chi\tilde\chi}$ is the invariant mass of the two LSPs
in the initial state, and $y_{A\tilde\chi\tilde\chi}$ is the coupling
of the $A$ to the two lightest neutralinos, given by
\BEQ
y_{A\tilde\chi\tilde\chi} = \edz \KL g_1 N_{11} - g_2 N_{12} \KR
                                 \KL \sin\be N_{13} - \cos\be N_{14} \KR~,
\label{eq:chichiAcoup}
\EEQ
with $N$ being the unitary matrix that diagonalizes the neutralino
mass matrix. 

The uncertainty of the experimental measurement of the CDM density as
shown in \refeq{cdmdensity} is currently at the 10\% level. How this
accuracy can be matched by experimental analyses and theoretical
evaluations will be discussed in the following subsections.

The next generation of CDM experiments like PLANCK~\cite{planck} will
improve the precision of $\Omega_{\rm CDM}$ down to about 2\%. This
will require additional efforts both from the experimental as well as
from the theoretical side.


\section{Experimental issues}
\label{sec:cosmo:exp}

The main quantities that enter the theoretical calculation of
\refeq{rapidannihilation} are the mass of the CP-odd Higgs boson,
$\MA$, and its width, $\Ga_A$, see \refeq{eq:chichiAXS}. 
They have to be determined with the highest possible accuracy. 

The ILC provides a unique tool to investigate the heavy MSSM Higgs
bosons via the channels
\BEA
e^+e^- &\to& AH \to b \bar b \; b \bar b \\
e^+e^- &\to& AH \to b \bar b \; \tau^+\tau^-~.
\EEA
Corresponding analyses can be found in
\citeres{eeAHdesch,eeAHbattaglia,eeAHmoroi}. 
The precision that can be reached in $\MA$ depends on the mass
splitting between $H$ and $A$. In most cases this splitting is too
small to be disentangled at the ILC. Assuming zero splitting a mass
determination better than $\sim 1 \gev$ can be
reached~\cite{eeAHdesch}. The Higgs boson width, for realistic values
of $\Ga_A \gsim 10 \gev$ could be determined at the 30-50\% level
only. 

In \citere{eeAHmoroi} two CMSSM points with 
$\MH \approx \MA \approx 400 \gev$ have been analyzed, assuming 
1 \iab of integrated luminosity at $\sqrt{s} = 1 \tev$.
Here the mass average, $1/2 (\MA + \MH)$ could be determined to 
$\sim \pm 2.5 \gev$, whereas the mass difference, $1/2 (\MA - \MH)$
could be measured to about $\pm 5 \gev$. The Higgs boson width was
determined better than about 25\%. Assuming that all other relevant
quantities, see \refse{sec:cosmo:theo}, are known with negligible
uncertainty, the authors used DarkSUSY~\cite{darksusy} to evaluate the
corresponding CDM density. The precision of $\Omega_{\rm CDM} h^2$ was
in the 20\% range, i.e., about a factor of two larger than the
current experimental precision. This would become even worse if all
other experimental uncertainties were included.

Other experimental aspects are the neutralino properties. In order to
determine $s_{\tilde\chi\tilde\chi}$, a measurement of the LSP mass is
necessary. However, this can be achieved at the $\sim 50 \mev$
level~\cite{tdr}. Furthermore the coupling of the lightest
neutralinos to the~$A$ has to be measured. Either
$y_{A\tilde\chi\tilde\chi}$ has to be measured directly (which would
be very difficult), or the various elements entering
\refeq{eq:chichiAcoup} have to be measured. 
Further experimental studies in this directions are required.

The current experimental uncertainty of $\sim 10\%$ in 
$\Omega_{\rm CDM} h^2$ requires a precise determination of $\MA$
and $\Ga_A$ down to the level of $\sim 1\%$ for the Higgs boson mass
and of about $\sim 10\%$ for the Higgs boson width. While the first
goal seems feasible~\cite{eeAHdesch}, the second one is very
challenging. The situation is complicated by the strong correlation 
which exists between the measured widths of the $H$ and $A$ bosons 
and the finite mass difference between these physical
states. A more precise determination of heavy Higgs widths
would require higher integrated luminosities as 
well as improved analyses. The latter should include the study of 
additional signal channels, e.g., the Yukawa process $e^+e^-\ra Ab\bar{b}$, 
and a more sophisticated treatment of correlation between $\Gamma_A$, 
$\Gamma_H$ and $\MA-\MH$. It was shown~\cite{eeAHdesch} that
by combining analyses in the $HA\ra b\bar{b}b\bar{b}$ and 
$HA\ra b\bar{b}b\tau^+\tau^-$ channels, the $A$ boson width can 
be determined with a relative accuracy of about 40\% for $\MA$ $\sim$
$\MH$ = 300 GeV, assuming 500 fb$^{-1}$ of integrated luminosity 
collected at 800 GeV centre-of-mass energy. 
Assuming a $1/\sqrt{N_{event}}$ dependence of the relative precision of this measurement, the accuracy can be 
improved to 20\% if 2000 fb$^{-1}$ of luminosity is accumulated at 800 GeV
centre-of-mass energy. Further improvement can be achieved by inclusion of 
the Yukawa process $e^+e^-\ra Ab\bar{b}(Hb\bar{b})$.
Combining the mass lineshape analysis in the $e^+e^-\ra Ab\bar{b}(Hb\bar{b})$ 
channel with the $HA\ra b\bar{b}b\bar{b}$ and 
$HA\ra b\bar{b}b\tau^+\tau^-$ channels would allow to improve the 
measurement of $\Gamma_A$, in particular at high $\tan\beta$ 
values where the $e^+e^-\ra Ab\bar{b}$ process is expected to have
an enhanced rate. On the other hand, $\Gamma_A$ can be determined 
by combining mesurements of BR$(A\ra b\bar{b})$ and the rate of the
$e^+e^-\ra Ab\bar{b}$ process. The first measurement can be performed 
by exploiting the $e^+e^-\ra HA$ process, while the second provides 
direct access to the partial width $\Gamma_{A\ra b\bar{b}}$. 
The total width is then calculated as 
$\Gamma_A = \Gamma_{A\ra b\bar{b}}/$BR$(A\ra b\bar{b})$. Again, 
the precision of measurement will crucially depend on the expected
rate in the $e^+e^-\ra Ab\bar{b}$ channel, which is governed by 
$\tan\beta$.
Clearly, given the relatively poor precision on $\Gamma_A$ predicted 
by currently existing analyses, the proposed ways to improve the experimental
accuracy on $\Gamma_A$ should be investigated 
by future studies.

In order to match the anticipated PLANCK accuracy for
$\Omega_{\rm CDM} h^2$ of about 2\%, a much better precision in $\MA$
and $\Ga_A$ will be necessary. The mass will have to be known at the 
$\sim 0.25\%$ level, whereas the width would require a $\sim 3\%$
accuracy. These goals are extremely ambitious and invite further 
investigations.


\section{Theoretical issues}
\label{sec:cosmo:theo}

The theoretical calculation of $\Omega_{\rm CDM} h^2$ also requires a
precise evaluation of \refeq{eq:chichiAXS}. The one-loop corrections
to \refeq{eq:chichiAXS} are expeted at the \order{{\rm few}\%}. 
Consequently, they can be omitted at the present level of
accuracy. However, in order to match the anticipated Planck precision
of about $\sim 2\%$ a calculation of the leading one-loop corrections
is necessary. Since for the `inverse' process, 
$e^+e^- \to \tilde\chi^0_1 \tilde\chi^0_1$, a full one-loop
calculation exists~\cite{eechichi}, the tools for the required
calculation appear to be available.

\bigskip
Even if the heavy Higgs boson are not observed at the LHC and the ILC,
the additional information from the ILC can be valuable:
the calculation of $\Omega_{\rm CDM} h^2$ can also incur large
uncertainties due to the lack of sufficiently stringent lower bounds 
on the masses of $A$ and $H$.  We show this fact in
\reffi{fig:CDMlowerMA}~\cite{Birkedal:2004talk,Birkedal:2005jq}.  Here
we display the 
%
\begin{figure}[htb!]
\centerline{
\includegraphics[width=.5\textwidth]{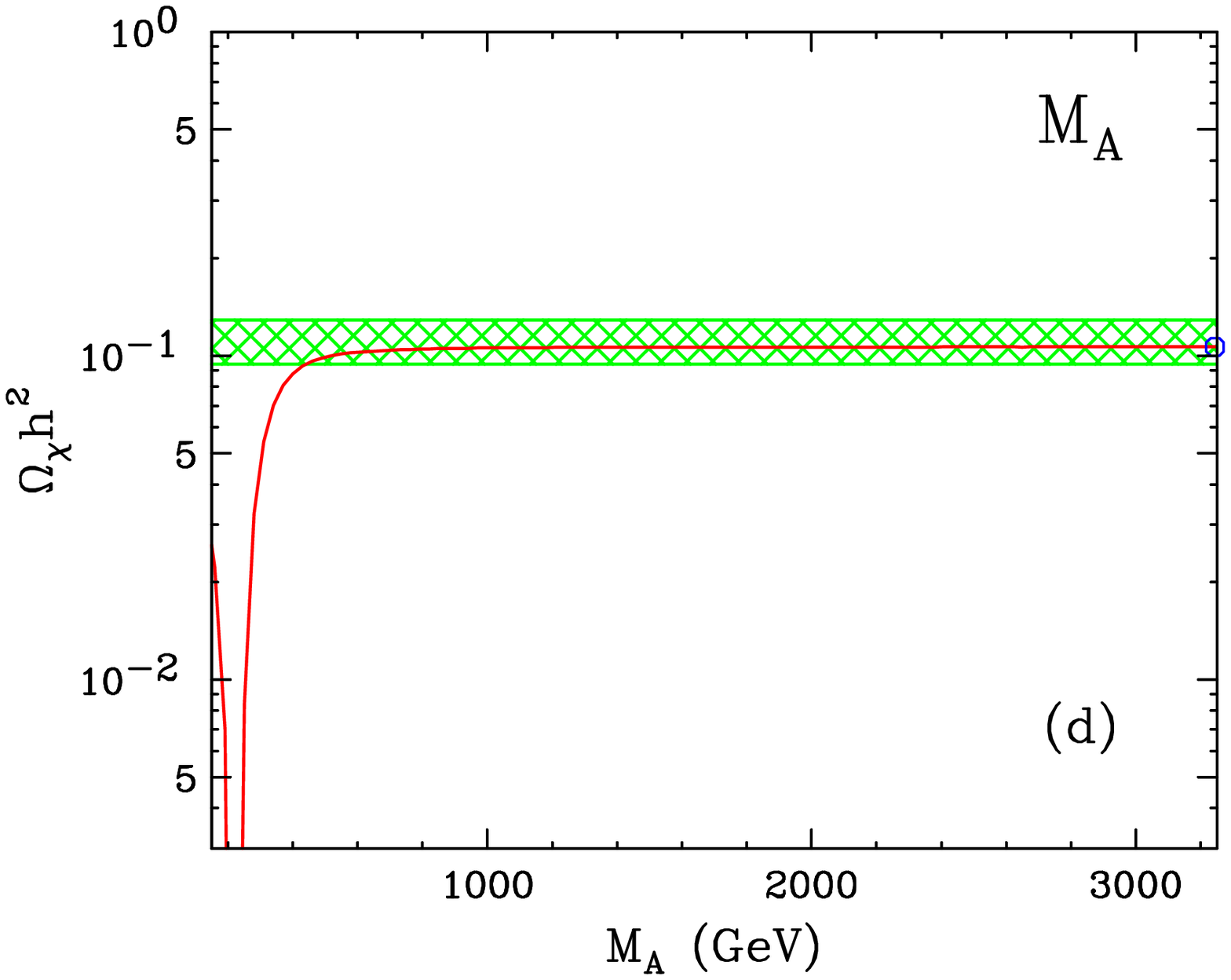}
\includegraphics[width=.5\textwidth]{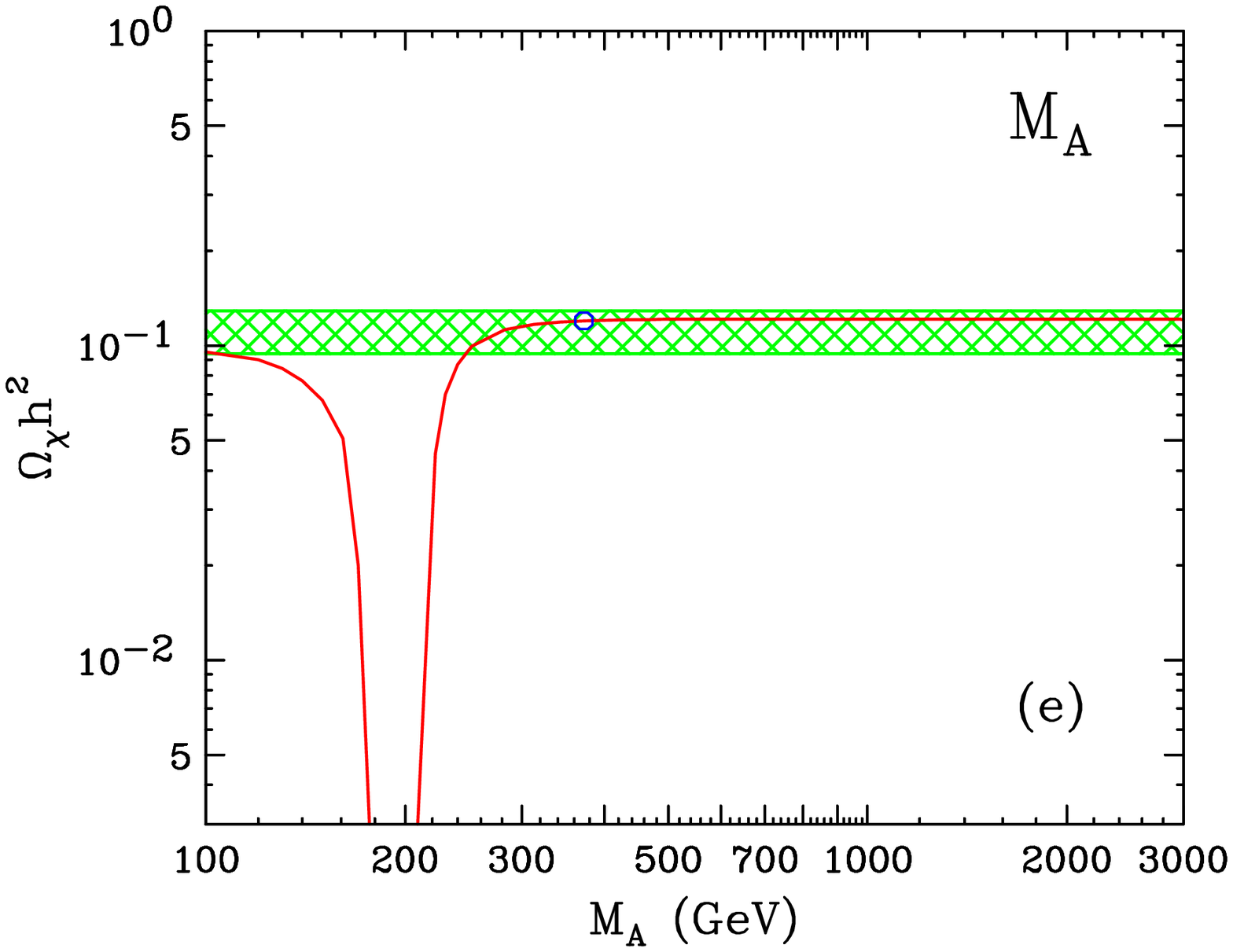}
}
\caption{
Effect on relic density of varying the SUSY mass parameter $M_A$ for point 
B$^{\prime}$ 
from the bulk region of mSUGRA (left plot) and the point LCC2 from the 
focus point region of mSUGRA (right plot).
The horizontal (green-shaded) region denotes
the $2\sigma$ WMAP limits on the dark matter relic density.  The red line 
shows the variation of the relic density as a function of $M_A$.
The blue dot in each plot denotes the nominal value for $M_A$ at point 
B$^{\prime}$ 
(left plot) and the point LCC2 (right plot).  Taken from 
\citeres{Birkedal:2004talk,Birkedal:2005jq}.
}
\label{fig:CDMlowerMA}
\end{figure}
%
variation in $\Omega_{\rm CDM} h^2$ caused only by a variation in
$M_A$.  In the left plot, we have chosen as our nominal point one from
the `bulk' region of mSUGRA parameter space.  The blue circle
identifies the nominal values of $\Omega_{\rm CDM} h^2$ and $M_A$ for
the point.  The green cross-hatched region denotes values consistent
with the WMAP determination of the relic density.  The red line
illustrates how the value of $\Omega_{\rm CDM} h^2$ changes as the
parameter $M_A$ is changed.  Since the slope of the red line is small
as it passes through the nominal value, $\Omega_{\rm CDM} h^2$ appears
rather insensitive to the exact value of $M_A$.  However, it is
possible that the value of $M_A$ cannot be determined at a collider
(if $\sqrt{s} = 500 \gev$ for example), and only a lower bound can be
attained.  If this is the case, when calculating $\Omega_{\rm CDM}
h^2$, one must allow $M_A$ to vary between its lower bound and
infinity.  From the plot, it is clear that this introduces large
uncertainties in $\Omega_{\rm CDM} h^2$ if the lower bound is not much
above $200 \gev$.  It is likely that this will be the best lower limit
available from the LHC at this point
~\cite{Birkedal:2004talk,Birkedal:2005jq}.  However, a $500 \gev$
linear collider will raise this lower limit to approximately $250
\gev$ (and higher energies can give rise to larger increases).  From
the plot, it can be seen that the extra $50 \gev$ gained 
from the ILC will vastly improve the uncertainty in the
calculation of $\Omega_{\rm CDM} h^2$ due to experimental uncertainty
in $M_A$.  The right plot shows the same story for the case of a
point~\cite{Gray:2005ci} from the `focus point' region~\cite{focus}.
If the value of $M_A$ can be 
determined, $\Omega_{\rm CDM} h^2$ will be very insensitive to its
exact numerical value.  This is unlikely to happen since $M_A \sim 3 \tev$.
However, $\Omega_{\rm CDM} h^2$ will be very sensitive to the
lower experimental bound on $M_A$ if it is around $200 \gev$.


\section{The NMSSM case}

\subsection{Cold dark matter and NMSSM Higgs physics}

Since the NMSSM has five neutralinos and two CP-odd Higgs bosons,
there are many new ways in which the relic density of the $\cnone$
could match the observed dark matter density; see,
e.g.,~\cite{Gunion:2005rw,Belanger:2005kh}. 
The latter group has made their code publicly available.
In the NMSSM the $\mai$ and $\maii$
masses are quite unconstrained by LEP data and theoretical model
structure, implying that $\cnone\cnone\to a_{1,2}\to X$ could be the
primary annihilation mechanism for large swaths of parameter space.
It is relatively easy to construct NMSSM models yielding
the correct relic density even 
for a very light neutralino, 100 MeV $< \mcnone <$
20 GeV. Even after including 
constraints from Upsilon decays, $b \rightarrow s \gamma$, $B_s
\rightarrow \mu^+ \mu^-$ and the magnetic moment of the muon, 
a light bino or singlino neutralino is allowed that can
generate the appropriate relic density.


The LSP is a mixture of the bino, neutral wino, neutral higgsinos
and singlino, which is the superpartner of the
singlet Higgs.  The
lightest neutralino therefore has a singlino component 
in addition to the four MSSM components; its eigenvector $\cnone$
can be written in terms of gauge eigenstates in the form
\begin{equation} 
    \cnone = \epsilon_u \tilde{H}^0_u + \epsilon_d
    \tilde{H}^0_d + \epsilon_W \tilde{W}^0   + \epsilon_B \tilde{B} +
    \epsilon_s \tilde{S}, 
\end{equation}
where $\epsilon_u$, $\epsilon_d$ are the up-type and down-type higgsino
components, $\epsilon_W$, $\epsilon_B$ are the wino and bino components
and $\epsilon_s$ is the singlet component of the lightest neutralino.
Similarly, the CP-even and CP-odd
Higgs states are mixtures of MSSM-like Higgses and singlets.
For the lightest CP-even Higgs state we can define (see also
\refse{sec:nmssmintro}) 
\begin{equation} 
    \hi = {1\over \sqrt 2}\left[\xi_u \Re (H_u^0 - v_u) + \xi_d \Re
    (H_d^0 - v_d) + \xi_s \Re (S - s)\right].
\end{equation}
Here, $\Re$ denotes the real component of the respective state. 
Lastly, the lightest CP-odd Higgs can be written as
(a similar formula also applies for the heavier $\aii$)
\begin{equation}
\label{as}
\ai = \cos \theta_{\ai} \ha_{\rm MSSM} + \sin \theta_{\ai} A_s, 
\end{equation}
where $A_s$ is the CP-odd piece of the singlet and $\ha_{\rm MSSM}$ is the state that would be
the MSSM pseudoscalar Higgs if the singlet were not present. 
Here, $\theta_{\ai}$ is the
mixing angle between these two states.  

In the NMSSM context, when annihilation proceeds via one
of the CP-odd Higgs bosons the calculation of the relic $\cnone$ density
is much more flexible than in the MSSM; see \refse{sec:mssmcosmo}.
The general formula for the thermally averaged cross section takes the
form 
%
\begin{eqnarray}
\label{sigmava}
\langle \sigma v\rangle &\approx& \frac{g^4_2 c_f
  m^2_f \cos^4 \theta_{a_1}  \tan^2 \beta}{8 \pi m^2_W}
\frac{m^2_{\cnone} \sqrt{1-m^2_f/m^2_{\cnone}}}{(4
  m^2_{\cnone}-m^2_{a_1})^2 + m^2_{a_1} \Gamma^2_{a_1}} \\ \nonumber 
&& \times \bigg[-\epsilon_u (\epsilon_W - \epsilon_B \tan \theta_W) \sin
  \beta + \epsilon_d (\epsilon_W-\epsilon_B \tan \theta_W) \cos \beta
  \\ \nonumber
&& + \sqrt{2} \frac{\lambda}{g_2}\epsilon_s (\epsilon_u \sin \beta +
  \epsilon_d \cos \beta) +\frac{\tan \theta_{a_1}}{g_2}\sqrt{2}
  (\lambda \epsilon_u \epsilon_d -\kappa \epsilon^2_s)  \bigg]^2, 
\label{sigmav}
\end{eqnarray}
where $c_f$ is a color factor, equal to 3 for quarks and 1 otherwise.
For this result, we have assumed that the final state fermions are
down-type. If they are instead up-type fermions, the 
$\tan^2\beta$ factor should be replaced by $\cot^2\beta$.
For medium to large values of $\tb$, neutralino dark matter can avoid
being overproduced for any $a_1$ mass below $\sim 20-60$ GeV, as long
as $m_{\cnone} > m_b$. For smaller values of $\tan \beta$, a lower
limit on $m_{a_1}$ can apply as well.

\begin{figure}[htb!]
\begin{center}
\includegraphics[scale=0.6]{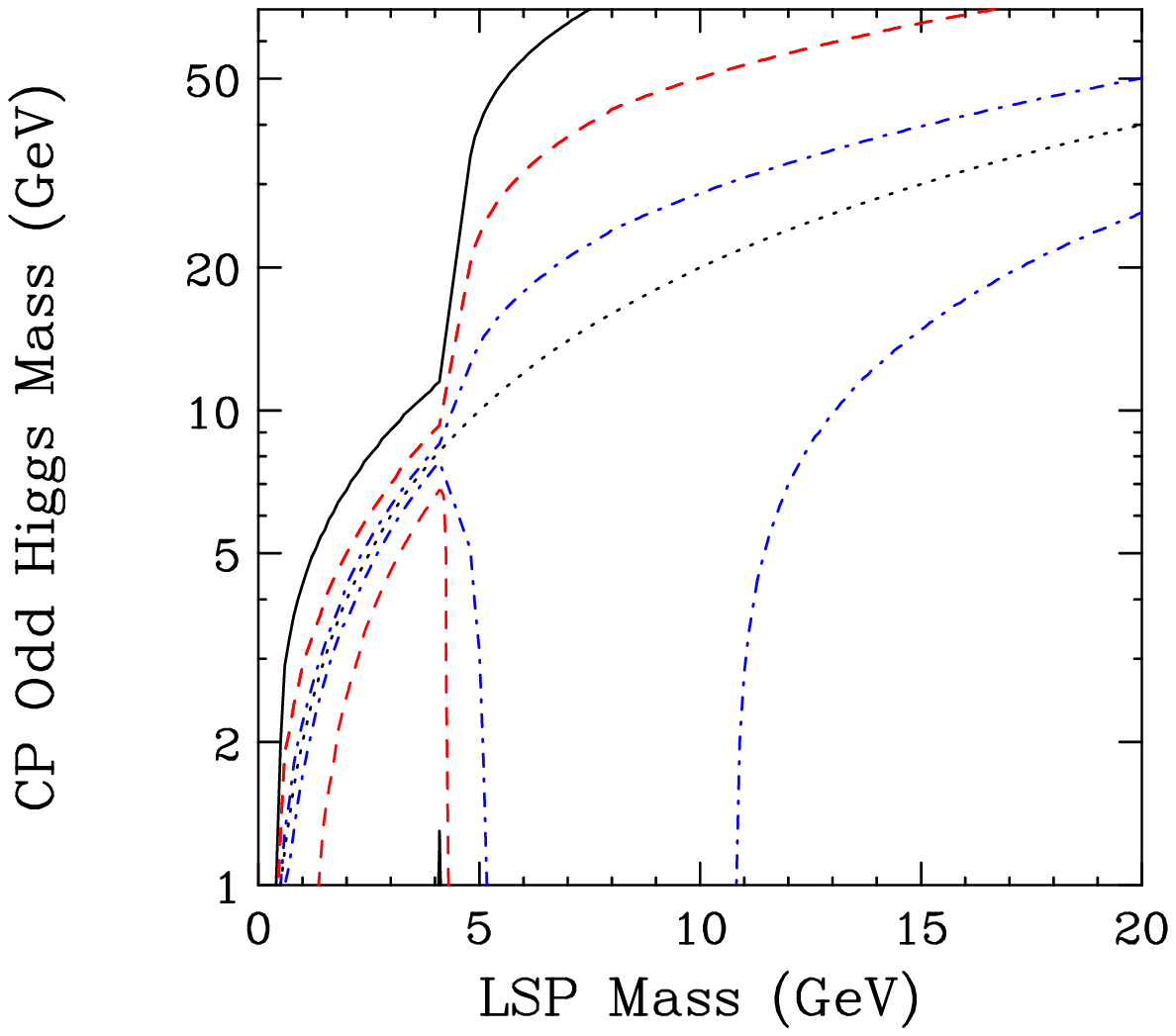}
\includegraphics[scale=0.6]{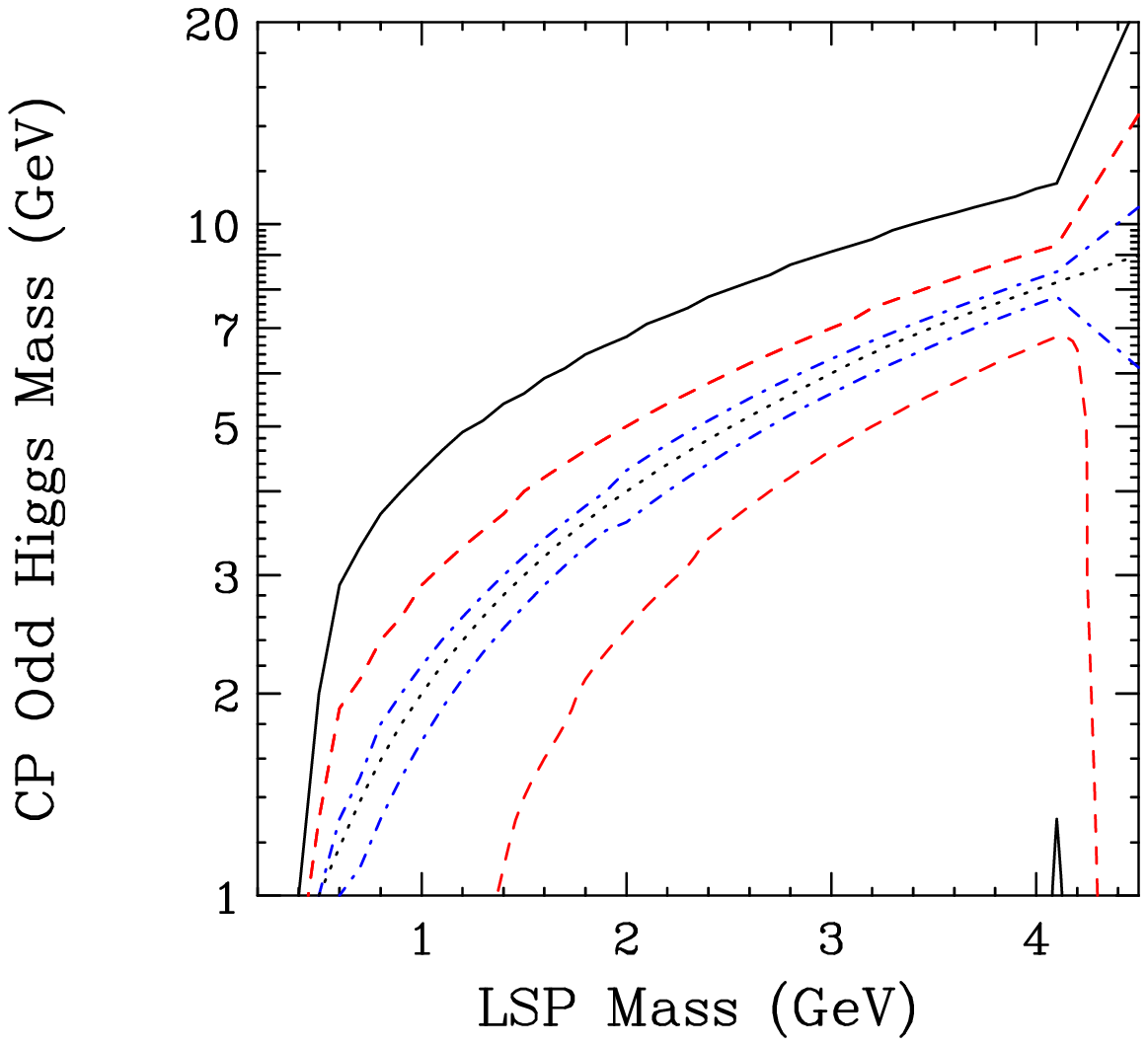}
\end{center}
\caption{
We display contours in $\mai$ -- $\mcnone$
parameter space for which $\Omega h^2=0.1$.
Points above or below each pair of curves produce more dark matter than
is observed; inside each set of curves less dark matter is produced than
is observed.  These results are for a bino-like neutralino with a small
higgsino admixture ($\epsilon^2_B = 0.94$, $\epsilon^2_u = 0.06$). Three
values of $\tan \beta$ (50, 15 and 3) have been used, shown as solid
black, dashed red, and dot-dashed blue lines, respectively. 
The dotted line is
the contour corresponding to $2 m_{\cnone}=\mai$. For each set of lines,
we have set $\cos^2 \theta_{\ai} =0.6$.}
\label{nmssmrelic}
\end{figure}
%
Fig.~\ref{nmssmrelic} shows typical NMSSM results.
There, the $\ai$ composition is chosen such that
$\cos^2 \theta_{\ai} = 0.6$ and the
neutralino composition is specified by $\eps_B^2=0.94$ and
$\eps_u^2=0.06$. These specific values are
representative of those that can be achieved for various NMSSM
parameter choices satisfying all constraints.  For each pair of
contours (solid black, dashed red, and dot-dashed blue), $\Omega
h^2=0.1$ along the contours and the region between the lines is the
space in which the neutralino's relic density obeys $\Omega h^2<0.1$.
The solid black, dashed red, and dot-dashed blue lines correspond to
$\tan \beta$=50, 15 and 3, respectively. Also shown as a dotted line
is the contour corresponding to the resonance condition, $2
m_{\cnone}=\mai$.
For the $\tan \beta$=50 or 15 cases, neutralino dark matter can avoid
being overproduced for any $\ai$ mass below $\sim 20-60$ GeV, as long
as $m_{\cnone} > m_b$. For smaller values of $\tan \beta$, a lower
limit on $\mai$ can apply as well.

For neutralinos lighter than the mass of the $b$-quark, annihilation is
generally less efficient. This region is shown in detail in the right
frame of Fig.~\ref{nmssmrelic}. In this funnel region, annihilations
to $c \bar{c}$, $\tau^+ \tau^-$ and $s \bar{s}$ all contribute
significantly. Despite the much smaller mass of the strange quark, its
couplings are enhanced by a factor proportional to $\tan \beta$ (as
with bottom quarks) and thus can play an important role in this mass
range. In this mass range, constraints from Upsilon and $J/\psi$
decays can be very important, often requiring fairly small values of
$\cos \theta_{\ai}$.

The above discussion focused on the case of a mainly bino LSP. If the
LSP is mostly singlino, it is also possible to generate the observed
relic abundance in the NMSSM. A number of features differ for the
singlino-like case in contrast to a bino-like LSP, however.  Most
importantly, an LSP mass that is chosen to be precisely at the Higgs
resonance, $\mai\simeq 2\mcnone$, is not possible for this case:
$\mai$ is always less than $2\mcnone$ by a significant amount.
Second, in models with a singlino-like LSP, the $\ai$ is generally
also singlet-like and the product of $\tan^2 \beta$ and $\cos^4
\theta_{\ai}$, to which annihilation rates are proportional, see
Eq.~(\ref{sigmava}), is typically 
very small. This limits the ability of a singlino-like LSP to generate
the observed relic abundance.  The result is that annihilation is too
inefficient for an LSP that is more than 80\% singlino. However,
there is no problem having $\mcnone\sim\mai/2$ so as
to  achieve the correct relic
density when the $\cnone$ is mainly bino while the $\ai$ is mainly singlet.

Finally, for the scenarios with a low $F$~value (see
\refeq{eq:finetuning}), one has
$\mai<2\mb$ and the $\ai$ is very singlet-like, with
$\cos^2\theta_{\ai}\lsim 0.015$. In this case, adequate annihilation
of a very light $\cnone$ via $\cnone\cnone\to\ai\to X$ occurs only if
$\mcnone\simeq \mai/2$.  This requires a rather fine adjustment of the
$M_1$ bino soft mass relative to $\mai$.
Because the $\ai$ is so light in the
low fine-tuning scenarios, if $\mcnone$ is significantly above $2\mb$
then consistency with relic abundance limits requires that
$\cnone\cnone$ annihilation proceed via one of the more conventional
co-annihilation channels or via $\cnone\cnone\to\aii\to X$.  The
latter case is only applicable if $\mcnone\gsim 200\gev$
(with typically $\maii\gsim 400 \gev$).


\subsection{Direct dark matter detection}

In \citere{Gunion:2005rw} it is estimated that the
neutralino-proton elastic scattering cross section is on the order of
$4\times 10^{-42}$ cm$^2$ ($4 \times 10^{-3}$ fb) for either a
bino-like or a singlino-like 
LSP. This value may be of interest to direct detection searches such as CDMS,
DAMA, Edelweiss, ZEPLIN and CRESST.  To account for the DAMA data, the cross
section would have to be enhanced by a local over-density of dark matter.

Prospects for direct detection of dark matter in the low fine-tuning
scenarios are rather constrained. Since the $\ai$ is so singlet in
nature, the only exchange 
of importance is $\hi$ exchange.  In the low-$F$ scenarios, $\hi$ is
almost entirely $H_u$.  In particular, the $H_d$ component
of the $\hi$ is $\xi_d\sim 0.1$, and correspondingly $\xi_u\sim 0.99$.
Whether or not $\mcnone$ is below $\mb$ ($\mcnone<\mb$ requiring
annihilation  via 
$\cnone\cnone\to\ai\to X$), the typical composition of a bino-like $\cnone$ is
such that $\eps_B>0.8$ and $\eps_u$ and $\eps_d$ take a range of
values from $0.1$ up to $0.5$.  Keeping only the $s$-quark
contribution and the dominant $\eps_d\xi_u$ piece in the external
factor, we obtain
\beq
\sigma_{\rm elastic}\sim 5\times 10^{-6} \eps_B^2\left({\eps_d\over
  0.25}\right)^2 
\left({100\gev \over \mhi}\right)^4 \left({\tanb\over 10}\right)^2~\fb\,.
\eeq  
Although $\mcnone$ can be chosen in the range $\gsim 20\gev$ (assuming
appropriate coannihilation to 
achieve proper relic density), which is relatively
optimal  for experiments like ZEPLIN and CRESST, the predicted cross
section is much below their expected sensitivity.  
If annihilation is via $\cnone\cnone\to \ai\to X$, the required mass
$\mcnone\simeq \mai/2<\mb$ is not very optimal and there would be no
hope of direct detection at such experiments.


\subsection{The role of the ILC}

In the case where $\mcnone \lsim 20\div 30 \gev$, and especially if
$\mcnone<\mb$, the best that a hadron collider can do will probably be
to set an upper limit on $\mcnone$.  Determining its composition is
almost certain to be very difficult.  Note that the $\mcntwo-\mcnone$
mass difference should be large, implying adequate room for $\cntwo\to
Z\cnone$ and a search for lepton kinematic edges and the like.  (Of
course, $\cntwo\to \hi\cnone$ will also probably be an allowed
channel, with associated implications for $\hi$ detection in SUSY
cascade decays.)  If the dominant annihilation channel is
$\cnone\cnone\to\ai\to X$, determination of the properties of the
$\ai$ would also be critical.  A light singlet-like $\ai$ is very hard
to detect.  At best, it might be possible to bound $\cos\theta_{\ai}$
by experimentally establishing an upper bound on the $WW\to \ai\ai$
rate (proportional to $\cos^4\theta_{\ai}$).  Thus, the ILC would be
absolutely essential.  Precise measurement of the $\cnone$ mass and
composition using the standard ILC techniques should be
straightforward.  A bigger question is how best to learn about the
$\ai$ at the precision level.  Of course, we will have lots of $\ai$'s
to study from $\hi\to\ai\ai$ decays. The events will give precise
measurements of $g_{ZZ\hi}^2\br(\hi\to\ai\ai)\br(\ai\to X)\br(\ai\to
Y)$, where $X,Y=b\anti b,\tauptaum$. The problem will be to unfold the
individual branching ratios so as to learn about the $\ai$ itself.
Particularly crucial would be some sort of determination of
$\cos\theta_{\ai}$ which enters so critically into the annihilation
rate.  (One can assume that a $\tanb$ measurement could come from
other supersymmetry particle production measurements and so take it as
given.)  There is some chance that $WW\to \ai\ai$ and $Z^* \to
Z\ai\ai$, with rates proportional to $\cos^4\theta_{\ai}$, could be
detected.  The $\cos^2\theta_{\ai}=1$ rates are very large, implying
that observation might be possible despite the fact that the low-$F$
scenarios have $\cos^2\theta_{\ai}\lsim 0.01$.  One could also
consider whether $\gam\gam \to\ai$ production would have an observable
signal despite the suppression due to the singlet nature of the $\ai$.
Hopefully, enough precision could be achieved for the $\ai$
measurements that they could be combined with the $\cnone$ precision
measurements so as to allow a precision calculation of the expected
$\cnone\cnone\to\ai\to X$ annihilation rate.  A study of the errors in the dark matter
density computation using the above measurements as compared to the
expected experimental error for the $\Omega h^2$ measurement is
needed.


\section{Electroweak Baryogenesis and quantum corrections 
to the triple Higgs boson coupling}

Baryogenesis is one of the fundamental cosmological problems. 
Electroweak baryogenesis~\cite{EWBG} is a scenario 
of baryogenesis at the electroweak phase transition, so that   
the Higgs sector plays an essentially important role.  
In the symmetric phase ($T > T_c$), where $T_c$ is the 
critical temperature, baryon number is generated by 
sphaleron processes.
For successful electroweak baryogenesis, it is required 
that the sphaleron decouples just after the phase transition. 
In the SM this condition implies $\varphi_c/T_c \gsim 1$, where 
$\varphi_c$ is the vacuum expectation value at $T_c$:  
i.e., the phase transition is strongly first order. 
This is qualitatively described in the high temperature 
expansion, where the effective potential is expressed by 
\begin{eqnarray}
V_{\rm eff} = D (T^2-T_0^2) \varphi^2 - E T \varphi^3 
+ \frac{\lambda_T}{4} \varphi^4 + \cdot\cdot\cdot ~. 
\label{eq:hte}
\end{eqnarray}
One can see that $\varphi_c=2 E T_c/\lambda_T$. 
Hence, the sphaleron decoupling condition yields 
$E/\lambda_T \gsim 0.5$. 
In the SM, where $E=\frac{1}{4\pi v^3}(2 m_W^3 + m_Z^3)$, 
this gives the upper bound for the mass of the Higgs boson 
$m_h \lsim 46$ GeV. 
A more quantitative analysis shows 
that the phase transition turns into a smooth cross-over 
for $m_h \gsim 80$ GeV. 
Therefore, the electroweak baryogenesis 
cannot be realized in the framework of the SM 
with the present experimental limit $m_h \gsim 114$ GeV~\cite{lep_higgs}.


\subsection{Strong first order phase transition by the 
additional bosonic loop contribution}

The condition of the sphaleron decoupling 
is compatible with the current experimental 
results when we consider extensions of the SM, which 
give larger values of the coefficient $E$ of the cubic 
term in Eq.~(\ref{eq:hte}) than in the SM prediction. 
Some additional bosons can contribute to enhance $E$ at one-loop level 
when they show the non-decoupling property: i.e., when
they receive masses mainly from the vacuum expectation value $v$. 
In the MSSM, light stops may be a candidate of such additional 
bosons\cite{ewbg-mssm}. 
An alternative viable model may be the Two-Higgs-Doublet Model 
(THDM)\cite{ewbg-thdm}. 

The phenomenological influence of such succesful baryogenesis scenarios 
on the Higgs physics has been discussed 
in the THDM and MSSM~\cite{hhh-ewbg}. 
It was shown that the condition 
on the finite temperature effective potential 
for a successful scenario
exactly corresponds to the condition of a large non-decoupling 
quantum effect due to extra Higgs boson loops 
on the $hhh$ coupling of the lightest Higgs boson $h$.
In the THDM
the leading contributions to the one-loop corrections to 
the $hhh$ coupling is given in the SM-like limit 
($\sin(\alpha-\beta) \simeq -1$) by\cite{hhh-THDM}
\BEA
\lambda_{hhh}^{\rm eff}({\rm THDM})\
&\simeq& \frac{3m_h^2}{v}
	\Big[ 
    1+\frac{m_H^4}{12\pi^2m_h^2v^2} \left(1-	
        \frac{M^2}{m_H^2}\right)^3
	+\frac{m_A^4}{12\pi^2m_h^2v^2} \left(1-\frac{M^2}{m_A^2}\right)^3
\non \\
&& \hspace{14mm} +\frac{m_{H^\pm}^4}{6\pi^2m_h^2v^2}
\left(1-\frac{M^2}{m_{H^\pm}^2}\right)^3-\frac{m_t^4}{\pi^2m_h^2v^2}\Big],
\label{eq:hhh-thdm}
\EEA
where $M$ is the soft-breaking mass parameter of the discrete $Z_2$
symmetry.

At finite temperatures, 
in the case with $M^2 \ll m_\Phi^2$  ($\Phi=H, A, H^\pm$), 
the coefficient $E$ of 
the cubic term in Eq.~(\ref{eq:hte}) is 
evaluated as $E=\frac{1}{12\pi v^3}
(6 m_W^3 + 3 m_Z^3 + m_H^3 + m_A^3 + 2 m_{H^\pm}^3)$, 
and the condition of sphaleron decoupling can be satisfied 
for  $m_h \gsim 115$ GeV for 
large values of heavy Higgs boson masses. 
In this case, from Eq.~(\ref{eq:hhh-thdm}) 
the effects of the heavy Higgs boson 
loops on the $hhh$ coupling are enhanced 
by $m_\Phi^4$.
These loop effects do not decouple in the large mass limit, 
and yield large deviations in the $hhh$ coupling from the SM prediction.
Such large non-decoupling effects can modify the $hhh$ coupling 
by more than 10\%~\cite{hhh-THDM} (maximally $\simeq$ a few times 100\%  
within 
the requirement from perturbative unitarity\cite{Kanemura:1993hm}), 
so that the deviation is expected to be detected at ILC\cite{hhh1,hhh2}.  
In the same scenario, the effects on the quartic Higgs boson 
coupling are also discussed in Ref.~\cite{Ham:2005ej}.

This connection between the condition for 
successful electroweak baryogenesis and 
the large non-decoupling effect on the $hhh$ coupling 
is a common feature, and can be seen in some other models 
such as the MSSM with light stops.   In the MSSM, 
the effect of additional stop loop contributions 
to the $hhh$ coupling is expressed by using the high temperature 
expansion as  
\begin{eqnarray}
 \frac{\Delta\lambda_{hhh}({\rm MSSM})}{\lambda_{hhh}({\rm SM})}
 \simeq \frac{2 v^4}{m_{t}^2 m_h^2} ({\Delta E_{\tilde{t}_1}})^2, 
\end{eqnarray}
where $\Delta E_{\tilde{t}_1}$ is the contribution of the light stop 
loop to the cubic term in the finite temperature effective potnetial 
in Eq.~(\ref{eq:hte}). 

Therefore, these scenarios of 
electroweak baryogenesis can be tested by precisely measuring the 
$hhh$ coupling at an ILC~(see Figure~\ref{fig:ewbg}). 

\begin{figure}[htb!]
\begin{minipage}[c]{\textwidth}
\psfig{figure=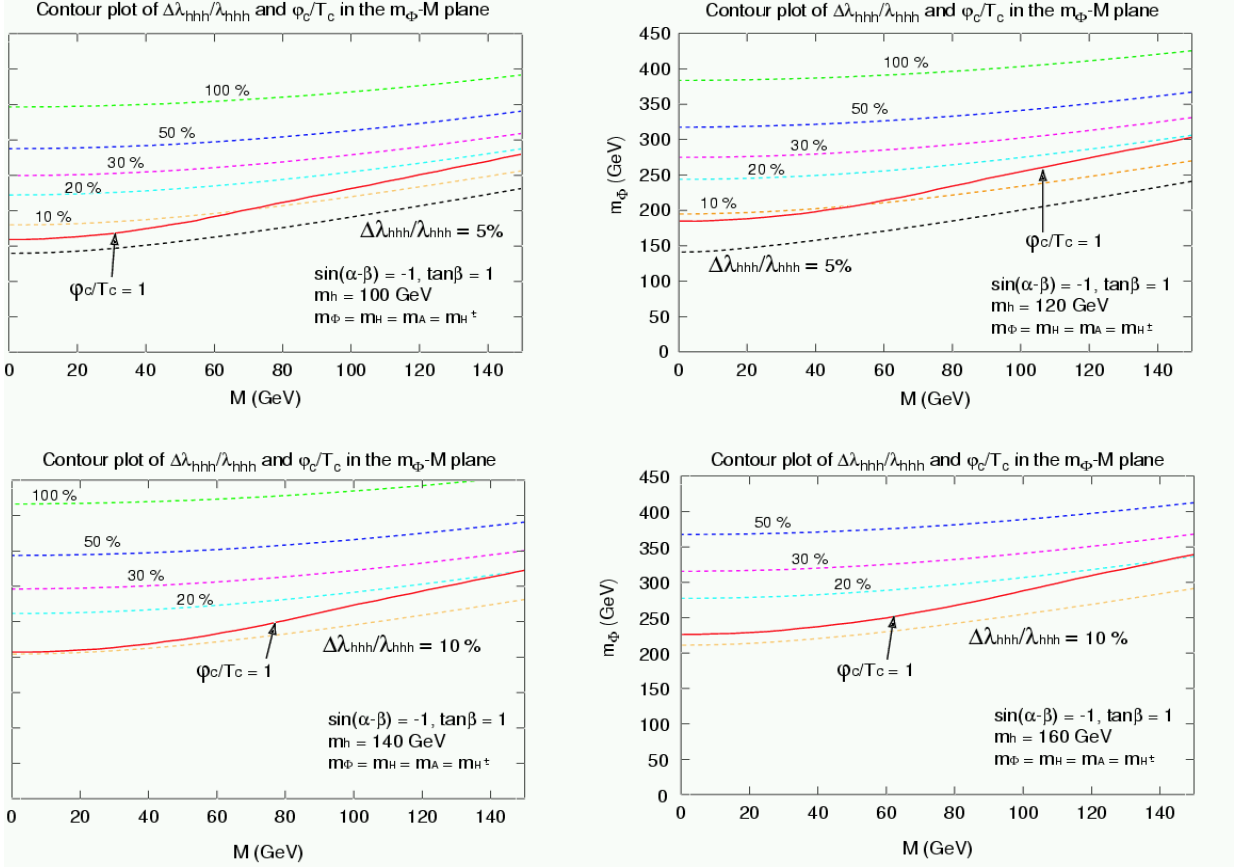,width=0.99\textwidth,height=11.5cm}
\end{minipage}
\caption{
The contour of the radiative correction of the triple Higgs boson coupling
constant overlaid with the line $\varphi_c/T_c=1$ in the $m_\Phi^{}$-$M$ plane
for $m_h$=100, 120, 140 and 160 GeV.
Above the critical line, the phase transition is strong enough for a
successful electroweak baryogenesis scenario\cite{hhh-ewbg}.
}
\label{fig:ewbg}
\end{figure}


\subsection{Strong first order phase transition in the 
SM with a low cutoff scenario}

Recently the possibility of electroweak baryogenesis 
has been studied in the SM with a low cut-off scale, by adding the 
higher dimensional Higgs boson operators \cite{Bodeker:2004ws,Grojean:2004xa}. 
In contrast to previous approaches which rely on 
large cubic Higgs interactions, 
they allow the possibility of a negative quartic 
coupling while the stability of the potential
is restored by higher dimensional operators\cite{Grojean:2004xa}: 
\begin{eqnarray}
V(\Phi)=\lambda \left(\Phi^\dagger\Phi - \frac{v^2}{2}\right)^2 
       + \frac{1}{\Lambda^2} 
          \left(\Phi^\dagger\Phi - \frac{v^2}{2}\right)^3,
\end{eqnarray}
where $\Phi$ is the SM electroweak Higgs doublet.
At zero temperature, the CP-even scalar state can be written
in terms of the zero temperature vacuum expectation value 
$\langle\varphi\rangle=v_0 \simeq 246$ GeV and the 
physical Higgs boson $H$: $\Phi=\varphi/\sqrt{2}=(H+v_0)/\sqrt{2}$.
One can minimize the zero temperature potential 
to find $\lambda=m_H^2/(2v_0^2)$ and $v=v_0$ in terms 
of physical parameters $m_H^{}$ and $v_0$.

The effective potential at finite temperatures 
is approximately obtained by adding the thermal mass to the 
zero temperature potential 
$V(\varphi,T)= c T^2\varphi^2/2+V(\varphi,0)$, where 
$c$ is generated by the quadratic terms, $T^2m_i^2$,    
in the high temperature expansion of the thermal potential.
The quartic term, which is neglected here, 
is also induced but its contribution 
is a few percent in the physically interesting 
region where a strongly first-order electroweak phase transition 
occurs. The cubic interaction is also induced at finite temperatures 
but it has a smaller 
role in this scenario, thus neglected here. 
From the requirements of the global minimum of the vacuum 
at zero temprature, a first order phase transition, 
and positivity of the thermal mass term, the cutoff scale
is constrained as 
\begin{eqnarray}
 {\rm max}\left(\frac{v_0^2}{m_H^2}, 
  \frac{\sqrt{3}v_0^2}{\sqrt{m_H^2+2m_c^2}}\right) 
 < \Lambda < \sqrt{3} \frac{v_0^2}{m_H^2}, 
\label{eq:Lambda}
\end{eqnarray}
where 
$m_c=v_0\sqrt{(4y_t^2 + 3 g^2 +g'^2)/8} \sim 200$ GeV, 
where $g$ and $g'$ are $SU(2)_L$ and $U(1)_{Y}$ gauge couplings, 
and $y_t$ is the top Yukawa coupling. 
The left plot of \reffi{fig:ewbg2} shows the contours of constant 
$v_c/T_c$ in the $\Lambda$ vs.~$m_H^{}$ plane, where 
$T_c$ is the critical temperature and $v_c$ 
is the vacuum expectation value at $T_c$.
The results are encouraging and motivate a full one-loop 
computation of the thermal potential embedded in a 
complete baryogenesis scenario. 

The value of the cutoff scale $\Lambda$ affects 
the prediction of the $HHH$ and $HHHH$ couplings. 
Describing the zero temperature Higgs self-interations by 
$V(H) = m_H^2 H^2/2 + \mu H^3/3! + \eta H^4/4! + \cdot\cdot\cdot$, 
one has 
\begin{eqnarray}
\mu=3\frac{m_H^2}{v_0} + 6 \frac{v_0^3}{\Lambda^2}, 
\hspace{1cm}
\eta=3\frac{m_H^2}{v_0^2} + 36 \frac{v_0^2}{\Lambda^2}~.
\end{eqnarray}
The right plot of \reffi{fig:ewbg2}
shows that the contours of $\mu/\mu_{\rm SM}-1$ 
in the $\Lambda$ vs.~$m_H^{}$ plane.

\begin{figure}[htb!]
\psfig{figure=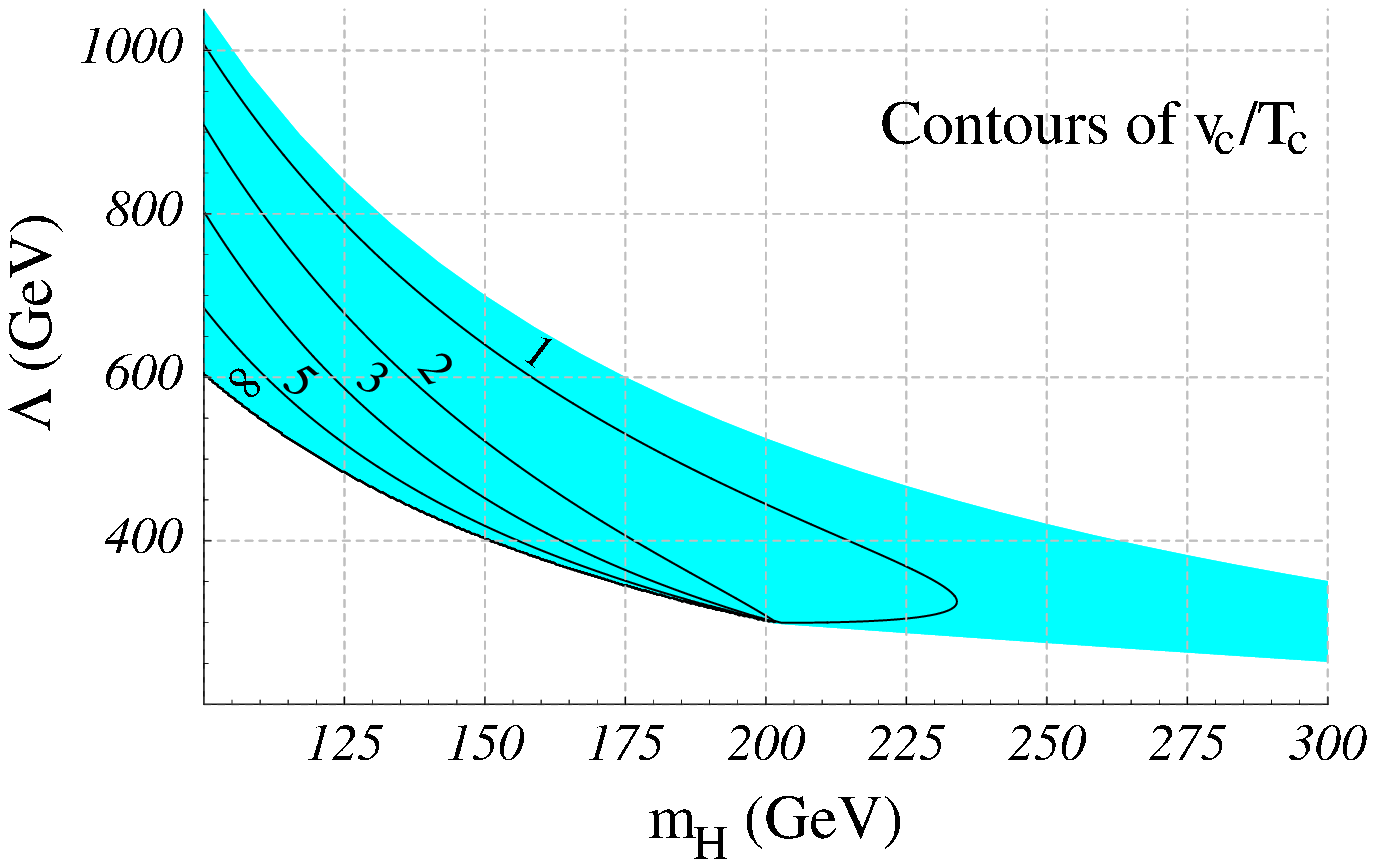,width=0.45\textwidth}
\psfig{figure=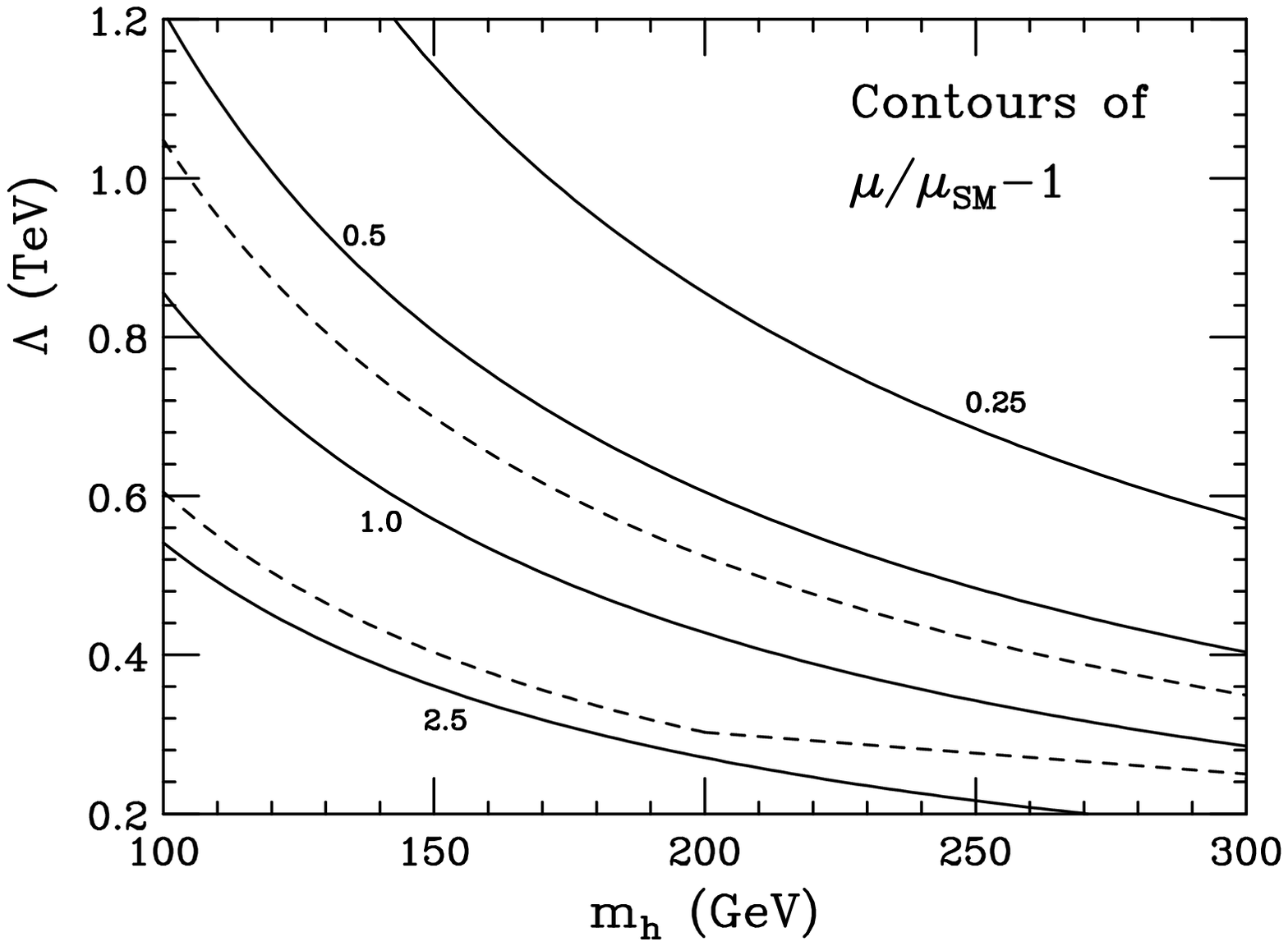,width=0.43\textwidth}
\caption{
[Left] Contours of constant $v_c/T_c$ from 1 to $\infty$. 
The shaded (blue) region satisfies the bounds of Eq.~(\ref{eq:Lambda}). 
[Right]
Contours of constant $\mu/\mu_{\rm SM}-1$ in the $\Lambda$ vs.~$m_H^{}$ 
plane. The dashed lines delimit the allowed region defined in 
Eq.~(\ref{eq:Lambda}). For details, see \cite{Grojean:2004xa}.}
\label{fig:ewbg2}
\end{figure}

\chapter{Impact of Machine and Detector Performance on Precision 
of Measurements in the Higgs Sector}
\label{sec:detector}

The physics processes at the ILC have been studied for about 15 years
and resulted in a variety of studies for discovery and precision
measurements. Many requirements on the detector and to a lesser extend
on the machine development have been established. The impact of the
detector (resolution, material budget, and coverage) and machine
aspects which directly influence the physics performance (beam energy
and luminosity spread, background induced in the detector) are
important aspects to be studied further. 
In this chapter we collect the existing results and discuss how the
detector could be optimized in view of Higgs boson physics.
Finally we present a strategy for future studies to
facilitate the final goal of Higgs precision measurements.

\bigskip
We start this section be recalling the main requirements for the ILC
detector~\cite{ILCdetectorAmsterdam}.

\begin{itemize}

\item
{\sl Momentum resolution:} 
$\si_{1/p} \sim 5 \times 10^{-5} \gev^{-1}$
(a factor of 10 better than that achieved at LEP). Good momentum
resolution is important for the reconstruction of the leptonic decays
of $Z$~bosons. This is particularly relevant for $\EEHZ$, which is
best suited to study the Higgs boson properties in detail. This works
especially well by investigating the Higgs boson recoiling against the
$Z$, with the clean decay $Z \to \mu^+\mu^-$.

\item
{\sl Impact parameter resolution:} 
Efficient $b$ and $c$ quark tagging is important, implying a good
impact parameter ($d_0$) resolution:
\BEQ
\si^2_{d_0} < (5.0 \ \mu{\rm m})^2 + 
\KL \frac{5.0 \ \mu{\rm m}}{p [{\rm GeV}] \sin^{3/2}\theta} \KR^2~,
\EEQ
where $p$ is particle momentum and $\theta$ is the polar angle.
This is a factor of three better than the resolution obtained at SLD.

\item
{\sl Jet energy resolution:}
$\si_E/E \sim 0.3/\sqrt{E [{\rm GeV}]}$ in order to be able to directly
reconstruct and identify gauge bosons from their hadronic decays to jets.

\item
{\sl Hermiticity:}
Hermiticity down to 5 mrad is needed for searches for missing energy
signals from new physics.

\item
{\sl High granularity:}
Events at the ILC will have high local 
particle densities due to the boosted
final states and the fact that many of the interesting physics
processes result in final states with six or more jets, especially in
the case of Higgs events. The ILC detector is required to have high
granularity calorimeters to enable efficient separation of close-by
showers and good pattern recognition in the tracking system.

\end{itemize}


\section{Existing Analyses}
\label{sec:existinganalyses}

A number of studies investigating the relation 
between machine and detector performance and the precision of measurements
in the Higgs sector have been presented at this Snowmass workshop and 
documented as LC Notes or e-prints.


\subsection{Beam related issues}
\label{subsec:beam}

In \citere{higgs_mass}, discussing the prospects 
of the Higgs boson mass measurements at ILC, the impact
of beam related systematic effects on the 
Higgs mass measurements have been evaluated.
The following experimental effects have been investigated: 
\begin{itemize}
\item{uncertainty in the beam energy measurement;}
\item{beam energy spread; and}
\item{uncertainty in the differential luminosity spectrum measurements.}
\end{itemize}

Since the analysis relies upon kinematic fits exploiting
energy-momentum conservation, a shift in center-of-mass energy
will inevitably introduce a bias in the measured Higgs mass.
This is illustrated in \reffi{fig:beam_error}. In order
to keep systematic uncertainty due to a bias in the beam 
energy measurement below the statistical error on the Higgs mass, 
the beam energy has to be controlled with relative precision 
better than 10$^{-4}$.  

To estimate the impact of the beam energy spread, Gaussian smearing 
of the beam energy has been applied at the stage of signal
event generation. It has been shown that for the case of 1\% energy spread
for both electron and positron beams, the statistical
precision in the Higgs boson mass measurement degrades from 45 to 50 MeV
in the $\HZbbqq$ channel and from 85 to 90 MeV in the 
$\HZqqll$ channel.
For a TESLA-like machine the expected energy spread amounts to 
0.15\% for the electron beam and 0.032\% for the positron beam~\cite{tdr}. 
For these energy spreads no significant degradation of the precision in 
the Higgs boson mass measurement is observed.

\begin{figure}[htb!]
\begin{minipage}[c]{0.48\textwidth}
\psfig{figure=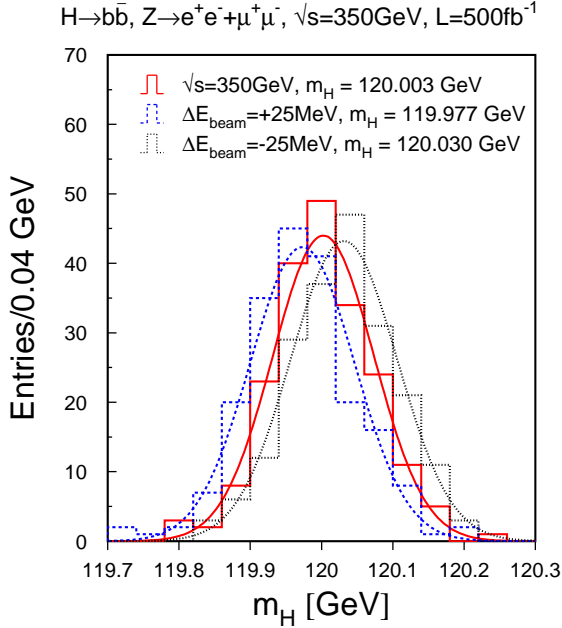,width=0.99\textwidth}
\end{minipage}
\begin{minipage}[c]{0.03\textwidth}
$\phantom{0}$
\end{minipage}
\begin{minipage}[c]{0.46\textwidth}
\caption{The spectrum of the fitted values of the Higgs 
boson mass as obtained from 200 independent signal 
samples in the $\HZqqll$ channel for the case when 
both electron and positron beam energies are overestimated 
by 25 MeV (dotted histogram), 
when they are underestimated by 25 MeV (dashed histogram) 
and when no shifts are introduced to the beam
energies (solid histogram).
\label{fig:beam_error}
}
\end{minipage}
\end{figure}

The energy spectra of the colliding electrons and positrons at the linear 
collider will be significantly affected
by photon radiation of an electron/positron in one bunch against the coherent 
field of the opposite bunch. This effect is referred to as beamstrahlung.
To have a fast simulation of beamstrahlung, the program CIRCE~\cite{CIRCE2} 
has been written. It assumes that the beamstrahlung in 
the two beams is equal and uncorrelated between the beams. The spectrum 
is parameterized according to
\BEQ
f(x) = a_0\delta(1-x)+a_1x^{a_2}(1-x)^{a_3},
\EEQ
where $x$ is the ratio of the colliding electron/positron energy to 
the initial energy of the undisrupted beam. The parameters $a_i$ 
depend on operational  
conditions of the linear collider. The normalization 
condition, $\int{f(x)dx=1}$,
fixes one of these parameters, leaving only three of them independent. 
The default parameters for the ILC (taken from TESLA)
operated at the centre-of-mass energy of 350 GeV are:
\BEQ
a_0=0.55,\hspace{2mm}a_1=0.59,\hspace{2mm}a_2=20.3,\hspace{2mm}a_3=-0.63.
\EEQ
It was shown that from the analysis of the acolinearity spectrum of Bhabha
scattering
events, the parameters $a_i$ can be determined with a precision of about 
1\%~\cite{lumi}. With such a precision on 
the parameters $a_i$, the uncertainty in 
the differential luminosity spectrum is expected to have 
a negligible effect on the Higgs mass measurement.

Beamstrahlung photons can interact and among others
produce hadronic final states, $\ga\ga\to {\rm hadrons}$. 
At a cold machine, the average number of these interactions with the 
center-of-mass energy of colliding photons in excess of 5 GeV
is expected to be about 0.25 (0.4) per bunch crossing in 
$\EE$ collisions
at $\sqrt{s}$ = 500 (800) GeV. This effect gets even stronger  
at a $\GG$ collider. About two $\ga\ga\to {\rm hadrons}$ events
are expected on average per bunch crossing at the ILC 
photon collider (taken from TESLA) 
for $\sqrt{s}_{ee}$ = 400 GeV at nominal 
luminosity\footnote{This effect was taken into consideration in the dedicated 
analyses related to Higgs boson studies at a photon 
collider~\cite{higgs_gg_p2,gg_hhiggs}}.
These parasitic interactions will inevitably contaminate 
physics events and deteriorate the performance of physics 
analyses. The impact of the hadronic $\ga\ga$ background  
on selected Higgs analyses has been investigated in 
Reference~\cite{gg_background}. Particles resulting from
$\ga\ga\to {\rm hadrons}$ interactions tend to be produced
at low polar angles. Therefore, for processes like $\EEHZ$,
where mainly central production of bosons is expected, the 
contamination from parasitic interactions can be reduced 
by rejecting particles produced at low polar angles. This 
allows for a partial recovery of the analysis performance. 
With special rejection criteria, allowing for suppression  
of $\ga\ga\to {\rm hadrons}$ background, 
the precision on the Higgs boson mass measurement in the 
$\HZbbqq$ channel
at $\sqrt{s}$ = 500 GeV with 500 fb$^{-1}$ degrades only 
by a few MeV. However, for the $WW$-fusion process the effect 
is expected to be more severe. For the cold machine, 
the presence of the $\ga\ga\to {\rm hadrons}$ background results in 
degradation of the signal-to-background ratio in the 
$e^+e^-\to W\nu\bar{\nu}\to b\bar{b}\nu\bar{\nu}$ 
channel by about 15\% at $\sqrt{s}$ = 1 TeV.
 


\subsection{Jet identification and resolution}

Many signal signatures at ILC will involve multi-jet final 
states. For their analysis, accurate jet identification and 
measurement is crucial. Efficient separation of $ZZ$ and $WW$ 
decay modes of a heavy Higgs boson (\reffi{fig:hhiggs_vv}) is possible 
with a jet energy resolution 
$\Delta E_{\rm jet}/E_{\rm jet} \leq 0.3/\sqrt{E_{\rm jet}}$.
A recent study has 
investigated the impact of jet energy resolution 
on the measurement of the Higgs
boson branching ratio to $W$ bosons in the 
Higgs-strahlung process~\cite{hww_jres}.
\begin{figure}[htb!]
\begin{minipage}[c]{0.47\textwidth}
\psfig{figure=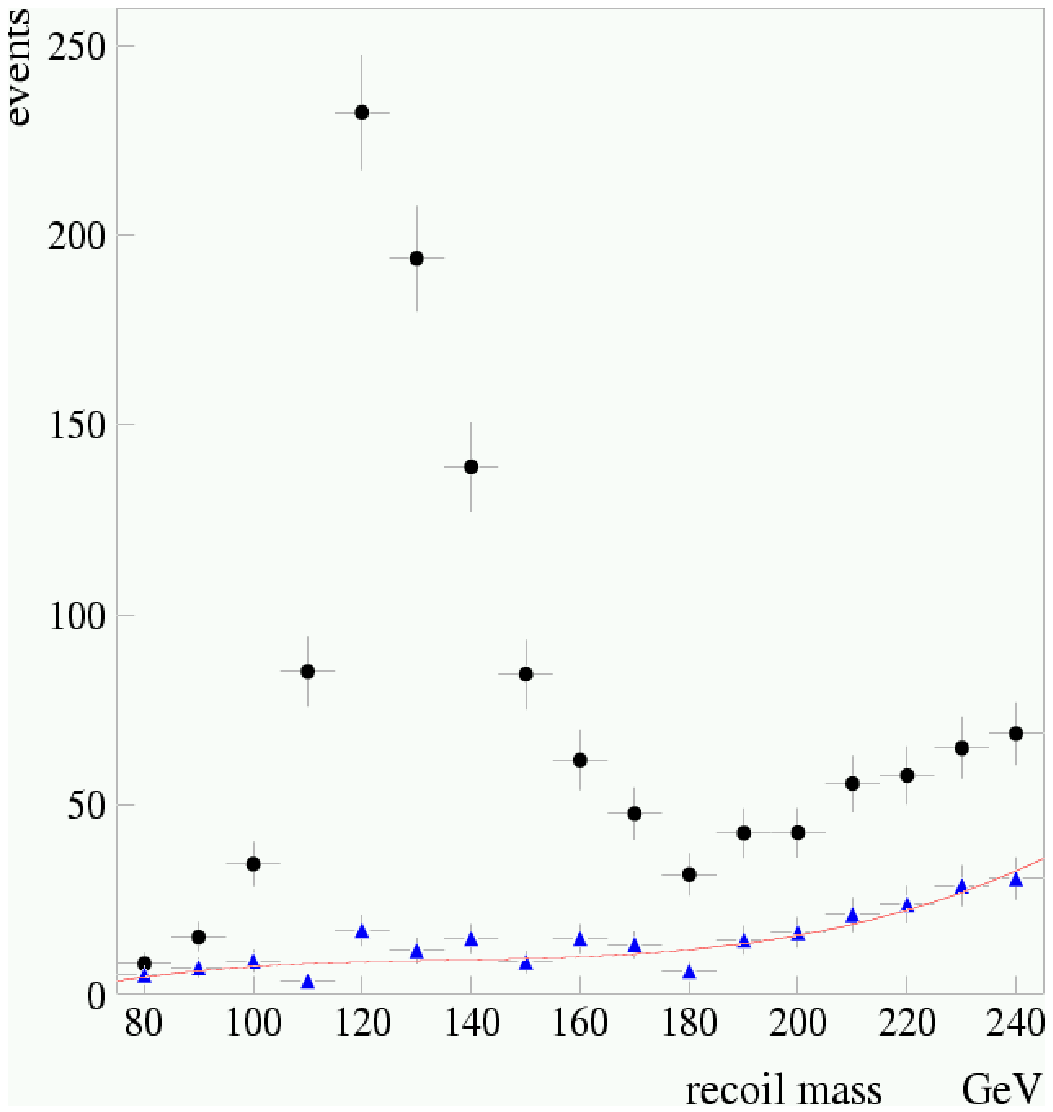,width=0.90\textwidth}\hspace{1em}
\end{minipage}
\begin{minipage}[c]{0.47\textwidth}
\psfig{figure=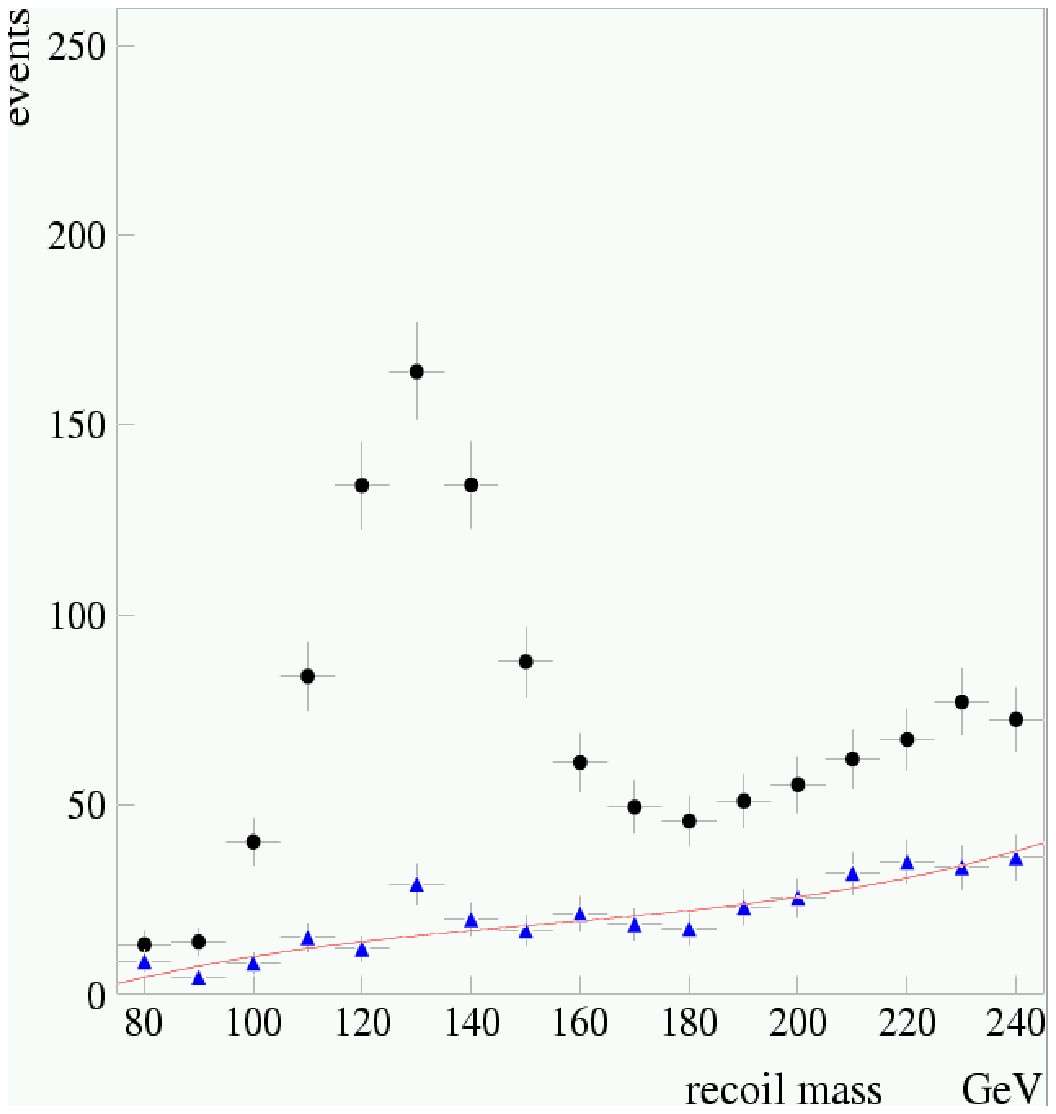,width=0.90\textwidth}
\end{minipage}
\caption{
Distribution of the reconstructed Higgs boson 
mass in the sample of selected Higgs-strahlung events
with subsequent decay of the Higgs into $W$ boson pairs and 
hadronic decays of the $Z$ boson. Only semileptonic decays of $W$ pairs
are considered. Left plot corresponds to jet 
energy resolution of $0.3/\sqrt{E_{\rm jet}}$, 
right plot corresponds to jet 
energy resolution of  $0.6/\sqrt{E_{\rm jet}}$. 
Selected sample corresponds to an integrated luminosity
of 500 fb$^{-1}$ collected at $\sqrt{s}$ = 360 GeV.
\label{fig:WW_jet}
}
\end{figure}

The analysis is based on selection of $\EEHZ$ events with subsequent 
semileptonic decays of $W$ pairs, $WW^*\to jj\ell\nu$, 
and hadronic decays of the $Z$ boson, $Z\to q\bar{q}$. 
The study is done for a light Higgs boson, $\mH$ = 120 GeV, assuming
500 fb$^{-1}$ of data collected at $\sqrt{s}$ = 360 GeV. Three
scenarios are considered, corresponding to assumed jet energy  
resolution of $0.3/\sqrt{E_{\rm jet}}$ 
(benchmark goal for the ILC detector), 
$0.6/\sqrt{E_{\rm jet}}$ (jet energy resolution for a 
typical LEP detector) and $0.4/\sqrt{E_{\rm jet}}$ 
(some intermediate value between the first two scenarios).
\reffi{fig:WW_jet} presents the 
reconstructed Higgs boson mass for the first and second scenarios. 
A considerable degradation in the signal-to-background ratio from
5 to 2.5 is observed for jet energy resolution of 
$0.6/\sqrt{E_{\rm jet}}$ compared to $0.3/\sqrt{E_{\rm jet}}$.
This results in deterioration of the precision in the 
$H\to WW^*$ branching fraction by 20\% which is equivalent 
to 45\% loss in the integrated luminosity.

A similar analysis~\cite{TBarklow_4jet} has been performed 
which investigated the impact of jet energy resolution on the 
Higgs boson mass measurement in the four-jet channel. It has
been shown that a deterioration in the jet energy resolution from 
30\%/$\sqrt{E}$ to 60\%/$\sqrt{E}$ results in a decrease in the 
Higgs mass accuracy from 42 MeV to 50 MeV. Like in the previous 
analysis, this roughly corresponds to an effective luminosity loss 
of 45\%. 

\begin{figure}[htb!]
\begin{minipage}[c]{0.55\textwidth}
\psfig{figure=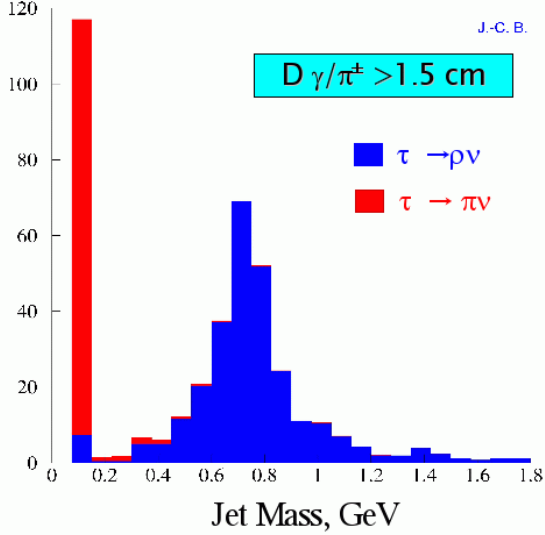,width=0.85\textwidth}
\end{minipage}
\begin{minipage}[c]{0.05\textwidth}
$\phantom{0}$
\end{minipage}
\begin{minipage}[c]{0.35\textwidth}
\caption{
The distribution of the measured tau-jet mass 
for the $\tau^\pm\ra \pi^\pm\nu$ and $\tau^\pm\ra \rho^\pm\nu \ra\pi^\pm\pi^0\nu$ channels. 
\label{fig:tau_separation}
}
\end{minipage}
\end{figure}

In \refse{sec:cp_higgs} the method to determine the CP quantum 
number of the Higgs  
boson by studying transverse spin correlation effects in the 
$H\ra\tau^+\tau^-$ decay has been described. 
Efficient identification of the tau decay mode is crucial for 
this measurement and sets stringent requirements on the 
performance of the electromagnetic calorimeter in combination 
with a dedicated pattern recognition algorithm~\cite{Brient_CPTau}. 
\reffi{fig:tau_separation} 
illustrates the separation of $\tau\ra \pi\nu$ and 
$\tau\ra \rho\nu$ modes in the distribution of the reconstructed
tau-jet mass. 
The result has been obtained with dedicated reconstruction tools 
aiming to identify and separate showers resulting from the charged pion 
and from the $\pi^0\ra\gamma\gamma$ decay. The simulation is performed
for a TESLA-like detector with a high granularity tungsten-silicon ECAL. 
The quality of separation between 
$\tau^\pm\ra \pi^\pm\nu$ and $\tau^\pm\ra \rho^\pm\nu \ra\pi^\pm\pi^0\nu$  
modes is quantified in \refta{tab:taudecays}. These results set one
of the benchmark conditions for the ECAL optimization studies.   

\begin{table}[htb!]
\begin{center}
\begin{tabular}{|c|c|c|}
\hline
Decay Mode & Jet mass $<$ 0.2 GeV & 0.2 GeV $<$ Jet mass $<$ 2 GeV\\ 
\hline
$\tau^\pm\ra \pi^\pm\nu$ & 82\% & 17\% \\
$\tau^\pm\ra \rho^\pm\nu \ra\pi^\pm\pi^0\nu$ & 2\% & 19\% \\
\hline
\end{tabular}
\end{center}
\caption{Fraction of the $\tau^\pm\ra \pi^\pm\nu$
and $\tau^\pm\ra \rho^\pm\nu \ra\pi^\pm\pi^0\nu$ final states 
with the tau-jet mass below 0.2 GeV and in the window 
0.2 $-$ 2 GeV. Distributions correspond to the samples
in which the distance between $\pi^+$ entering calorimeter and 
the nearest photon is greater that 1.5 cm, one and a half
calorimeter cell size.
\label{tab:taudecays}
}
\end{table}


\subsection{Identification of $b$- and $c$-quarks}

For the study of the hadronic branching fractions, an excellent 
$b$- and $c$-tagging capability of the detector is crucial. 
Tools have been developed which implement in detail 
the procedure of tagging $b$- and $c$-hadrons~\cite{bc_tagging}.
%
\begin{figure}[htb!]
\begin{minipage}[c]{0.47\textwidth}
\psfig{figure=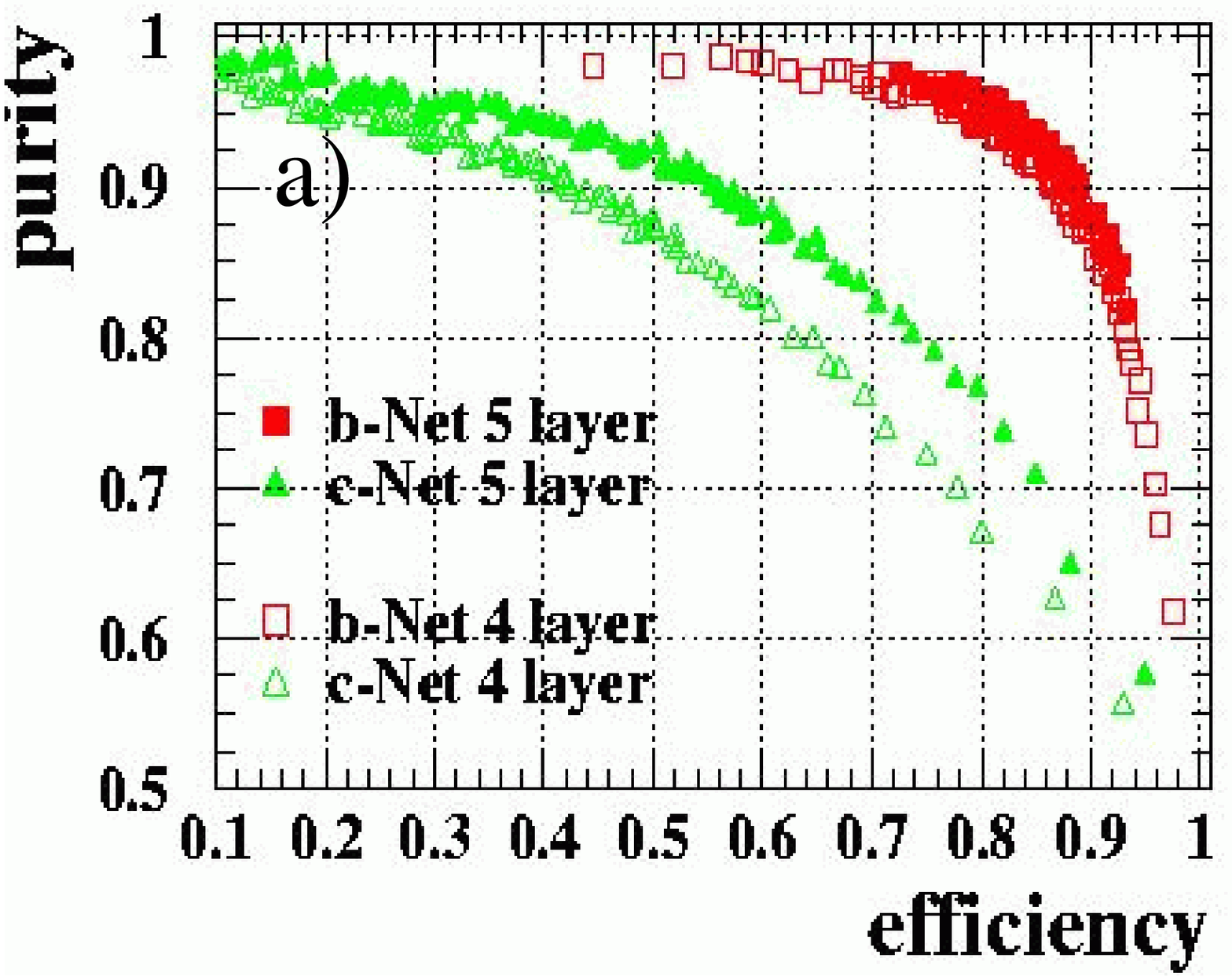,width=0.90\textwidth}
\end{minipage}
\begin{minipage}[c]{0.47\textwidth}
\psfig{figure=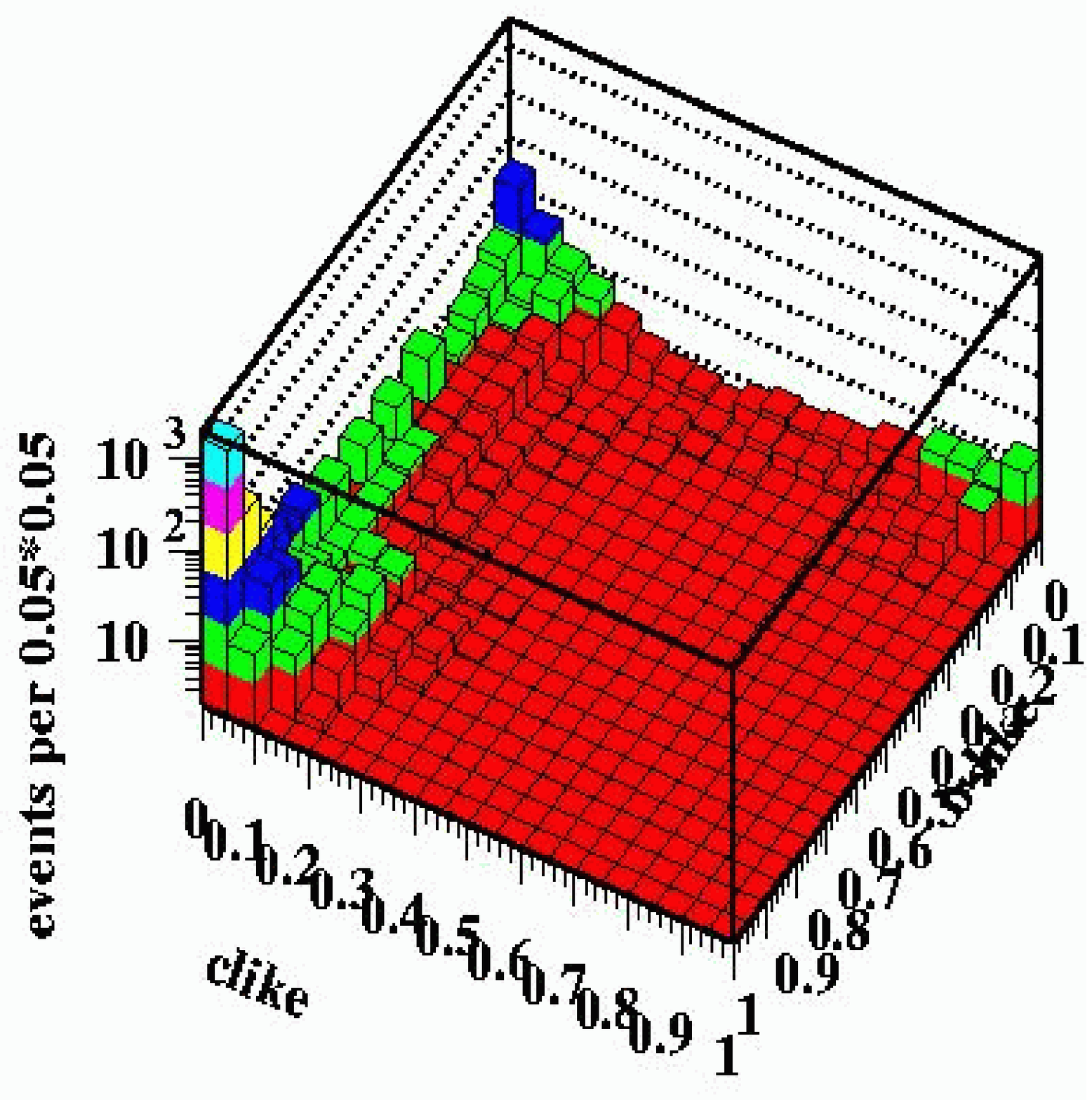,width=0.90\textwidth}
\end{minipage}
\begin{minipage}[c]{0.47\textwidth}
\caption{
The performance of $b$- and $c$-tagging as envisaged for 
a vertex device of the TESLA detector. Performance 
is quantified in terms of $b$/$c$ purity as a function of 
$b$/$c$ efficiency. 
\label{fig:bc_tag}
}
\end{minipage}
\begin{minipage}[c]{0.03\textwidth}
$\phantom{0}$
\end{minipage}
\begin{minipage}[c]{0.47\textwidth}
\caption{
The 2D distribution of event $b$-tag vs.\ $c$-tag in the 
sample of the selected Higgs-strahlung events with 
$H \to {\rm hadrons}$ and $Z \to q\bar{q}$~\cite{ee_hqq}.
\label{fig:bc_fit}
}
\end{minipage}
\end{figure}
%
\reffi{fig:bc_tag} illustrates performance of $b$- and $c$-tagging 
in terms of $b$ ($c$) purity as a function of $b$ ($c$) efficiency 
as obtained with dedicated tools for the micro-vertex detector
configuration foreseen for a TESLA-like detector.
Recently, a dedicated analysis based on these detailed 
flavor-tagging tools has been performed~\cite{ee_hqq} 
to evaluate the ILC potential for the measurement
of hadronic decays of a light Higgs boson with $\mH$ = 120 GeV. 
The study is based on the the selection of an inclusive 
sample of hadronic Higgs decays using the Higgs-strahlung
process and exploiting all possible decay modes of the $Z$~boson. 
Each selected event is assigned quantities that
quantify the probability to contain $b$- or $c$-jets. These quantities
are referred to as $b$- and $c$-tag variables. The branching 
ratios $\Hbb$, $\Hcc$ and $\Hgg$ are determined
from the fit of 2D distribution of $b$-tag vs.\ $c$-tag variables 
(\reffi{fig:bc_fit})
with three free parameters, which are
normalization factors quantifying the fractions of 
$\Hbb$, $\Hcc$ and $\Hgg$ events in the final 
selected samples. Results of this study are presented in 
\refta{tab:hqq_branchings}.

\begin{table}[htb!]
\begin{minipage}[c]{0.55\textwidth}
\renewcommand{\arraystretch}{1.2}
\begin{tabular}{|c|c|c|c|c|}
\hline
Decay  & $\Zll$ & $\Zvv$ & $\Zqq$ & Combined \\
\hline
       & \multicolumn{4}{|c|}{Relative precision (\%)} \\
\hline
$\Hbb$ &  3.0 &   2.1 &  1.5 &  1.1 \\
$\Hcc$ & 33.0 &  20.5 & 17.5 & 12.1 \\
$\Hgg$ & 18.5 &  12.3 & 14.4 &  8.3 \\
\hline
\end{tabular}
\renewcommand{\arraystretch}{1}
\end{minipage}
\begin{minipage}[c]{0.05\textwidth}
$\phantom{0}$
\end{minipage}
\begin{minipage}[c]{0.35\textwidth}
\caption{
Relative precision on the measurement of the 
hadronic branching ratios of Higgs for $\mH$ = 120 GeV.
The analysis is performed for $\sqrt{s}$ = 350 GeV, 
assuming an integrated luminosity of 500~fb$^{-1}$.
\label{tab:hqq_branchings}
}
\end{minipage}
\end{table}

Another study (reported at Snowmass) has evaluated the impact 
of the vertex detector parameters on the measurement of the hadronic 
branching fractions of the Higgs boson~\cite{ciborowsky_snowmass}. 
The analysis is performed for the SM Higgs boson with 
a mass of 127 GeV and for one representative MSSM parameter point
giving the same mass for the light supersymmetric Higgs.
The detector response is simulated with the 
SGV program~\cite{SGV}. Two configurations of the vertex detector 
are considered. The first configuration has  
five layers, located at the distance 15, 26, 37, 48, 60 mm away 
from the beam axis. In the second configuration the innermost  
layer is removed. For both configurations, the precision on 
$\Gamma(H\ra c\bar{c})$ and $\Gamma(H\ra b\bar{b})$ has been 
investigated as a function of the layer thickness and
spatial resolution of a single layer. 
The analysis is based on the selection 
of Higgs-strahlung events. The prospective measurement is 
simulated assuming an integrated luminosity of 500 fb$^{-1}$ 
collected at 350 GeV center-of-mass energy. 
The study revealed strong dependence of the relative
precision of the $\Gamma(H\ra c\bar{c})$ measurement on the 
%
\begin{figure}[htb!]
\begin{center}
\psfig{figure=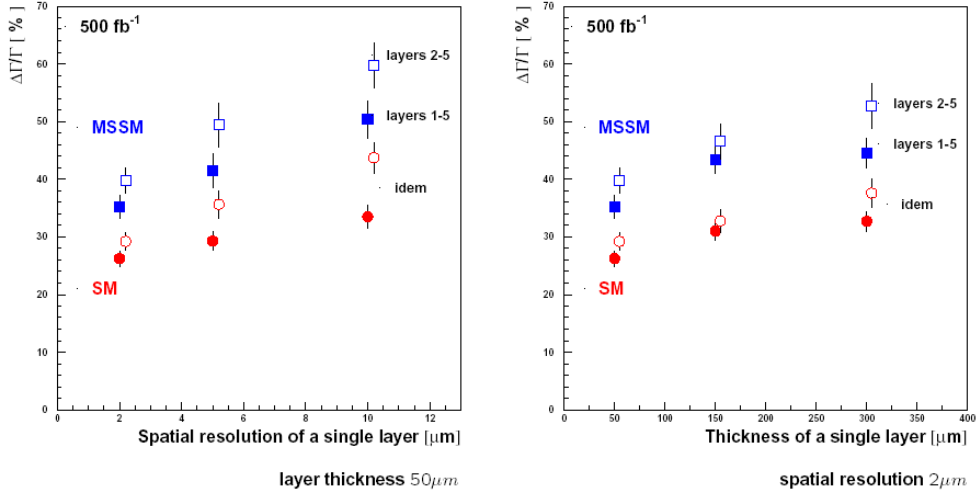, width=0.8\textwidth}
\end{center}
\caption{
The dependence of relative precision on $\Gamma(H\ra c\bar{c})$
as a function of spatial resolution of single layer (left plot) 
and layer thickness (right plot).
In the left plot layer thickness is fixed to 50 $\mu$m. In the 
right plot spatial resolution is fixed to 2$\mu$m. Filled dots 
correspond to the vertex configuration with five layers.
Open dots show results for the case when the innermost layer is removed.
Results for the SM Higgs are indicated with circles, for MSSM 
Higgs with square markers.
\label{fig:hqq_vertex}
}
\end{figure}
%
vertex detector characteristics as illustrated in  
\reffi{fig:hqq_vertex}.  
The dependence is found to be less pronounced for the 
$H\ra b\bar{b}$ channel.  



\section{Higgs Physics and Detector Optimization}

The International Linear Collider project enters now the phase of 
detector design and optimization. At this stage, the detailed analyses
studying the impact of the detector design and event reconstruction procedure
on the precision of the measurements at ILC
are of crucial importance. The studies discussed in previous sections
have been performed using fast parametric Monte Carlo programs, in which
the detector response and event reconstruction performance are 
simulated in an idealized manner for a specific 
detector configuration option.  
Such an approach is illustrated by 
the {\tt Simdet}~\cite{simdet} 
program, which found wide acceptance within the ECFA studies.
{\tt Simdet} parameterizes the response of 
the ILC detector with fixed detector geometry and therefore 
is not appropriate for studies relating various 
detector configurations to the attainable precision of 
measurements in various physics channels. 
{\tt Simdet} performs a fast Monte Carlo by
smearing the particle momenta and energies according 
to subdetector resolutions obtained from studies with 
full simulation programs based on the {\tt Geant3} or {\tt Geant4} packages.
Parameterization of the detector response  
in {\tt Simdet} relies on Monte Carlo studies performed with the 
full {\tt Geant3}-based simulation program {\tt Brahms} on single particle
samples. Results of these studies are then extrapolated 
to events with complex topology, e.g., multi-jet final
states, assuming highly efficient pattern recognition in 
the tracking system and calorimeters in a high 
local particle density environment. Clearly, the physics
potential of the ILC will depend on both \underline{detector performance}
and \underline{performance of the corresponding reconstruction software},
and the issue is still open whether the performance of the 
detector and event reconstruction are realistically 
implemented in the fast parametric Monte Carlo programs. 
In order to answer this question and to perform detailed 
and reliable detector 
optimization studies, future analyses related to Higgs 
physics at the ILC should be done with a full detector simulation based on  
{\tt Geant3} or {\tt Geant4} implementations of existing detector designs 
and with consistent and realistic reconstruction software.


\subsection{Tools for Full Detector Simulation}

Currently, activities are ongoing worldwide which aim at 
the development of realistic simulation tools based 
on the {\tt Geant4} package as well as reconstruction software
which can be used for the detector optimization studies. A big 
progress in the development of such tools has been reported
at Snowmass. Most of these tools use the common output format  
realized in the LCIO data model~\cite{LCIO}. This 
feature facilitates direct comparison between the various detector 
models implemented in the different simulators.
The list of available full simulation programs
for the linear collider detector is given below:

\begin{itemize}

\item{{\bf{Brahms}}~\cite{Brahms}:
{\tt Geant3} based package which implements 
the TESLA TDR geometry in a hard-coded manner. The 
program reads HEPEVT files and produces an output 
in the ASCII or LCIO format. Since the detector 
geometry is fixed, the {\tt Brahms} package cannot be utilized 
for detector optimization studies. However, it may still 
be useful for revision of currently existing Higgs analyses 
with the full realistic simulation.}

\item{{\bf{SLIC}}~\cite{SLIC}: 
{\tt Geant4} based simulation package that uses an XML 
geometry input format called LCDD. This package 
includes various detector models developed within the 
compact detector concept with the silicon tracker (SiD) as 
well as large detector (LDC) and huge detector (GLD) 
concepts with the TPC as the main tracking device. The program
reads a StdHep file for the event input and produces an output in the form
of a LCIO file.
}

\item{{\bf{Mokka}}~\cite{Mokka}: 
{\tt Geant4} simulator that retrieves geometry information 
from a MySQL database. It reads binary StdHep or ASCII HEPEVT files and 
writes a LCIO file. Although originally {\tt Mokka} was 
designed for simulation of the large detector with the TPC as the main 
tracking device, the latest release of {\tt Mokka}
contains also models with the silicon tracker. The package
provides an option for coherent scaling of various detector
components such as the TPC, ECAL and HCAL, which allows to optimize 
the length and radius of the tracker, the depth and sampling frequency of 
calorimeters, and other geometrical parameters of the detector.}
\item{{\bf{Jupiter}}~\cite{Jupiter}: 
{\tt Geant4} based program for simulation of the GLD 
detector. This program uses as input StdHep or HEPEVT files.
The geometry of the detector is specified using XML interfaces. 
The data flow controller is based on ROOT. However,
work is ongoing to implement output in the LCIO format to enable
comparison with other detector concepts.}

\end{itemize}


\subsection{Tools for Event Reconstruction}

The most promising strategy for event reconstruction at the ILC is
based on the particle flow concept, implying reconstruction of the
four-vectors of all particles produced in an event. The particle flow
algorithm works best at moderate energies of individual 
particles, below about 100 GeV. In this regime, the 
tracking system reconstructs the momentum 
of the charged particles with an accuracy better than 
that achievable for the energy measurements
with calorimeters. This is illustrated in Figure~\ref{fig:tpc_calo}.
%
\begin{figure}[htb!]
\begin{minipage}[c]{0.45\textwidth}
\psfig{figure=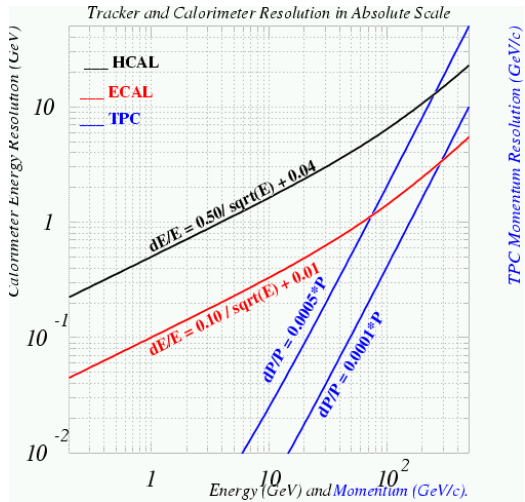,width=0.92\textwidth}
\end{minipage}
\begin{minipage}[c]{0.05\textwidth}
$\phantom{0}$
\end{minipage}
\begin{minipage}[c]{0.45\textwidth}
\caption{The track momentum resolution for charged
objects and the energy resolution for photons and hadrons
as a function of particle energy. Plotted curves correspond
to the resolutions envisaged for the TESLA detector.
\label{fig:tpc_calo}
}
\end{minipage}
\end{figure}
%
Hence, in order to attain a better reconstruction of 
events, the charged particle measurement must be solely based 
on the tracking information. The crucial step of the 
particle flow algorithm is the correct assignment of calorimeter
hits to the charged particles,  and efficient separation of 
close-by showers produced by charged and neutral particles.
A clustering algorithm is then performed on calorimetric hits 
not assigned to the charged particles to reconstruct 
photons and neutral hadrons. The reconstruction of photons
and neutral hadrons buried in jets is nearly impossible 
with a typical LEP detector. This explains why LEP detectors
could reconstruct jet energy with something like 60\%/$\sqrt{E}$ 
(where $E$ represents the jet energy).
The goals of the physics program at the linear collider 
requires however much better jet energy resolution in order to 
efficiently separate, e.g., $ZZ$ and $WW$ final states. 
Monte Carlo studies have shown that an ideal reconstruction
algorithm~\cite{pflow_morgunov}, 
which finds each particle and measures its energy and 
direction with the detector resolution expected for single particles,
could reach a jet energy resolution of 14\%/$\sqrt{E}$.
Over the years a jet energy resolution of 30\%/$\sqrt{E}$ 
has become accepted as a good compromise between the theoretically
possible and practically achievable resolution.
The RERECO package~\cite{Brahms}, developed at DESY, 
represents the first complete 
software implementation of the event reconstruction following
the particle flow concept. A good 
performance of this implementation was demonstrated in a number of 
Monte Carlo studies\cite{pflow_morgunov,top_quark}. The software 
includes pattern recognition in the vertex detector, TPC, and forward 
tracking system, followed by a track fitting procedure. The calorimeter 
clustering, track-cluster association and particle identification 
is done with the SNARK~\cite{snark} package. The output is produced in the 
form of the reconstructed particle flow objects stored in 
the LCIO file. Despite the main advantage of being a complete and consistent 
implementation of the particle flow algorithm,
the RERECO package has serious 
shortcomings. It is heavily bound to the TESLA TDR geometry implemented 
in {\tt Brahms} and, therefore, cannot be used for detector optimization.
Nevertheless, the revision of the Higgs analyses in the key channels 
can be done with this package for the TESLA TDR geometry. 

The {\tt MarlinReco} package~\cite{MarlinReco}, meant as a successor of RERECO,
is currently being developed at DESY. Presently, this package
includes track reconstruction and fitting in the TPC, 
calorimeter clustering, track-cluster matching and particle 
identification. Unlike RERECO, {\tt MarlinReco} is not bound to any 
specific detector geometry and can be used for the detector
optimization studies. 
Other reconstruction packages presented at the Snowmass workshop
include {\tt org.lcsim}~\cite{orglcsim} designed
primarily for the SiD detector and the package designed for 
the GLD detector~\cite{GLD}.
Both packages, {\tt MarlinReco} and {\tt org.lcsim}, 
use the same LCIO data model,
which serves as an interface between the detector simulation and 
reconstruction software. The LCIO produced by full simulation 
contains a collection of hits recorded in various subdetectors. 
This information is used as an input for the reconstruction 
software. The reconstructed objects, namely tracks, calorimeter clusters
and particle flow objects, are also stored in the LCIO file 
and can be used for further analysis. 
Currently, the event reconstruction 
package used for the GLD detector stores information on the 
reconstructed objects in a ROOT file, 
but efforts are ongoing to make this package also compatible 
with the LCIO data format. 
 
Existing reconstruction packages provide an option to disentangle detector 
effects from the inefficiencies of the reconstruction algorithm.
This is done by emulating perfect pattern recognition in the 
tracking system and calorimeters and constructing 
true Monte Carlo tracks and calorimeter clusters. This approach, 
referred to as ``perfect particle flow'', is very important at the 
stage of detector optimization since it allows to trace
and eliminate possible systematic effects coming solely from 
the reconstruction algorithms.


\subsection{Benchmarking Detector Performance with Measurements 
in the Higgs Sector}

One of the main issues that has to be addressed by future studies 
is the relation between detector design and precision of measurements
in the Higgs sector, going beyond the existing studies summarized in
\refse{sec:existinganalyses}. We propose the following list of 
questions to be addressed by dedicated studies along with 
Higgs channels which can be used to benchmark the performance
of the different detector components.

\begin{itemize}
\item
{\bf{\underline{Vertex Detector}}}

How does the precision of Higgs hadronic branching ratio measurements 
depend on the specific configuration of the micro-vertex detector
(number of layers, position of the innermost layer, etc.)?
How is the analysis performance in other channels, e.g.\
$HA\to b\bar{b}b\bar{b}$, influenced by 
the micro-vertex detector configuration? The first study 
addressing these questions is presented in
\citere{ciborowsky_snowmass}. Efficient determination 
of the charge of secondary vertices is important for 
the reduction of combinatorial background in the $HHZ$ channel
with the $H\ra b\bar{b}$ decay and therefore crucial for the measurement
of the Higgs self-coupling.
How does the efficiency of the reconstruction of the vertex 
charge depend on the configuration of the vertex detector? 

\item
{\bf{\underline{Tracking System}}}

What charged particle momentum resolutions can be achieved with 
the currently considered options for the main tracking device 
(TPC, silicon tracker)? How does the geometry of the tracking 
system (radius and length of TPC, number and position of layers
in the silicon tracker, etc.) affect the charged
particle momentum resolution?
What impact will the charged particle momentum resolution have
on the performance of analyses which utilize the inclusive 
$\HZXll$ channel? How will the pattern recognition capabilities 
provided by different configurations of the tracking system 
influence the reconstruction of multi-jet final states 
with high track multiplicity, 
e.g.\ $ZHH\to 6{\rm ~jets}$ or $HZ\to WW(ZZ)qq \to 6{\rm ~jets}$?

\item
{\bf{\underline{Calorimeters}}}

How does the precision on the Higgs photonic branching ratio 
measurement depend
on the energy and angular resolution of the electromagnetic calorimeter?
What impact will a specific option for electromagnetic and hadronic
calorimeters (absorber material, active media, sampling fraction, 
transverse granularity) have on the efficiency of the reconstruction 
of multi-jet final states with high particle multiplicity 
in various Higgs channels? 
What jet energy resolution can be achieved with a given 
option for the calorimeter system? How does the efficiency of $\tau$ lepton
identification via its hadronic decays depend on the configuration 
of the calorimeter system. What effects will it have on the 
identification of $H\ra\tau\tau$ decays, which is crucial 
for the determination of the CP quantum numbers 
of Higgs bosons?

\item
{\bf{\underline{Muon System}}}

What improvement can be achieved in the identification of high
energy the muons if tracking and calorimeter information 
is complemented by the muon system? Should the muon system be realized
in a conventional way by instrumenting the magnet return yoke or
are alternative ways of implementing the muon detector more favorable?
What active element (Resistive Plate Chambers, scintillating strips/tiles)
is preferable for the muon system? These studies should be based on the
analysis of final states involving high energy muons 
($\HZXmm$, $e^+e^-\to H\nu\bar{\nu}\to \mu^+\mu^-\nu\bar{\nu}$). 

\item
{\bf{\underline{Overall Detector Performance and Performance of Reconstruction Software}}}

The overall detector and reconstruction software performance can be 
expressed in terms of the jet energy resolution. Hence, it would be
desirable to investigate the dependence of jet energy resolution 
on a given detector configuration and also to study 
the impact of the jet energy resolution on
the precision of measurements in the various 
Higgs channels involving multi-jet final states.

\end{itemize}

The ultimate goal of these studies would be to establish
a mapping between performance/configurations of different 
subdetectors and the precision of measurements in the Higgs sector.


\section{Proposed Strategy}

The detector optimization for the Higgs channels can be 
done in two steps. The first step is based on 
the simple MC smearing of the 4-momentum of isolated particles, 
jet energies and angles, and a parameterization of the heavy flavor tagging
procedure. The resolution functions 
are extracted separately for each type of particle by running a 
full simulation program on single particle samples. For charged
particles it implies the study of the performance of the tracking
system and the parameterization of the momentum resolution and 
reconstruction efficiency as a function of particle momentum and 
polar angle $\theta$. For neutral particles, 
detailed investigation of the calorimeter response 
has to be performed in order to quantify  
the energy, position and angular resolutions 
separately for photons and neutral hadrons.  
The Higgs analyses relying on flavor-tagging will need as
an input the parameterization of the vertex detector performance 
in terms of $b$- and $c$- tagging efficiency versus purity.
Realistic reconstruction packages can be employed to evaluate 
the jet energy resolutions for the case of perfect particle flow
as well as for realistic particle flow algorithms.
Once the main characteristics quantifying the detector response 
are extracted from the full simulation and reconstruction 
tools, fast Monte Carlo smearing can be applied to the 
generated background and signal samples relevant for a given Higgs channel.
The program is supposed to emulate
the measurements of the four-momenta of isolated particles 
and jets, heavy-flavor tagging, etc.,  using as an input 
the detector resolution functions extracted from the Monte Carlo studies
with the full simulation. Results of the relevant  
analyses will directly reflect the potential of a specific detector
model for measurements in the Higgs sector. 

On the other hand, with this simple parametric approach, 
one can study the precision of measurements as a function 
of the main detector performance characteristics, such as 
jet energy resolution, momentum resolution of the tracking system,
etc., without referring to a specific detector model.   
The main advantage of the parametric approach is that 
it provides a fast procedure for mapping the detector performance 
(described in terms of generic resolution functions and 
inefficiencies) into the precision of measurements in the key 
physics channels. 

However, this strategy has a major shortcoming. 
Lacking detailed simulation of the detector response and 
realistic event reconstruction, it fails to predict in an accurate 
way the analysis performance in the physics channels where details 
of the pattern recognition in the detector play a crucial role.
This situation can be illustrated by two examples.
Successful determination of Higgs CP quantum numbers
in the $H\ra\tau^+\tau^-$ channel will require efficient 
pattern recognition in the electromagnetic calorimeter in order 
to identify the tau decay mode and to measure 
precisely the decay products of the tau leptons. Accurate analysis
in this channel is possible only with the full detector simulation and 
realistic reconstruction tools. The second example is given 
by measurements of Higgs self-couplings
in the double Higgs-strahlung process, $e^+e^-\ra HHZ$. 
The large combinatorial background in this channel can be 
reduced by identifying the vertex charge, which 
would allow to discriminate between $b$ and $\bar{b}$ jets. 
Again, reliable prediction of the vertex charge identification 
efficiency requires a detailed study of the vertex detector 
performance with the full simulation and realistic 
reconstruction tools.   

At this point one can conclude that it is reasonable to 
employ fast parametric 
simulation at the early stage of the detector optimization 
to reject those detector configurations that result in 
unacceptable deterioration of the precision of measurements 
in the Higgs sector. 

For detector configurations passing the first step 
of optimization with the fast parametric simulation, a more 
detailed study is needed that employs the full 
detector simulation and reconstruction. The large-scale
analyses including background estimates represent a time-consuming 
procedure in this case. Therefore, we propose first to run full simulation 
programs, implementing a chosen detector geometry,
only on HEPEVT or StdHep files of signal samples. Performing
event reconstruction, selection and analysis of only the signal
samples, one can get a rough idea about  
the dependence of the analysis performance on a specific detector
configuration. Variations in analyzing power of variables crucial for 
a given Higgs channel will serve as a major criterion to discriminate
between different detector designs. 
An example of such an approach is illustrated in
\reffi{fig:hvv_full}. It presents reconstructed Higgs boson 
mass spectra in the $e^+e^-\ra H\nu\bar{\nu}\ra b\bar{b}\nu\bar{\nu}$ channel 
after performing the full simulation of the TESLA TDR detector with 
{\tt Mokka}. 
The figure compares results of the  
realistic reconstruction based on the {\tt MarlinReco} package  
and the perfect particle flow algorithm. 
It is very important that the results of
the realistic event reconstruction are cross 
checked with the simulation of perfect reconstruction.
This is necessary to disentangle detector effects
from the inefficiencies in pattern recognition algorithms and 
to guide the efforts
of the developers of the reconstruction software.
In this particular example, the obtained resolution on the 
reconstructed Higgs boson mass plays the role of a parameter 
evaluating the performance of the given detector model  
and event reconstruction package. 
Similar studies can be done in other channels for 
various detector options which may emerge in the future as 
a result of the detector R\&D program for the ILC.

\begin{figure}[htb!]
\begin{minipage}[c]{0.65\textwidth}
\psfig{figure=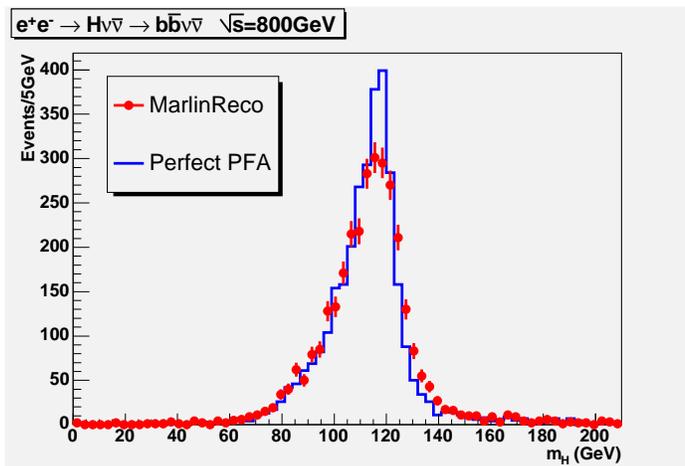,width=0.85\textwidth}
\end{minipage}
\begin{minipage}[c]{0.30\textwidth}
\caption{
The distribution of the dijet mass in the 
$e^+e^-\ra H\nu\bar{\nu}\ra b\bar{b}\nu\bar{\nu}$ channel.
Distribution is obtained after performing 
full simulation of the TESLA TDR detector with {\tt Mokka}.
The result of the realistic event reconstruction with 
the {\tt MarlinReco} package (dots) is compared with 
perfect reconstruction (line).
\label{fig:hvv_full}
}
\end{minipage}
\end{figure}



\end{document}